%% file: Manuscript.tex
\documentclass{article}
\usepackage{template, bm, hyperref, natbib, graphicx, verbatim, amsmath, amsthm, thmtools}
\usepackage[normalem]{ulem}
\usepackage{subcaption}
\usepackage{graphicx}
\usepackage{tikz}
\usepackage{tikz-3dplot}
\usetikzlibrary{arrows.meta,calc,decorations.pathreplacing}
\usepackage{contour}
\contourlength{0.6pt}
\title{Identifiability and Spectral Estimation of Mixed Memberships in Heterogeneous Multilayer Networks}
\author{Taeseok Park, Jes\'us Arroyo}
\author{Taeseok Park\thanks{\href{mailto:tspark@tamu.edu}{tspark@tamu.edu}}, Jes\'us Arroyo\thanks{\href{mailto:jarroyo@tamu.edu}{jarroyo@tamu.edu}}\\
\normalsize{Department of Statistics, Texas A\&M University}}

\begin{document}

\maketitle
\begin{abstract}

    Multilayer network data, arising from multiple sources, repeated measurements, or multi-view interactions, are increasingly common, highlighting the need for statistical methods that jointly leverage information across layers while allowing for layer heterogeneity and computational scalability. Here, we study a multilayer degree-corrected mixed membership model (ML-DCMM), in which nodes share a common mixed membership across layers while degree effects and block connectivity patterns vary by layer. This model encompasses several existing models for multilayer data while preserving identifiability and interpretation. We develop a spectral estimation procedure that integrates layer-wise embeddings with corner estimation for membership recovery. A further refinement that connects the resulting representation to a ParaTuck-2 tensor decomposition is also introduced. We establish theoretical guarantees for the proposed spectral method, showing that the probability of near-corner recovery increases with the number of layers, and we provide the convergence rate for the resulting membership estimator. We establish a minimax lower bound for membership estimation, which does not improve with increasing number of layers. Simulation studies show that ML-DCMM can outperform other approaches under strong layer heterogeneity. In an analysis of London bike-sharing networks, the methodology recovers interpretable memberships and achieves lower held-out edge prediction error under edge-sampling cross-validation than competing models.

\end{abstract}

\tableofcontents

\input{Sections/1-Introduction}

\section{Methodology}
\input{Sections/2-Methodology}

\section{Theory}\label{sec:Theory}
\input{Sections/3-Theory}
\section{Simulations}\label{sec:Simulations}
\input{Sections/4-Simulations} 
\section{Data analysis}
\input{Sections/5-Data_analysis}\label{sec:data_analysis}
\section{Discussion}

\input{Sections/6-Discussion}

\section*{Acknowledgment}
Jes\'us Arroyo acknowledges support from the National Science Foundation under grant DMS-2413553. The authors also acknowledge that portions of this research were conducted with the advanced computing resources provided by Texas A\&M Department of Statistics Arseven Computing Cluster.

\section*{Declaration of generative AI use}
The authors used ChatGPT 5.6 to partially assist with language editing, constructing examples in Theorem~\ref{thm:minimax-membership-lower-bound} and improving computational efficiency of the code implementation. All AI-assisted outputs were reviewed by the authors, who take full responsibility for the final manuscript.

\bibliographystyle{plainnat_nodoi}
\bibliography{Reference}    

\appendix
\input{Sections/Appendix-closeness}
\input{Sections/Appendix-identifiability}
\input{Sections/Appendix-bi_Lipschitz}
\input{Sections/Appendix-corner_estimation}
\input{Sections/Appendix-consistency_membership}
\input{Sections/Appendix-minimax}
\input{Sections/Appendix-cv}
\end{document}

%% file: Sections/1-Introduction.tex
\section{Introduction}
Network data provide a natural framework for representing pairwise relationships among a collection of entities. A single network is commonly encoded by an adjacency matrix, whose entries record whether, or to what extent, pairs of nodes are connected. Examples include social networks, where nodes represent individuals and edges represent friendship or communication ties \citep{BALLNEWMAN2013}, and structural brain networks, where nodes represent brain regions and edges represent anatomical connections estimated from diffusion MRI tractography \citep{Hagmann2008}. Multilayer network data, represented as collections of adjacency matrices on a common set of nodes, have become increasingly common in modern applications \citep{Kivela2014, BOCCALETTI2014}. Such layers may arise from repeated measurements over time \citep{mucha2010community}, different experimental or cognitive conditions \citep{Lukemire2021}, or distinct types of relations among the same entities \citep{Szell2010}. They may also arise from multimodal measurements. For example, in neuroimaging, a common anatomical parcellation can be used to define a shared set of nodes, while different layers may encode distinct forms of structural or functional connectivity derived from different imaging modalities \citep{DeDomenico2017}.

A central task in network data analysis is to uncover latent community or membership structure from the observed pattern of edges \citep{FORTUNATO201075}. The \emph{stochastic block model} (SBM) provides a canonical probabilistic framework for this task \citep{HollandLaskeyLeinhardt1983}. Its extensions include the \emph{mixed-membership stochastic block model} (MMSBM), which allows nodes to participate in multiple communities simultaneously \citep{airoldi2008mixed}, the \emph{degree-corrected stochastic block model} (DCBM), which accounts for node-specific degree heterogeneity \citep{karrer2011stochastic}, and the \emph{degree-corrected mixed-membership model} (DCMM), which combines these two features \citep{zhang2020detecting,jin2024mixed}. Because the DCMM jointly encodes degree effects, mixed memberships, and block connectivity in the same probability matrix, identifiability is a nontrivial issue. A wide range of methods have been developed for community detection, including likelihood-based and Bayesian procedures, modularity maximization, matrix factorization approaches, and spectral methods. The latter are especially relevant to the present work, as they provide computationally efficient estimators based on low-rank structure in the population adjacency matrix. The \emph{random dot product graph} (RDPG) \citep{Young2007, Athreya2017} framework provides a latent-position interpretation of such embeddings by representing edge probabilities through inner products of latent node positions.

For multilayer network data, block models are often extended by assuming that the layers share a latent membership structure \citep{Peixoto2015}. For example, the \emph{multilayer degree-corrected stochastic block model} (ML-DCBM) extends the DCBM by assuming common non-overlapping community assignments across layers, together with layer-specific degree-correction parameters and block connectivity matrices. The \emph{multilayer degree-corrected mixed-membership model} (ML-DCMM), a multilayer extension of the single-layer DCMM, assumes the layers share a common mixed-membership matrix while the degree-correction parameters and block connectivity matrices are allowed to vary across layers. From the perspective of membership estimation, it is important to use information across layers without collapsing the multilayer structure into a single averaged adjacency matrix. While averaging can improve signal strength, it may also discard layer-specific connectivity patterns and degree heterogeneity when the layers are heterogeneous. This has motivated joint multilayer approaches that retain the layer index, including tensor-based methods that stack adjacency matrices into a third-order array and use tensor-decomposition techniques to extract shared node-level structure across layers \citep{Jing2021, Aguiar2024}. 

In this paper, we study the ML-DCMM, provide identifiability conditions and propose algorithms for community detection under this model. The main contributions of this paper are as follows.
\begin{itemize}
    \item We establish identifiability conditions for the ML-DCMM, clarifying how the shared mixed-membership matrix can be separated from layer-specific degree corrections and block connectivity matrices. We also show that other latent space models for multilayer network data can be described as a special case of this model, showing that this model provides a general framework with interpretable latent structure.
    \item We propose an efficient spectral algorithm for estimation of the common node memberships via a corner estimation procedure. This method consists of the concatenation of normalized adjacency spectral embeddings (CNASE) followed by vertex hunting. We also establish a connection of these embeddings with the ParaTuck-2 tensor factorization and propose a refinement of the estimator based on this decomposition.    
    \item We establish error bounds for spectral estimation of node memberships in terms of the two-to-infinity norm of the mixed membership matrix, providing row-wise control of the community-detection error. We also establish a minimax lower bound for membership estimation that does not improve with increasing number of layers. This lower bound match the error bounds of the proposed method up to logarithmic factors under homogeneous degree-correction parameters, showing that our method is nearly minimax optimal.
    \item We demonstrate the practical usefulness of the proposed method through simulations and real data. In a multilayer bike-sharing network analysis, the ML-DCMM is shown to provide better generalization error than other models as well as interpretable latent memberships.
\end{itemize}

\subsection{Related work}

The \emph{stochastic block model} (SBM) assumes that nodes of a graph are divided into $K$ communities, and edges are independently generated according to a block-structured probability matrix. In matrix form, this probability matrix can be written as $\mathbf P = \mathbf Z \mathbf B \mathbf Z^\top,$ where \(\mathbf Z\) is an \(n\times K\) binary membership matrix whose \(i\)th row is the canonical basis vector corresponding to the community label of node \(i\), and \(\mathbf B\in[0,1]^{K\times K}\) specifies the connection probabilities between communities \citep{HollandLaskeyLeinhardt1983}. A large body of work has studied estimation and theory under the SBM, including likelihood- and modularity-based approaches \citep{Bickel2009} and spectral methods based on graph Laplacians \citep{Rohe2011}, adjacency spectral embeddings \citep{Sussman2012}, and adjacency spectral clustering \citep{Lei2015}. While this parsimonious model can approximate a wide range of random networks, it has notable limitations. First, nodes assigned to the same community are stochastically equivalent so the model cannot capture node-level degree heterogeneity. \citet{karrer2011stochastic} proposed the DCBM to overcome this limitation by introducing node-specific degree-correction parameters. Several spectral methods have been proposed including regularized graph Laplacian \citep{Qin2013}, SCORE \citep{jin2015fastcommunity} as well as theoretical works including consistency \citep{Zhao2011} and minimax optimality \citep{Gao2018}. Second, each node is assigned to a single community, which can be restrictive when communities overlap or when nodes participate in multiple communities simultaneously. The MMSBM relaxes this restriction by allowing overlapping communities for each node \citep{airoldi2008mixed}. \citet{Anandkumar2013} proposed a tensor spectral method and \citet{Mao2017} connected MMSBM estimation to symmetric nonnegative matrix factorization. \citet{Mao2021} exploited the row-wise simplex geometry of the leading population eigenspace under the MMSBM, where pure nodes correspond to simplex vertices and mixed nodes are represented as convex combinations of these vertices. 

The \emph{degree-corrected mixed-membership model} (DCMM) allows both node-specific degree correction and mixed memberships. Existing simplex-based spectral methods for DCMM-type models exploit a normalized spectral embedding in which degree effects are factored out and membership information is encoded by a simplex geometry. Estimating the corners of this simplex provides the anchor points needed for membership reconstruction. Examples of these methods include sketched vertex search or successive projection on the ratios of eigenvectors \citep{jin2015fastcommunity}, and $k$-medians clustering or one-class SVM on row-wise $\ell_{2}$ normalized spectral embeddings \citep{zhang2020detecting,mao2018overlapping}.

For multilayer networks, most existing work on community detection focuses on extensions that typically assume a shared community structure across layers, with the intention of capturing joint structure. In the multilayer SBM, estimators of shared community structure includes likelihood-based approaches \citep{Han2015, PaulChen2016}, least-squares \citep{Lei2020},  spectral methods \citep{Battacharyya2018, arroyo2021inference,lei2023biasadjusted}, and nonparametric Bayesian methods \citep{Peixoto2015}. Extensions to other multilayer models are often non-trivial due to the specific identifiability and estimation constraints that arise. For instance, identifiability the \emph{multilayer degree-corrected mixed-membership model} (ML-DCMM) is more challenging than in the single-layer DCMM because the model introduces additional layer-specific parameters. \citet{Xie2025} developed a spectral-assisted Gaussian variational inference method for the ML-DCMM under a direct layer-wise extension of the single-layer DCMM identifiability condition. 

Spectral methods are a particularly appealing approach in modern data science due to their computational scalability and theoretical guarantees \citep{chen2021spectral}. In multilayer community detection, these methods require constructing a joint spectral embedding matrix that captures community information across the layers. While several of these approaches have been developed for the ML-SBM, these ideas do not generalize to other models due to the more complex geometry of the embedding spaces. For instance, \citet{Agterberg2025} proposed a spectral clustering method for the ML-DCBM based on a joint spectral embedding obtained from the singular value decomposition of concatenated spectral embeddings. In this paper, we employ a similar concatenation idea, but in the ML-DCMM model with mixed memberships, this approach does not yield consistent estimators of the node memberships due to heterogeneity across the layer, motivating the use of a different embedding matrix to estimate the corners of the simplex that identify the communities of the model.

Tensor-based approaches have also been proposed for multilayer community detection, applying Tucker-type or nonnegative Tucker decompositions \citep{Jing2021, Aguiar2024}. These tensor decompositions are natural in the SBM and some of its extensions as the parameters of these models can be expressed as low-rank tensors. However, the ML-DCMM structure does not allow us to apply these decompositions to recover the model parameters. Instead, we establish a connection between the parameters of the ML-DCMM and the ParaTuck-2 decomposition \citep{Harshman_Lundy_1996}. Specifically, for a third-order tensor $\mathcal{U} \in \bbR^{I\times J\times L}$, let $\bU_{\cdot, \cdot, l}\in\mathbb{R}^{I\times J}$ be its $l$-th frontal slice. The ParaTuck-2 decomposition represents the $l$-th slice as 
\begin{equation} \label{eq:paratuck2-definition}
\bU_{\cdot, \cdot, l} = \bA \, \mathbf{D}_{A}^{(l)} \, \bH \, \mathbf{D}_{B}^{(l)} \, \bB^{\top},    
\end{equation} where $\bA \in \mathbb{R}^{I\times R_1}$ and $\bB \in \mathbb{R}^{J\times R_2}$ are shared loading matrices independent of $l$, $\mathbf{D}_{A}^{(l)} \in \mathbb{R}^{R_1 \times R_{1}}$ and $\mathbf{D}_{B}^{(l)} \in \mathbb{R}^{R_2 \times R_{2}}$ are diagonal slice-specific scaling matrices, and $\bH \in \mathbb{R}^{R_1\times R_2}$ captures the interactions between the two factor spaces. In particular, when $\bA$ and $\bB$ are the identity matrix, the entry $(i,j,l)$ of this tensor is $\bU_{ijl} = \bC_{il}\bH_{ij} \bD_{jl}$ for some matrices $\bC\in\real^{I\times L}, \bD\in\real^{J\times L}$, which is known as the no three-way interaction model \citep{diaconis1998algebraic}, and we show that it correspond to a factorization of the parameters of the ML-DCMM.
Several algorithms to fit this decomposition have been proposed, including alternating least squares (ALS) \citep{Bro1998}, with additional stochastic ALS step \citep{zniyed2025} and a lifting-based algebraic solver \citep{usevich2023}.

\subsection{Notation}
Let $a$, $\mathbf{a}$, and $\mathbf{A}$ denote a scalar, column vector, and matrix, respectively. For any matrix $\bA \in \bbR^{m \times n}$, $\bA_{i, \cdot}$, $\bA_{\cdot, j}$, and $\bA_{ij}$ denote the $i$-th row, $j$-th column of $\bA$ and $(i, j)$-th entry of $\bA$, respectively. For a positive integer \(m\), let $[m]:=\{1,\ldots,m\}.$ Let $\mathbf{1}_{K}$ and $\mathbf{I}_{K}$ denote the vector of ones in $\bbR^{K}$ and identity matrix in $\bbR^{K \times K}$, respectively. Let $\bI_{p, q}$ denote the diagonal matrix whose diagonal entries consist of $p$ ones followed by $q$ negative ones. Let $\mathcal{Q}_{K}$ denote the set of \(K\times K\) permutation matrices. Let $\lambda_{\max}(\bA), \lambda_{\min}(\bA), \sigma_{\max}(\bA), \sigma_{\min}(\bA)$ denote the maximum/minimum eigenvalues/singular values of $\bA$, respectively. Define $\bbO(K)$ be the set of $K \times K$ orthogonal matrices and $\bbO(p, q) = \{ \bA \in \bbR^{K \times K} \mid \bA \bI_{p, q} \bA^{\top} = \bI_{p, q}, p + q = K \}$ be the indefinite orthogonal group. For any symmetric positive definite matrix \(\mathbf A\), let \(\mathbf A=\mathbf V\mathbf\Lambda\mathbf V^\top\) be a spectral decomposition and define \(\mathbf A^{\pm1/2}:=\mathbf V\mathbf\Lambda^{\pm1/2}\mathbf V^\top\), where, for \(\mathbf\Lambda:=\operatorname{diag}(\lambda_1,\ldots,\lambda_d)\), \(\mathbf\Lambda^{\pm1/2}:=\operatorname{diag}(\lambda_1^{\pm1/2},\ldots,\lambda_d^{\pm1/2})\). Let $\Delta^{K-1} := \{ \mathbf{x} \in \mathbb{R}^{K}: \sum_{i=1}^{K} \mathbf{x}_{i} = 1, \mathbf{x}_{i} \geq 0 \}$ denote the standard simplex in $\mathbb{R}^{K}$. Let $\mathbb{S}^{K-1} := \{ \mathbf{x} \in \mathbb{R}^{K}: \| \mathbf{x}\|_{2} = 1 \}$ denote the unit $\ell_{2}$-norm ball in $\mathbb{R}^{K}$. For any matrix $\mathbf{A} \in \mathbb{R}^{n \times K}$ and an index set $\mathcal{J} \subset [n]$ with $|\mathcal{J}| = K$ such that $\mathbf{A}_{\mathcal{J}, \cdot}$ is nonsingular, let $\mathcal{SS}(\mathbf{A}, \mathcal{J}) := \left \{ \frac{\sum_{j \in \mathcal{J}} a_{j} \mathbf{A}_{j, \cdot}}{\| \sum_{j \in \mathcal{J}} a_{j} \mathbf{A}_{j, \cdot} \|_{2}}: \sum_{j \in \mathcal{J}}a_{j} = 1, a_{j} \geq 0 \right \}$.
denote the spherical simplex generated by the rows of $\mathbf{A}$ indexed by $\mathcal{J}$.

%% file: Sections/2-Methodology.tex
\subsection{The Multilayer Degree-Corrected Mixed Membership model}
In this paper, we consider the degree-corrected mixed-membership stochastic block model in the multilayer setting (ML-DCMM). Concretely, consider a collection of $L$ matrices $\bP^{(1)}, \cdots, \bP^{(L)}\in\real^{n\times n}$. From these matrices, we observe $L$ adjacency matrices $\bA^{(1)}, \cdots, \bA^{(L)}$, where $\bA_{ij}^{(l)} \overset{indep}{\sim} \text{Bernoulli} (\bP_{ij}^{(l)})$. For simplicity of the analysis, we assume undirected, binary graphs allowing self-loops. Thus, each $\bA^{(l)}$ is symmetric, and its diagonal entries are treated in the same way as the off-diagonal entries. This convention is mainly technical, as it avoids a separate treatment of the diagonal elements. The loopless setting can be handled by setting the diagonal entries to zero; this modification does not affect the main methodology, although it would require minor notational changes in the analysis. In the ML-DCMM, we assume $\bP^{(l)}$ of the form
\begin{equation}\label{eq:model_def}
    \bP^{(l)} = \bTheta^{(l)} \bZ \bB^{(l)} \bZ^{\top} \bTheta^{(l)}, \quad 1 \leq l \leq L,
\end{equation}
where $\bTheta^{(l)}=\operatorname{diag}(\btheta^{(l)}) \in \bbR^{n \times n}$ is a diagonal matrix that consists of positive degree-correction parameters $\btheta^{(l)} = (\theta_{1}^{(l)}, \cdots, \theta_{n}^{(l)})$, $\bZ \in \bbR^{n \times K}$ is a membership matrix with $\norm{\bZ_{i, \cdot}}_{1} = 1$ and $\bZ_{ij} \geq 0$, and $\bB^{(l)} \in \bbR^{K \times K}$ is a {full-rank} symmetric encoding the connectivity across blocks. 
Specifically, the element in the $i$-th row and $j$-th column of $\bP^{(l)}$ is computed as $\bP_{ij}^{(l)} = \theta_{i}^{(l)} \theta_{j}^{(l)} \bZ_{i, \cdot}^{\top} \bB^{(l)} \bZ_{j.}$.
Note that the membership matrix $\bZ$ is shared across all layers, indicating that the community structure of a given node is the same across layers, whereas $(\btheta^{(l)}, \bB^{(l)})$ are layer-specific. This modeling structure is natural in many real-world multilayer networks, where the latent roles of nodes are relatively stable but the edge formation patterns may vary across layers. For example, in transportation networks, a station or location may have a stable role in the system, while traffic intensity and connectivity patterns may change across time periods, days, or operating conditions. The shared matrix $\bZ$ represents these stable node-level roles, whereas the layer-specific parameters $\btheta^{(l)}$ and $\bB^{(l)}$ capture variation in node activity levels and block-level interaction strengths across layers.

The model in Equation~\eqref{eq:model_def}
extends several existing community detection and low-rank latent space models to multilayer networks by allowing a common mixed-membership matrix and layer-specific degree-correction and block-connectivity parameters. In particular, when \(L=1\), it reduces to the single-layer DCMM \citep{zhang2020detecting,jin2024mixed}. When each row of \(\bZ\) is a canonical basis vector in \(\mathbb R^K\), it reduces to the ML-DCBM, as in \citet{Agterberg2025}. When both restrictions hold, it reduces to the DCBM of \citet{karrer2011stochastic}. When the matrices $\bB^{(l)}$ have unit diagonal, it reduces to the version of this model studied in \cite{Xie2025}.

More generally, the ML-DCMM model extends network models with a common invariant subspace shared across the parameters of all the layers \citep{arroyo2021inference,jones2020multilayer, wang2026multilayer}, which extend the notion of latent spaces to multilayer settings. The next proposition formalizes this result. 

\begin{restatable}{proposition}{COSIE}
\label{prop:cosie_graphs}
 Consider a collection of symmetric matrices $\bP^{(1)}, \ldots, \bP^{(L)}\in[0,1]^{n\times n}$ such that $\bP^{(l)} = \bV \bR^{(l)}\bV^\top$ for full-rank matrices $\bV\in \real^{n\times d}$ and symmetric $\bR^{(l)}\in\real^{d\times d}, l\in[L]$. Then, there exist a diagonal matrix $\bTheta\in\real^{n\times n}$, full-rank square symmetric matrices $\bB^{(l)}$, $l\in[L]$ and a membership matrix $\bZ$ with $\|\bZ_{i,\cdot}\|_1=1$ and $\bZ_{ij}\geq 0, i\in[n],j\in[d]$ such that $\bP^{(l)} = \bTheta \bZ \bB^{(l)}\bZ^\top \bTheta$ for $l\in[L]$.
\end{restatable}

The previous proposition establishes the connection between a shared latent space across the layers encoded in the matrix $\bV$ and 
a common mixed-membership matrix $\bZ$. While the latter provides a more interpretable representation of the latent factors through cluster memberships, the matrix $\bZ$ is generally not unique, as additional conditions are needed to ensure that the memberships are identifiable. We provide some sufficient conditions in Section~\ref{sec:identifiability} through a latent space representation of the model.

\subsection{Latent Space Geometry of the ML-DCMM}

We now examine the latent geometry induced by the ML-DCMM. This geometry links the common membership matrix to the population spectral embedding and therefore motivates our use of spectral methods to fit this model. To this end, consider the 
eigendecomposition of $\mathbf P^{(l)} = \mathbf V^{(l)}\mathbf \Lambda^{(l)}\mathbf V^{(l)\top},$ where $\mathbf V^{(l)} = [\mathbf v_1^{(l)},\ldots,\mathbf v_K^{(l)}] \in\mathbb R^{n\times K}$ contains orthonormal eigenvectors corresponding to the nonzero eigenvalues of \(\mathbf P^{(l)}\), and $\mathbf \Lambda^{(l)} = \operatorname{diag}\{\lambda_1^{(l)},\ldots,\lambda_K^{(l)}\} \in\mathbb R^{K\times K}.$ For each \(l\in[L]\), let \(p^{(l)}\) and \(q^{(l)}\) denote the numbers of positive and negative eigenvalues of \(\mathbf P^{(l)}\), respectively, so that \(p^{(l)}+q^{(l)}=K\).
We order the eigenvalues so that $\lambda_1^{(l)}\ge\cdots\ge\lambda_{p^{(l)}}^{(l)}>0$ and $|\lambda_{p^{(l)}+1}^{(l)}| \ge\cdots\ge |\lambda_K^{(l)}|>0,$
where \(\lambda_{p^{(l)}+1}^{(l)},\ldots,\lambda_K^{(l)}<0\).

The \textit{generalized random dot product graph} (GRDPG) model \citep{Young2007,Rubin2022} provides a model-based interpretation of adjacency spectral embedding as latent positions in a low-dimensional space. The following proposition formalizes the connection between spectral embeddings in the GRDPG and the ML-DCMM.
\begin{restatable}[Spectral representation of the ML-DCMM]{proposition}{representation}
\label{prop:embedding_aligned_representation}
Consider the ML-DCMM model in Equation~\eqref{eq:model_def}. 
Let $\mathbf X^{(l)} = \mathbf V^{(l)}|\mathbf \Lambda^{(l)}|^{1/2}$ be the population spectral embedding of \(\mathbf P^{(l)}\), and let $\mathbf I_{p^{(l)},q^{(l)}} = \operatorname{sgn}\!\left(\mathbf\Lambda^{(l)}\right)$
be the signature matrix associated with the ordered eigendecomposition above. Then there exists a matrix $\mathbf R^{(l)}\in\mathbb R^{K\times K}$ such that
\begin{equation}\label{eq:pop_ebd_model_repre}
    \bX^{(l)} = \bTheta^{(l)} \bZ \bR^{(l)}
\end{equation}
Moreover, with the same choice of \(\mathbf R^{(l)}\), it follows that
\begin{equation}\label{eq:relationship_bw_B_and_R}
    \bB^{(l)} = \bR^{(l)} \bI_{p^{(l)}, q^{(l)}} \bR^{(l)\top}.
\end{equation}
\end{restatable}

\begin{remark}
Proposition~\ref{prop:embedding_aligned_representation} establishes two useful identities relating the population spectral embeddings to the ML-DCMM model parameters. Equation \eqref{eq:pop_ebd_model_repre} shows that the spectral embedding, although defined through the eigendecomposition of \(\mathbf P^{(l)}\), can be written exactly in the model-based form. The matrix \(\mathbf R^{(l)}\) is not an arbitrary auxiliary factor introduced only for this representation: equation \eqref{eq:relationship_bw_B_and_R} shows that the same \(\mathbf R^{(l)}\) also encodes the layer-specific block connectivity matrix \(\mathbf B^{(l)}\) through the indefinite inner product.
\end{remark}

Observe that each row of \(\mathbf X^{(l)}\) can be written as $\mathbf X_{i,\cdot}^{(l)} = \theta_i^{(l)} \bR^{(l)\top} \bZ_{i, \cdot}.$
That is, each row of $\bX^{(l)}_{i, \cdot}$ is a convex combination of the rows of $\bR^{(l)}$, with coefficients $\bZ_{i, \cdot}$ shared across all layers,multiplied by its degree-correction parameter $\theta_{i}^{(l)}$. In particular, for a pure node \(i\) in community \(k\), we have $\mathbf X_{i,\cdot}^{(l)} = \theta_i^{(l)}\mathbf R_{k,\cdot}^{(l)}$.
Therefore, the rows of \(\mathbf R^{(l)}\) determine the community-specific directions of the population embedding, while \(\theta_i^{(l)}\) controls the radial position along the corresponding ray. This fact motivates estimation approaches for $\bZ$ via row-normalizing \(\mathbf X^{(l)}\) \citep{jin2024mixed,zhang2020detecting}, which removes the radial effect of the degree-correction parameter and reveals the a geometry that contains information about the communities. However, in the multilayer setup, the different parameters on each layer have a different effect on the normalized embeddings from each layer, complicating estimation of these memberships directly from the normalized embeddings. We thus handle the different layers separately in order to construct an appropriate estimator. 

\subsection{Identifiability of multilayer mixed memberships with degree corrections}\label{sec:identifiability}

Identifiability of the DCMM parameters generally consists of two problems: global scale and community membership-specific scaling. In the single-layer model, the former issue can be resolved by assuming $\| \mathbf{Z}_{i, \cdot} \|_{1} = 1$ for all $i \in [n]$, and an additional constraint to distinguish the degree parameters from the global connectivity, such as $\sum_{i=1}^{n} \theta_{i} = n$ or $\| \mathbf{B} \|_{\max} = 1$. The second identifiability problem implies that for any positive diagonal matrix $\mathbf{D} \in \mathbb{R}^{K \times K}$, there exists a positive diagonal matrix $\mathbf{N}_{\mathbf{D}} \in \mathbb{R}^{n \times n}$ such that $(\mathbf{\Theta}, \mathbf{Z}, \mathbf{B})$ and $(\mathbf{\Theta} \mathbf{N}_{\mathbf{D}}^{-1}, \mathbf{N}_{\mathbf{D}} \mathbf{Z} \mathbf{D}, \mathbf{D}^{-1} \mathbf{B} \mathbf{D}^{-1})$ yield the same probability matrix $\mathbf{P}$ \citep{jin2024mixed}. One can address this nonidentifiability issue by imposing a constraint on the entries of $\bB$, such as $\operatorname{diag}(\mathbf{B}) = \mathbf{1}_{K}$. Note that this nonidentifiability of the membership matrix is specific to mixed-membership models and does not arise under non-overlapping memberships.

Moving onto multilayer models, identifiability conditions must allow sufficient flexibility on each layer to capture heterogeneous patterns. Directly imposing the identifiability conditions for single-layer block models on every layer can unnecessarily restrict the ML-DCMM parameter space. For example, imposing a unit-diagonal constraint on each \(\mathbf B^{(l)}\) generally requires layer-specific rescalings of the membership matrix, which need not preserve a common \(\mathbf Z\) across layers  (see Appendix~\ref{sec:appendix_example_unit_diagonal}). Instead, for the ML-DCMM, we designate one layer as a \textit{reference layer} to fix the parameterization of the shared membership matrix, and represent the remaining layers in terms of this reference. Without loss of generality, we take layer \(1\) to be the reference layer. We state identifiability conditions of the ML-DCMM.

\begin{restatable}[Identifiability conditions]{condition}{identifiabilitycond}
\label{cond:identifiability}
The following conditions hold.

\begin{enumerate}[(a)]
\item \textbf{Membership matrix.}
Each row of $\bZ$ has unit $\ell_1$-norm, and there exists at least one pure node for each community. 

\item \textbf{Community-overlap connectivity.} Define a community graph $\mathcal{G}_{Z} = ([K], \mathcal{E}_{Z})$, where $$\{a, b\} \in \mathcal{E}_{Z} \Leftrightarrow \exists i \in [n] \text{ such that } \mathbf{Z}_{ia} > 0, \mathbf{Z}_{ib} > 0, \quad a\neq b.$$ Then, $\mathcal{G}_{Z}$ is connected.

\item \textbf{Block connectivity matrices.}
The matrix $\bB^{(l)}$ is full rank for every $1\le l\le L$. Moreover, for the reference layer $l=1$ the matrix $\bR^{(1)}$ defined in Equation~\eqref{eq:relationship_bw_B_and_R} satisfies $\|(\bR^{(1)})_{k,\cdot}\|_2=1$ for all $ k \in [K]$.
Lastly, $\|\bB^{(l)}\|_{\max}=1$ for all $l \neq 1$.
\end{enumerate}
\end{restatable}

Under these conditions, the ML-DCMM parameters are identifiable up to a common permutation of the community labels as shown next.
\begin{restatable}[Identifiability of ML-DCMM]{proposition}{identifiabilityprop}
\label{prop:identifiability}
    Suppose that two parameterizations
$
\{\bTheta^{(l)}\}_{l=1}^L$, $\bZ$, $\{\bB^{(l)}\}_{l=1}^L$ and 
$\{\widetilde\bTheta^{(l)}\}_{l=1}^L, \widetilde \bZ, \{\widetilde \bB^{(l)}\}_{l=1}^L
$
both satisfy Condition~\ref{cond:identifiability} and generate the same population probability matrices, that is, $\bP^{(l)}
=
\bTheta^{(l)} \bZ \bB^{(l)} \bZ^\top \bTheta^{(l)}
=
\widetilde\bTheta^{(l)} \widetilde \bZ \widetilde \bB^{(l)} \widetilde \bZ^\top \widetilde\bTheta^{(l)}$ for all $l \in [L]$.
Then the model parameters are identifiable up to a common permutation of the community labels. More precisely, there exists a $K\times K$ permutation matrix $\bPi$ such that, for every $1\le l\le L$,
\[
\widetilde\bTheta^{(l)}=\bTheta^{(l)},\quad
\widetilde \bZ = \bZ\bPi,
\quad
\widetilde \bB^{(l)}=\bPi^\top \bB^{(l)}\bPi.
\]
\end{restatable}

\begin{remark}[Multilayer-specific identifiability considerations]
Compared with the single-layer setting, the multilayer model introduces two additional identifiability considerations. First, the community-overlap connectivity condition is additionally required to identify $\{\bTheta^{(l)}, \bB^{(l)}\}_{l \neq 1}$. Once \(\mathbf Z\) is identified from the reference layer, a non-reference layer may still have community-specific scale ambiguities between \(\mathbf\Theta^{(l)}\) and \(\mathbf B^{(l)}\). Mixed nodes with overlapping supports link the scale ratios of the corresponding communities. See Figure~\ref{fig:community-overlap-connectivity} in the Appendix. A sufficient condition is the existence of a fully mixed node $i$ with $\bZ_{ik} > 0$ for all $k \in [K]$. When condition (b) does not hold, additional constraints on the connectivity matrices across all the layers are required to distinguish $\bTheta^{(l)}$ and $\bB^{(l)}$. For example, when all nodes are pure so that it reduces to the ML-DCBM, this condition can instead be replaced by $\| (\bR^{(l)})_{k, \cdot} \|_{2} = 1$ for all $l \in [L]$ and $k \in [K]$.
Second, only the reference layer satisfies $\|(\bR^{(1)})_{k,\cdot}\|_2=1$ for the common membership matrix $\mathbf{Z}$ across all layers to exist. Under ML-DCBM, however, one can assume $\| (\mathbf{R}^{(l)})_{k, \cdot} \|_{2} = 1$ for all $l$ and $k$ since the community-specific scaling does not change the membership matrix $\mathbf{Z} \in \{0, 1 \}^{n \times K}$. Note that, when $\bB^{(1)}$ is positive definite, the unit-row-norm condition of $\mathbf{R}^{(1)}$ is equivalent to the unit-diagonal condition $\mathrm{diag}(\bB^{(1)})=\mathbf 1_K.$
\end{remark}

\subsection{Proposed estimation procedure}\label{sec:2_proposed}
Estimation of the mixed memberships in the model can be done by identifying the indices of the pure nodes, which characterize the communities. From a latent space point of view, this reduces to estimating the corners of the space in which the latent positions lie. Single-layer methods typically resolve this problem by normalizing the latent positions to remove the degree heterogeneity effect, followed by a corner estimation procedure \citep{zhang2020detecting,jin2024mixed,mao2018overlapping}. In the multilayer setup, these methods are not directly applicable due to the potential heterogeneity across layers that needs to be individually handled.

To remove the degree heterogeneity effect, in this paper we use an $\ell_2$ normalization of the latent positions for each layer. At the population level, consider the matrices $\bX^{(l)}$ and $\bR^{(l)}$ from Proposition~\ref{prop:embedding_aligned_representation}. Define a diagonal matrix $\mathbf{M}^{(l)} := \operatorname{diag}(m_1^{(l)},\ldots,m_K^{(l)}) \in \mathbb{R}^{K \times K}$ with $m_k^{(l)} := \|\mathbf R_{k,\cdot}^{(l)}\|_2$ so that $\mathbf{S}^{(l)} := (\mathbf{M}^{(l)})^{-1} \mathbf{R}^{(l)}$ denotes the $\ell_{2}$-norm normalized corner matrix. Let $\mathbf{Y}^{(l)} \in \mathbb{R}^{n \times K}$ denote the $\ell_{2}$-norm normalized spectral embedding matrix defined by
\begin{equation}
    \mathbf{Y}_{i, \cdot}^{(l)} = \frac{\mathbf{X}_{i, \cdot}^{(l)}}{\| \mathbf{X}_{i, \cdot}^{(l)} \|_{2}} = \frac{(\mathbf Z_{i,\cdot}^{\top}\mathbf M^{(l)}\mathbf S^{(l)})^{\top} }{ \|(\mathbf Z_{i,\cdot}^{\top} \mathbf M^{(l)}\mathbf S^{(l)})^{\top}\|_2 } \in \mathcal{SS}(\mathbf{S}^{(l)}, [K]) , \quad 1 \leq i \leq n. \label{eq:Y-definition}
\end{equation}
Consequently, the rows of $\mathbf Y^{(l)}$ lie on a spherical simplex in $\mathbb S^{K-1}$, whose corners are given by the rows of $\mathbf S^{(l)}$. Moreover, $\mathbf Y_{i,\cdot}^{(l)}=\mathbf S_{k,\cdot}^{(l)}$ if and only if node $i$ is pure in community $k$, which implies that the corners of the spherical simplex correspond precisely to pure nodes in each community, and these are the same across every layer. This observation motivates our approach for clustering.

Our proposed method for corner estimation constructs a matrix that concatenates row-normalized adjacency spectral embeddings (CNASE) across layers to jointly utilize layer-wise information. For each \(l\in[L]\), let \(\widehat{\bV}^{(l)}\in\mathbb R^{n\times K}\) contain the leading eigenvectors of \(\bA^{(l)}\) corresponding to its \(K\) largest eigenvalues in magnitude, and let \(\widehat{\bLambda}^{(l)}\) be the diagonal matrix containing the corresponding eigenvalues. The rank-\(K\) adjacency spectral embedding \citep{Sussman2012,Rubin2022} is defined as \(\widehat{\bX}^{(l)} := \widehat{\bV}^{(l)} |\widehat{\bLambda}^{(l)}|^{1/2}\).  We then estimate the shared corner indices using the concatenated normalized embeddings $\widehat{\bY}^{(l)}$ computed as in Equation~\eqref{eq:Y-definition}, which combine the \(L\) layer-wise spherical-simplex structures. This procedure is summarized in Algorithm~\ref{alg:step1}.
\begin{algorithm}[!ht]
\caption{Corner and membership estimation on concatenated normalized adjacency spectral embeddings (CNASE)}
\label{alg:step1}
\begin{algorithmic}[1]
\Input Adjacency matrices $\{\bA^{(1)},\ldots,\bA^{(L)}\}$ and number of communities $K$.
\Ensure Membership estimate $\widehat{\bZ}_{0}$ and matrices $\{\widehat{\bU}^{(1)},\ldots,\widehat{\bU}^{(L)}\}$ 

\For{$l = 1$ to $L$}
    \State Obtain the adjacency spectral embedding \(\widehat{\bX}^{(l)}\) of $\bA^{(l)}$.
    \State Normalize the rows of \(\widehat{\bX}^{(l)}\) to obtain \(\widehat{\bY}^{(l)}\), where \(\widehat{\bY}_{i,\cdot}^{(l)} = \widehat{\bX}_{i,\cdot}^{(l)} / \|\widehat{\bX}_{i,\cdot}^{(l)}\|_2\) for all \(i\in[n]\).
\EndFor
\State Concatenate the normalized adjacency spectral embeddings and form $\widehat{\bY} = \left[ \widehat{\bY}^{(1)} \cdots \widehat{\bY}^{(L)} \right]$.
\State Apply a corner estimation method to the rows of \(\widehat{\bY}\) and obtain a set of \(K\) corner indices \(\widehat{\mathcal J}\).
\State For each \(l\in[L]\), set \(\widehat{\bS}^{(l)} = \widehat{\bY}^{(l)}_{\widehat{\mathcal J},\cdot}\), and form \(\widehat{\bS} = [\widehat{\bS}^{(1)}\cdots\widehat{\bS}^{(L)}] \in\mathbb R^{K\times KL}\).
\For{$l = 1$ to $L$}
    \State Obtain $\widehat{\bU}^{(l)} = \widehat{\bX}^{(l)} \left( \widehat{\bS}^{(l)} \right)^{-1}$.
\EndFor
\State For the reference layer, normalize $\widehat{\bU}^{(1)}$ by $\widetilde{\bU}^{(1)} = \text{diag}(\widehat{\bU}^{(1)} \mathbf{1}_{K})^{-1} \widehat{\bU}^{(1)}$ and obtain membership estimate $\widehat{\bZ}_{0}$ by $(\widehat{\bZ}_{0})_{i, \cdot} = \Pi_{\Delta^{K-1}}\left(\widetilde{\bU}^{(1)}_{i, \cdot} \right)$.
\end{algorithmic}
\end{algorithm}

The corner-estimation step in Algorithm~\ref{alg:step1} can be treated as a plug-in step. That is, the proposed procedure is not tied to a particular vertex-hunting routine; any suitable simplex corner-finding method can be used. For any corner candidate set $\mathcal{I} \subseteq [n]$ with $| \mathcal{I} | \geq K$, define the averaged vertex-hunting objective
\begin{equation}\label{eq:def_of_vh_objective}
    \widetilde{\Phi}_{L, \mathcal{I}}(\mathcal{J}) := \max_{i \in \mathcal{I}} \frac{1}{L} \sum_{l=1}^{L} d\left( \widehat{\mathbf{Y}}_{i, \cdot}^{(l)}, \mathcal{SS}(\widehat{\mathbf{Y}}^{(l)}, \mathcal{J}) \right), \quad \mathcal{J} \subseteq \mathcal{I}, \quad |\mathcal{J}|=K.
\end{equation}
Here, $d(\cdot, \cdot)$ denotes the Euclidean distance between a point and a set. We use a variant of \textit{Sketched Vertex Search} \citep{jin2024mixed} equipped with the $k$-medoids algorithm for the corner estimation as follows:
\begin{enumerate}
    \item Denoise step (\textit{optional}). For a value $m > K$, apply the \(k\)-medoids algorithm with \(m\) medoids to the rows of the concatenated normalized embedding \(\widehat{\mathbf Y}\) from Algorithm~\ref{alg:step1}, and let $\widehat{\mathcal M}_m$ denote the resulting index set of medoids.
    \item Corner search step. Select $\widehat{\mathcal{J}} \in \arg\min_{\substack{ \mathcal{J} \subseteq\widehat{\mathcal M}_m\\  |\mathcal{J}|=K }} \widetilde{\Phi}_{L,\widehat{\mathcal M}_m}(\mathcal{J}).$
\end{enumerate}
The first denoise step is optional but recommended for computational efficiency. Otherwise, one can find corners directly from the set of $n$ concatenated adjacency spectral embeddings instead of $m$ medoids. In practice, we choose the number of medoids $m$ in a data-driven manner (see Appendix~\ref{sec:tensor_refinement} for details).

\begin{remark}[Computational scalability of corner estimation] For the denoising step, we use \(k\)-medoids rather than \(k\)-means. Since each medoid is an observed row of \(\widehat{\mathbf Y}\), it remains on the Cartesian product of \(L\) unit spheres. In contrast, averaging within a cluster generally moves a \(k\)-means centroid off this product space. This geometric property is used in the subsequent vertex-hunting analysis. 
Although exact optimization of the \(k\)-medoids objective is NP-hard in general, Theorem~\ref{thm:VH_almost_pure_recovery} does not require a global minimizer. Its proof only requires one-swap local optimality which implies no single swap between a selected medoid and an unselected node decreases the $k$-medoids objective. This is precisely the local optimality condition targeted by the swap phase of standard PAM-type algorithms \citep{KaufmanRousseeuw1990}. 
\end{remark}

\subsection{Connection with a tensor decomposition}

Several approaches for multilayer clustering and joint embedding often rely in joint spectral decompositions that recover informative signal about the communities or the shared latent space \citep{PaulChen2016,lei2023biasadjusted,Agterberg2025}. In contrast, the structure of the ML-DCMM does not admit a joint spectral decomposition to recover the mixed-membership matrix due to individual-specific parameters. However, here we show that the parameters of the model can be expressed as the solution of the ParaTuck-2 decomposition \citep{Harshman_Lundy_1996}. This connection motivate us to propose an algorithmic refinement of the parameter estimates based on computing this decomposition.

Recall that the ParaTuck-2 decomposition \eqref{eq:paratuck2-definition} with $\mathbf{A} = \mathbf{I}_{n}$ and $\mathbf{B} = \mathbf{I}_{K}$ induces two diagonal matrices $\mathbf{C}^{(l)} \in \mathbb{R}^{n \times n}$ and $\mathbf{D}^{(l)} \in \mathbb{R}^{K \times K}$ such that the $l$-th frontal slice of $\mathcal{U}$ can be written as $\mathbf{U}^{(l)} = \mathbf{C}^{(l)} \mathbf{H} \mathbf{D}^{(l)}$ for $l \in [L]$. Under the ML-DCMM, define a corner-coordinate matrix $\mathbf{U}^{(l)} := \mathbf{X}^{(l)} (\mathbf{S}^{(l)})^{-1}$. By the definition of $\mathbf{S}^{(l)}$, we obtain the exact ParaTuck-2 decomposition for the corner-coordinate matrix $\mathbf{U}^{(l)} = \bm{\Theta}^{(l)} \mathbf{Z} \mathbf{M}^{(l)}$. Note that, by Condition~\ref{cond:identifiability}(c), we have $\mathbf{M}^{(1)} = \mathbf{I}_{K}$ and hence $\mathbf{U}^{(1)} = \bm{\Theta}^{(1)} \mathbf{Z}$. Consequently, row-wise $\ell_{1}$-norm normalization of $\mathbf{U}^{(1)}$ recovers the common membership matrix $\mathbf{Z}$. Note that applying the same normalization to non-reference layers does not necessarily recover $\mathbf{Z}$. Additional Euclidean projection onto the standard simplex $\Delta^{K-1}$, denoted by $\Pi_{\Delta^{K-1}}$ in Algorithm~\ref{alg:step1}, ensures the membership estimates to lie on the standard simplex. 

This equivalent expression motivates the following refinement step: Construct $\widehat{\mathbf U}^{(l)} := \widehat{\mathbf X}^{(l)} (\widehat{\mathbf S}^{(l)})^{-1}$ for all $l\in[L]$, with $\widehat{\mathbf X}^{(l)}$ and $\widehat{\mathbf S}^{(l)}$ as defined in Algorithm~\ref{alg:step1}. Define a mean square error loss $\mathcal{L}(\{\bm{\Theta}^{(l)}\}_{l=1}^{L}, \mathbf{Z}, \{ \mathbf{M}^{(l)} \}_{l=1}^{L}):= \sum_{l=1}^{L} \| \widehat{\mathbf{U}}^{(l)} - \bm{\Theta}^{(l)} \mathbf{Z} \mathbf{M}^{(l)} \|_{F}^{2}$. Implement ALS \citep{Bro1998} with random initiates $\{ \bm{\Theta}_{0}^{(l)}, \mathbf{M}_{0}^{(l)} \}_{l=1}^{L}$ and spectral membership estimates $\widehat{\mathbf{Z}}_{0}$ defined in algorithm~\ref{alg:step1}. We can obtain a refined membership estimate $\widehat{\mathbf{Z}}$ as the output of the refinement step. See Appendix~\ref{sec:tensor_refinement} for the algorithm.

%% file: Sections/3-Theory.tex
In this section, we establish the non-asymptotic properties of mixed-membership estimation following our proposed method. We state regularity assumptions for theoretical analysis. For each layer $l$, define the normalized degree-weighted membership Gram matrix $\bG^{(l)}\in\real^{K\times K}$ as $\bG^{(l)} := \frac{\sqrt{K}}{\|\btheta^{(l)}\|_{2}} \left\{ \bZ^{\top}(\bTheta^{(l)})^{2}\bZ \right\}^{1/2}$.
\begin{assumption}\label{asm:gram_matrix}
There exists a constant $C_G>0$, independent of $l$, such that $$\max\bigl\{\|\bG^{(l)}\|_{\mathrm{op}},\|(\bG^{(l)})^{-1}\|_{\mathrm{op}}\bigr\}\le C_G, \quad 1 \leq l \leq L.$$
\end{assumption}
Assumption~\ref{asm:gram_matrix} requires the eigenvalues of $\bG^{(l)}$ to be uniformly bounded above and away from zero across layers. This implies the degree-weighted membership vectors $(\bm{\Theta}^{(l)} \mathbf{Z})_{i, \cdot}$ to remain sufficiently diverse across all community directions. In particular, each community must carry a nonvanishing amount of degree-weighted squared membership mass.

\begin{assumption}\label{asm:min_singular_value_of_B}
For each layer $l$, let $\sigma_B^{(l)} \in (0,1)$ be a constant independent of $n$, with $\inf_{l\geq 1}\sigma_B^{(l)} = \sigma_{B, \inf}>0$, such that $\sigma_{\min}(\bB^{(l)}) \geq \sigma_B^{(l)}$ for all $l \in [L]$.
\end{assumption}
Assumption~\ref{asm:min_singular_value_of_B} prevents the layer-specific block connectivity matrices from becoming asymptotically singular. Next, for each layer $l$, define $$\beta_{n}^{(l)} = \left( \frac{\theta_{\max}^{(l)}}{\theta_{\min}^{(l)}} \right)\frac{K^{8} \theta_{\max}^{(l)} \|\bm{\theta}^{(l)}\|_{1} \log(n)}{(\sigma_{B}^{(l)})^{3} \|\bm{\theta}^{(l)}\|_{2}^{4}}.$$
We refer to $\beta_n^{(l)}$ as the layer-wise spectral difficulty parameter, which is analogous to the corresponding quantity in the analysis of multilayer degree-corrected block models \citep{Agterberg2025}.
\begin{assumption}\label{asm:embedding_regular}
There exist constants \(c_{\deg}>0\) and \(c_{\beta}>0\), independent of \(l\), where \(c_{\deg}\) is sufficiently large and \(c_{\beta}\) is sufficiently small, such that $\theta_{\min}^{(l)}\|\bm{\theta}^{(l)}\|_1 \ge c_{\deg}\log (n)$ and $\max_{1\le l\le L}\beta_n^{(l)} \le c_{\beta}$ for all $ l\in [L]$.
\end{assumption}

For the theoretical analysis, we work with aligned empirical embeddings. For each layer \(l\), let $\bW_*^{(l)}$ be the orthogonal Procrustes alignment matrix between the empirical and population embeddings and define corresponding aligned embeddings $\widetilde{\bX}^{(l)} := \widehat{\bX}^{(l)}\bW_{*}^{(l)}$ and $\widetilde{\bY}^{(l)}_{i,\cdot} := \frac{\widetilde{\bX}^{(l)}_{i,\cdot}} {\|\widetilde{\bX}^{(l)}_{i,\cdot}\|_2}.$
Since \(\bW_{*}^{(l)}\) is orthogonal, we have $\widetilde{\bY}^{(l)} = \widehat{\bY}^{(l)}\bW_{\ast}^{(l)}$.
Let $\widetilde{\bY} := \left[ \widetilde{\bY}^{(1)} \cdots \widetilde{\bY}^{(L)} \right]$ denote the matrix of concatenated normalized adjacency spectral embeddings.
The matrices \(\widetilde{\bX}^{(l)}\) and \(\widetilde{\bY}^{(l)}\) are used only for analysis; the estimator in Algorithm~\ref{alg:step1} is computed from the observable embeddings \(\widehat{\bX}^{(l)}\) and \(\widehat{\bY}^{(l)}\). It can be verified that this alignment does not change the vertex-hunting objective. Indeed, for any candidate set \(\mathcal I\subset[n]\) and any \(\mathcal J\subset\mathcal I\),  $d\bigl( \widehat \bY_{i, \cdot}^{(l)}, {\mathcal{SS}}(\widehat{\bY}^{(l)}, \mathcal J) \bigr) = d\bigl( \widetilde \bY_{i, \cdot}^{(l)}, {\mathcal{SS}}(\widetilde{\bY}^{(l)}), \mathcal J \bigr)$ for all \(i\), \(l\), and \(\mathcal J\). Hence the minimizers of the observable objective based on \(\widehat{\bY}\) are identical to those of the aligned objective based on \(\widetilde{\bY}\).

\subsection{Geometry and Stability of the Spherical Normalization}
We introduce the layer-wise population transformations $T^{(l)}:\Delta^{K-1}\to\mathbb S^{K-1}$ defined by $T^{(l)}(\mathbf z) = \frac{(\bR^{(l)})^\top \mathbf z} {\|(\bR^{(l)})^\top \mathbf z\|_2}$ for all $l \in [L]$.
For a node with membership vector \(\mathbf Z_{i,\cdot}\), this gives the population row-normalized spectral embedding in layer \(l\) by $T^{(l)}(\mathbf Z_{i,\cdot}) = \mathbf Y_{i,\cdot}^{(l)}.$ We then introduce the concatenated transformation $T:\Delta^{K-1}\to \prod_{l=1}^{L}\mathbb S^{K-1}$ defined by $T(\mathbf z) = \left( T^{(1)}(\mathbf z)^{\top},\ldots,T^{(L)}(\mathbf z)^{\top} \right)^{\top}$.
The transformation $T$ maps membership vectors to concatenated row-normalized spectral embeddings by $T(\bZ_{i,\cdot})=\left((\bY_{i,\cdot}^{(1)})^\top,\ldots,(\bY_{i,\cdot}^{(L)})^\top\right)^\top =: \bY_{i, \cdot}.$
The following theorem establishes bi-Lipschitz bounds for the concatenated population map \(T\), showing that the concatenated row-normalized spectral embedding preserves distances on the membership simplex up to layer-dependent constants.
\begin{theorem}[Bi-Lipschitz geometry of the concatenated row-normalized spectral embedding]
\label{thm:concat_embedding_geometry}
For any
\(\mathbf z_1,\mathbf z_2\in\Delta^{K-1}\), there exist constants $c_{\mathrm{Lip}}$, $C_{\mathrm{Lip}} >0$ such that
\begin{equation*}\label{eq:theory_bi_Lipschitz_of_T}
c_{\mathrm{Lip}} \left( \sum_{l=1}^L \frac{(\sigma_{B}^{(l)})^2}{K^3} \right)^{1/2} \|\mathbf z_2-\mathbf z_1\|_2 \leq \|T(\mathbf z_2)-T(\mathbf z_1)\|_2 \leq C_{\mathrm{Lip}} K \left( \sum_{l=1}^L \frac{1}{\sigma_{B}^{(l)}} \right)^{1/2} \|\mathbf z_2-\mathbf z_1\|_2.
\end{equation*}
\end{theorem}
Consequently, since \(\sigma_B^{(l)}>0\) for all \(1\le l\le L\), the map \(T\) is bi-Lipschitz on \(\Delta^{K-1}\) for each fixed finite \(L\). Moreover, under Assumption~\ref{asm:min_singular_value_of_B} and fixed $K$, the rescaled map $L^{-1/2}T$ is bi-Lipschitz with constants independent of $L$. We next state maximum perturbation bounds for both layer-wise and concatenated row-normalized spectral embeddings. This result extends the analysis of \citet{Agterberg2025} from the ML-DCBM to the mixed membership setup.
\begin{theorem}[Layer-wise and concatenated row-wise perturbation bounds]\label{thm:perturbation_bounds}
    Suppose that Assumptions~\ref{asm:gram_matrix}, \ref{asm:min_singular_value_of_B}, and \ref{asm:embedding_regular}, together with Condition~\ref{cond:identifiability}, hold. Then there exist constants $c_{\mathrm{E}}$, $c_{\mathrm{Y}} > 0$, independent of $n$, $L$, and $l$, such that, for each $l \in [L]$, the following bound holds with probability at least $1 - c_{\mathrm{E}}n^{-18}$:
    \begin{equation}\label{eq:layerwise_perturbation_bound}
        \hat{h}^{(l)}:=\max_{1 \leq i \leq n} \| \widetilde{\bY}_{i, \cdot}^{(l)} - \bY_{i, \cdot}^{(l)} \|_{2} \leq c_{\mathrm{Y}} \sqrt{\beta_{n}^{(l)}}.
    \end{equation}
    Consequently, by the union bound, with probability at least $1 - c_{\mathrm{E}}Ln^{-18}$, 
    \begin{equation*}
        \max_{1 \leq i \leq n} \frac{1}{\sqrt{L}} \| \widetilde{\bY}_{i, \cdot} - \bY_{i, \cdot} \|_{2} \leq \left( \frac{c_{\mathrm{Y}}}{L}\sum_{l=1}^{L} \beta_{n}^{(l)} \right)^{1/2}.
    \end{equation*}
\end{theorem}
To illustrate the upper bound in the theorem, consider the homogeneous degree correction parameters $\theta_{i}^{(l)} \asymp \sqrt{\rho_{n}}$. In this case, Assumption~\ref{asm:gram_matrix} and Assumption~\ref{asm:embedding_regular} require $n_{k} \asymp \frac{n}{K}$ for all $k \in [K]$ and $\rho_n \gtrsim \frac{\log(n)}{n}$, respectively. Moreover, Equation~\eqref{eq:layerwise_perturbation_bound} implies that $\max_{i} \| \widetilde{\bY}_{i, \cdot}^{(l)} - \bY_{i, \cdot}^{(l)} \|_{2} \lesssim \sqrt{\beta_{n}^{(l)}} \asymp \sqrt{\frac{\log n}{n \rho_{n}}}$, which is also the rate of the scaled concatenated embedding.

\subsection{Near-Pure Recovery via an Averaged Vertex-Hunting Objective}

We now show that the nodes recovered by the corner estimation procedure are nearly-pure, and this corner-recovery error is controlled by the average layerwise embedding error $\bar{h} := \sum_{l=1}^{L} \hat{h}^{(l)}/L$, reflecting the averaging structure of the vertex-hunting objective defined in \eqref{eq:def_of_vh_objective}. This result uses similar arguments as the vertex-hunting scheme of \citet{jin2024mixed}, but the use of a different normalization strategy and the heterogeneity across layers require special care. In particular, for denoising we use $k$-medoids instead of $k$-means, which ensures the recovery of a point near the spherical simplex. 

We first state a high-probability result on the recovery of pure nodes via corner estimation. Independence across layers allows us to obtain a high-probability bound on $\bar{h}$ with a concentration margin that decreases as $L$ increases. For a deterministic version of this result, see Appendix~\ref{sec:appendix_corner_estimation}. Let
\begin{equation*}\label{eq:def_of_margin_r}
    r_{n,L}(x):= C_{\mathrm{VH}}\left( \frac1L\sum_{l=1}^L\sqrt{\beta_n^{(l)}} + n^{-18} + \frac{x}{L} \right)
\end{equation*}
denote a radius of the estimated corner from every pure node, where $C_{\mathrm{VH}}$ is a constant independent of $n$ and $L$. We state the corner estimation theorem that allows a general choice of $x$ in the probability.
\begin{restatable}[High-probability almost-pure recovery]{theorem}{VHalmostpure}
\label{thm:VH_almost_pure_recovery}
Suppose that $K$ is fixed, the layers are independent, and Assumptions~\ref{asm:gram_matrix}, \ref{asm:min_singular_value_of_B}, \ref{asm:embedding_regular} and Condition~\ref{cond:identifiability} hold. For any candidate set $\mathcal I\subseteq[n]$ containing at least one pure node from each community, let $\widehat{\mathcal J}(\mathcal I) \in \arg\min_{\substack{\mathcal J\subseteq\mathcal I\\ |\mathcal J|=K}} \widetilde\Phi_{L,\mathcal I}(\mathcal J).$ Then for any $x>0$, with probability at least $1-e^{-x}$, every candidate set $\mathcal I\subseteq[n]$ containing at least one pure node from each community satisfies $\min_{k\in[K]}\max_{j\in\widehat J(\mathcal I)} \mathbf{Z}_{jk} \ge 1-r_{n,L}(x)$.
\end{restatable}
\begin{remark}
    When $x = c \log L$ for some $c > 0$, Theorem~\ref{thm:VH_almost_pure_recovery} yields an almost-pure recovery guarantee with probability at least $1-L^{-c}$ and an error radius $$r_{n, L}(c \log L) = C_{\mathrm{VH}}\left(\frac{1}{L} \sum_{l=1}^{L} \sqrt{\beta_{n}^{(l)}} + n^{-18} + \frac{\log L}{L} \right).$$ As the number of layers increases, the last concentration error term vanishes, while the failure probability decays polynomially in $L$. On the other hand, when $x = \kappa L$ for some $\kappa > 0$, we have an error radius $r_{n, L}(\kappa L) = C_{\mathrm{VH}} \left(\frac{1}{L} \sum_{l=1}^{L} \sqrt{\beta_{n}^{(l)}} + n^{-18} + \kappa \right)$ with probability $1 - O(e^{-\kappa L})$. Hence, we have a nonvanishing error radius with exponentially decreasing probability in $L$.
\end{remark}
Since Theorem~\ref{thm:VH_almost_pure_recovery} applies to any corner candidate set containing at least one pure node from each community, it can be naturally applied to the corner-search step without denoising by taking $\mathcal{I}=[n]$ under Condition~\ref{cond:identifiability}. When the denoise step is performed, the same guarantee also applies with $\mathcal{I}=\widehat{\mathcal{M}}_{m}$ if $\widehat{\mathcal{M}}_{m} \cap \mathcal{P}_{k} \neq \varnothing$ for all $k \in [K]$, where $\mathcal{P}_{k} := \{ i \in [n] : \mathbf{Z}_{i, \cdot} = \mathbf{e}_{k} \}$ denote the set of pure nodes in community $k$. We next introduce an additional conditions under which this property holds with high probability.

\begin{condition}\label{cond:denoising_purity_separation}
The following conditions hold.
\begin{enumerate}[(a)]
    \item \textbf{Pure node abundance.} There exists a constant $p_{\mathrm{pure}} > 0$, independent of $n$ and $L$, such that $\min_{k \in [K]} \frac{|\mathcal{P}_{k}|}{n} \geq p_{\mathrm{pure}}$.
    \item \textbf{Mixed node separation.} There exists a constant $\delta_{\mathrm{mix}} \in (0, 1)$, independent of $n$ and $L$, such that for every mixed node $i$, $\min_{k \in [K]} \| \mathbf{Z}_{i, \cdot} - \mathbf{e}_{k} \|_{2} \geq \delta_{\mathrm{mix}}$.
\end{enumerate}
\end{condition}
Condition~\ref{cond:denoising_purity_separation} ensures that each community contains a nonvanishing fraction of pure nodes and that mixed nodes are separated from the pure vertices. We provide a sufficient assumption for the one-swap local minimizer $m$-medoids set $\widehat{\mathcal{M}}_{m}$ to contain a pure node from each community.
\begin{restatable}{proposition}{medoidpreservation}\label{prop:medoid_preservation}
    Suppose Assumptions~\ref{asm:gram_matrix}, \ref{asm:min_singular_value_of_B}, \ref{asm:embedding_regular} and Condition~\ref{cond:identifiability}, \ref{cond:denoising_purity_separation} are satisfied. Let $\widehat{\mathcal{M}}_{m}$ be a one-swap local minimizer of the $m$-medoids objective, where $K \leq m \leq n$. Define the population separation $\Delta_{0} := \frac{1}{\sqrt{L}} \min_{k \in [K]} \min_{i \in [n] \setminus \mathcal{P}_{k}} \| T(\mathbf{Z}_{i, \cdot}) - T(\mathbf{e}_{k}) \|_{2}$.
    There exists a constant $C>0$, independent of $n,L,m,x$, such that, for any $x>0$, if 
    \begin{equation*}
        4 C\left( \frac{1}{L} \sum_{l=1}^{L} \beta_{n}^{(l)} + n^{-18} + \frac{x}{L} \right)^{1/2} + \frac{2}{m p_{\mathrm{pure}}} < \Delta_{0},
    \end{equation*}
    then $\widehat{\mathcal{M}}_{m} \cap \mathcal{P}_{k} \neq \varnothing$ for all $k \in [K]$ with probability at least $1 - e^{-x}$.
\end{restatable}
Theorem~\ref{thm:VH_almost_pure_recovery} guarantees near-pure corner recovery for any candidate set containing at least one pure node from each community. Under the mixed-node separation condition, this guarantee can be strengthened to exact pure-node recovery when the recovery radius $r_{n, L}(x)$ is sufficiently small.
\begin{restatable}[Exact pure-node recovery]{corollary}{exactrecovery}\label{cor:VH_exact}
    Suppose that the assumptions of Theorem~\ref{thm:VH_almost_pure_recovery} and Condition~\ref{cond:denoising_purity_separation}(b) hold. Fix $x>0$ such that $r_{n,L}(x)<\frac{\delta_{\mathrm{mix}}}{\sqrt{2}}.$
Then, with probability at least $1-e^{-x}$, $\widehat J(\mathcal I)$ contains exactly one pure node from each community for any corner candidate set $\mathcal{I}$ with $\mathcal{I} \cap \mathcal{P}_{k} \neq \varnothing$ for all $k \in [K]$.
\end{restatable}
In particular, the corollary applies directly to $\mathcal I=[n]$. It also applies to the $m$-medoid candidate set $\mathcal I=\widehat{\mathcal M}_m$ if the separation condition in Proposition~\ref{prop:medoid_preservation} holds for the same $x$, together with $r_{n,L}(x)<\delta_{\mathrm{mix}}/\sqrt{2}$.

\subsection{Rate of the Membership Estimator}
Given the estimated corners of the $L$ spherical simplices, we construct a membership estimator using the reference layer, as described in Algorithm~\ref{alg:step1}. The following corollary establishes its estimation error bound.

\begin{restatable}[Rate of the spectral membership estimator]
{corollary}{thmInitMembershipConsistency}
\label{cor:init_membership_consistency}
Suppose the conditions of Theorem~\ref{thm:VH_almost_pure_recovery} hold. Let $\mathcal I\subseteq[n]$ be any corner candidate set containing at least one pure node from each community.
Let \(\widehat{\bZ}_0\) be the membership estimator obtained from Algorithm~\ref{alg:step1}. Then there exist constants \(C>0\), $c_{x} > 0$ and $n_{0} \geq 1$ such that, for every $n \geq n_{0}$ and $0 < x \leq c_{x}L$, with probability at least \(1-e^{-x} - c_{E}n^{-18}\), there exists a permutation matrix \(\mathbf\Pi\in\mathcal Q_K\) satisfying
\[
\frac{1}{\sqrt n}\|\widehat{\bZ}_0\mathbf\Pi-\bZ\|_F\le\|\widehat{\bZ}_0\mathbf\Pi-\bZ\|_{2,\infty} \le C\left(\sqrt{\beta_{n}^{(1)}} + r_{n, L}(x) \right).
\]
Moreover, if Condition~\ref{cond:denoising_purity_separation}(b)
holds with $r_{n, L}(x) < \delta_{\mathrm{mix}}/\sqrt{2}$, then on the same event
\begin{align*}
    \frac{1}{\sqrt n}\|\widehat{\bZ}_0\mathbf\Pi-\bZ\|_F\le\|\widehat{\bZ}_0\mathbf\Pi-\bZ\|_{2,\infty} \le C\sqrt{\beta_n^{(1)}}.
\end{align*}
\end{restatable}
The above corollary implies that consistency of the membership estimator is ensured when the reference-layer error $\beta_n^{(1)}$ and the average layer-wise error rate $\frac{1}{L}\sum_{l=1}^L\sqrt{\beta_n^{(l)}}$ both vanish, without requiring $\beta_n^{(l)}\to0$ for every non-reference layer.

\subsection{Minimax lower bound for membership estimation}
An important feature of the ML-DCMM is that additional layers need not always provide additional information for membership estimation. Under the normalization imposed by Condition~\ref{cond:identifiability}, community-specific rescaling can alter the membership vectors of mixed nodes while its effect is absorbed by the layer-specific $\bm{\Theta}^{(l)}$ and $\mathbf{B}^{(l)}$ in non-reference layers. Consequently, there exist distinct membership matrices that induce identical edge-probability matrices $\mathbf{P}^{(l)}$ in all non-reference layers. Therefore, even as additional non-reference layers are observed, they provide no further information for distinguishing between such membership matrices. In such configurations, the distinguishability of the membership matrices, and hence a component of the membership estimation error, is determined entirely by the reference layer and cannot be improved by increasing $L$. This particular reparametrization phenomenon does not arise in the analogous non-overlapping membership model, since community-specific rescaling followed by row normalization leaves pure membership vectors unchanged. A concrete construction illustrating this bottleneck is provided in Appendix~\ref{sec:Appendix-minimax-example}. The following theorem formalizes its statistical consequence.

\begin{restatable}[Minimax lower bound for membership estimation]
{theorem}{thmMinimaxLowerbound}
\label{thm:minimax-membership-lower-bound}
    Fix $K \geq 2$. Suppose that $n\ge n_{0}(K)$ and $\rho_n\in(0,1]$ satisfy $n\rho_n\ge C_\rho(K)\log n$, where $n_{0}(K)$ and $C_\rho(K)>0$ are sufficiently large constants. Let $\mathcal F_{n, K, L}$ denote the parameter space of the ML-DCMM with $n$ nodes, $K$ communities and $L$ layers satisfying Assumptions~\ref{asm:gram_matrix}, \ref{asm:min_singular_value_of_B}, \ref{asm:embedding_regular} and Condition~\ref{cond:identifiability}. Then there exists a universal constant $c>0$, independent of $n$, $K$, $L$, and $\rho_n$, such that
    \begin{equation}\label{eq:minimax_lower_bound}
        \inf_{\widehat{\mathbf Z}} \sup_{\bm{\vartheta}\in\mathcal F_{n,K, L}} \mathbb E_{\bm{\vartheta}} \left[ \min_{\bm{\Pi}\in\mathcal Q_K} \|\widehat{\mathbf Z}\bm{\Pi}-\mathbf Z\|_{2,\infty} \right] \ge \frac{c}{\sqrt{n\rho_n}}.
    \end{equation}
\end{restatable}
\begin{remark}
Consider homogeneous degree-correction parameter setting in the reference layer $\theta_{i}^{(1)} \asymp \sqrt{\rho_{n}}$. When $K$ is fixed, our minimax lower bound of ML-DCMM \eqref{eq:minimax_lower_bound} has the same order as the minimax lower bound of DCMM in \citet{ke2025optimal}. Moreover, if $r_{n,L}(x)<\delta_{\mathrm{mix}}/\sqrt{2}$,  Corollary~\ref{cor:init_membership_consistency} gives
$
\min_{\bm{\Pi}\in\mathcal Q_K} \|\widehat{\mathbf Z}_0\bm{\Pi}-\mathbf Z\|_{2,\infty} = O_p\left( \sqrt{\frac{\log n}{n\rho_n}} \right).
$
Combining this upper bound with Theorem~\ref{thm:minimax-membership-lower-bound}, the proposed spectral membership estimator is minimax optimal up to a factor of $\sqrt{\log n}$ under the $2\to\infty$ loss.
\end{remark}
These results also highlight the distinct roles of additional layers in our procedure. Increasing \(L\) can improve corner recovery through information aggregation across layers, whereas the subsequent membership estimation rate can remain bottlenecked by the reference layer once the corners are sufficiently well estimated.

%% file: Sections/4-Simulations.tex
We conduct simulation studies to evaluate the proposed method. The first experiment studies the effect of the number of layers on the performance of the proposed methodology to validate our theoretical results. The second experiment compares the performance in estimating the node memberships across different data-generating settings and other alternative methodologies.

\noindent\textbf{Experiment 1 (Corner recovery).}
We investigate whether observing more layers improves the recovery of spherical simplex corners. 
Fix \((n,n_0,K,L_{\max})=(300,30,3,30)\), where \(n_0\) denotes the number of pure nodes per community. 
The remaining $n - K n_{0}$ mixed nodes are independently generated from \(\mathrm{Dirichlet}(\mathbf{1}_K)\). 
We set \(\bTheta^{(l)}=\bI_n\) for all \(l\), and use a common block connectivity matrix
$\bB^{(l)}=\bB$ with
$\bB_{kk}=1$ and
$\bB_{jk}=0.2$ for $j\neq k$
for all layers. The probability matrices are scaled so that the expected average degree is 30. 
We generate \(L_{\max}=30\) independent layers from this model.

We apply Algorithm~\ref{alg:step1} to a collection of layers and report the result of step 6 (corner estimation) using the first $L'$ layers for each $L' \in \{2,4,6,8,10,15,20,30\}$.
Let $\widehat{\mathcal J}^{(L')}$ denote the estimated corner index set obtained from these $L'$ layers. 
For each community $k$, define the recovered purity $q_k^{(L')} := \max_{j\in\widehat{\mathcal J}^{(L')}} \mathbf{Z}_{jk}.$
Observe that larger values of $q_k^{(L')}$ indicate that the true membership of a node from community $k$ is closer to a pure node with $\mathbf{Z}_{jk} = 1$.
We summarize corner recovery by the minimum purity score $Q^{(L')} := \min_{1\le k\le K} q_k^{(L')} = \min_{1\le k\le K}\max_{j\in\hat{\mathcal J}^{(L')}} \mathbf{Z}_{jk},$
which is only equal to 1 when each community is represented by a pure node on the estimated set of corners. For $r_0\in\{0.05,0.1\}$, we also define the success indicator $\mathbf{1}\left[ Q^{(L')} \ge 1-r_0 \right]$.
We report the empirical averages of $Q^{(L')}$ and the success indicators over 100 independent repetitions. 

\begin{figure}[!htbp]
\centering

\begin{subfigure}{0.4\textwidth}
\centering
\includegraphics[width=\linewidth]{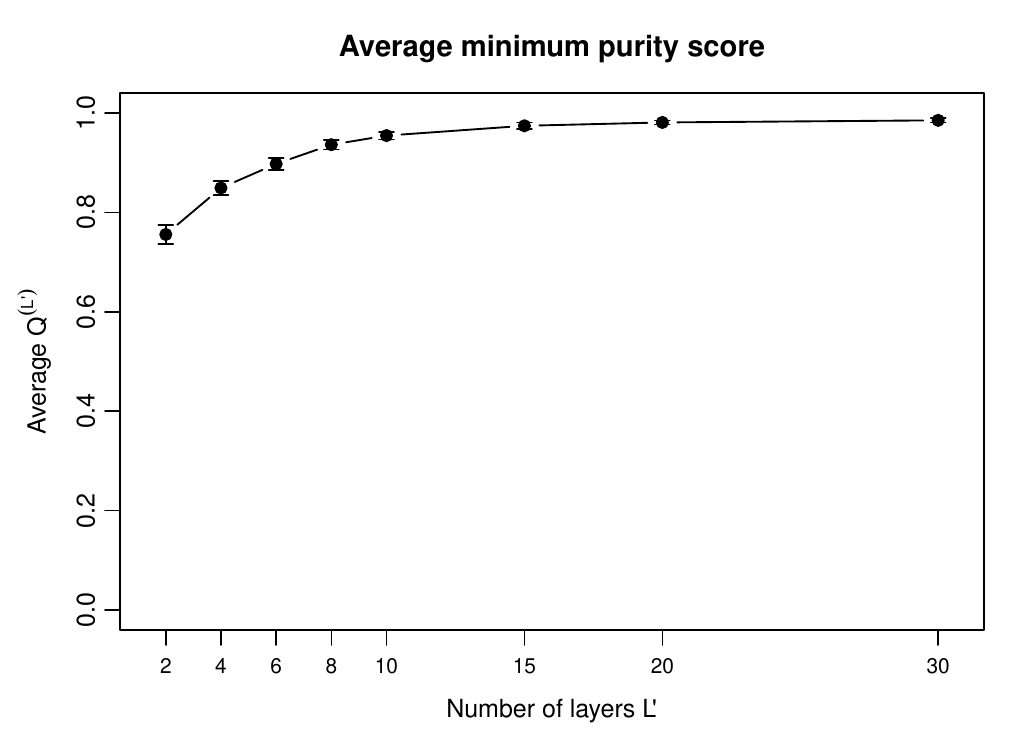}
\end{subfigure}
\begin{subfigure}{0.4\textwidth}
\centering
\includegraphics[width=\linewidth]{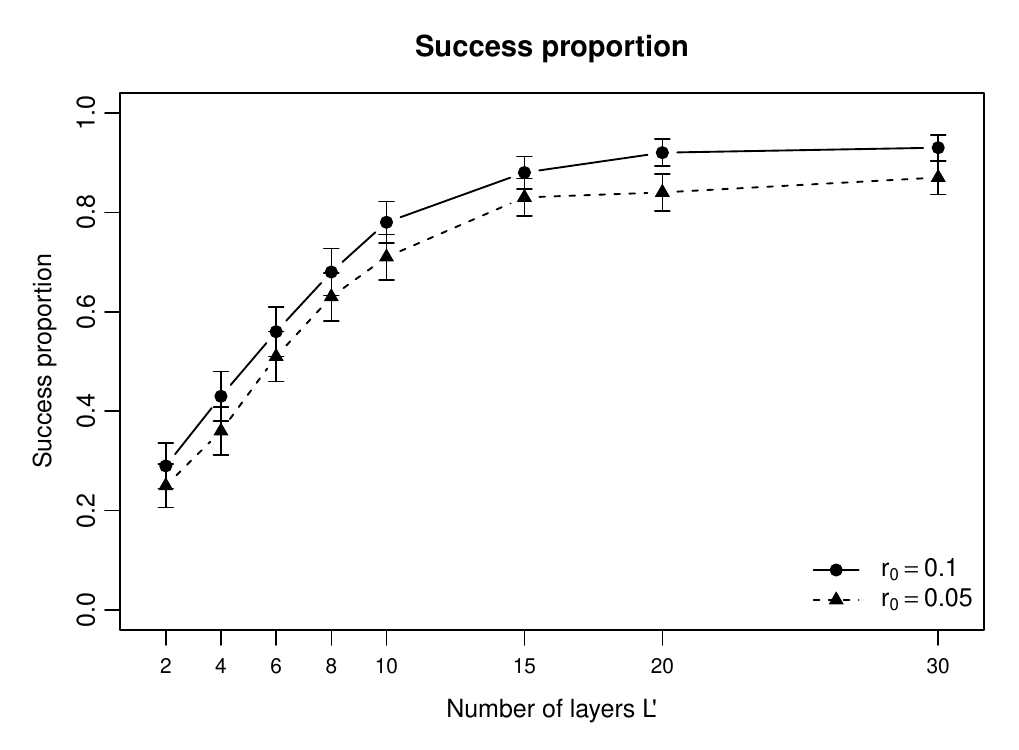}
\end{subfigure}

\caption{
Vertex-hunting performance as the number of layers \(L'\) increases. Left: average minimum purity score \(Q^{(L')}\). Right: success proportion for \(r_0=0.05\) and \(r_0=0.10\). Error bars indicate one standard error over 100 independent repetitions.
}
\label{fig:sim10_vh}
\end{figure}

Figure~\ref{fig:sim10_vh} reports the corner recovery results. The left panel shows that the average minimum purity score \(Q^{(L')}\) improves as the number of layers used for vertex hunting increases. The right panel shows that the success proportion also increases with \(L'\) for both \(r_0=0.05\) and \(r_0=0.10\). These empirical trends are consistent with our theoretical results, suggesting that additional informative layers improve the reliability of corner recovery.

\noindent\textbf{Experiment 2 (Membership recovery).}
We compare the proposed method with competing approaches across six simulation settings, formed by varying the layer-specific block connectivity matrices and the purity of mixed memberships. Fix \((n,n_0,K,L)=(300,30,3,10)\). 
The six settings are obtained by combining one of the following three block-connectivity regimes with one of the following two membership-purity regimes.
\begin{itemize}
    \item \textit{Same connectivity:} the matrices \(\bB^{(l)}=\bB\) are the same for all layers, with unit diagonal entries \(\bB_{kk}=1\) and off-diagonal entries \(\bB_{jk}=0.2\) for \(j\neq k\).

    \item \textit{Different connectivity:} the matrices \(\bB^{(l)}\) vary across layers. Each \(\bB^{(l)}\) has unit diagonal entries \(\bB_{kk}^{(l)}=1\); for \(j<k\), the off-diagonal entries are generated independently as \(\bB_{jk}^{(l)}\sim U(0.1,0.3)\), and we set \(\bB_{kj}^{(l)}=\bB_{jk}^{(l)}\).

    \item \textit{Alternating connectivity:} odd layers have an assortative community structure, with \(\bB_{kk}^{(l)}=1\) and \(\bB_{jk}^{(l)}=0.2\) for \(j\neq k\), whereas even layers have a disassortative structure, with \(\bB_{kk}^{(l)}=0.2\) and \(\bB_{jk}^{(l)}=1\) for \(j\neq k\).
\end{itemize}
For the membership vectors, the remaining \(n-Kn_0\) mixed nodes are evenly split among the following six membership patterns supported on pairs of communities:
\[
(1-r,r,0),\ (r,1-r,0),\ (1-r,0,r),\ (r,0,1-r),\ (0,1-r,r),\ (0,r,1-r).
\]
We consider two values of \(r\): \(r=0.2\), corresponding to \textit{higher-purity} mixed memberships, and \(r=0.4\), corresponding to \textit{lower-purity} mixed memberships. 
We use the same degree-correction parameters across layers $\bTheta^{(l)}=\bTheta=\operatorname{diag}(\theta_1,\ldots,\theta_n)$, $1\le l\le L,$
where \(1/\theta_i\overset{\mathrm{iid}}{\sim}U(1,3)\). For each target density
$\rho\in\{0.04,0.06,0.08,0.10,0.12,0.14,0.16\},$
the probability matrices are scaled so that the expected average degree is $(n-1)\rho$.

We compare the proposed method with three types of baselines. First, we include \textsc{Mixed-SCORE (Avg)}, an aggregation-based baseline that applies the single-layer \textsc{Mixed-SCORE} method of \citet{jin2024mixed} to the average adjacency matrix \(L^{-1}\sum_{l=1}^L A^{(l)}\). This baseline represents a natural way to extend a single-layer mixed-membership method to multilayer data by network aggregation. Second, we include \textsc{DC-MASE} of \citet{Agterberg2025} and \textsc{Bias-adjusted SoS} of \citet{lei2023biasadjusted}, two multilayer methods modeling non-overlapping community structure. This comparison separates two sources of improvement: the benefit of using multilayer information and the benefit of explicitly modeling mixed memberships. For each method, we report the membership estimation error $\operatorname{Err}(\widehat{\bZ},\bZ)
:=
\min_{\bPi\in\mathcal{Q}_K}
\frac{1}{n}\|\bZ-\widehat{\bZ}\bPi\|_F^2$
averaged over 50 independent repetitions.

\begin{figure}[!ht]
    \centering
    \includegraphics[width=0.6\textwidth]{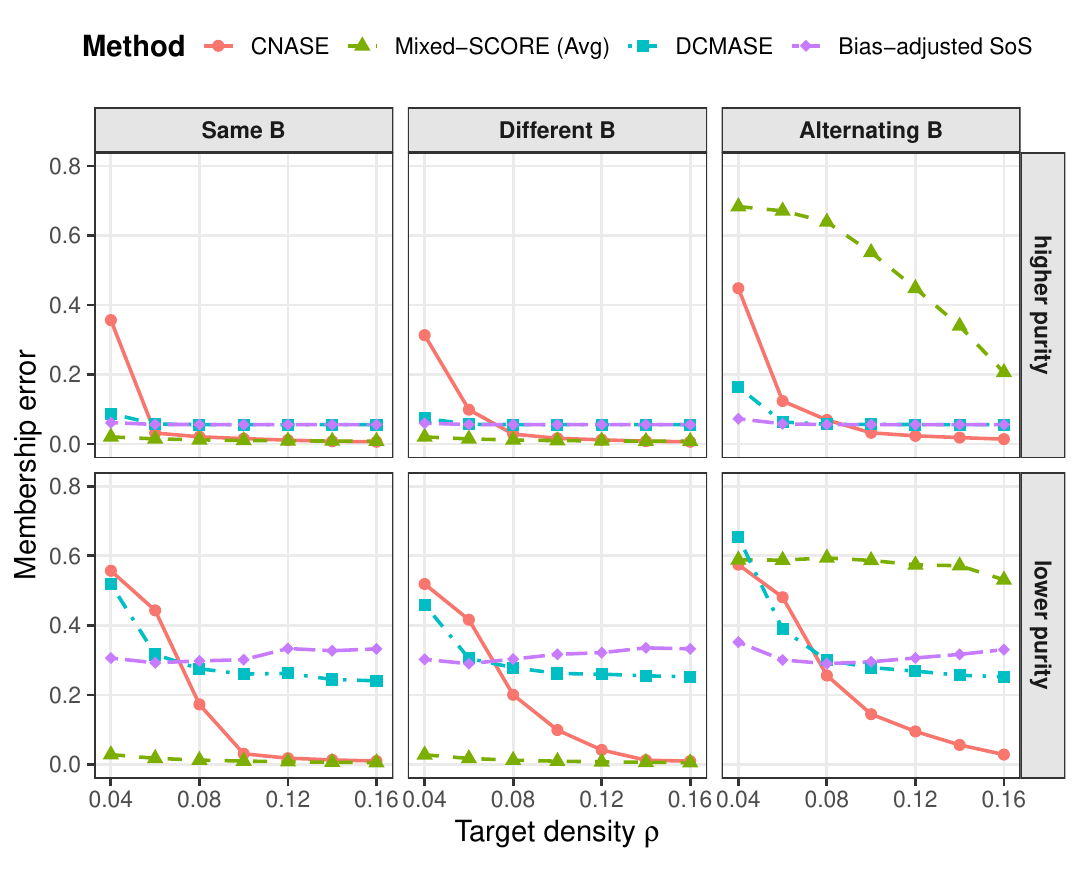}
    \caption{Membership estimation error across six simulation settings. Columns correspond to the three block-connectivity regimes, and rows correspond to the two membership-purity regimes. Each curve reports the average membership error over 50 independent repetitions as the target density \(\rho\) varies. Lower values indicate better membership recovery.}
    \label{fig:membership_error}
\end{figure}

Figure~\ref{fig:membership_error} shows the membership estimation errors across the six simulation settings. The relative performance depends strongly on both the layer-specific block connectivity structure and the purity of the membership vectors. In the Same \(\bB\) and Different \(\bB\) settings, \textsc{Mixed-SCORE (Avg)} performs very well, and in several cases it outperforms \textsc{CNASE}. This behavior is expected because the layers share a common assortative structure, or have block connectivity matrices that are similar in expectation, so averaging the adjacency matrices preserves most of the membership signal. The picture changes substantially in the Alternating \(\bB\) setting. In this case, the assortative and disassortative layer-specific connectivity patterns can cancel each other when the adjacency matrices are averaged, leading to poor performance of \textsc{Mixed-SCORE (Avg)}. By contrast, \textsc{CNASE} preserves the layer-specific connectivity information, and its error decreases rapidly as the target density \(\rho\) increases. This illustrates the advantage of using the multilayer structure when the connectivity patterns differ substantially across layers. 

The comparison with the multilayer hard-clustering methods, \textsc{DC-MASE} and \textsc{Bias-adjusted SoS}, highlights a different aspect. These methods can use information across layers, but they target hard community assignments rather than mixed memberships. As a result, they are competitive in some higher-purity settings, where most nodes are close to pure, but their errors remain bounded away from zero in the lower-purity settings. In contrast, \textsc{CNASE} explicitly models mixed memberships and therefore achieves substantially smaller errors in the lower-purity regimes once the networks are sufficiently dense. Overall, these results suggest that \textsc{CNASE} is particularly beneficial when multilayer networks exhibit substantial layer-to-layer variation in connectivity patterns together with non-negligible mixed membership structure.

%% file: Sections/5-Data_analysis.tex
We analyze London bicycle-sharing trip data originally provided by Transport for London (TfL) and preprocessed and distributed through the \texttt{bikecommclust} GitHub repository by \citet{Munoz-Mendez2018}. 
The data contain 1,469,851 bike trips from June 22, 2014 to July 31, 2014 among \(n=750\) bike stations.

We construct four layers by crossing day type (weekday versus weekend) with time period (AM versus PM), where AM and PM refer to trips starting before and after noon, respectively. 
Throughout this analysis, the layers are ordered as weekday AM, weekday PM, weekend AM, and weekend PM for \(l=1,2,3,4\), respectively. 
For each layer, we first symmetrize the directed trip counts by summing trips in the two directions between each pair of stations, and then divide by the number of days in the corresponding day-type category. 
Let $\mathbf{W}^{(l)}\in\mathbb{R}^{750\times750}$ denote the resulting symmetric weighted adjacency matrix, where \(\mathbf{W}_{ij}^{(l)}\) is the average number of trips per day between stations \(i\) and \(j\) in layer \(l \in \{1, 2, 3, 4 \}\). 
We then threshold these weighted matrices to obtain binary symmetric adjacency matrices \(\mathbf{A}^{(1)},\ldots,\mathbf{A}^{(4)}\), defined by $\mathbf{A}_{ij}^{(l)} = \mathbf{1}\{\mathbf{W}_{ij}^{(l)}\ge 0.1\}$ for every $i < j$ with \(\mathbf{A}_{ii}^{(l)}=0\). 
Summary statistics for the four thresholded layers are reported in Table~\ref{tab:bike_summary}. 
The layer densities vary substantially, suggesting non-negligible heterogeneity across the four temporal layers.

\begin{table}[!ht]
\centering
\caption{Summary statistics for the four thresholded layers of the London bike-sharing network.}
\label{tab:bike_summary}
\small
\begin{tabular}{lcccc}
\hline
Layer & Days & Density & Average degree & Median degree \\
\hline
Weekday AM & 29 & 0.1206 & 90.32 & 80.5 \\
Weekday PM & 29 & 0.2032 & 152.21 & 139.0 \\
Weekend AM & 11 & 0.0644 & 48.23 & 43.0 \\
Weekend PM & 11 & 0.1534 & 114.92 & 104.0 \\
\hline
\end{tabular}
\end{table}

To evaluate the model and the proposed ML-DCMM-based membership estimator, we compare our methodology with other baseline models and membership estimation methods. The competing models include DCBM, DCMM and ML-DCBM, while the estimation methods include \textsc{SCORE} \citep{jin2015fastcommunity}, \textsc{Mixed-SCORE} \citep{jin2024mixed} and \textsc{DC-MASE} \citep{Agterberg2025}. For the ML-DCMM, we include \textsc{Mixed-SCORE (Avg)}, introduced in Section~\ref{sec:Simulations}, to compare estimation performance of the proposed \textsc{CNASE} under the same model.  Note that these baselines include both overlapping and non-overlapping communities, as well as same or different connectivity parameters across layers. See Table~\ref{tab:baseline_methods} in Appendix~\ref{sec:appendix_baseline_models} for a summary of the compared models and methods.

Following the node-pair cross-validation scheme of \citet{Li2020}, we use a 10-fold cross-validation procedure adapted to multilayer networks. See Appendix~\ref{sec:appendix-cv} for details. 
\begin{table}[!ht]
\centering
\caption{
$10$-fold node-pair cross-validation errors for the London bike-sharing network.
Entries are mean CV errors across 10 folds, with empirical standard errors in parentheses.
For each \(K\), the smallest mean CV error is highlighted in bold. Each column corresponds to a different model, with the method to fit the communities indicated in parentheses.
}
\label{tab:bike_cv_K}
\small
\renewcommand{\arraystretch}{1.15}
\begin{tabular*}{0.98\textwidth}{@{\extracolsep{\fill}}cccccc@{}}
\toprule
\(K\)
& \textsc{\makecell{ML-DCMM\\(CNASE)}}
& \textsc{\makecell{DCMM\\(Mixed-SCORE)}}
& \textsc{\makecell{ML-DCMM\\(Mixed-SCORE)}}
& \textsc{\makecell{ML-DCBM\\(DC-MASE)}}
& \textsc{\makecell{DCBM\\(SCORE)}} \\
\midrule
4
& \(\mathbf{0.0751}\ (0.0002)\)
& \(0.0833\ (0.0003)\)
& \(0.0757\ (0.0002)\)
& \(0.0882\ (0.0006)\)
& \(0.0976\ (0.0007)\) \\
5
& \(\mathbf{0.0727}\ (0.0002)\)
& \(0.0812\ (0.0003)\)
& \(0.0736\ (0.0002)\)
& \(0.0849\ (0.0002)\)
& \(0.0971\ (0.0009)\) \\
6
& \(\mathbf{0.0721}\ (0.0002)\)
& \(0.0832\ (0.0002)\)
& \(0.0758\ (0.0002)\)
& \(0.0856\ (0.0008)\)
& \(0.1008\ (0.0011)\) \\
7
& \(\mathbf{0.0708}\ (0.0003)\)
& \(0.0815\ (0.0005)\)
& \(0.0740\ (0.0005)\)
& \(0.0848\ (0.0005)\)
& \(0.1034\ (0.0008)\) \\
8
& \(\mathbf{0.0756}\ (0.0011)\)
& \(0.0833\ (0.0007)\)
& \(0.0762\ (0.0007)\)
& \(0.0860\ (0.0010)\)
& \(0.1080\ (0.0011)\) \\
\bottomrule
\end{tabular*}
\end{table}
Table~\ref{tab:bike_cv_K} summarizes the 10-fold cross-validation errors. The proposed \textsc{CNASE} method achieves the smallest error for every \(K\in\{4,5,6,7,8\}\), with the minimum attained at \(K=7\). The comparison of cross-validation errors under the DCMM and ML-DCMM using the same \textsc{Mixed-SCORE} method is informative. Both procedures estimate the shared membership matrix from the aggregated network, but the ML-DCMM additionally refits the layer-specific degree-correction parameters and block connectivity matrices. Across all values of \(K\), this refitting step substantially reduces the cross-validation error relative to the DCMM. This suggests that layer-specific degree effects and connectivity patterns play an important role in explaining the bike-sharing networks. At the same time, \textsc{Mixed-SCORE} still has larger cross-validation errors than \textsc{CNASE} for every \(K\). Since both procedures allow layer-specific \(\bTheta^{(l)}\) and \(\bB^{(l)}\) in the reconstruction step, this remaining gap points to the estimation of the shared membership structure. In particular, estimating the membership matrix from the concatenated multilayer embeddings appears to yield a more predictive shared community representation than estimating it from the averaged adjacency matrix alone.

\begin{figure}[!ht]
    \centering
    \includegraphics[width=0.95\textwidth]{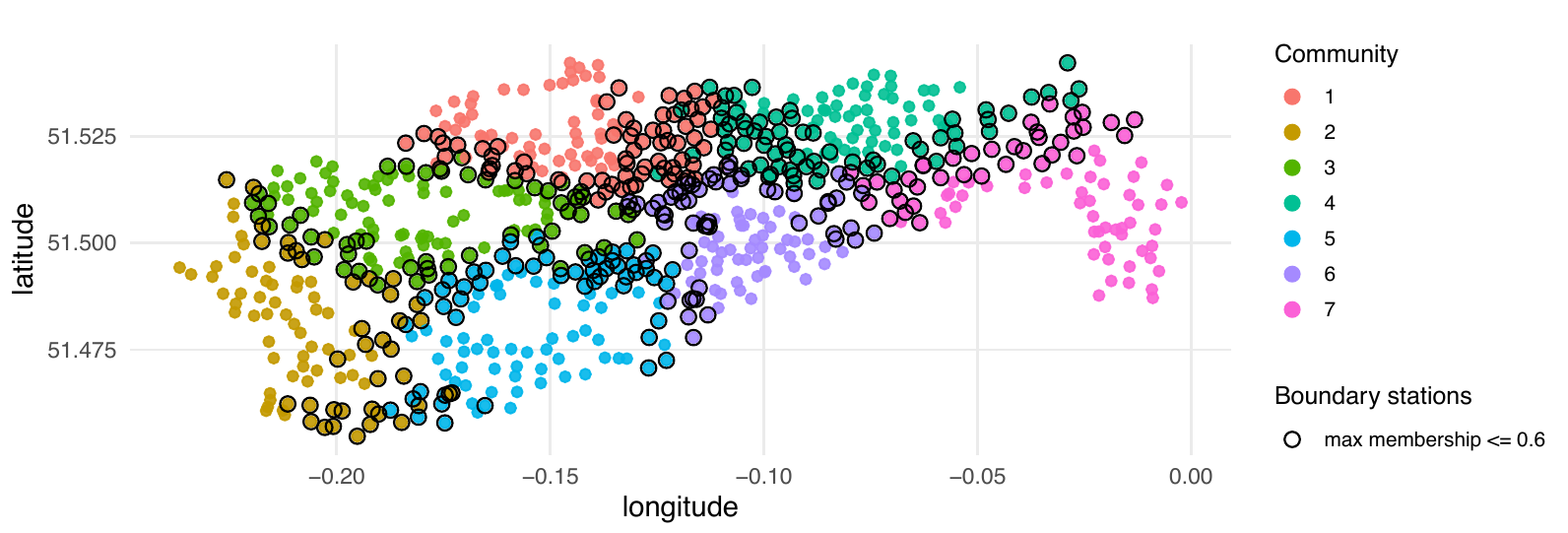}
    \caption{Estimated community structure of the London bike-sharing stations. Stations are colored by their dominant estimated community, \(\arg\max_k \widehat \bZ_{ik}\). Black outlines indicate stations with maximum membership at most \(0.6\), corresponding to relatively mixed-membership stations.}
    \label{fig:London_bike}
\end{figure}

Figure~\ref{fig:London_bike} shows dominant estimated community structure by our method assuming $K=7$ communities. The map suggests a spatially coherent community structure, in which stations assigned to the same dominant community are geographically clustered. The station coordinates are not used in fitting the model, so this spatial coherence provides an external validation of the estimated community structure. Stations located near transition regions between neighboring estimated communities are expected to have more ambiguous community assignments. Consistent with this interpretation, many stations with maximum estimated membership at most \(0.6\) appear near such transition regions. In contrast to hard-clustering models such as ML-DCBM, which assign each station to a single community, ML-DCMM allows these boundary stations to retain non-negligible memberships in multiple communities. This provides a more flexible representation of the spatial community structure.

%% file: Sections/6-Discussion.tex
In this work, we studied the multilayer degree-corrected mixed-membership model. We established identifiability conditions and a minimax lower bound for estimating the membership matrix under the ML-DCMM. We developed a spectral method for estimating the shared membership matrix, and we established a connection with the ParaTuck-2 tensor decomposition. Unlike approaches that collapse the multilayer structure into a single averaged network, the proposed method jointly uses information across layers while retaining layer-specific variation. Accurate corner estimation is crucial for membership estimation, and our theory shows that incorporating more layers can improve corner recovery. Simulation studies show that the proposed method performs well, and it is particularly suitable when the layers are heterogeneous. The bike-sharing network analysis further demonstrates that the method can produce meaningful mixed-membership estimates and improve predictive performance in real multilayer network data, suggesting that the ML-DCMM model equipped with CNASE for membership estimation provides a better fit to the real data.

One limitation is that our theoretical analysis assumes the number of communities $K$ is known, a common simplifying assumption in the community detection literature. In practice, however, $K$ is typically unknown and data-driven estimation requires additional substantial work. Extension of existing methods under several single-layer block models can address this limitation \citep{Jin2023optimal, Han2023, Fan2022}. A second limitation is algorithmic. The ParaTuck-2 refinement step leads to a structured nonconvex optimization problem, and our implementation uses an alternating least-squares procedure because it is easy to implement and performs well empirically. Nevertheless, a more systematic study of initialization, optimization strategies, and convergence behavior for this constrained tensor decomposition would strengthen the practical reliability of the method. 

Future work could further extend the proposed framework in several directions. A particularly interesting direction is to develop methods that combine complementary partial signals across layers, so that the full community structure can be recovered even when individual layers are insufficient . It would also be useful to study dependent multilayer networks, including temporal networks, and to extend the framework to directed or weighted multilayer networks.

%% file: Sections/Appendix-closeness.tex
\section{Algorithmic details}\label{sec:tensor_refinement}
In this section, we provide details of the proposed estimation procedure, including choosing the number of medoids in the corner search step and random initialization in the refinement step. 

\subsection{Choosing the number of medoids}
The number of medoids is selected in a data-driven manner from the candidate set $\mathcal M_{\mathrm{med}}:=\{2K,2K+1,\ldots,3K\}.$
For each \(m\in\mathcal M_{\mathrm{med}}\), we run the vertex-hunting procedure with \(m\) medoids and obtain a candidate index set \(\widehat{\mathcal J}_m\). 
We then compute the corresponding objective value $\widetilde\Phi_{L,\widehat{\mathcal{M}}_{m}}(\widehat{\mathcal J}_m).$ Let $m^\ast \in \operatorname*{arg\,min}_{m\in\mathcal M_{\mathrm{med}}} \widetilde\Phi_{L,\widehat{\mathcal{M}}_{m}}(\widehat{\mathcal J}_m).$
Rather than selecting \(m^\ast\) directly, we allow a small tolerance around the minimum objective value. 
Specifically, for a fixed tolerance parameter \(\tau>0\), we define
\[
\mathcal M_{\tau}
:=
\left\{
m\in\mathcal M_{\mathrm{med}}:
\widetilde\Phi_{L,\widehat{\mathcal{M}}_{m}}(\widehat{\mathcal J}_m)
\le
(1+\tau)
\widetilde\Phi_{L,\widehat{\mathcal{M}}_{m}^\ast}(\widehat{\mathcal J}_{m^\ast})
\right\}.
\]
Among these near-optimal choices, we select the value that gives the best-conditioned estimated corner matrices:
\[
\widehat m \in \operatorname*{arg\,max}_{m\in\mathcal M_{\tau}} \min_{1\le l\le L} \sigma_K\!\left(\widehat{\bS}_{m}^{(l)}\right),
\]
where \(\widehat{\bS}_m^{(l)} = \widehat{\bY}^{(l)}_{\widehat{\mathcal J}_m,\cdot}\) is the corner matrix in layer \(l\) formed by the \(K\) medoids selected from the \(m\) candidates. In our numerical studies, we set \(\tau=0.1\). This aims to obtain numerically stable inverse matrix $(\widehat{\mathbf{S}}_{m}^{(l)})^{-1}$ among possible candidates in $\mathcal{M}_{\tau}$. 

If $\max_{m\in\mathcal M_{\mathrm{med}}}\binom{m}{K} = \binom{3K}{K} \leq 5000$, we enumerate all $K$-subsets and obtain an exact minimizer of $\widetilde\Phi_{L,\widehat{\mathcal{M}}_{m}}$ for every $m\in\mathcal M_{\mathrm{med}}$. Otherwise, we apply the following same greedy search procedure for every candidate \(m\), followed by a local-swap refinement. Our greedy search is inspired by the successive projection algorithm \citep{Araujo001}. For each $m \in \mathcal{M}_{\mathrm{med}}$,
\begin{enumerate}
    \item Choose two farthest medoids and let $\hat{i}_{1}, \hat{i}_{2}$ denote the index of the medoids. 
    \item For each $3 \leq b < K$, let $\widehat{\mathcal{I}}_{b-1} = \{ \hat{i}_{1}, \cdots, \hat{i}_{b-1}\}$ denote the selected $b-1$ medoids index set. Let $P_{\widehat{\mathcal{I}}_{b-1}}^{(l)}$ denote the orthogonal projection matrix onto the linear span of the medoids indexed by $\widehat{\mathcal{I}}_{b-1}$ in layer $l$. Select $$\hat{i}_{b} \in \arg \max_{j \in \mathcal{M}_{\mathrm{med}} \setminus \widehat{\mathcal{I}}_{b-1}} \frac{1}{L} \sum_{l=1}^{L} \| (\mathbf{I} - P_{\widehat{\mathcal{I}}_{b-1}}^{(l)}) \mathbf{y}_{j}^{(l)} \|_{2}^{2},$$
    and set $\widehat{\mathcal{I}}_{b} = \widehat{\mathcal{I}}_{b-1} \cup \{ \hat{i}_{b} \}$.
    \item For $b = K$, select
    \begin{equation*}
        \hat{i}_{K} \in \arg \min_{j \in \mathcal{M}_{\mathrm{med}} \setminus \widehat{\mathcal{I}}_{K-1}} \widetilde{\Phi}_{L, \mathcal{M}_{\mathrm{med}}}(\widehat{\mathcal{I}}_{K-1} \cup \{j\}),
    \end{equation*}
    and set $\widehat{\mathcal{I}}_{K} = \widehat{\mathcal{I}}_{K-1} \cup \{ \hat{i}_{K} \}$. That is, we select $\hat{i}_{K}$ that minimize the averaged vertex-hunting objective defined in \eqref{eq:def_of_vh_objective}.
    \item Refine the selected medoid set $\widehat{\mathcal{I}}_{K}$ to one-swap local optimality.
\end{enumerate}

The resulting solution need not be a global minimizer of \(\widetilde\Phi_{L,\widehat{\mathcal{M}}_{m}}\); consequently, our current recovery theory, which is stated for global minimizers of the corner-search objective, does not directly apply to this scalable implementation.

\subsection{Estimation refinement via a tensor decomposition}
Following the spectral membership estimation, the second step refines $\widehat{\bZ}_{0}$ through the ParaTuck-2 representation. We estimate the ParaTuck-2 factors from 
$\{\widehat{\bU}^{(l)}\}_{l=1}^L$. 
The ParaTuck-2 decomposition may be computed by different numerical or algebraic algorithms; throughout our experiments, we use alternating least squares (ALS). The refinement step is summarized in Algorithm~\ref{alg:step2}.

\begin{algorithm}[htbp]
\caption{ParaTuck-2 refinement}
\label{alg:step2}
\begin{algorithmic}[1]
\Input List of matrices $\{ \widehat{\bU}^{(1)}, \ldots, \widehat{\bU}^{(L)} \}$, number of clusters $K$, spectral membership estimates $\widehat{\bZ}_{0}$, maximum number of iterations $T_{\max}$, convergence criteria $\epsilon$
\Ensure Estimated membership matrix $\widehat{Z}$

\State Obtain random initialization $\{ \widehat{\bTheta}_{0}^{(1)}, \ldots, \widehat{\bTheta}_{0}^{(L)} \}$ and $\{ \widehat{\bM}_{0}^{(1)}, \ldots, \widehat{\bM}_{0}^{(L)} \}$ by $(\widehat{\bTheta}^{(l)})_{ii} \sim U(0.5, 1.5)$ and $(\widehat{\bM}_{0})_{ii} \sim U(0.1, 0.5)$.
\For{$t=1$ to $T_{\max}$}
    \State Perform one ALS iteration to update 
    $\{\widehat{\bTheta}^{(l)}\}_{l=1}^L$, $\widehat{\bZ}$, and 
    $\{\widehat{\bM}^{(l)}\}_{l=1}^L$.
    \If{the convergence criterion is satisfied}
        \State \textbf{break}
    \EndIf
\EndFor
\end{algorithmic}
\end{algorithm}

\section{Spectral Embedding Error Bounds}\label{sec:closeness}
Let $f:\bbR^{d} \setminus \{ \mathbf{0} \} \mapsto \bbR^{d}$ be the $\ell_{2}$ normalization function defined as: $$f(\bx) = \frac{\bx}{\norm{\bx}_{2}}.$$ In this section, we derive row-wise error bounds for the adjacency spectral embedding $\widehat{\bX}_{i, \cdot}^{(l)}$ relative to its population counterpart $\bX_{i, \cdot}^{(l)}$. Using the elementary bound,
\begin{equation*}
    \| f(\bx) - f(\by) \| \leq \frac{2 \| \bx - \by \|}{\| \by \|},
\end{equation*}
we further derive an upper bound for $\| (\widehat{\bY}^{(l)} \bW_{\ast}^{(l)})_{i, \cdot} - \bY_{i, \cdot}^{(l)} \|$. 

To this end, we use the following decomposition from \citet{Agterberg2025}:
\begin{equation}\label{eq:ASE_decomposition}
    \widehat{\bX}^{(l)} \bW_{\ast}^{(l)} - \bX^{(l)} = (\bA^{(l)} - \bbE \bA^{(l)}) \bV^{(l)} \abs{\bLambda^{(l)}}^{-\frac{1}{2}} \bI_{p, q}^{(l)} + \bR_{\bX}^{(l)},
\end{equation}
where the residual term $\bR_{\bX}^{(l)}$ consists of the following 6 terms:
\begin{equation*}
    \begin{aligned}
        \bR_{\bX}^{(l)} &= \bR_{1}^{(l)} \bW_{\ast}^{(l)} + \bR_{2}^{(l)} \bW_{\ast}^{(l)} + \bR_{3}^{(l)} \bW_{\ast}^{(l)} + \bR_{4}^{(l)} + \bR_{5}^{(l)} + \bR_{6}^{(l)}, \\
        \bR_{1}^{(l)} &= - \bV^{(l)} (\bV^{(l)})^{\top} (\bA^{(l)} - \bbE \bA^{(l)}) \widehat{\bV}^{(l)} \abs{\widehat{\bLambda}^{(l)}}^{-\frac{1}{2}} \bI_{p, q}^{(l)}; \\
        \bR_{2}^{(l)} &= \bV^{(l)} \left( (\bV^{(l)})^{\top} \widehat{\bV}^{(l)} \abs{\widehat{\bLambda}^{(l)}}^{\frac{1}{2}} - \abs{\bLambda^{(l)}}^{\frac{1}{2}} (\bV^{(l)})^{\top} \widehat{\bV}^{(l)} \right); \\
        \bR_{3}^{(l)} &= \bV^{(l)} \abs{\bLambda^{(l)}}^{\frac{1}{2}} \left( (\bV^{(l)})^{\top} \widehat{\bV}^{(l)} - (\bW_{\ast}^{(l)})^{\top} \right); \\
        \bR_{4}^{(l)} &= (\bA^{(l)} - \bbE \bA^{(l)}) (\widehat{\bV}^{(l)} (\widehat{\bV}^{(l)})^{\top} \bV^{(l)} - \bV^{(l)}) \abs{\bLambda^{(l)}}^{-\frac{1}{2}} \bI_{p, q}^{(l)}; \\
        \bR_{5}^{(l)} &= -(\bA^{(l)} - \bbE \bA^{(l)}) \widehat{\bV}^{(l)} \left( (\widehat{\bV}^{(l)})^{\top} \bV^{(l)} \abs{\bLambda^{(l)}}^{-\frac{1}{2}} \bI_{p, q}^{(l)} - \abs{\widehat{\bLambda}^{(l)}}^{-\frac{1}{2}} \bI_{p, q}^{(l)} (\widehat{\bV}^{(l)})^{\top} \bV^{(l)} \right); \\
        \bR_{6}^{(l)} &= (\bA^{(l)} - \bbE \bA^{(l)}) \widehat{\bV}^{(l)} \abs{\widehat{\bLambda}^{(l)}}^{-\frac{1}{2}} \bI_{p, q}^{(l)} \left( \bW_{\ast}^{(l)} - (\widehat{\bV}^{(l)})^{\top} \bV^{(l)} \right).
    \end{aligned}
\end{equation*}
We will bound the leading term and each residual term following the approach of \citet{Agterberg2025}. 

\subsection{Preliminary lemmas}\label{sec:C1_prelim_lemmas}
We begin by presenting several lemmas used in the proof. 
\begin{lemma}\label{lem:bound_of_spectral}
    Suppose Assumption \ref{asm:gram_matrix} and \ref{asm:min_singular_value_of_B} hold, and let $\sigma_{r}(\bP^{(l)})$ denote the $r$-th singular value of $\mathbf{P}^{(l)}$ and let $\sigma_{r}(\mathbf{B}^{(l)})$ denote the $r$-th singular value of $\mathbf{B}^{(l)}$ in descending order. Then for all $1 \leq r \leq K$ and $1 \leq i \leq n$,
    \begin{gather}
        \theta_{i}^{(l)} \frac{\sigma_{B}^{(l)}}{K} \lesssim \| \mathbf{X}_{i, \cdot}^{(l)} \| \lesssim \theta_{i}^{(l)} \sqrt{K}; \label{eq:lemma_A.1_lbd&ubd_of_SE}\\
        \| \mathbf{V}_{i}^{(l)} \| \leq C_{G} \sqrt{K } \frac{\theta_{i}^{(l)}}{\| \bm{\theta}^{(l)} \|}; \label{eq:lemma_A.1_ubd_of_eigenvectors}\\
        \sigma_r\bigl(\mathbf P^{(l)}\bigr) \asymp \frac{\|\bm{\theta}^{(l)}\|^2}{K} \sigma_r\bigl(\mathbf B^{(l)}\bigr). \label{eq:lemma_A.1_singular_values}
    \end{gather}
\end{lemma}

\begin{proof}
First, we prove \eqref{eq:lemma_A.1_singular_values}. Define
\[
\bQ^{(l)} := \bTheta^{(l)} \bZ \left( \bZ^\top (\bTheta^{(l)})^2 \bZ \right)^{-1/2}.
\]
Then,
\[
(\bQ^{(l)})^\top \bQ^{(l)} = \bI_K,
\]
so $\bQ^{(l)}$ has orthonormal columns. By the definition of $\bG^{(l)} = \frac{\sqrt{K}}{\norm{\btheta^{(l)}}} \left( \bZ^{\top} \left( \bTheta^{(l)} \right)^{2} \bZ \right)^{\frac{1}{2}}$, we have
\[
\left( \bZ^\top (\bTheta^{(l)})^2 \bZ \right)^{1/2} = \frac{\|\btheta^{(l)}\|}{\sqrt{K}} \bG^{(l)}.
\]
Therefore,
\[
\bP^{(l)} = \frac{\|\btheta^{(l)}\|^2}{K} \bQ^{(l)} \bG^{(l)} \bB^{(l)} \bG^{(l)} (\bQ^{(l)})^\top.
\]
Since $\bQ^{(l)}$ has orthonormal columns, multiplication by $\bQ^{(l)}$ and $(\bQ^{(l)})^\top$ preserves the nonzero singular values. Hence, for every $1 \leq r \leq K$,
\[
\sigma_r(\bP^{(l)}) = \frac{\|\btheta^{(l)}\|^2}{K} \sigma_r\left(\bG^{(l)} \bB^{(l)} \bG^{(l)}\right).
\]
Using the standard singular-value inequality and Assumption~\ref{asm:gram_matrix},
\[
\sigma_r\left(\bG^{(l)} \bB^{(l)} \bG^{(l)}\right) \leq \|\bG^{(l)}\|^2 \sigma_r(\bB^{(l)}) \leq C_G^2 \sigma_r(\bB^{(l)}).
\]
Conversely,
\[
\bB^{(l)} = (\bG^{(l)})^{-1} \left\{ \bG^{(l)} \bB^{(l)} \bG^{(l)} \right\} (\bG^{(l)})^{-1},
\]
and hence
\[
\sigma_r(\bB^{(l)}) \leq \|(\bG^{(l)})^{-1}\|^2 \sigma_r\left(\bG^{(l)} \bB^{(l)} \bG^{(l)}\right) \leq C_G^2 \sigma_r\left(\bG^{(l)} \bB^{(l)} \bG^{(l)}\right).
\]
It follows that
\[
C_G^{-2} \sigma_r(\bB^{(l)}) \leq \sigma_r\left(\bG^{(l)} \bB^{(l)} \bG^{(l)}\right) \leq C_G^2 \sigma_r(\bB^{(l)}).
\]
Therefore,
\[
\frac{\|\btheta^{(l)}\|^2}{K C_G^2} \sigma_r(\bB^{(l)}) \leq \sigma_r(\bP^{(l)}) \leq \frac{C_G^2 \|\btheta^{(l)}\|^2}{K} \sigma_r(\bB^{(l)}).
\]
Thus,
\[
\sigma_r(\bP^{(l)}) \asymp \frac{\|\btheta^{(l)}\|^2}{K} \sigma_r(\bB^{(l)}),
\]
which proves \eqref{eq:lemma_A.1_singular_values}.

Next, we show \eqref{eq:lemma_A.1_ubd_of_eigenvectors}. First, we show $\bP^{(l)}$ and $\left( \bZ^{\top} \left( \bTheta^{(l)} \right)^{2} \bZ \right)^{\frac{1}{2}} \bB^{(l)} \left( \bZ^{\top} \left( \bTheta^{(l)} \right)^{2} \bZ \right)^{\frac{1}{2}}$ have the same non-zero eigenvalues. We can rewrite $\bP^{(l)}$ in terms of $\left( \bZ^{\top} \left( \bTheta^{(l)} \right)^{2} \bZ \right)^{\frac{1}{2}} \bB^{(l)} \left( \bZ^{\top} \left( \bTheta^{(l)} \right)^{2} \bZ \right)^{\frac{1}{2}}$ by following:
\begin{align*}
    \bP^{(l)} &= \bTheta^{(l)} \bZ \bB^{(l)} \bZ^{\top} \bTheta^{(l)} \\
    &= \bTheta^{(l)} \bZ \left( \bZ^{\top} \left( \bTheta^{(l)} \right)^{2} \bZ \right)^{-\frac{1}{2}} \left( \bZ^{\top} \left( \bTheta^{(l)} \right)^{2} \bZ \right)^{\frac{1}{2}} \bB^{(l)} \left( \bZ^{\top} \left( \bTheta^{(l)} \right)^{2} \bZ \right)^{\frac{1}{2}} \left( \bZ^{\top} \left( \bTheta^{(l)} \right)^{2} \bZ \right)^{-\frac{1}{2}} \bZ^{\top} \bTheta^{(l)}.
\end{align*}
Since $\bTheta^{(l)} \bZ \left( \bZ^{\top} \left( \bTheta^{(l)} \right)^{2} \bZ \right)^{-\frac{1}{2}}$ is full column rank matrix with orthonormal columns, $\bP^{(l)}$ and $\left( \bZ^{\top} \left( \bTheta^{(l)} \right)^{2} \bZ \right)^{\frac{1}{2}} \bB^{(l)} \left( \bZ^{\top} \left( \bTheta^{(l)} \right)^{2} \bZ \right)^{\frac{1}{2}}$ have the same non-zero eigenvalues. Using the same non-zero eigenvalues, let
\begin{equation}\label{eq:C_prelim_eigendecomp_middle_matrix}
\left( \bZ^{\top} \left( \bTheta^{(l)} \right)^{2} \bZ \right)^{\frac{1}{2}}
\bB^{(l)}
\left( \bZ^{\top} \left( \bTheta^{(l)} \right)^{2} \bZ \right)^{\frac{1}{2}}
=
\widetilde \bU^{(l)} \bLambda^{(l)} \widetilde \bU^{(l){\top}}
\end{equation}
be an eigendecomposition, where the columns of \(\widetilde{\mathbf U}^{(l)}\) are ordered consistently with the diagonal entries of \(\mathbf\Lambda^{(l)}\). Then it holds that
\begin{equation}\label{eq:svdofP}
    \bP^{(l)} = \bTheta^{(l)} \bZ \left( \bZ^{\top} \left( \bTheta^{(l)} \right)^{2} \bZ \right)^{-\frac{1}{2}} \widetilde{\bU}^{(l)} \bLambda^{(l)} \left( \widetilde{\bU}^{(l)} \right)^{\top} \left( \bZ^{\top} \left( \bTheta^{(l)} \right)^{2} \bZ \right)^{-\frac{1}{2}} \bZ^{\top} \bTheta^{(l)}.
\end{equation}
Note that the matrix $\bTheta^{(l)} \bZ \left( \bZ^{\top} \left( \bTheta^{(l)} \right)^{2} \bZ \right)^{-\frac{1}{2}} \widetilde{\bU}^{(l)}$ has orthonormal columns. That is,
\begin{equation*}
\begin{aligned}
    &\left( 
        \bTheta^{(l)} \bZ 
        \left( \bZ^{\top} \left( \bTheta^{(l)} \right)^2 \bZ \right)^{-1/2}
        \widetilde{\bU}^{(l)} 
    \right)^{\top}
    \bTheta^{(l)} \bZ 
    \left( \bZ^{\top} \left( \bTheta^{(l)} \right)^2 \bZ \right)^{-1/2} 
    \widetilde{\bU}^{(l)} \\[0.5ex]
    &= 
    \left( \widetilde{\bU}^{(l)} \right)^{\top}
    \left( \bZ^{\top} \left( \bTheta^{(l)} \right)^2 \bZ \right)^{-1/2}
    \bZ^{\top} \left( \bTheta^{(l)} \right)^2 \bZ 
    \left( \bZ^{\top} \left( \bTheta^{(l)} \right)^2 \bZ \right)^{-1/2}
    \widetilde{\bU}^{(l)} \\[0.5ex]
    &= 
    \left( \widetilde{\bU}^{(l)} \right)^{\top} \widetilde{\bU}^{(l)} 
    = \bI_{K}.
\end{aligned}
\end{equation*}
Therefore, the decomposition \eqref{eq:svdofP} is also an eigen-decomposition of $\bP^{(l)}$. That is, $\bV^{(l)}$ can be written as
\begin{equation}\label{eq:form_of_V}
    \bV^{(l)} = \bTheta^{(l)} \bZ \left( \bZ^{\top} \left( \bTheta^{(l)} \right)^{2} \bZ \right)^{-\frac{1}{2}} \widetilde{\bU}^{(l)} \bO^{(l)},
\end{equation}
up to orthogonal transformation $\bO^{(l)}$ satisfying $\bO^{(l)} \bLambda^{(l)} \bO^{(l)\top} = \bLambda^{(l)}$.
Now, we can obtain the upper bound of $\norm{\bV_{i, \cdot}^{(l)}}$ as
\begin{equation*}
    \begin{aligned}
        \|\bV_{i, \cdot}^{(l)}\|
        &= \Bigl\| \be_i^\top \bTheta^{(l)} \bZ \bigl( \bZ^\top (\bTheta^{(l)})^2 \bZ \bigr)^{-1/2}\widetilde \bU^{(l)} \Bigr\| \\
        &= \theta_i^{(l)} \Bigl\| \bZ_{i, \cdot}^\top \bigl( \bZ^\top (\bTheta^{(l)})^2 \bZ \bigr)^{-1/2}\widetilde \bU^{(l)} \Bigr\| \\
        &\leq \theta_i^{(l)}
        \Bigl\| \bigl( \bZ^\top (\bTheta^{(l)})^2 \bZ \bigr)^{-1/2} \Bigr\|
        \|\widetilde \bU^{(l)}\|
        \|\bZ_{i, \cdot}\| \\
        &\leq C_{G} \frac{\theta_{i}^{(l)} \sqrt{K}}{\norm{\btheta^{(l)}}},
    \end{aligned}
\end{equation*}

where the last inequality comes from the assumption \ref{asm:gram_matrix}, $$\norm{\left( \bG^{(l)} \right)^{-1}} = \frac{\norm{\btheta^{(l)}}}{\sqrt{K}}\norm{\left( \bZ^{\top} \left( \bTheta^{(l)} \right)^{2} \bZ \right)^{-\frac{1}{2}}} \leq C_{G}.$$ This proves \eqref{eq:lemma_A.1_ubd_of_eigenvectors}. 

Lastly, we show \eqref{eq:lemma_A.1_lbd&ubd_of_SE}. The upper bound of $\norm{\bX_{i, \cdot}^{(l)}}$ can be obtained by
\begin{equation*}
    \begin{aligned}
        \norm{\bX_{i, \cdot}^{(l)}} &= \Bigl \| \bV_{i, \cdot}^{(l)} \abs{\bLambda^{(l)}}^{\frac{1}{2}} \Bigr \| \\
        &\leq \Bigl \| \abs{\bLambda^{(l)}}^{\frac{1}{2}} \Bigr \| \norm{\bV_{i, \cdot}^{(l)}} \\
        &= \sqrt{\sigma_{1}(\bP^{(l)})} \norm{\bV_{i, \cdot}^{(l)}} \\
        &\leq C_{G} \sqrt{\sigma_{1}(\bP^{(l)})} \sqrt{K} \frac{\theta_{i}^{(l)}}{\norm{\btheta^{(l)}}} \\
        &\lesssim \frac{\norm{\btheta^{(l)}}}{\sqrt{K}} \sqrt{\sigma_{1}(\bB^{(l)})} \sqrt{K} \frac{\theta_{i}^{(l)}}{\norm{\btheta^{(l)}}} \\
        &= \sqrt{\sigma_{1}(\bB^{(l)})} \theta_{i}^{(l)} \\
        &\leq\sqrt{K} \theta_{i}^{(l)}, 
    \end{aligned}
\end{equation*}
where the last inequality comes from the equivalence property between max norm and spectral norm of a matrix. 
Next, we derive a lower bound for $\norm{\bX_{i, \cdot}^{(l)}}$. Using the expression of $\bV^{(l)}$ by \eqref{eq:form_of_V}, $\bX^{(l)}$ can be written as
\begin{equation*}
    \bX^{(l)} = \bV^{(l)} \abs{\bLambda^{(l)}}^{\frac{1}{2}} = \bTheta^{(l)} \bZ \left( \bZ^{\top} \left( \bTheta^{(l)} \right)^{2} \bZ \right)^{-\frac{1}{2}} \widetilde{\bU}^{(l)} \abs{\bLambda^{(l)}}^{\frac{1}{2}}.
\end{equation*}
Recall that by \eqref{eq:C_prelim_eigendecomp_middle_matrix}, we have
\begin{equation*}
    \left( \bZ^{\top} \left( \bTheta^{(l)} \right)^{2} \bZ \right)^{-\frac{1}{2}} \widetilde{\bU}^{(l)} = \bB^{(l)} \left( \bZ^{\top} \left( \bTheta^{(l)} \right)^{2} \bZ \right)^{\frac{1}{2}} \widetilde{\bU}^{(l)} \left( \bLambda^{(l)} \right)^{-1},
\end{equation*}
which implies that
\begin{equation*}
    \begin{aligned}
        \bV^{(l)} \abs{\bLambda^{(l)}}^{\frac{1}{2}} &= \bTheta^{(l)} \bZ \bB^{(l)} \left( \bZ^{\top} \left( \bTheta^{(l)} \right)^{2} \bZ \right)^{\frac{1}{2}} \widetilde{\bU}^{(l)} \left( \bLambda^{(l)} \right)^{-1} \abs{\bLambda^{(l)}}^{\frac{1}{2}} \\
        &= \bTheta^{(l)} \bZ \bB^{(l)} \left( \bZ^{\top} \left( \bTheta^{(l)} \right)^{2} \bZ \right)^{\frac{1}{2}} \widetilde{\bU}^{(l)} \abs{\bLambda^{(l)}}^{-\frac{1}{2}} \bW^{(l)},
    \end{aligned}
\end{equation*}
where $\bW^{(l)}$ is the diagonal matrix of sign of $\bLambda^{(l)}$. Then,
\begin{equation*}
    \begin{aligned}
        \norm{\bX_{i, \cdot}^{(l)}} &= \theta_{i}^{(l)} \norm{\bZ_{i, \cdot}^{\top} \bB^{(l)} \left( \bZ^{\top} \left( \bTheta^{(l)} \right)^{2} \bZ \right)^{\frac{1}{2}} \widetilde{\bU}^{(l)} \abs{\bLambda^{(l)}}^{-\frac{1}{2}} \bW^{(l)}} \\
        &\geq \theta_{i}^{(l)} \norm{\bZ_{i, \cdot}^{\top} \bB^{(l)}} \sigma_{\min} \left( \left( \bZ^{\top} \left( \bTheta^{(l)} \right)^{2} \bZ \right)^{\frac{1}{2}} \widetilde{\bU}^{(l)} \abs{\bLambda^{(l)}}^{-\frac{1}{2}} \right) \\
        &\geq \theta_{i}^{(l)} \norm{\bZ_{i, \cdot}} \sigma_{\min}(\bB^{(l)}) \sigma_{\min} \left( \left( \bZ^{\top} \left( \bTheta^{(l)} \right)^{2} \bZ \right)^{\frac{1}{2}} \widetilde{\bU}^{(l)} \abs{\bLambda^{(l)}}^{-\frac{1}{2}} \right) \\
        &\geq \theta_{i}^{(l)} \frac{1}{\sqrt{K}} \sigma_{\min}(\bB^{(l)}) \sigma_{\min} \left( \left( \bZ^{\top} \left( \bTheta^{(l)} \right)^{2} \bZ \right)^{\frac{1}{2}} \right) \sigma_{\min}\left( \abs{\bLambda^{(l)}}^{-\frac{1}{2}} \right) ,
    \end{aligned}
\end{equation*}
where the last inequality comes from $\sigma_{\min}(\bA \bB) \geq \sigma_{\min}(\bA) \sigma_{\min}(\bB)$ for any invertible matrix $\bA, \bB$ and $\sigma_{\min}(\bW) = 1$ for any orthogonal matrix $\bW$. We provide a lower bound for each minimum singular value. We know $\sigma_{\min} \left( \left( \bZ^{\top} \left( \bTheta^{(l)} \right)^{2} \bZ \right)^{\frac{1}{2}} \right) = \frac{1}{\sigma_{\max} \left( \left( \bZ^{\top} \left( \bTheta^{(l)} \right)^{2} \bZ \right)^{-\frac{1}{2}} \right)}$ and by the assumption~\ref{asm:gram_matrix}, 
\begin{equation*}
    \begin{aligned}
        \norm{\left( \bG^{(l)} \right)^{-1}} &= \frac{\norm{\btheta^{(l)}}}{\sqrt{K}} \norm{\left( \bZ^{\top} \left( \bTheta^{(l)} \right)^{2} \bZ \right)^{-\frac{1}{2}}} \leq C_{G} \\
        &\Leftrightarrow \sigma_{\max} \left( \left( \bZ^{\top} \left( \bTheta^{(l)} \right)^{2} \bZ \right)^{-\frac{1}{2}} \right) \leq C_{G} \frac{\sqrt{K}}{\norm{\btheta^{(l)}}} \\
        &\Leftrightarrow \frac{1}{\sigma_{\max} \left( \left( \bZ^{\top} \left( \bTheta^{(l)} \right)^{2} \bZ \right)^{-\frac{1}{2}} \right)} \geq \frac{\norm{\btheta^{(l)}}}{C_{G}\sqrt{K}}.
    \end{aligned}
\end{equation*}
Also, it can be shown that
\begin{equation*}
    \sigma_{\min}\left( \abs{\bLambda^{(l)}}^{-\frac{1}{2}} \right) = \lambda_{\min} \left( \abs{\bLambda^{(l)}}^{-\frac{1}{2}} \right) = \frac{1}{\lambda_{\max} \left( \abs{\bLambda^{(l)}}^{\frac{1}{2}} \right)} = \frac{1}{\sqrt{\sigma_{1}(\bP^{(l)})}}.
\end{equation*}
Therefore, 
\begin{equation*}
    \begin{aligned}
        \norm{\bX_{i, \cdot}^{(l)}} &\geq \theta_{i}^{(l)} \frac{1}{\sqrt{K}} \sigma_{\min}(\bB^{(l)}) \sigma_{\min} \left( \left( \bZ^{\top} \left( \bTheta^{(l)} \right)^{2} \bZ \right)^{\frac{1}{2}} \right) \sigma_{\min} \left( \abs{\bLambda^{(l)}}^{-\frac{1}{2}} \right) \\
        &\geq \theta_{i}^{(l)} \frac{1}{\sqrt{K}} \sigma_{\min}(\bB^{(l)}) \frac{\norm{\btheta^{(l)}}}{C_{G} \sqrt{K}} \frac{1}{\sqrt{\sigma_{1}(\bP^{(l)})}}\\
        &\gtrsim \theta_{i}^{(l)} \frac{1}{\sqrt{K}} \sigma_{\min}(\bB^{(l)}) \frac{\norm{\btheta^{(l)}}}{\sqrt{K}} \frac{\sqrt{K}}{\norm{\btheta^{(l)}}}\frac{1}{\sqrt{\sigma_{1}(\bB^{(l)})}} \\
        &\geq \theta_{i}^{(l)} \frac{1}{\sqrt{K}} \frac{1}{\sqrt{K}} \sigma_{\min}(\bB^{(l)}).
    \end{aligned}
\end{equation*}
\end{proof}

We use the following key lemmas from \citet{Agterberg2025} (Lemmas C.3--C.7) without proof, which are crucial for our proof. The proofs of Lemmas C.3 – C.7 depend only on concentration bounds for $\bA^{(l)} - \bP^{(l)}$, the eigendecomposition of $\bP^{(l)}$, and algebraic identities involving $\bV^{(l)}$ and $\widehat{\bV}^{(l)}$. In particular, they do not use the hard-clustering structure of $\bZ$. Hence the same arguments extend to the MLDCMM setting.
\begin{lemma}\label{lem:C3}
    When $\theta_{\max}^{(l)} \norm{\btheta^{(l)}}_{1} \geq \log(n),$ it holds that
    \begin{equation*}
        \begin{aligned}
            \norm{\bA^{(l)} - \bP^{(l)}} &\lesssim \sqrt{\theta_{\max}^{(l)} \norm{\btheta^{(l)}}_{1}} \\
            \norm{(\bV^{(l)})^{\top} (\bA^{(l)} - \bP^{(l)}) \bV^{(l)}} &\lesssim \sqrt{K} + \sqrt{\log(n)},
        \end{aligned}
    \end{equation*}
    with probability at least $1 - O(n^{-20})$.
\end{lemma}

\begin{lemma}\label{lem:C4}
    When $\sigma_{K}(\bP^{(l)}) \geq \sqrt{\theta_{\max}^{(l)} \norm{\btheta^{(l)}}_{1} \log(n)}$ and $\min_{i} \theta_{i}^{(l)} \norm{\btheta^{(l)}}_{1} \gtrsim \log(n),$ the following bounds hold with probability at least $1 - O(n^{-20}):$
    \begin{equation*}
        \begin{aligned}
            \norm{(\bV^{(l)})^{\top} \widehat{\bV}^{(l)} - (\bW_{\ast}^{(l)})^{\top}} &\lesssim \frac{\theta_{\max}^{(l)} \norm{\btheta^{(l)}}_{1}}{(\sigma_{K}(\bP^{(l)}))^{2}} \\
            \norm{(\widehat{\bV}^{(l)})^{\top} \bV^{(l)} \abs{\bLambda^{(l)}}^{\frac{1}{2}} - \abs{\widehat{\bLambda}^{(l)}}^{\frac{1}{2}} (\widehat{\bV}^{(l)})^{\top} \bV^{(l)}} &\lesssim \frac{K^{2}}{(\sigma_{K}(\bP^{(l)}))^{\frac{1}{2}}} \left( \sqrt{K \log(n)} + \frac{\theta_{\max}^{(l)} \norm{\btheta^{(l)}}_{1}}{\sigma_{K}(\bP^{(l)})} \right) \\
            \norm{(\widehat{\bV}^{(l)})^{\top} \bV^{(l)} \abs{\bLambda^{(l)}}^{-\frac{1}{2}} \bI_{p, q}^{(l)} - \abs{\widehat{\bLambda}^{(l)}}^{-\frac{1}{2}} \bI_{p, q}^{(l)} (\widehat{\bV}^{(l)})^{\top} \bV^{(l)}} &\lesssim \frac{K^{2}}{(\sigma_{K}(\bP^{(l)}))^{\frac{3}{2}}} \left( \sqrt{K \log(n)} + \frac{\theta_{\max}^{(l)} \norm{\btheta^{(l)}}_{1}}{\sigma_{K}(\bP^{(l)})} \right).
        \end{aligned}
    \end{equation*}
\end{lemma}

\begin{lemma}\label{lem:C5}
    When $\sigma_{K}(\bP^{(l)}) \gtrsim \sqrt{\theta_{\max}^{(l)} \norm{\btheta^{(l)}}_{1} \log(n)}$ and $\min_{i} \theta_{i}^{(l)} \norm{\btheta^{(l)}}_{1} \gtrsim \log(n)$, the following bound holds with probability at least $1 - O(n^{-20}):$
    \begin{equation*}
        \norm{\be_{i}^{\top} (\bA^{(l)} - \bbE \bA^{(l)}) \widehat{\bV}^{(l)}} \lesssim \sqrt{\theta_{i}^{(l)} \norm{\btheta^{(l)}}_{1} \log(n)} \norm{\widehat{\bV}^{(l)}}_{2, \infty}.
    \end{equation*}
\end{lemma}

\begin{lemma}\label{lem:C6}
    When $\sigma_{K}(\bP^{(l)}) \gtrsim \sqrt{\theta_{\max}^{(l)} \norm{\btheta^{(l)}}_{1} \log(n)}$ and $\min_{i} \theta_{i}^{(l)} \norm{\btheta^{(l)}}_{1} \gtrsim \log(n)$, the following bounds holds with probability at least $1 - O(n^{-19}):$
    \begin{equation*}
        \begin{aligned}
            \norm{\widehat{\bV}^{(l)} - \bV^{(l)} (\bV^{(l)})^{\top} \widehat{\bV}^{(l)}}_{2, \infty} &\lesssim \frac{\sqrt{\theta_{\max}^{(l)} \norm{\btheta^{(l)}}_{1} \log(n)}}{\sigma_{K}(\bP^{(l)})} \norm{\bV^{(l)}}_{2, \infty} \\
            \norm{\widehat{\bV}^{(l)} \bW_{\ast}^{(l)} - \bV^{(l)}}_{2, \infty} &\lesssim \frac{\sqrt{\theta_{\max}^{(l)} \norm{\btheta^{(l)}}_{1} \log(n)}}{\sigma_{K}(\bP^{(l)})} \norm{\bV^{(l)}}_{2, \infty} \\
            \norm{\widehat{\bV}^{(l)} (\widehat{\bV}^{(l)})^{\top} \bV^{(l)} - \bV^{(l)}}_{2, \infty} &\lesssim \frac{\sqrt{\theta_{\max}^{(l)} \norm{\btheta^{(l)}}_{1} \log(n)}}{\sigma_{K}(\bP^{(l)})} \norm{\bV^{(l)}}_{2, \infty}.
        \end{aligned}
    \end{equation*}
\end{lemma}

\begin{lemma}\label{lem:C7}
    When $\sigma_{K}(\bP^{(l)}) \gtrsim \sqrt{\theta_{\max}^{(l)} \norm{\btheta^{(l)}}_{1} \log(n)}$ and $\min_{i} \theta_{i}^{(l)} \norm{\btheta^{(l)}}_{1} \gtrsim \log(n)$, with probability at least $1 - O(n^{-19}),$ it holds that
    \begin{equation*}
        \norm{\be_{i}^{\top} (\bA^{(l)} -\bbE \bA^{(l)}) (\widehat{\bV}^{(l)} (\widehat{\bV}^{(l)})^{\top} \bV^{(l)} - \bV^{(l)})} \lesssim \sqrt{\theta_{i}^{(l)} \norm{\btheta^{(l)}}_{1} \log(n)} \frac{\sqrt{\theta_{\max}^{(l)} \norm{\btheta^{(l)}}_{1} \log(n)}}{\sigma_{K}(\bP^{(l)})} \norm{\bV^{(l)}}_{2, \infty}.
    \end{equation*}
\end{lemma}

Lastly, we derive useful bounds relating \(\widehat{\mathbf \Lambda}^{(l)}\) and \(\mathbf \Lambda^{(l)}\) as a consequence of Weyl's inequality.

\begin{lemma}[Weyl consequence]\label{lem:weyl_consequence}
Suppose the conditions of Lemma \ref{lem:C3} hold. In addition, assume that
\[
\sigma_K(\mathbf P^{(l)})
\gtrsim
\sqrt{
\theta_{\max}^{(l)}
\|\boldsymbol{\theta}^{(l)}\|_1
\log (n)
}.
\]
Then, on the event of Lemma \ref{lem:C3}, for all sufficiently large \(n\),
\[
\|\mathbf A^{(l)}-\mathbf P^{(l)}\|
\le
\frac{1}{2}\sigma_K(\mathbf P^{(l)}).
\]
Consequently,
\[
\sigma_K(\mathbf A^{(l)})
\ge
\frac{1}{2}\sigma_K(\mathbf P^{(l)})
\]
and
\[
\sigma_1(\mathbf A^{(l)})
\le
\frac{3}{2}\sigma_1(\mathbf P^{(l)}).
\]
Therefore,
\[
\left\| |\widehat{\mathbf \Lambda}^{(l)}|^{-1} \right\|
\le
2
\left\| |\mathbf \Lambda^{(l)}|^{-1} \right\|,
\]
\[
\left\| |\widehat{\mathbf \Lambda}^{(l)}|^{-1/2} \right\|
\le
\sqrt{2}
\left\| |\mathbf \Lambda^{(l)}|^{-1/2} \right\|,
\]
and
\[
\left\| |\widehat{\mathbf \Lambda}^{(l)}|^{1/2} \right\|
\le
\sqrt{\frac{3}{2}}
\left\| |\mathbf \Lambda^{(l)}|^{1/2} \right\|.
\]
\end{lemma}

\begin{proof}
On the event of Lemma \ref{lem:C3}, there exists a constant \(C>0\) such that
\begin{equation}\label{lem:C1_C7_lemC2}
    \|\mathbf A^{(l)}-\mathbf P^{(l)}\| \le C \sqrt{ \theta_{\max}^{(l)} \|\boldsymbol{\theta}^{(l)}\|_1 }.
\end{equation}
Moreover, by assumption, we have 
\begin{equation}\label{lem:C1_C7_lower_bound_of_min_sigma}
    \sigma_{K}(\bP^{(l)}) \gtrsim \sqrt{\theta_{\max}^{(l)} \| \btheta^{(l)} \|_{1} \log(n)}.
\end{equation}
By combining \eqref{lem:C1_C7_lemC2} and \eqref{lem:C1_C7_lower_bound_of_min_sigma}, we have
\[
\frac{
\|\mathbf A^{(l)}-\mathbf P^{(l)}\|
}{
\sigma_K(\mathbf P^{(l)})
}
\lesssim
\frac{1}{\sqrt{\log n}} \to 0.
\]
Hence, for all sufficiently large \(n\),
\[
\|\mathbf A^{(l)}-\mathbf P^{(l)}\|
\le
\frac{1}{2}\sigma_K(\mathbf P^{(l)}).
\]
By Weyl's inequality,
\[
\sigma_K(\mathbf A^{(l)})
\ge
\sigma_K(\mathbf P^{(l)})
-
\|\mathbf A^{(l)}-\mathbf P^{(l)}\|
\ge
\frac{1}{2}\sigma_K(\mathbf P^{(l)}).
\]
Similarly,
\[
\sigma_1(\mathbf A^{(l)})
\le
\sigma_1(\mathbf P^{(l)})
+
\|\mathbf A^{(l)}-\mathbf P^{(l)}\|
\le
\frac{3}{2}\sigma_1(\mathbf P^{(l)}),
\]
because \(\sigma_K(\mathbf P^{(l)})\le \sigma_1(\mathbf P^{(l)})\). The displayed bounds for
\(|\widehat{\mathbf \Lambda}^{(l)}|^{-1}\),
\(|\widehat{\mathbf \Lambda}^{(l)}|^{-1/2}\), and
\(|\widehat{\mathbf \Lambda}^{(l)}|^{1/2}\)
follow directly from these two singular-value inequalities.
\end{proof}

\subsection{\texorpdfstring{Bound of the residual terms $\bR_{\bX}^{(l)}$}{Bound of residual terms}}\label{sec:bound_of_ASE_residual}
For each \(l\in[L]\) and \(i\in[n]\), let \(\mathcal{E}_{res, i}^{(l)}\) denote the event on which the conclusions of Lemmas \ref{lem:C3}--\ref{lem:C7}, the Weyl consequence in Lemma \ref{lem:weyl_consequence}, and the Davis--Kahan bound
\[
\|(\mathbf V_{\perp}^{(l)})^\top \widehat{\mathbf V}^{(l)}\|
\le
C
\frac{\|\mathbf A^{(l)}-\mathbf P^{(l)}\|}{\sigma_K(\mathbf P^{(l)})}
\]
hold simultaneously.
By a union bound, there exists a constant \(C>0\), independent of \(l\), such that
\[
\mathbb P(\mathcal{E}_{res, i}^{(l)})
\ge
1-Cn^{-19}.
\]
Define
\[
\mathcal{E}_{res}^{(l)}
:=
\bigcap_{i=1}^n
\mathcal{E}_{res, i}^{(l)}.
\]
Then
\[
\mathbb P(\mathcal{E}_{res}^{(l)})
\ge
1-Cn^{-18}.
\]
In the residual-bound subsection, all row-wise inequalities for layer \(l\) are understood to hold on \(\mathcal{E}_{res}^{(l)} \).

We provide an upper bound of the residual term in the decomposition \eqref{eq:ASE_decomposition}. We will apply the lemmas from Section \ref{sec:C1_prelim_lemmas} to their corresponding terms throughout the proof.

\subsubsection*{The term $\bR_{1}^{(l)}$}
We start with the first term $\bR_{1}^{(l)} = - \bV^{(l)} (\bV^{(l)})^{\top} (\bA^{(l)} - \bP^{(l)}) \widehat{\bV}^{(l)} \abs{\widehat{\bLambda}^{(l)}}^{-\frac{1}{2}} \bI_{p, q}^{(l)}.$ 
\begin{equation*}
    \begin{aligned}
        \norm{\be_{i}^{\top} \bR_{1}^{(l)}} &= \norm{\be_{i}^{\top} \bV^{(l)} (\bV^{(l)})^{\top} (\bA^{(l)} - \bP^{(l)}) \widehat{\bV}^{(l)} \abs{\widehat{\bLambda}^{(l)}}^{-\frac{1}{2}} \bI_{p, q}^{(l)}} \\
        &= \norm{\be_{i}^{\top} \bV^{(l)} (\bV^{(l)})^{\top} (\bA^{(l)} - \bP^{(l)}) (\bV^{(l)} (\bV^{(l)})^{\top} + \bI - \bV^{(l)} (\bV^{(l)})^{\top}) \widehat{\bV}^{(l)} \abs{\widehat{\bLambda}^{(l)}}^{-\frac{1}{2}} \bI_{p, q}^{(l)}} \\
        &\leq \norm{\be_{i}^{\top} \bV^{(l)} (\bV^{(l)})^{\top} (\bA^{(l)} - \bP^{(l)}) \bV^{(l)} (\bV^{(l)})^{\top} \widehat{\bV}^{(l)} \abs{\widehat{\bLambda}^{(l)}}^{-\frac{1}{2}} \bI_{p, q}^{(l)}} \\
        & \quad + \norm{\be_{i}^{\top} \bV^{(l)} (\bV^{(l)})^{\top} (\bA^{(l)} - \bP^{(l)}) \bV_{\perp}^{(l)} (\bV_{\perp}^{(l)})^{\top} \widehat{\bV}^{(l)} \abs{\widehat{\bLambda}^{(l)}}^{-\frac{1}{2}} \bI_{p, q}^{(l)}} \\
        &\leq \norm{\be_{i}^{\top} \bV^{(l)}} \norm{(\bV^{(l)})^{\top} (\bA^{(l)} - \bP^{(l)}) \bV^{(l)}} \norm{\abs{\widehat{\bLambda}^{(l)}}^{-\frac{1}{2}}} \\
        & \quad + \norm{\be_{i}^{\top} \bV^{(l)}} \norm{\bA^{(l)} - \bP^{(l)}} \norm{(\bV_{\perp}^{(l)})^{\top} \widehat{\bV}^{(l)}} \norm{\abs{\widehat{\bLambda}^{(l)}}^{-\frac{1}{2}}} \\
        &\lesssim \frac{\norm{\be_{i}^{\top} \bV^{(l)}}}{(\sigma_{K}(\bP^{(l)}))^{\frac{1}{2}}} \left( \norm{(\bV^{(l)})^{\top} (\bA^{(l)} - \bP^{(l)}) \bV^{(l)}} + \frac{\norm{\bA^{(l)} - \bP^{(l)}}^{2}}{\sigma_{K}(\bP^{(l)})} \right) \\
        &\lesssim \frac{\norm{\be_{i}^{\top} \bV^{(l)}}}{(\sigma_{K}(\bP^{(l)}))^{\frac{1}{2}}} \left( \sqrt{K} + \sqrt{\log(n)} + \frac{\theta_{\max}^{(l)}\norm{\btheta^{(l)}}_{1}}{\sigma_{K}(\bP^{(l)})} \right) \\
        &\lesssim \frac{\sqrt{K}\frac{\theta_{i}^{(l)}}{\norm{\btheta^{(l)}}}}{\frac{\norm{\btheta^{(l)}} (\sigma_{B}^{(l)})^{\frac{1}{2}}}{\sqrt{K}}} \left( \sqrt{K \log(n)} + \frac{\theta_{\max}^{(l)} \norm{\btheta^{(l)}}_{1}}{\frac{\norm{\btheta^{(l)}}^{2} \sigma_{B}^{(l)}}{K}} \right) \\
        &\asymp \frac{ K \theta_{i}^{(l)}}{\norm{\btheta^{(l)}}^{2} (\sigma_{B}^{(l)})^{\frac{1}{2}}} \left( \sqrt{K \log(n)} + \frac{K \theta_{\max}^{(l)} \norm{\btheta^{(l)}}_{1}}{\norm{\btheta^{(l)}}^{2} \sigma_{B}^{(l)}} \right).
    \end{aligned}
\end{equation*}

\subsubsection*{The term $\bR_{2}^{(l)}$}
Restate the term $\bR_{2}^{(l)} = \bV^{(l)} \left( (\bV^{(l)})^{\top} \widehat{\bV}^{(l)} \abs{\widehat{\bLambda}^{(l)}}^{\frac{1}{2}} - \abs{\bLambda^{(l)}}^{\frac{1}{2}} (\bV^{(l)})^{\top} \widehat{\bV}^{(l)} \right)$ for convenience. 
\begin{equation*}
    \begin{aligned}
        \norm{\be_{i}^{\top} \bR_{2}^{(l)}} &= \norm{\be_{i}^{\top} \bV^{(l)} \left( (\bV^{(l)})^{\top} \widehat{\bV}^{(l)} \abs{\widehat{\bLambda}^{(l)}}^{\frac{1}{2}} - \abs{\bLambda^{(l)}}^{\frac{1}{2}} (\bV^{(l)})^{\top} \widehat{\bV}^{(l)} \right)} \\
        &\leq \norm{\be_{i}^{\top} \bV^{(l)}} \norm{(\bV^{(l)})^{\top} \widehat{\bV}^{(l)} \abs{\widehat{\bLambda}^{(l)}}^{\frac{1}{2}} - \abs{\bLambda^{(l)}}^{\frac{1}{2}} (\bV^{(l)})^{\top} \widehat{\bV}^{(l)}} \\
        &\lesssim \sqrt{K} \frac{\theta_{i}^{(l)}}{\norm{\btheta^{(l)}}} \frac{K^{2}}{(\sigma_{K}(\bP^{(l)}))^{\frac{1}{2}}} \left( \sqrt{K \log(n)} + \frac{\theta_{\max}^{(l)} \norm{\btheta^{(l)}}_{1}}{\sigma_{K}(\bP^{(l)})} \right) \\
        &\lesssim \sqrt{K} \frac{\theta_{i}^{(l)}}{\norm{\btheta^{(l)}}} \frac{K^{\frac{5}{2}}}{\norm{\btheta^{(l)}} (\sigma_{B}^{(l)})^{\frac{1}{2}}} \left( \sqrt{K \log(n)} + \frac{K \theta_{\max}^{(l)} \norm{\btheta^{(l)}}_{1}}{\norm{\btheta^{(l)}}^{2} \sigma_{B}^{(l)}} \right) \\
        &\asymp \frac{\theta_{i}^{(l)} K^{3}}{\norm{\btheta^{(l)}}^{2} (\sigma_{B}^{(l)})^{\frac{1}{2}}} \left( \sqrt{K \log(n)} + \frac{K \theta_{\max}^{(l)} \norm{\btheta^{(l)}}_{1}}{\norm{\btheta^{(l)}}^{2} \sigma_{B}^{(l)}} \right).
    \end{aligned}
\end{equation*}

\subsubsection*{The term $\bR_{3}^{(l)}$}
Restate the term $\bR_{3}^{(l)} = \bV^{(l)} \abs{\bLambda^{(l)}}^{\frac{1}{2}} \left( (\bV^{(l)})^{\top} \widehat{\bV}^{(l)} - (\bW_{\ast}^{(l)})^{\top} \right)$ for convenience. 
\begin{equation*}
    \begin{aligned}
        \norm{\be_{i}^{\top} \bR_{3}^{(l)}} &= \norm{\be_{i}^{\top} \bV^{(l)} \abs{\bLambda^{(l)}}^{\frac{1}{2}} \left( (\bV^{(l)})^{\top} \widehat{\bV}^{(l)} - (\bW_{\ast}^{(l)})^{\top} \right)} \\
        &\lesssim \norm{\be_{i}^{\top} \bX^{(l)}} \norm{(\bV^{(l)})^{\top} \widehat{\bV}^{(l)} - (\bW_{\ast}^{(l)})^{\top}} \\
        &\lesssim \sqrt{K} \theta_{i}^{(l)} \frac{\theta_{\max}^{(l)} \norm{\btheta^{(l)}}_{1}}{(\sigma_{K}(\bP^{(l)}))^{2}} \\
        &\lesssim \frac{K^{\frac{5}{2}} \theta_{i}^{(l)} \theta_{\max}^{(l)} \norm{\btheta^{(l)}}_{1}}{\norm{\btheta^{(l)}}^{4} (\sigma_{B}^{(l)})^{2}}.
    \end{aligned}
\end{equation*}

\subsubsection*{The term $\bR_{4}^{(l)}$}
Restate the term $\bR_{4}^{(l)} = (\bA^{(l)} - \bP^{(l)}) (\widehat{\bV}^{(l)} (\widehat{\bV}^{(l)})^{\top} \bV^{(l)} - \bV^{(l)}) \abs{\bLambda^{(l)}}^{-\frac{1}{2}} \bI_{p, q}^{(l)}$ for convenience. 
\begin{equation*}
    \begin{aligned}
        \norm{\be_{i}^{\top} \bR_{4}^{(l)}} &= \norm{\be_{i}^{\top} (\bA^{(l)} - \bP^{(l)}) (\widehat{\bV}^{(l)} (\widehat{\bV}^{(l)})^{\top} \bV^{(l)} - \bV^{(l)}) \abs{\bLambda^{(l)}}^{-\frac{1}{2}} \bI_{p, q}^{(l)}} \\
        &\lesssim \norm{\be_{i}^{\top} (\bA^{(l)} - \bP^{(l)}) (\widehat{\bV}^{(l)} (\widehat{\bV}^{(l)})^{\top} \bV^{(l)} - \bV^{(l)})} \frac{1}{(\sigma_{K}(\bP^{(l)}))^{\frac{1}{2}}} \\
        &\lesssim \sqrt{\theta_{i}^{(l)} \norm{\btheta^{(l)}}_{1} \log(n)} \frac{\sqrt{\theta_{\max}^{(l)} \norm{\btheta^{(l)}}_{1} \log(n)}}{\sigma_{K}(\bP^{(l)})} \norm{\bV^{(l)}}_{2, \infty} \frac{1}{(\sigma_{K}(\bP^{(l)}))^{\frac{1}{2}}} \\
        &\lesssim \sqrt{\theta_{i}^{(l)} \norm{\btheta^{(l)}}_{1} \log(n)} \frac{K\sqrt{\theta_{\max}^{(l)} \norm{\btheta^{(l)}}_{1} \log(n)}}{\norm{\btheta^{(l)}}^{2} \sigma_{B}^{(l)}} \sqrt{K} \frac{\theta_{\max}^{(l)}}{\norm{\btheta^{(l)}}} \frac{\sqrt{K}}{\norm{\btheta^{(l)}} (\sigma_{B}^{(l)})^{\frac{1}{2}}} \\
        &\asymp \frac{K^{2} (\theta_{i}^{(l)})^{\frac{1}{2}} (\theta_{\max}^{(l)})^{\frac{3}{2}} \norm{\btheta^{(l)}}_{1} \log(n)}{\norm{\btheta^{(l)}}^{4} (\sigma_{B}^{(l)})^{\frac{3}{2}}}.
    \end{aligned}
\end{equation*}

\subsubsection*{The term $\bR_{5}^{(l)}$}
Restate the term $\bR_{5}^{(l)} = -(\bA^{(l)} - \bP^{(l)}) \widehat{\bV}^{(l)} \left( (\widehat{\bV}^{(l)})^{\top} \bV^{(l)} \abs{\bLambda^{(l)}}^{-\frac{1}{2}} \bI_{p, q}^{(l)} - \abs{\widehat{\bLambda}^{(l)}}^{-\frac{1}{2}} \bI_{p, q}^{(l)} (\widehat{\bV}^{(l)})^{\top} \bV^{(l)} \right)$ for convenience. 
\begin{equation*}
    \begin{aligned}
        \norm{\be_{i}^{\top} \bR_{5}^{(l)}} &= \norm{\be_{i}^{\top} (\bA^{(l)} - \bP^{(l)}) \widehat{\bV}^{(l)} \left( (\widehat{\bV}^{(l)})^{\top} \bV^{(l)} \abs{\bLambda^{(l)}}^{-\frac{1}{2}} \bI_{p, q}^{(l)} - \abs{\widehat{\bLambda}^{(l)}}^{-\frac{1}{2}} \bI_{p, q}^{(l)} (\widehat{\bV}^{(l)})^{\top} \bV^{(l)} \right)} \\
        &\leq \norm{\be_{i}^{\top} (\bA^{(l)} - \bP^{(l)}) \widehat{\bV}^{(l)}} \norm{(\widehat{\bV}^{(l)})^{\top} \bV^{(l)} \abs{\bLambda^{(l)}}^{-\frac{1}{2}} \bI_{p, q}^{(l)} - \abs{\widehat{\bLambda}^{(l)}}^{-\frac{1}{2}} \bI_{p, q}^{(l)} (\widehat{\bV}^{(l)})^{\top} \bV^{(l)}} \\
        &\lesssim \sqrt{\theta_{i}^{(l)} \norm{\btheta^{(l)}}_{1} \log(n)} \norm{\widehat{\bV}^{(l)}}_{2, \infty} \frac{K^{2}}{(\sigma_{K}(\bP^{(l)}))^{\frac{3}{2}}} \left( \sqrt{K \log(n)} + \frac{\theta_{\max}^{(l)} \norm{\btheta^{(l)}}_{1}}{\sigma_{K}(\bP^{(l)})} \right) \\
        &\lesssim \sqrt{\theta_{i}^{(l)} \norm{\btheta^{(l)}}_{1} \log(n)} \norm{\bV^{(l)}}_{2, \infty} \frac{K^{\frac{7}{2}}}{\norm{\btheta^{(l)}}^{3} (\sigma_{B}^{(l)})^{\frac{3}{2}}} \left( \sqrt{K \log(n)} + \frac{K \theta_{\max}^{(l)} \norm{\btheta^{(l)}}_{1}}{\norm{\btheta^{(l)}}^{2} \sigma_{B}^{(l)}} \right) \\
        &\lesssim \sqrt{\theta_{i}^{(l)} \norm{\btheta^{(l)}}_{1} \log(n)} \sqrt{K} \frac{\theta_{\max}^{(l)}}{\norm{\btheta^{(l)}}} \frac{K^{\frac{7}{2}}}{\norm{\btheta^{(l)}}^{3} (\sigma_{B}^{(l)})^{\frac{3}{2}}} \left( \sqrt{K \log(n)} + \frac{K \theta_{\max}^{(l)} \norm{\btheta^{(l)}}_{1}}{\norm{\btheta^{(l)}}^{2} \sigma_{B}^{(l)}} \right) \\
        &\asymp \frac{K^{4} \theta_{\max}^{(l)} \sqrt{\theta_{i}^{(l)} \norm{\btheta^{(l)}}_{1} \log(n)}}{\norm{\btheta^{(l)}}^{4} (\sigma_{B}^{(l)})^{\frac{3}{2}}} \left( \sqrt{K \log(n)} + \frac{K \theta_{\max}^{(l)} \norm{\btheta^{(l)}}_{1}}{\norm{\btheta^{(l)}}^{2} \sigma_{B}^{(l)}} \right).
    \end{aligned}
\end{equation*}

\subsubsection*{The term $\bR_{6}^{(l)}$}
Restate the term $\bR_{6}^{(l)} = (\bA^{(l)} - \bP^{(l)}) \widehat{\bV}^{(l)} \abs{\widehat{\bLambda}^{(l)}}^{-\frac{1}{2}} \bI_{p, q}^{(l)} \left( \bW_{\ast}^{(l)} - (\widehat{\bV}^{(l)})^{\top} \bV^{(l)} \right)$ for convenience. 
\begin{equation*}
    \begin{aligned}
        \norm{\be_{i}^{\top} \bR_{6}^{(l)}} &= \norm{\be_{i}^{\top} (\bA^{(l)} - \bP^{(l)}) \widehat{\bV}^{(l)} \abs{\widehat{\bLambda}^{(l)}}^{-\frac{1}{2}} \bI_{p, q}^{(l)} \left( \bW_{\ast}^{(l)} - (\widehat{\bV}^{(l)})^{\top} \bV^{(l)} \right)} \\
        &\leq \norm{\be_{i}^{\top} (\bA^{(l)} - \bP^{(l)}) \widehat{\bV}^{(l)}} \norm{\abs{\widehat{\bLambda}^{(l)}}^{-\frac{1}{2}}} \norm{\bW_{\ast}^{(l)} - (\widehat{\bV}^{(l)})^{\top} \bV^{(l)}} \\
        &\lesssim \sqrt{\theta_{i}^{(l)} \norm{\btheta^{(l)}}_{1} \log(n)} \norm{\widehat{\bV}^{(l)}}_{2, \infty} \frac{1}{(\sigma_{K}(\bP^{(l)}))^{\frac{1}{2}}} \frac{\theta_{\max}^{(l)} \norm{\btheta^{(l)}}_{1}}{(\sigma_{K}(\bP^{(l)}))^{2}} \\
        &\lesssim \sqrt{\theta_{i}^{(l)} \norm{\btheta^{(l)}}_{1} \log(n)} \norm{\bV^{(l)}}_{2, \infty} \frac{K^{\frac{5}{2}} \theta_{\max}^{(l)} \norm{\btheta^{(l)}}_{1}}{\norm{\btheta^{(l)}}^{5} (\sigma_{B}^{(l)})^{\frac{5}{2}}} \\
        &\lesssim \sqrt{\theta_{i}^{(l)} \norm{\btheta^{(l)}}_{1} \log(n)} \sqrt{K} \frac{\theta_{\max}^{(l)}}{\norm{\btheta^{(l)}}} \frac{K^{\frac{5}{2}} \theta_{\max}^{(l)} \norm{\btheta^{(l)}}_{1}}{\norm{\btheta^{(l)}}^{5} (\sigma_{B}^{(l)})^{\frac{5}{2}}} \\
        &\asymp \frac{\sqrt{\theta_{i}^{(l)} \log(n)} \norm{\btheta^{(l)}}_{1}^{\frac{3}{2}} (\theta_{\max}^{(l)})^{2} K^{3}}{\norm{\btheta^{(l)}}^{6} (\sigma_{B}^{(l)})^{\frac{5}{2}}}.
    \end{aligned}
\end{equation*}

Combining the bounds for \(\mathbf R_1^{(l)},\ldots,\mathbf R_6^{(l)}\), on the event \(\mathcal E_l^{\mathrm{res}}\), we obtain
\begin{equation}\label{eq:bound_of_residual_X}
    \begin{aligned}
        \norm{\be_{i}^{\top} \bR_{\bX}^{(l)}} \lesssim \theta_{i}^{(l)} \frac{K^{5/2} \theta_{\max}^{(l)} \norm{\btheta^{(l)}}_{1}}{\norm{\btheta^{(l)}}^{4} (\sigma_{B}^{(l)})^{2}} + \theta_{i}^{(l)} \left( \frac{\theta_{\max}^{(l)}}{\theta_{\min}^{(l)}} \right)^{\frac{1}{2}} \left( \frac{K^{\frac{9}{2}} \theta_{\max}^{(l)} \norm{\btheta^{(l)}}_{1} \log(n)}{\norm{\btheta^{(l)}}^{4} (\sigma_{B}^{(l)})^{\frac{3}{2}}} \right)
    \end{aligned}
\end{equation}
Denote $$\alpha_{\bR}^{(l)} = \frac{K^{5/2} \theta_{\max}^{(l)} \norm{\btheta^{(l)}}_{1}}{\norm{\btheta^{(l)}}^{4} (\sigma_{B}^{(l)})^{2}} + \left( \frac{\theta_{\max}^{(l)}}{\theta_{\min}^{(l)}} \right)^{\frac{1}{2}} \left( \frac{K^{\frac{9}{2}} \theta_{\max}^{(l)} \norm{\btheta^{(l)}}_{1} \log(n)}{\norm{\btheta^{(l)}}^{4} (\sigma_{B}^{(l)})^{\frac{3}{2}}} \right).$$
Then, $\norm{\be_{i}^{\top} \bR_{\bX}^{(l)}} \lesssim \theta_{i}^{(l)} \alpha_{\bR}^{(l)}$.

\subsection{\texorpdfstring{Closeness of $\widehat{\bX}^{(l)}$ to $\bX^{(l)}$}{Closeness of Xhat to X}}\label{sec:Closeness_Xhat}
For each \(l\in[L]\) and \(i\in[n]\), let \(\mathcal{E}_{X, i}^{(l)}\) denote the event
\[
\left\{
\left\|
\mathbf e_i^\top(\mathbf A^{(l)}-\mathbf P^{(l)})
\mathbf V^{(l)}
|\mathbf \Lambda^{(l)}|^{-1/2}\mathbf I_{p,q}^{(l)}
\right\|_2
\le
C
\frac{
\sqrt{\theta_i^{(l)}\|\boldsymbol{\theta}^{(l)}\|_1\log n}
}{
\sqrt{\sigma_K(\mathbf P^{(l)})}
}
\|\mathbf V^{(l)}\|_{2,\infty}
\right\}.
\]
By a Bernstein inequality for sums of independent random vectors,
\[
\mathbb P(\mathcal{E}_{X, i}^{(l)})
\ge
1-Cn^{-20}.
\]
Define
\[
\mathcal{E}_{X}^{(l)}
:=
\bigcap_{i=1}^n
\mathcal{E}_{X, i}^{(l)}.
\]
Then
\[
\mathbb P(\mathcal{E}_{X}^{(l)})
\ge
1-Cn^{-19}.
\]
Let $\mathcal E^{(l)}
:=
\mathcal{E}_{res}^{(l)}
\cap
\mathcal{E}_{X}^{(l)}.$
Then
\[
\mathbb P(\mathcal{E}^{(l)})
\ge
1-Cn^{-18}.
\]
In the rest of this subsection and in the normalized-embedding subsection, all inequalities are understood to hold on \(\mathcal{E}^{(l)}\).

Note that
\begin{equation}\label{eq:decomposition_of_X}
    \be_{i}^{\top} (\widehat{\bX}^{(l)} \bW_{\ast}^{(l)} - \bX^{(l)}) = \be_{i}^{\top} (\bA^{(l)} - \bP^{(l)}) \bV^{(l)} \abs{\bLambda^{(l)}}^{-\frac{1}{2}} \bI_{p, q}^{(l)} + \be_{i}^{\top} \bR_{\bX}^{(l)},
\end{equation}
and we have $\norm{\be_{i}^{\top} \bR_{\bX}^{(l)}} \lesssim \theta_{i}^{(l)} \alpha_{\bR}^{(l)}$ from Section \ref{sec:bound_of_ASE_residual}. Now, we bound the main term $\be_{i}^{\top} (\bA^{(l)} - \bP^{(l)}) \bV^{(l)} \abs{\bLambda^{(l)}}^{-\frac{1}{2}} \bI_{p, q}^{(l)}$. Matrix Bernstein inequality gives
\begin{align}\label{eq:ASE_matrix_bernstein}
    \norm{\be_{i}^{\top} (\bA^{(l)} - \bP^{(l)}) \bV^{(l)} \abs{\bLambda^{(l)}}^{-\frac{1}{2}} \bI_{p, q}^{(l)}} &= \norm{\sum_{j=1}^{n} (\bA_{ij}^{(l)} - \bP_{ij}^{(l)}) (\bV^{(l)} \abs{\bLambda^{(l)}}^{-\frac{1}{2}} \bI_{p, q}^{(l)})_{j.}} \notag \\
        &\lesssim \frac{\sqrt{\theta_{i}^{(l)} \norm{\btheta^{(l)}}_{1} \log(n)}}{(\sigma_{K}(\bP^{(l)}))^{\frac{1}{2}}} \norm{\bV^{(l)}}_{2, \infty} \notag \\
        &\lesssim \frac{\sqrt{K \theta_{i}^{(l)} \norm{\btheta^{(l)}}_{1} \log(n)}}{\norm{\btheta^{(l)}} (\sigma_{B}^{(l)})^{\frac{1}{2}}} \sqrt{K} \frac{\theta_{\max}^{(l)}}{\norm{\btheta^{(l)}}} \notag \\
        &\asymp \frac{K \theta_{\max}^{(l)} \sqrt{\theta_{i}^{(l)} \norm{\btheta^{(l)}}_{1} \log(n)}}{\norm{\btheta^{(l)}}^{2} (\sigma_{B}^{(l)})^{\frac{1}{2}}} \notag \\
        &\asymp \theta_{i}^{(l)} \left( \frac{\theta_{\max}^{(l)}}{\theta_{i}^{(l)}} \right)^{\frac{1}{2}} \frac{K \sqrt{\theta_{\max}^{(l)} \norm{\btheta^{(l)}}_{1} \log(n)}}{\norm{\btheta^{(l)}}^{2} (\sigma_{B}^{(l)})^{\frac{1}{2}}}.
\end{align}
Combining \eqref{eq:bound_of_residual_X} and \eqref{eq:ASE_matrix_bernstein}, 
\begin{align}
    \norm{\be_{i}^{\top} (\widehat{\bX}^{(l)} \bW_{\ast}^{(l)} - \bX^{(l)})}
    &= \norm{\be_{i}^{\top} (\bA^{(l)} - \bP^{(l)}) \bV^{(l)} \abs{\bLambda^{(l)}}^{-\frac{1}{2}} \bI_{p, q}^{(l)} + \be_{i}^{\top} \bR_{\bX}^{(l)}} \notag\\
    &\leq \norm{\be_{i}^{\top} (\bA^{(l)} - \bP^{(l)}) \bV^{(l)} \abs{\bLambda^{(l)}}^{-\frac{1}{2}} \bI_{p, q}^{(l)}} + \norm{\be_{i}^{\top} \bR_{\bX}^{(l)}} \notag\\
    &\lesssim \theta_{i}^{(l)} \left( \left( \frac{\theta_{\max}^{(l)}}{\theta_{i}^{(l)}} \right)^{\frac{1}{2}} \frac{K \sqrt{\theta_{\max}^{(l)} \norm{\btheta^{(l)}}_{1} \log(n)}}{\norm{\btheta^{(l)}}^{2} (\sigma_{B}^{(l)})^{\frac{1}{2}}} + \alpha_{\bR}^{(l)} \right) \notag\\
    &\asymp \theta_{i}^{(l)} \frac{\sigma_{B}^{(l)}}{K} \left( \left( \frac{\theta_{\max}^{(l)}}{\theta_{i}^{(l)}} \right)^{\frac{1}{2}} \frac{K^{2} \sqrt{\theta_{\max}^{(l)} \norm{\btheta^{(l)}}_{1} \log(n)}}{\norm{\btheta^{(l)}}^{2} (\sigma_{B}^{(l)})^{\frac{3}{2}}} + \frac{K}{\sigma_{B}^{(l)}} \alpha_{\bR}^{(l)} \right)
    \label{eq:C3_bound_of_diffX}
\end{align}
We rewrite \eqref{eq:C3_bound_of_diffX} in terms of $\beta_{n}^{(l)}$. Recall that
\begin{equation*}
    \beta_n^{(l)}
:=
\left( \frac{\theta_{\max}^{(l)}}{\theta_{\min}^{(l)}} \right)
\frac{K^{8}\theta_{\max}^{(l)} \|\btheta^{(l)}\|_{1} \log (n)}
{(\sigma_{B}^{(l)})^{3}\|\btheta^{(l)}\|^{4}}.
\end{equation*}
We bound the first term in \eqref{eq:C3_bound_of_diffX}:
\begin{equation}\label{eq:C3_bound_of_first_term}
    \left( \frac{\theta_{\max}^{(l)}}{\theta_i^{(l)}} \right)^{1/2}
    \frac{K^{2} \sqrt{\theta_{\max}^{(l)} \|\theta^{(l)}\|_{1} \log n}}
    {\|\theta^{(l)}\|^{2} (\sigma_{B}^{(l)})^{3/2}}
    \le
    \frac{1}{K^{2}} \sqrt{\beta_n^{(l)}}.
\end{equation}
Next, we bound the second term in \eqref{eq:C3_bound_of_diffX}:
\begin{equation}\label{eq:C3_bound_of_second_term}
    \frac{K}{\sigma_{B}^{(l)}} \alpha_R^{(l)}
    =
    \beta_n^{(l)}
    \left[
    \left( \frac{\theta_{\min}^{(l)}}{\theta_{\max}^{(l)}} \right)
    \frac{1}{K^{9/2}\log n}
    +
    \left( \frac{\theta_{\min}^{(l)}}{\theta_{\max}^{(l)}} \right)^{1/2}
    \frac{(\sigma_{B}^{(l)})^{1/2}}{K^{5/2}}
    \right]
    \lesssim
    \beta_n^{(l)}.
\end{equation}
Substituting \eqref{eq:C3_bound_of_first_term} and \eqref{eq:C3_bound_of_second_term} into \eqref{eq:C3_bound_of_diffX}, we obtain
\begin{equation*}
    \left\|
    \mathbf e_i^\top (\widehat{\mathbf X}^{(l)} \mathbf W_{\ast}^{(l)} - \mathbf X^{(l)})
    \right\|_2
    \lesssim
    \theta_i^{(l)} \frac{\sigma_{B}^{(l)}}{K}
    \left(
    \sqrt{\beta_n^{(l)}} + \beta_n^{(l)}
    \right).
\end{equation*}
If, in addition,
\[
\max_{1 \le l \le L} \beta_n^{(l)} \le C_{\beta}
\]
for some fixed constant \(C_{\beta}>0\), then
\[
\beta_n^{(l)} \le \sqrt{C_{\beta}}\,\sqrt{\beta_n^{(l)}},
\]
and hence
\[
\sqrt{\beta_n^{(l)}} + \beta_n^{(l)}
\le
(1+\sqrt{C_{\beta}})\sqrt{\beta_n^{(l)}}.
\]
Therefore, on \(\mathcal E_l\), there exists a constant \(C_X>0\), depending only on \(C_\beta\) and fixed model constants, such that for all \(i\in[n]\),
\begin{equation}\label{eq:C3_final_bound_of_diffX}
    \left\|
    \mathbf e_i^\top (\widehat{\mathbf X}^{(l)} \mathbf W_{\ast}^{(l)} - \mathbf X^{(l)})
    \right\|_2
    \le
    C_X\, \theta_i^{(l)} \frac{\sigma_{B}^{(l)}}{K} \sqrt{\beta_n^{(l)}}.
\end{equation}

\subsection{\texorpdfstring{Closeness of $\widehat{\mathbf Y}^{(l)}$ to $\mathbf Y^{(l)}$}{Closeness of Yhat to Y}}\label{sec:Closeness_Yhat}

Recall that
\[
\mathbf Y_{i,\cdot}^{(l)} = \frac{\mathbf X_{i,\cdot}^{(l)}}{\|\mathbf X_{i,\cdot}^{(l)}\|_2},
\qquad
\widehat{\mathbf Y}_{i,\cdot}^{(l)} = \frac{\widehat{\mathbf X}_{i,\cdot}^{(l)}}{\|\widehat{\mathbf X}_{i,\cdot}^{(l)}\|_2}.
\]
Let
\[
f(\bx) := \frac{\bx}{\|\bx\|_2},
\quad \bx \in \mathbb R^K \setminus \{\mathbf 0\}.
\]
Since \(\mathbf W_{\ast}^{(l)}\) is orthogonal, we have
\[
f(\bx \mathbf W_{\ast}^{(l)}) = f(\bx)\mathbf W_{\ast}^{(l)}
\]
for every \(\bx \in \mathbb R^K \setminus \{\mathbf 0\}\). Hence
\[
(\widehat{\mathbf Y}^{(l)} \mathbf W_{\ast}^{(l)})_{i,\cdot}
=
f\bigl((\widehat{\mathbf X}^{(l)} \mathbf W_{\ast}^{(l)})_{i,\cdot}\bigr).
\]

We use the elementary inequality
\begin{equation}\label{eq:C4_stability_of_f}
\left\|
f(\hat \bx) - f(\bx)
\right\|_2
\le
\frac{2\|\hat \bx - \bx\|_2}{\|\bx\|_2},
\quad
\bx,\hat \bx \in \mathbb R^K \setminus \{\mathbf 0\}.
\end{equation}
Applying \eqref{eq:C4_stability_of_f} with
\[
\bx = \mathbf X_{i,\cdot}^{(l)},
\qquad
\hat \bx = (\widehat{\mathbf X}^{(l)} \mathbf W_{\ast}^{(l)})_{i,\cdot},
\]
we obtain
\begin{equation}\label{eq:C4_rowwise_norm_stability}
\left\|
(\widehat{\mathbf Y}^{(l)} \mathbf W_{\ast}^{(l)})_{i,\cdot} - \mathbf Y_{i,\cdot}^{(l)}
\right\|_2
\le
\frac{
2\left\|
(\widehat{\mathbf X}^{(l)} \mathbf W_{\ast}^{(l)})_{i,\cdot} - \mathbf X_{i,\cdot}^{(l)}
\right\|_2
}{
\|\mathbf X_{i,\cdot}^{(l)}\|_2
}.
\end{equation}

By Lemma \ref{lem:bound_of_spectral}, there exists a constant \(c_X>0\) such that
\begin{equation}\label{eq:C4_lower_bound_of_denom}
\|\mathbf X_{i,\cdot}^{(l)}\|_2
\ge
c_X \,\theta_i^{(l)} \frac{\sigma_{B}^{(l)}}{K}.
\end{equation}
Moreover, by \eqref{eq:C3_final_bound_of_diffX}, there exists a constant \(C_X>0\) such that
\[
\left\|
(\widehat{\mathbf X}^{(l)} \mathbf W_{\ast}^{(l)})_{i,\cdot} - \mathbf X_{i,\cdot}^{(l)} 
\right\|_2
\le
C_X \,\theta_i^{(l)} \frac{\sigma_{B}^{(l)}}{K}\sqrt{\beta_n^{(l)}}.
\]
Substituting this and \eqref{eq:C4_lower_bound_of_denom} into \eqref{eq:C4_rowwise_norm_stability}, we obtain
\[
\left\|
(\widehat{\mathbf Y}^{(l)} \mathbf W_{\ast}^{(l)})_{i,\cdot} - \mathbf Y_{i,\cdot}^{(l)} 
\right\|_2
\le
\frac{2C_X}{c_X}\sqrt{\beta_n^{(l)}}.
\]
Define
\[
C_Y := \frac{2C_X}{c_X}.
\]
Then
\begin{equation}\label{eq:C4_fine_bound_Y}
\left\|
(\widehat{\mathbf Y}^{(l)} \mathbf W_{\ast}^{(l)})_{i,\cdot} - \mathbf Y_{i,\cdot}^{(l)} 
\right\|_2
\le
C_Y \sqrt{\beta_n^{(l)}}.
\end{equation}

Since \(\mathbf Y_{i,\cdot}^{(l)}\) and \((\widehat{\mathbf Y}^{(l)}\mathbf W_*^{(l)})_{i,\cdot}\) both lie on the unit sphere,
\[
\left\|
(\widehat{\mathbf Y}^{(l)} \mathbf W_{\ast}^{(l)})_{i,\cdot} - \mathbf Y_{i,\cdot}^{(l)} 
\right\|_2
\le 2.
\]
Combining this with \eqref{eq:C4_fine_bound_Y}, on the same event \(\mathcal E^{(l)}\), we conclude that
\begin{equation}\label{eq:C4_final_bound_Y}
\left\|
(\widehat{\mathbf Y}^{(l)} \mathbf W_{\ast}^{(l)})_{i,\cdot} - \mathbf Y_{i,\cdot}^{(l)}
\right\|_2
\le
\min\{ C_Y \sqrt{\beta_n^{(l)}},\, 2\}.
\end{equation}

%% file: Sections/Appendix-identifiability.tex
\section{Population Embedding Representation and Identifiability}

This section collects auxiliary results for the identifiability analysis of the MLDCMM model. We begin with Proposition~\ref{prop:embedding_aligned_representation}, which expresses the population spectral embeddings in terms of the model parameters. Next, we show that the reference-layer $\ell_{2}$-unit-row-norm condition of $\bR^{(\mathrm{ref})}$ in Condition~\ref{cond:identifiability} can always be achieved through a canonical reparametrization, and hence should be viewed as a normalization rather than an additional model restriction. With these results, we prove the identifiability of MLDCMM (Proposition~\ref{prop:identifiability}).

\subsection{Model-Based Representation of Population Embeddings}
Although the population spectral embedding is defined as
\begin{equation*}
    \bX^{(l)}=\bV^{(l)}|\bLambda^{(l)}|^{1/2},
\end{equation*}
via the eigendecomposition of $\bP^{(l)}$, it can also be represented directly in terms of the MLDCMM parameters. Specifically, we show that
\begin{equation*}
    \bX^{(l)}=\bTheta^{(l)}\bZ\bR^{(l)}
\end{equation*}
for some invertible matrix $\bR^{(l)} \in \mathbb{R}^{K \times K}$. This model-based representation links the spectral embedding to the membership matrix, and plays a key role both in the identifiability argument and in characterizing the spherical simplex structure induced by the embeddings. We prove this representation in Proposition~\ref{prop:embedding_aligned_representation}. 
\representation*
\begin{proof}
By the eigenvector representation in \eqref{eq:form_of_V}, there exists an orthogonal matrix
\(\mathbf O^{(l)}\) satisfying
\[
\mathbf O^{(l)}\mathbf \Lambda^{(l)}\mathbf O^{(l)\top} = \mathbf \Lambda^{(l)}
\]
such that
\[
\mathbf V^{(l)} = \mathbf \Theta^{(l)} \mathbf Z \left( \bZ^{\top} \left( \bTheta^{(l)} \right)^{2} \bZ \right)^{-\frac{1}{2}} \widetilde{\mathbf U}^{(l)} \mathbf O^{(l)}.
\]
Define
\[
\mathbf R^{(l)} := \left( \bZ^{\top} \left( \bTheta^{(l)} \right)^{2} \bZ \right)^{-\frac{1}{2}} \widetilde{\mathbf U}^{(l)} \mathbf O^{(l)} |\mathbf \Lambda^{(l)}|^{1/2}.
\]
Then
\[
\mathbf X^{(l)} = \mathbf V^{(l)}|\mathbf \Lambda^{(l)}|^{1/2} = \mathbf \Theta^{(l)} \mathbf Z \mathbf R^{(l)}.
\]

Now we show $\bB^{(l)} = \bR^{(l)} \bI_{p, q}^{(l)} \bR^{(l)\top}$ using the following relation from \eqref{eq:C_prelim_eigendecomp_middle_matrix}:
\begin{equation*}
    \left( \bZ^{\top} \left( \bTheta^{(l)} \right)^{2} \bZ \right)^{\frac{1}{2}}
\bB^{(l)}
\left( \bZ^{\top} \left( \bTheta^{(l)} \right)^{2} \bZ \right)^{\frac{1}{2}}
=
\widetilde \bU^{(l)} \bLambda^{(l)} \widetilde \bU^{(l){\top}}.
\end{equation*}
Since
\[
\mathbf \Lambda^{(l)} = |\mathbf \Lambda^{(l)}|^{1/2} \mathbf I_{p,q}^{(l)} |\mathbf \Lambda^{(l)}|^{1/2},
\]
we have
\[
\begin{aligned}
\mathbf R^{(l)} \mathbf I_{p,q}^{(l)} \mathbf R^{(l)\top} &= \left( \bZ^{\top} \left( \bTheta^{(l)} \right)^{2} \bZ \right)^{-\frac{1}{2}} \widetilde{\mathbf U}^{(l)} \mathbf O^{(l)} \mathbf \Lambda^{(l)} \mathbf O^{(l)\top} \widetilde{\mathbf U}^{(l)\top} \left( \bZ^{\top} \left( \bTheta^{(l)} \right)^{2} \bZ \right)^{-\frac{1}{2}} \\
&= \left( \bZ^{\top} \left( \bTheta^{(l)} \right)^{2} \bZ \right)^{-\frac{1}{2}} \widetilde{\mathbf U}^{(l)} \mathbf \Lambda^{(l)} \widetilde{\mathbf U}^{(l)\top} \left( \bZ^{\top} \left( \bTheta^{(l)} \right)^{2} \bZ \right)^{-\frac{1}{2}} \\
&= \mathbf B^{(l)}. 
\end{aligned}
\]
\end{proof}

\subsection{\texorpdfstring{Singular-Value Relations Between $\bR$ and $\bB$}{Singular-Value Relations Between R and B}}\label{sec:Prop_of_R}
The model-based representation in Proposition~\ref{prop:embedding_aligned_representation} introduces the direction matrix $\bR^{(l)}$ by
\[
\bR^{(l)} = \left( \bZ^{\top} \left( \bTheta^{(l)} \right)^{2} \bZ \right)^{-\frac{1}{2}}
\widetilde{\mathbf U}^{(l)}
\mathbf O^{(l)}
|\mathbf \Lambda^{(l)}|^{\frac{1}{2}}.
\]
The direction matrices $\bR^{(l)}$ map membership vectors in $\Delta^{K-1}$ to population spectral embeddings under the special case $\bTheta^{(l)}=\bI_n$. After row normalization, these linear images form spherical simplices whose hemisphere poles are determined by $(\bR^{(l)})^{-1}\mathbf 1_K$. Thus, the maximum and minimum singular values of $\bR^{(l)}$ are central to the bi-Lipschitz analysis of the induced embedding map. See Section~\ref{sec:bi-Lipschitz_of_T} for details.

Since our assumptions are stated in terms of the block connectivity matrix $\bB^{(l)}$, we study relationships between the singular values of $\bB^{(l)}$ and $\bR^{(l)}$. Note that $$\sigma_{k}(\bB^{(l)}) \neq (\sigma_{k}(\bR^{(l)}))^{2}$$ unless $\bI_{p, q}^{(l)} = \bI_{K}^{(l)}$. Nevertheless, standard singular-value inequalities yield the following bounds:
\begin{align}
\sigma_{\max}(\bB^{(l)}) &= \sigma_{\max}\!\left(\bR^{(l)} \bI_{p,q}^{(l)} \bR^{(l)\top}\right) \notag\\
&\leq \sigma_{\max}(\bR^{(l)}) \sigma_{\max}(\bI_{p,q}^{(l)}) \sigma_{\max}(\bR^{(l)\top}) \notag\\
&= \left(\sigma_{\max}(\bR^{(l)})\right)^2, \label{eq:smax_B_R} \\[0.5em] \sigma_{\min}(\bB^{(l)})
&= \sigma_{\min}\!\left(\bR^{(l)} \bI_{p,q}^{(l)} \bR^{(l)\top}\right) \notag\\
&\geq \sigma_{\min}(\bR^{(l)}) \sigma_{\min}(\bI_{p,q}^{(l)}) \sigma_{\min}(\bR^{(l)\top}) \notag\\
&= \left(\sigma_{\min}(\bR^{(l)})\right)^2.
\label{eq:smin_B_R}
\end{align}
The bounds in \eqref{eq:smax_B_R} and \eqref{eq:smin_B_R} control the singular values of $\bR^{(l)}$ in terms of those of $\bB^{(l)}$. We now derive bounds in the opposite direction, controlling $\sigma_{\max}(\bR^{(l)})$ from above and $\sigma_{\min}(\bR^{(l)})$ from below in terms of the singular values of $\bB^{(l)}$.

\subsubsection*{Upper bound of $\sigma_{\max}(\bR^{(l)})$}
We provide an upper bound for $\sigma_{\max}(\bR^{(l)})$. Recall that $\norm{\bG^{(l)}} = \norm{\frac{\sqrt{K}}{\norm{\btheta^{(l)}}} \left( \bZ^{\top} (\bTheta^{(l)})^{2} \bZ \right)^{\frac{1}{2}}} \leq C$ and $\norm{(\bG^{(l)})^{-1}} = \norm{\frac{\norm{\btheta^{(l)}}}{\sqrt{K}} \left( \bZ^{\top} (\bTheta^{(l)})^{2} \bZ \right)^{-\frac{1}{2}}} \leq C$ by assumption \ref{asm:gram_matrix}.
\begin{align}\label{eq:D_upper_bound_of_smaxR}
    \sigma_{\max}(\bR^{(l)}) &= \sigma_{\max} \left( \left( \bZ^{\top} (\bTheta^{(l)})^{2} \bZ \right)^{-\frac{1}{2}} \widetilde{\bU}^{(l)} \bO^{(l)} \abs{\bLambda^{(l)}}^{\frac{1}{2}} \right) \nonumber \\
    &= \sigma_{\max} \left( \frac{\norm{\btheta^{(l)}}}{\sqrt{K}} ( \bZ^{\top} (\bTheta^{(l)})^{2} \bZ )^{-\frac{1}{2}} \frac{\sqrt{K}}{\norm{\btheta^{(l)}}} \widetilde{\bU}^{(l)} \bO^{(l)} \abs{\bLambda^{(l)}}^{\frac{1}{2}} \right) \nonumber \\
    &= \frac{\sqrt{K}}{\norm{\btheta^{(l)}}} \sigma_{\max} \left( (\bG^{(l)})^{-1} \widetilde{\bU}^{(l)} \bO^{(l)} \abs{\bLambda^{(l)}}^{\frac{1}{2}} \right) \nonumber \\
    &\leq \frac{\sqrt{K}}{\norm{\btheta^{(l)}}} \sigma_{\max}((\bG^{(l)})^{-1}) \sigma_{\max}(\widetilde{\bU}^{(l)}) \sigma_{\max}(\bO^{(l)}) \sigma_{\max}(\abs{\bLambda^{(l)}}^{\frac{1}{2}}) \nonumber \\
    &\leq C \frac{\sqrt{K}}{\norm{\btheta^{(l)}}} \sqrt{\sigma_{\max}(\bP^{(l)})} \nonumber \\
    &\lesssim C \frac{\sqrt{K}}{\norm{\btheta^{(l)}}} \frac{\norm{\btheta^{(l)}}}{\sqrt{K}} \sqrt{\sigma_{\max}(\bB^{(l)})} \nonumber \\
    &= C \sqrt{\sigma_{\max}(\bB^{(l)})}.
\end{align}

\subsubsection*{Lower bound of $\sigma_{\min}(\bR^{(l)})$}
Similarly, we provide a lower bound of $\sigma_{\min}(\bR^{(l)})$.
\begin{align}\label{eq:D_lower_bound_of_sminR}
    \sigma_{\min}(\bR^{(l)}) &= \sigma_{\min} \left(\left( \bZ^{\top} (\bTheta^{(l)})^{2} \bZ \right)^{-\frac{1}{2}} \widetilde{\bU}^{(l)} \bO^{(l)} \abs{\bLambda^{(l)}}^{\frac{1}{2}} \right) \nonumber \\
    &= \frac{1}{\sigma_{\max} \left( \abs{\bLambda^{(l)}}^{-\frac{1}{2}} (\bO^{(l)})^{\top} (\widetilde{\bU}^{(l)})^{\top} ( \bZ^{\top} (\bTheta^{(l)})^{2} \bZ )^{\frac{1}{2}}\right)} \nonumber \\
    &= \frac{1}{\frac{\norm{\btheta^{(l)}}}{\sqrt{K}} \sigma_{\max}\left( \abs{\bLambda^{(l)}}^{-\frac{1}{2}} (\bO^{(l)})^{\top} (\widetilde{\bU}^{(l)})^{\top} \frac{\sqrt{K}}{\norm{\btheta^{(l)}}} ( \bZ^{\top} (\bTheta^{(l)})^{2} \bZ )^{\frac{1}{2}} \right)} \nonumber \\
    &\geq \frac{\sqrt{K}}{\norm{\btheta^{(l)}}} \frac{1}{\sigma_{\max} \left( \abs{\bLambda^{(l)}}^{-\frac{1}{2}} \right)} \frac{1}{\norm{\bG^{(l)}}} \nonumber \\
    &\geq \frac{\sqrt{K}}{\norm{\btheta^{(l)}}} \sigma_{\min}\left( \abs{\bLambda^{(l)}}^{\frac{1}{2}} \right) \frac{1}{C} \nonumber \\
    &\gtrsim \frac{\sqrt{K}}{\norm{\btheta^{(l)}}} \frac{\norm{\btheta^{(l)}}}{\sqrt{K}} \sqrt{\sigma_{\min}(\bB^{(l)})} \nonumber \\
    &\asymp \sqrt{\sigma_{\min}(\bB^{(l)})}.
\end{align}

\subsection{Reparametrization of MLDCMM model}
We restate the identifiability conditions for convenience. 
\identifiabilitycond*

Condition~\ref{cond:identifiability} includes a reference-layer normalization: there exists a layer, denoted by $\mathrm{ref}$, such that
\begin{equation*}
    \|\mathbf{R}_{k,\cdot}^{(\mathrm{ref})}\|_2=1, \quad 1 \leq k \leq K.
\end{equation*}
We show that this condition should be viewed as a normalization rather than a substantive model restriction. Specifically, any parametrization satisfying the remaining structural identifiability conditions can be reparametrized into a canonical form
\begin{equation*}
    \left(\{\bTheta_{\mathrm{can}}^{(l)}\}_{l=1}^{L}, \bZ_{\mathrm{can}},\{\bB_{\mathrm{can}}^{(l)}\}_{l=1}^{L}\right)
\end{equation*}
that yields the same population matrices and satisfies the required normalization conditions.

Without loss of generality, assume that the first layer is the reference layer. Let $\bD_{1}^{(1)} \in \bbR^{K \times K}$ be a diagonal matrix with $(\bD_{1}^{(1)})_{kk} = \norm{(\bR^{(1)})_{k, \cdot}}_{2}$ for $1 \leq k \leq K$. Then, we obtain row-normalized matrix $\bR_{\mathrm{can}}^{(1)}$:
\begin{equation*}
    \begin{aligned}
        \bP^{(1)} &= \bTheta^{(1)} \bZ \bB^{(1)} \bZ^{\top} \bTheta^{(1)} \\
        &= \bTheta^{(1)} \bZ \bD_{1}^{(1)} (\bD_{1}^{(1)})^{-1} \bR^{(1)} \bI_{p, q}^{(1)} (\bR^{(1)})^{\top} (\bD_{1}^{(1)})^{-1} \bD_{1}^{(1)} \bZ^{\top} \bTheta^{(1)} \\
        &\equiv \bTheta^{(1)} \bZ_{\dagger} \bR_{\mathrm{can}}^{(1)} \bI_{p, q}^{(1)} (\bR_{\mathrm{can}}^{(1)})^{\top} \bZ_{\dagger}^{\top} \bTheta^{(1)},
    \end{aligned}
\end{equation*}
where 
\[
\bZ_{\dagger} := \bZ \bD_{1}^{(1)}, 
\quad 
\bR_{\mathrm{can}}^{(1)} := (\bD_{1}^{(1)})^{-1}\bR^{(1)}.
\]
By construction, $\bR_{\mathrm{can}}^{(1)}$ has unit $\ell_2$-norm rows, while $\bZ_{\dagger}$ may not have unit $\ell_1$-norm rows. Next, let $\bD_{2}^{(1)} \in \bbR^{n \times n}$ be the diagonal matrix with $(\bD_{2}^{(1)})_{ii} = \norm{(\bZ_{\dagger})_{i, \cdot}}_{1}$ for $1 \leq i \leq n$.     Then we obtain a matrix with unit $\ell_1$-norm rows $\bZ_{\mathrm{can}}$:
\begin{equation*}
    \begin{aligned}
        \bP^{(1)} &= \bTheta^{(1)} \bZ_{\dagger} \bR_{\mathrm{can}}^{(1)} \bI_{p, q}^{(1)} (\bR_{\mathrm{can}}^{(1)})^{\top} \bZ_{\dagger}^{\top} \bTheta^{(1)} \\
        &= \bTheta^{(1)} \bD_{2}^{(1)} (\bD_{2}^{(1)})^{-1} \bZ_{\dagger} \bR_{\mathrm{can}}^{(1)} \bI_{p, q}^{(1)} (\bR_{\mathrm{can}}^{(1)})^{\top} \bZ_{\dagger}^{\top} (\bD_{2}^{(1)})^{-1} \bD_{2}^{(1)} \bTheta^{(1)} \\
        &\equiv \bTheta_{\mathrm{can}}^{(1)} \bZ_{\mathrm{can}} \bR_{\mathrm{can}}^{(1)} \bI_{p, q}^{(1)} (\bR_{\mathrm{can}}^{(1)})^{\top} \bZ_{\mathrm{can}}^{\top} \bTheta_{\mathrm{can}}^{(1)},
    \end{aligned}
\end{equation*}
where 
\[
\bZ_{\mathrm{can}} := (\bD_{2}^{(1)})^{-1}\bZ_{\dagger}, \quad \bTheta_{\mathrm{can}}^{(1)} := \bTheta^{(1)}\bD_{2}^{(1)}.
\]
Equivalently, the reference layer can be written as
\[
\bP^{(1)} = \bTheta_{\mathrm{can}}^{(1)} \bZ_{\mathrm{can}} \bB_{\mathrm{can}}^{(1)} \bZ_{\mathrm{can}}^\top \bTheta_{\mathrm{can}}^{(1)},
\]
where
\[
\bB_{\mathrm{can}}^{(1)} := \bR_{\mathrm{can}}^{(1)} \bI_{p,q}^{(1)} (\bR_{\mathrm{can}}^{(1)})^\top .
\]
This representation satisfies the membership row-normalization and the reference-layer unit-row-norm condition.

Now, we represent $\bP^{(l)} = \bTheta^{(l)} \bZ \bB^{(l)} \bZ^{\top} \bTheta^{(l)}$ in terms of $\bZ_{\mathrm{can}}$ for $l \geq 2$. 
\begin{equation*}
    \begin{aligned}
        \bP^{(l)} &= \bTheta^{(l)} \bZ \bB^{(l)} \bZ^{\top} \bTheta^{(l)} \\
        &= \bTheta^{(l)} \bD_{2}^{(1)} (\bD_{2}^{(1)})^{-1} \bZ \bD_{1}^{(1)} (\bD_{1}^{(1)})^{-1} \bB^{(l)} (\bD_{1}^{(1)})^{-1} \bD_{1}^{(1)} \bZ^{\top} (\bD_{2}^{(1)})^{-1} \bD_{2}^{(1)} \bTheta^{(l)} \\
        &\equiv \bTheta_{\dagger}^{(l)} \bZ_{\mathrm{can}} \bB_{\dagger}^{(l)} \bZ_{\mathrm{can}}^{\top} \bTheta_{\dagger}^{(l)},
    \end{aligned}
\end{equation*}
where $\bTheta_{\dagger}^{(l)} := \bTheta^{(l)} \bD_{2}^{(1)}$ and $\bB_{\dagger}^{(l)} := (\bD_{1}^{(1)})^{-1} \bB^{(l)} (\bD_{1}^{(1)})^{-1}$. Note that $\bB_{\dagger}^{(l)}$ need not satisfy
\[
\|\bB_{\dagger}^{(l)}\|_{\max}=1.
\]
To this end, let $b^{(l)} = \norm{\bB_{\dagger}^{(l)}}_{\max}$. Then, 
\begin{equation*}
    \begin{aligned}
        \bP^{(l)} &= \bTheta_{\dagger}^{(l)} \bZ_{\mathrm{can}} \bB_{\dagger}^{(l)} \bZ_{\mathrm{can}}^{\top} \bTheta_{\dagger}^{(l)} \\
        &= \left( \sqrt{b^{(l)}} \bTheta_{\dagger}^{(l)} \right) \bZ_{\mathrm{can}} \, \frac{\bB_{\dagger}^{(l)}}{b^{(l)}} \, \bZ_{\mathrm{can}}^{\top} \left( \sqrt{b^{(l)}} \bTheta_{\dagger}^{(l)} \right) \\
        &\equiv \bTheta_{\mathrm{can}}^{(l)} \bZ_{\mathrm{can}} \bB_{\mathrm{can}}^{(l)} \bZ_{\mathrm{can}}^{\top} \bTheta_{\mathrm{can}}^{(l)},
    \end{aligned}
\end{equation*}
where $\bTheta_{\mathrm{can}}^{(l)} = \sqrt{b^{(l)}} \bTheta_{\dagger}^{(l)}$, and $\bB_{\mathrm{can}}^{(l)} = \frac{\bB_{\dagger}^{(l)}}{b^{(l)}}$. Since $\bD_1^{(1)}$ and $\bD_2^{(1)}$ are positive diagonal matrices, this reparametrization preserves the support pattern of the membership vectors. Since \(\mathbf D_1^{(1)}\) and \(\mathbf D_2^{(1)}\) are positive diagonal matrices, this reparametrization preserves the support pattern of the membership vectors. In particular, pure nodes remain pure, and the community-overlap graph \(\mathcal G_Z\) is unchanged. Hence the community-overlap connectivity condition is preserved. Moreover, full rank of the block connectivity matrices is preserved under multiplication by nonsingular diagonal matrices. Finally, by construction,
\[
\|(\bR_{\mathrm{can}}^{(1)})_{k,\cdot}\|_2=1, \quad 1\le k\le K,
\]
and
\[
\|\bB_{\mathrm{can}}^{(l)}\|_{\max}=1, \quad l\ge2.
\]
Therefore, the canonical parametrization
\[
\left(\{\bTheta_{\mathrm{can}}^{(l)}\}_{l=1}^{L}, \bZ_{\mathrm{can}}, \{\bB_{\mathrm{can}}^{(l)}\}_{l=1}^{L}\right)
\]
satisfies all required identifiability conditions and yields the same population matrices $\{\bP^{(l)}\}_{l=1}^L$.

\subsection{Identifiability of MLDCMM}\label{sec:identifiability_proof}
We now establish identifiability of the MLDCMM model under Condition~\ref{cond:identifiability}. Compared with the single-layer DCMM setting, the main additional difficulty is to identify the layer-specific scaling in the non-reference layers. The reference layer identifies the shared membership matrix $\bZ$ up to a common permutation of community labels. For each non-reference layer $l\ge2$, the mixed-node richness condition ensures that the mixed nodes link the community coordinates through overlapping membership supports; these overlap constraints determine the relative row norms of $\bR^{(l)}$. Once these relative row-norm ratios are identified, the remaining layer-specific parameters can be recovered. 

We restate Proposition~\ref{prop:identifiability} for convenience. 
\identifiabilityprop*   
\begin{proof}
    Assume the permutation of the communities is fixed. Let $\mathcal{J}_{1}$ and $\mathcal{J}_{2}$ be pure node index sets such that $\bZ_{\mathcal{J}_{1}} = \widetilde{\bZ}_{\mathcal{J}_{2}} = \bI_{K}$. By the proof of Lemma B.1 in \citet{mao2018overlapping}, it can be shown that pure nodes match up. By Proposition~\ref{prop:embedding_aligned_representation}, we can define two population spectral embeddings $\bX^{(l)}$ and $\widetilde{\bX}^{(l)}$ such that
    \begin{equation*}
        \bX^{(l)} = \bTheta^{(l)} \bZ \bR^{(l)} \quad \text{and} \quad \widetilde{\bX}^{(l)} = \widetilde{\bTheta}^{(l)} \widetilde{\bZ} \widetilde{\bR}^{(l)}.
    \end{equation*}
    Since the two parametrizations generate the same \(\bP^{(l)}\), we have
    \begin{equation*}
        \begin{aligned}
            \bX^{(l)} (\bX^{(l)})^{\top} &= \widetilde{\bX}^{(l)} (\widetilde{\bX}^{(l)})^{\top} \\
            \Rightarrow \: \bTheta^{(l)} \bZ \bR^{(l)} (\bR^{(l)})^{\top} \bZ^{\top} \bTheta^{(l)} &= \widetilde{\bTheta}^{(l)} \widetilde{\bZ} \widetilde{\bR}^{(l)} (\widetilde{\bR}^{(l)})^{\top} \widetilde{\bZ}^{\top} \widetilde{\bTheta}^{(l)}
        \end{aligned}
    \end{equation*}
    Selecting the rows and columns corresponding to the indices in $\mathcal{J}_{1}$ on both sides,
    \begin{equation*}
        \bTheta_{\mathcal{J}_{1}}^{(1)} \bR^{(1)} (\bR^{(1)})^{\top} \bTheta_{\mathcal{J}_{1}}^{(1)} = \widetilde{\bTheta}_{\mathcal{J}_{1}}^{(1)} \bPi^{(1)} \widetilde{\bR}^{(1)} (\widetilde{\bR}^{(1)})^{\top} (\bPi^{(1)})^{\top} \widetilde{\bTheta}_{\mathcal{J}_{1}}^{(1)},
    \end{equation*}
    where $\bPi^{(1)} := \widetilde{\bZ}_{\mathcal{J}_{1}}$ is a permutation matrix. Note that $\bR^{(1)} (\bR^{(1)})^{\top}$ and $\widetilde{\bR}^{(1)} (\widetilde{\bR}^{(1)})^{\top}$ have unit diagonals since $\bR^{(1)}$ and $\widetilde{\bR}^{(1)}$ have unit $\ell_{2}$-norm rows. By comparing diagonal elements of both sides, it can be shown that 
    \begin{equation}\label{eq:D3_Theta_P_eq}
        \bTheta_{\mathcal{J}_{1}}^{(1)} = \widetilde{\bTheta}_{\mathcal{J}_{1}}^{(1)}.
    \end{equation}
    We use the following key result from Lemma 2.1. in \citet{mao2018overlapping}, 
    \begin{equation}\label{eq:Lemma2.1._Mao}
        \bV^{(l)} = \bTheta^{(l)} \bZ (\bTheta_{P}^{(l)})^{-1} \bV_{P}^{(l)}
    \end{equation}
    Substituting $P$ with $\mathcal{J}_{1}$ and multiplying $(\bV_{\mathcal{J}_{1}}^{(1)})^{-1}$ to \eqref{eq:Lemma2.1._Mao} give
    \begin{equation}\label{eq:D3_ThetaZTheta}
        \bTheta^{(1)} \bZ (\bTheta_{\mathcal{J}_1}^{(1)})^{-1} = \widetilde{\bTheta}^{(1)} \widetilde{\bZ} (\widetilde{\bTheta}_{\mathcal{J}_1}^{(1)})^{-1}.
    \end{equation}
    By combining \eqref{eq:D3_Theta_P_eq} and \eqref{eq:D3_ThetaZTheta}, we have
    \begin{equation}\label{eq:D3_ThetaZ}
        \bTheta^{(1)} \bZ = \widetilde{\bTheta}^{(1)} \widetilde{\bZ}.
    \end{equation}
    By comparing the row-wise $\ell_{1}$-norms on both sides of \eqref{eq:D3_ThetaZ}, we have $\bTheta^{(1)} = \widetilde{\bTheta}^{(1)}$. By pre-multiplying $(\bTheta^{(1)})^{-1}$ to \eqref{eq:D3_ThetaZ}, we have $\bZ = \widetilde{\bZ}$. Note that $\bTheta^{(1)} \bZ$ is full column rank matrix. By multiplying $(\bZ^{\top} \bTheta^{(1)} \bTheta^{(1)} \bZ)^{-1} \bZ^{\top} \bTheta^{(1)}$ on the left and its transpose on the right of $\bP^{(1)}$, we prove $\bB^{(1)} = \widetilde{\bB}^{(1)}$.

    Now we show $\bTheta^{(l)} = \widetilde{\bTheta}^{(l)}$, and $\bB^{(l)} = \widetilde{\bB}^{(l)}$ for all $2 \leq l \leq L$. Let $\mathcal{J} = \{ i_{1}, \cdots, i_{K} \}$ be a set of pure nodes such that $\bZ_{\mathcal{J}} = \bI_{K}$. Using the relationship \eqref{eq:Lemma2.1._Mao}, we have the following:
    \begin{equation}\label{eq:D3_ThetaZThetainv_l}
        \bTheta^{(l)} \bZ (\bTheta_{\mathcal{J}}^{(l)})^{-1} = \widetilde{\bTheta}^{(l)} \bZ (\widetilde{\bTheta}_{\mathcal{J}}^{(l)})^{-1}.
    \end{equation}
    The $i$-th row of both sides are calculated by
    \begin{equation*}
        \theta_{i}^{(l)} \left( \frac{\bZ_{i1}}{\theta_{i_{1}}^{(l)}}, \cdots, \frac{\bZ_{iK}}{\theta_{i_{K}}^{(l)}} \right) = \widetilde{\theta}_{i}^{(l)} \left( \frac{\bZ_{i1}}{\widetilde{\theta}_{i_{1}}^{(l)}}, \cdots, \frac{\bZ_{iK}}{\widetilde{\theta}_{i_{K}}^{(l)}} \right).
    \end{equation*}
    By the community-overlap connectivity condition, the ratios are propagated across all communities. Hence
\[
\frac{\theta_{i_1}^{(l)}}{\widetilde\theta_{i_1}^{(l)}} = \cdots = \frac{\theta_{i_K}^{(l)}}{\widetilde\theta_{i_K}^{(l)}} =:c^{(l)}.
\]
Thus
\[
\bTheta_{\mathcal J}^{(l)}
=
c^{(l)}\widetilde\bTheta_{\mathcal J}^{(l)}.
\]
Multiplying \eqref{eq:D3_ThetaZThetainv_l} on the right by \(\bTheta_{\mathcal J}^{(l)}\), we obtain
\[
\bTheta^{(l)} \bZ
=
c^{(l)}\widetilde\bTheta^{(l)} \bZ.
\]
Since each row of \(Z\) has unit \(\ell_1\)-norm, comparing the row-wise \(\ell_1\)-norms gives
\[
\bTheta^{(l)}
=
c^{(l)}\widetilde\bTheta^{(l)}.
\]

Substituting this relation into the equality of the probability matrices,
\[
\bTheta^{(l)} \bZ \bB^{(l)} \bZ^\top\bTheta^{(l)}
=
\widetilde\bTheta^{(l)} \bZ \widetilde \bB^{(l)} \bZ^\top \widetilde\Theta^{(l)},
\]
yields
\[
(c^{(l)})^{2}
\widetilde\bTheta^{(l)}\bZ \bB^{(l)} \bZ^\top \widetilde\bTheta^{(l)}
=
\widetilde\bTheta^{(l)} \bZ \widetilde \bB^{(l)} \bZ^\top \widetilde\bTheta^{(l)}.
\]
Since \(\widetilde\bTheta^{(l)} \bZ\) has full column rank, it follows that
\[
\widetilde \bB^{(l)}
=
(c^{(l)})^{2}\bB^{(l)}.
\]
Both parametrizations satisfy
\[
\|\bB^{(l)}\|_{\max}
=
\|\widetilde \bB^{(l)}\|_{\max}
=
1
\]
for every non-reference layer \(l\). Therefore,
\[
1
=
\|\widetilde \bB^{(l)}\|_{\max}
=
(c^{(l)})^{2}\|\bB^{(l)}\|_{\max}
=
(c^{(l)})^{2}.
\]
Since \(c^{(l)}>0\), we have \(c^{(l)}=1\). Hence
\[
\bTheta^{(l)}
=
\widetilde\bTheta^{(l)}
\quad\text{and}\quad
\bB^{(l)}
=
\widetilde \bB^{(l)}.
\]
\end{proof}

\subsection{Illustrative Examples of the Identifiability Conditions}
We provide two illustrative examples showing the roles of community-overlap connectivity and unit $\ell_{2}$-norm condition on $\| \mathbf{R}_{k, \cdot}^{(1)} \|_{2}$ only on the reference layer. Specifically, we show how nonidentifiability can arise when community-overlap connectivity condition fails and how imposing the unit-norm constraint $\| \mathbf{R}_{k, \cdot}^{(l)} \|_{2}$ on every layer can be restrictive.

\subsubsection{Nonidentifiability Without Community-Overlap Connectivity}
\begin{figure}[!ht]
\centering

\begin{subfigure}[t]{0.35\textwidth}
    \centering
    \includegraphics[width=\linewidth]
        {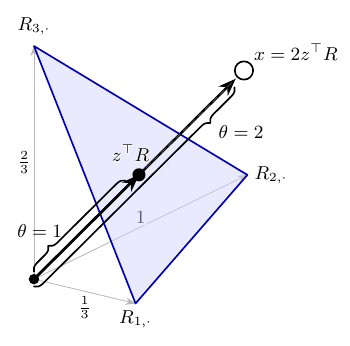}
    \caption{Connected overlap.}
    \label{fig:connected-overlap}
\end{subfigure}%
\hspace{0.03\textwidth}%
\begin{subfigure}[t]{0.35\textwidth}
    \centering
    \includegraphics[width=\linewidth]
        {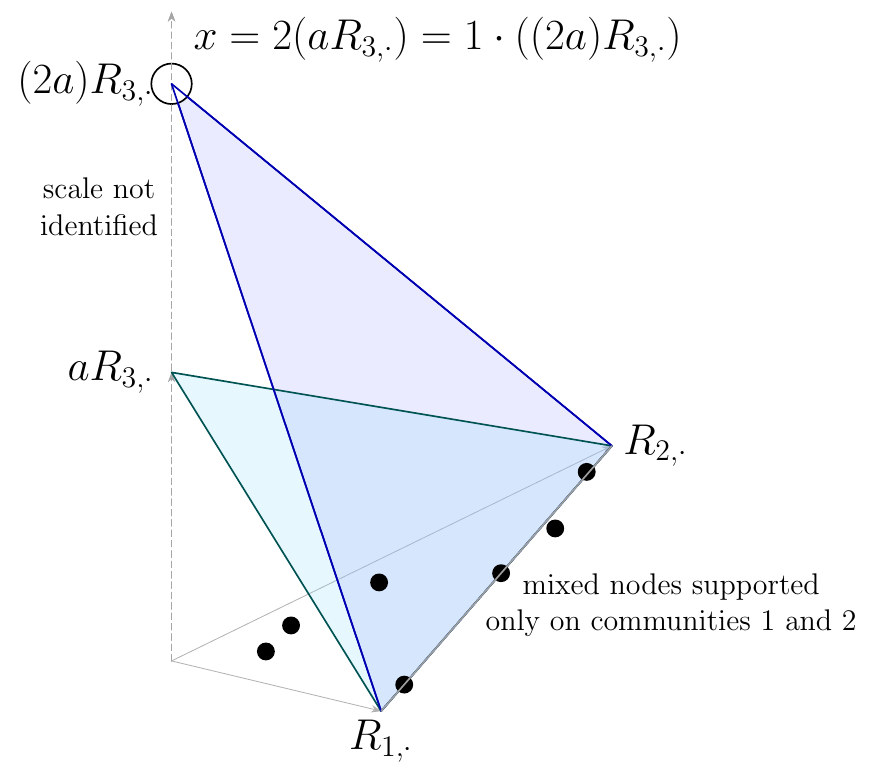}
    \caption{Disconnected overlap.}
    \label{fig:disconnected-overlap}
\end{subfigure}
\caption{
Geometric intuition for the community-overlap connectivity condition.
This figure is drawn in the unnormalized embedding
$\bX = \bTheta \bZ \bR$.
(a) In the connected case, mixed-membership nodes link the community-specific rays, making their relative scales identifiable.
(b) In the disconnected case, mixed nodes are supported only on communities 1 and 2, so the scale of the third ray is not identified. 
}
\label{fig:community-overlap-connectivity}
\end{figure}

We provide an illustrative example of the role of the Community-overlap connectivity condition in Condition~\ref{cond:identifiability} when $K=3$. A set of mixed nodes satisfying the condition determines the relative ratio of $\| \mathbf{R}_{k, \cdot}^{(l)} \|_{2}$ for $k \in [K]$. Specifically, once the membership matrix $\mathbf{Z}$ is identified in the reference layer, we can recover the relative ratio of $\| \mathbf{R}_{k, \cdot}^{(l)} \|_{2}$ by solving $\mathbf{X}^{(l)} = \bm{\Theta}^{(l)} \mathbf{Z} \mathbf{R}^{(l)}$. We can fix the relative ratio by assuming $\| \mathbf{B}^{(l)} \|_{\max} = 1$ and then determine the remaining $\bm{\Theta}^{(l)}$ for every non-reference layer $l$. If the condition is not satisfied, we have a nonidentifiability between $\bm{\Theta}^{(l)}$ and $\mathbf{R}^{(l)}$ even though $\mathbf{Z}$ is identified from the reference layer. Figure~\ref{fig:community-overlap-connectivity} illustrates the phenomenon. In (a), given the membership of the black point node is identified from the reference layer, for example, $\mathbf{z} = (1/3, 1/3, 1/3)^{\top}$, and $x = \theta \mathbf{z}^{\top} \mathbf{R}$, we can identify $\theta$ and $\mathbf{R}$. In (b), however, since community 3 is not connected to the other communities, we have a scale nonidentifiability between $\theta$ and $\mathbf{R}$. Therefore, one can represent the same white point node $x$ by assuming $(\theta, \mathbf{R}_{3, \cdot}) = (1, 2a \mathbf{R}_{3, \cdot}) = (2, a \mathbf{R}_{3, \cdot})$. 

\subsubsection{Restrictions of Uniform Unit Diagonal Assumptions on $\mathbf{B}^{(l)}$}\label{sec:appendix_example_unit_diagonal}
\begin{figure}[!ht]
    \centering

    \includegraphics[
        width=0.6\textwidth,
        keepaspectratio
    ]{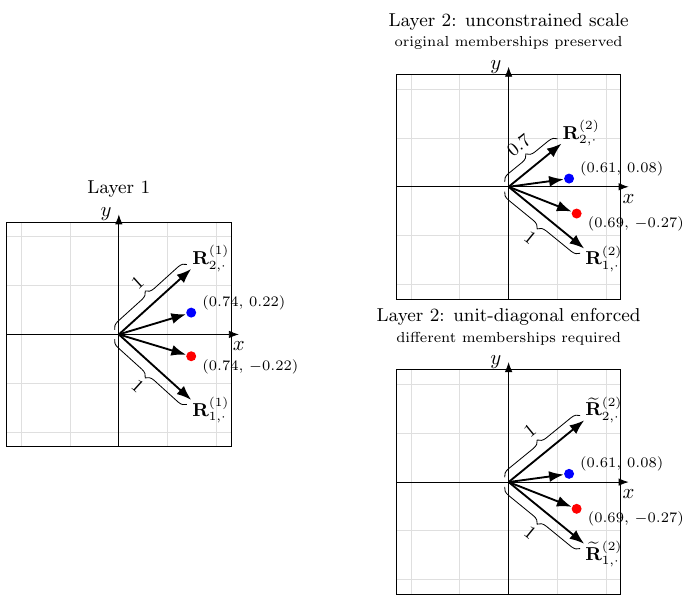}

    \caption{
    Illustration of the effect of imposing a unit-diagonal
    constraint across layers.
    The left panel shows the latent positions of two nodes
    in Layer~1.
    In Layer~2, allowing non-unit diagonal entries in
    $\mathbf{B}^{(2)}$ preserves the original memberships
    (upper right), whereas enforcing unit diagonal requires
    a different reparametrization and hence different
    memberships for the same latent positions (lower right).
    }

    \label{fig:identifiability}
\end{figure}

We illustrate how imposing unit-diagonal constraints on $\mathbf{B}^{(l)}$ across all layers can affect the shared membership structure. Consider a two-layer ML-DCMM with the following parameters. Let $\mathbf{P}^{(l)} = \bm{\Theta}^{(l)} \mathbf{Z} \mathbf{B}^{(l)} \mathbf{Z}^{\top} \bm{\Theta}^{(l)} \in \mathbb{R}^{n \times n}$ be given with $\bm{\Theta}^{(l)} = \mathbf{I}_{n}$ and $\mathbf{B}^{(1)} = 
    \begin{pmatrix}
        1 & 0.1 \\
        0.1 & 1
    \end{pmatrix}, \mathbf{B}^{(2)} = 
    \begin{pmatrix}
        1 & 0.14 \\
        0.14 & 0.49
    \end{pmatrix}$ for $1 \leq l \leq 2$. Assume $\bZ_{1, \cdot} = \left( \frac{2}{3}, \frac{1}{3} \right)^{\top}, \bZ_{2, \cdot} = \left( \frac{1}{3}, \frac{2}{3} \right)^{\top}$. Although the unit-diagonal constraint can be imposed on each layer through reparameterization, the resulting membership vectors differ across layers. See Figure~\ref{fig:identifiability}. Thus, applying the normalization separately to each layer need not preserve the shared membership structure. This motivates imposing the normalization only on the reference layer. 

\subsection{Proof of Proposition~\ref{prop:cosie_graphs}}\label{sec:appendix_COSIE}
We restate Proposition~\ref{prop:cosie_graphs} for convenience. 
\COSIE*

\begin{proof}
    Without loss of generality, we fix layer 1 as the reference layer. Let $\mathbf{U}^{(1)} \bm{\Lambda}^{(1)} \mathbf{U}^{(1)\top}$ denote a spectral decomposition of $\mathbf{R}^{(1)}$. Following the GRDPG framework \citep{Rubin2022}, let $\mathbf{X}^{(1)} := \mathbf{V} \mathbf{U}^{(1)} |\bm{\Lambda}^{(1)}|^{1/2} \in \mathbb{R}^{n \times d}$ denote a population spectral embedding such that $\mathbf{X}^{(1)} \mathbf{I}_{p^{(1)}, q^{(1)}} \mathbf{X}^{(1)\top} = \mathbf{P}^{(1)}$, where $p^{(1)}$ and $q^{(1)}$ denote the number of positive and negative eigenvalues of $\mathbf{R}^{(1)}$. Since $\mathrm{rank}(\mathbf{V})=d$, we can define $\mathcal{J}$ an index set of $d$ independent rows of $\mathbf{V}$. Then $\mathbf{W} := \mathbf{P}_{\cdot, \mathcal{J}}^{(1)} = \mathbf{X}^{(1)} \mathbf{I}_{p^{(1)}, q^{(1)}} \mathbf{X}_{\mathcal{J}, \cdot}^{(1)\top} \in \mathbb{R}^{n \times d}$ is a nonnegative matrix since $\mathbf{P}_{ij}^{(1)} \geq 0$ for all $i$ and $j$. Moreover, we have the following relationships:
    \begin{equation}\label{eq:COSIE_expression}
        \mathbf{P}^{(1)} = \mathbf{V} \mathbf{R}^{(1)} \mathbf{V}^{\top} = \mathbf{X}^{(1)} \mathbf{I}_{p^{(1)}, q^{(1)}} \mathbf{X}^{(1)\top} = \mathbf{W} (\mathbf{X}_{\mathcal{J}, \cdot}^{(1)})^{-\top} \mathbf{I}_{p^{(1)}, q^{(1)}} (\mathbf{X}_{\mathcal{J}, \cdot}^{(1)})^{-\top} \mathbf{W}^{\top}.
    \end{equation}
    Let $\bm{\Theta} := \operatorname{diag}(\mathbf{W} \mathbf{1}_{d})$ and $\mathbf{Z} := \bm{\Theta}^{-1} \mathbf{W}$. Then we define a diagonal matrix $\bm{\Theta}$ and a membership matrix $\mathbf{Z}$ such that $\mathbf{W} = \bm{\Theta} \mathbf{Z}$. Substituting $\mathbf{W}$ with $\bm{\Theta} \mathbf{Z}$ in \eqref{eq:COSIE_expression}, we have
    \begin{equation*}
        \mathbf{P}^{(1)} = \bm{\Theta} \mathbf{Z} (\mathbf{X}_{\mathcal{J}, \cdot}^{(1)})^{-\top} \mathbf{I}_{p^{(1)}, q^{(1)}} (\mathbf{X}_{\mathcal{J}, \cdot}^{(1)})^{-\top} \mathbf{Z}^{\top} \bm{\Theta}.
    \end{equation*}
    By taking $\mathbf{B}^{(1)} := (\mathbf{X}_{\mathcal{J}, \cdot}^{(1)})^{-\top} \mathbf{I}_{p^{(1)}, q^{(1)}} (\mathbf{X}_{\mathcal{J}, \cdot}^{(1)})^{-\top}$, we have an equivalent expression for the reference layer. 

    For non-reference layers, let $l > 1$ be given. Then $\mathbf{P}^{(l)} = \mathbf{V} \mathbf{R}^{(l)} \mathbf{V}^{\top}$. By definition of $\mathbf{X}^{(1)}$, it follows that
    \begin{equation*}
        \mathbf{W} = \mathbf{X}^{(1)} \mathbf{I}_{p^{(1)}, q^{(1)}} \mathbf{X}_{\mathcal{J}, \cdot}^{(1)\top} = \mathbf{V} \mathbf{U}^{(1)} | \bm{\Lambda}^{(1)}|^{1/2} \mathbf{I}_{p^{(1)}, q^{(1)}} \mathbf{X}_{\mathcal{J}, \cdot}^{(1)\top}.
    \end{equation*}
    Since $\mathbf{U}^{(1)}$, $\bm{\Lambda}^{(1)}$, $\mathbf{I}_{p^{(1)}, q^{(1)}}$ and $\mathbf{X}_{\mathcal{J}, \cdot}^{(1)\top}$ are nonsingular, we have $\mathrm{col}(\mathbf{W}) = \mathrm{col}(\mathbf{V})$ and hence there exists a nonsingular matrix $\mathbf{C}$ such that $\mathbf{V} = \mathbf{W} \mathbf{C}$. By substituting $\mathbf{V}$ with $\mathbf{W} \mathbf{C}$, we have
    \begin{equation*}
        \mathbf{P}^{(l)} = \mathbf{V} \mathbf{R}^{(l)} \mathbf{V}^{\top} = \mathbf{W} \mathbf{C} \mathbf{R}^{(l)} \mathbf{C}^{\top} \mathbf{W}^{\top} = \bm{\Theta} \mathbf{Z} \mathbf{C} \mathbf{R}^{(l)} \mathbf{C}^{\top} \mathbf{Z}^{\top} \bm{\Theta}.
    \end{equation*}
    By taking $\mathbf{B}^{(l)} := \mathbf{C} \mathbf{R}^{(l)} \mathbf{C}^{\top}$, we obtain an equivalent expression of the common subspace model. This completes the proof. 
\end{proof}

%% file: Sections/Appendix-bi_Lipschitz.tex
\section{Bi-Lipschitz Continuity of the Concatenated Embedding Map}
\label{sec:bi-Lipschitz_of_T}

The error bounds in Section~\ref{sec:closeness} control the norm of difference between the population and empirical row-normalized spectral embeddings. To translate these embedding-space bounds into guarantees for membership recovery, we need to relate distances in the standard simplex to distances between the corresponding normalized embeddings. This section establishes such a relationship through a concatenated embedding map.

For $\bz\in\Delta^{K-1}$, define
\[
T(\bz) := \left( T^{(1)}(\bz)^\top,\ldots,T^{(L)}(\bz)^\top \right)^\top \in\mathbb{R}^{KL},
\]
where
\[
T^{(l)}(\bz) := \frac{\bR^{(l)\top} \bz}{\| \bR^{(l)\top} \bz\|_2}.
\]
In particular, for the membership vector $\bZ_{i,\cdot}$ of node $i$, this map satisfies
\[
T(\bZ_{i,\cdot})=\left((\bY_{i,\cdot}^{(1)})^\top,\ldots,(\bY_{i,\cdot}^{(L)})^\top\right)^\top .
\]
The goal of this section is to establish the bi-Lipschitz continuity of $T$ on $\Delta^{K-1}$, which allows us to convert spectral embedding error bounds into membership recovery guarantees and to analyze the vertex hunting step. The argument proceeds in two steps: we first obtain bi-Lipschitz bounds for each layer-wise map $T^{(l)}$, and then combine the layer-wise bounds to obtain the desired result for the concatenated map $T$. 

\paragraph{Definition of the layer-wise map $T^{(l)}$}
For each $l \in [L]$, define
\[
\mathcal{C}_{R}^{(l)} := \operatorname{conv}\left\{ \bR_{1,\cdot}^{(l)}, \bR_{2,\cdot}^{(l)}, \ldots, \bR_{K,\cdot}^{(l)} \right\} \subset \mathbb{R}^{K}.
\]
When $\bR^{(l)}$ is nonsingular, $\mathcal{C}_{R}^{(l)}$ is a $(K-1)$-simplex whose vertices are the rows of $\bR^{(l)}$. Then, there exists a normal vector 
\begin{equation}\label{eq:def_of_nl}
    \bn^{(l)} = (\bR^{(l)})^{-1} \mathbf{1}_{K}
\end{equation}
such that $\mathcal{C}_{R}^{(l)}$ is a subset of the hyperplane of $\mathcal{H}_{R}^{(l)}$ defined by
\begin{equation}\label{eq:def_of_Hl}
    \mathcal{H}_{R}^{(l)} = \left\{ (x_{1}, x_{2}, \cdots, x_{K}) \in \bbR^{K}: \langle \bn^{(l)}, \bx \rangle= 1 \right\}.
\end{equation}

Define a linear transformation $$T_{1}^{(l)}: \Delta^{K-1} \rightarrow \mathcal{C}_{R}^{(l)} \text{ by } T_{1}(\bx) = (\bR^{(l)})^{\top} \bx.$$ Also define the $\ell_{2}$-normalized image of $\mathcal{C}_{R}^{(l)}$ by
\begin{equation}\label{eq:def_of_S2}
    \mathcal{S}_{R}^{(l)} = \left \{ (x_{1}, x_{2}, \cdots, x_{K}) \in \bbR^{K}: \bx = \frac{\by}{\norm{\by}_{2}}, \by \in \mathcal{C}_{R}^{(l)} \right \} \subset \mathbb{S}^{K-1}.
\end{equation}
Let $T_{2}^{(l)}: \mathcal{C}_{R}^{(l)} \to \mathcal{S}_{R}^{(l)}$ denote the $\ell_{2}$-norm normalization function. Then we can define the layer-wise map $T^{(l)}$ by following:
\begin{equation*}
    T^{(l)} = T_{2}^{(l)}\circ T_{1}^{(l)}.
\end{equation*}
We derive lower and upper Lipschitz bounds for $T_1^{(l)}$ and $T_2^{(l)}$ separately. These bounds imply corresponding bi-Lipschitz bounds for $T^{(l)}$ by composition. Finally, the bi-Lipschitz constants for the concatenated map $T$ are obtained by combining the layer-wise bounds across $l=1,\dots,L$. 

\paragraph{Bi-Lipschitz continuity of $T_{1}^{(l)}$}
Note that $T_{1}^{(l)} (\bx) = \left( \bR^{(l)} \right)^{\top} \bx$, where $\bR^{(l)}$ is a full rank matrix in $\bbR^{K \times K}$. Then, for any $\bx, \by \in \bbR^{K}$, 
\begin{equation}\label{eq:bi-Lipschitz_constants_T1}
    \sigma_{\min}\left( (\bR^{(l)})^{\top} \right) \norm{\bx - \by}_{2} \leq \norm{T_{1}^{(l)}(\bx) - T_{1}^{(l)}(\by)} \leq \sigma_{\max} \left( (\bR^{(l)})^{\top} \right) \norm{\bx - \by}_{2}.
\end{equation}
Since $\bR^{(l)}$ is a full rank matrix, $\sigma_{\min} (\bR^{(l)}) > 0$. As shown in Section \ref{sec:Prop_of_R}, we have the following:
\begin{equation*}
    \sqrt{\sigma_{B}^{(l)}} \norm{\bx - \by}_{2} \lesssim \norm{T_{1}^{(l)}(\bx) - T_{1}^{(l)}(\by)} \lesssim \sqrt{K} \norm{\bx - \by}_{2}.
\end{equation*}

\paragraph{Bi-Lipschitz continuity of $T_{2}^{(l)}$}
First, we show the transformation $T_{2}^{(l)}$ is $C^{1}$ on $\bbR^{K} \setminus \{ \mathbf{0} \}$. It is known that $T_{2}^{(l)}$ is differentiable at a point $\bx$ if and only if the component functions $$\left( T_{2}^{(l)} (\bx)\right)_{k} = \frac{x_{k}}{\norm{\bx}_{2}}$$ is differentiable for all $1 \leq k \leq K$. For each component functions, we obtain partial derivatives by
\begin{equation*}
    \frac{\partial \left( T_{2}^{(l)} (\bx)\right)_{k}}{\partial x_{j}} =
    \begin{cases}
        -\frac{x_{k} x_{j}}{\norm{\bx}^{3}} \quad & \text{if } k \neq j \\
        \frac{1}{\norm{\bx}} - \frac{x_{j}^{2}}{\norm{\bx}^{3}} \quad & \text{if } k = j.
    \end{cases}
\end{equation*}
Since all partial derivatives are well-defined and continuous, $T_{2}^{(l)}$ is $C^{1}$ on $\bbR^{K} \setminus \{ \mathbf{0} \}$. Next, we show Lipschitz continuity of $T_{2}^{(l)}$ by finding a constant $L_{2, U}^{(l)} > 0$ such that 
\begin{equation*}
    \norm{T_{2}^{(l)}(\bx) - T_{2}^{(l)}(\by)} \leq L_{2, U}^{(l)} \norm{\bx - \by}. 
\end{equation*}
Recall one useful equation \eqref{eq:FTC} from the fundamental theorem of calculus:
\begin{equation}\label{eq:FTC}
    T_{2}^{(l)}(\by) - T_{2}^{(l)}(\bx) = \int_{0}^{1} DT_{2}^{(l)}(\bx + t(\by - \bx))(\by - \bx) dt,
\end{equation}
where $$DT_{2}^{(l)}(\bx) = \frac{\bI_{K}}{\norm{\bx}} - \frac{\bx \bx^{\top}}{\norm{\bx}^{3}}$$ is the Jacobian matrix of the transformation $T_{2}^{(l)}$ at the point $\bx$. Let $$\sigma_{U}^{(l)} = \underset{\bx \in \mathcal{C}_{R}^{(l)}}{\sup} \sigma_{\max} \left( DT_{2}^{(l)}(\bx) \right)$$ and $\bx_{t} = \bx + t(\by - \bx)$ for $t \in [0, 1]$. Note that $\sigma_{U}^{(l)}$ is finite by the extreme value theorem. Now we bound $\sigma_{U}^{(l)}$:
\begin{equation*}
    \begin{aligned}
        \sigma_{U}^{(l)} &= \underset{\bx \in \mathcal{C}_{R}^{(l)}}{\sup} \sigma_{\max} \left( DT_{2}^{(l)}(\bx) \right) \\
        &= \underset{\bx \in \mathcal{C}_{R}^{(l)}}{\sup} \sigma_{\max} \left( \frac{\bI_{K}}{\norm{\bx}} - \frac{\bx \bx^{\top}}{\norm{\bx}^{3}} \right) \\
        &= \underset{\bx \in \mathcal{C}_{R}^{(l)}}{\sup} \frac{1}{\norm{\bx}} \\
        &= \frac{1}{\underset{\bx \in \mathcal{C}_{R}^{(l)}}{\inf} \norm{\bx}}.
    \end{aligned}
\end{equation*}
For any $\bx \in \mathcal{C}_{R}^{(l)}$, by Cauchy-Schwarz inequality, we obtain
\begin{equation*}
    \begin{aligned}
        \langle \bx, \bn^{(l)} \rangle \leq \| \bx \|_{2} \| \bn^{(l)} \|_{2} \Leftrightarrow \frac{1}{\| \bx \|_{2}} \leq \frac{\| \bn^{(l)} \|_{2}}{\langle \bx, \bn^{(l)} \rangle} = \| \bn^{(l)} \|_{2}.
    \end{aligned}
\end{equation*}
Hence, by \eqref{eq:D_lower_bound_of_sminR}, we obtain
\begin{align}\label{eq:C_upper_bound_of_n}
    \| \bn^{(l)} \| &= \| (\bR^{(l)})^{-1} \mathbf{1}_{K} \| \notag \\
    &\leq \| (\bR^{(l)})^{-1} \| \| \mathbf{1}_{K} \|_{2} \notag \\
    &\lesssim \frac{\sqrt{K}}{\sqrt{\sigma_{B}^{(l)}}}.
\end{align}
We show $T_{2}^{(l)}$ is Lipschitz continuous. Since \(\mathcal C_R^{(l)}\) is convex and \(0\notin\mathcal C_R^{(l)}\), the line segment between any two points of \(\mathcal C_R^{(l)}\) lies in the domain of \(T_2^{(l)}\). Hence, by the fundamental theorem of calculus and \eqref{eq:C_upper_bound_of_n},
\begin{align}\label{eq:C_Lipschitz_of_T2}
    \norm{T_{2}^{(l)}(\by) - T_{2}^{(l)}(\bx)} &= \norm{\int_{0}^{1} DT_{2}^{(l)}(\bx + t(\by - \bx))(\by - \bx) dt} \notag \\
        &\leq \int_{0}^{1} \norm{DT_{2}^{(l)}(\bx + t(\by - \bx))(\by - \bx)} dt \notag \\
        &\leq \int_{0}^{1} \sigma_{\max} \left( DT_{2}^{(l)}(\bx + t(\by - \bx)) \right) \norm{\by - \bx}_{2} dt \notag \\
        &\leq \int_{0}^{1} \sigma_{U}^{(l)} \norm{\by - \bx}_{2} dt \notag \\
        &= \sigma_{U}^{(l)} \norm{\by - \bx}_{2} \notag \\
        &\lesssim \frac{\sqrt{K}}{\sqrt{\sigma_{B}^{(l)}}} \| \by - \bx \|_{2}
\end{align}

Similarly, we show $\left( T_{2}^{(l)} \right)^{-1}$ is Lipschitz continuous. Let $\bx \in \mathcal{C}_{R}^{(l)}$ and $\by = \frac{\bx}{\| \bx \|_{2}}$ be given. Since $\bx \in \mathcal{C}_{R}^{(l)} \subset \mathcal{H}_{R}^{(l)}$, we have $\langle \bn^{(l)}, \bx \rangle = 1$. Hence,
\begin{equation}\label{eq:C_inner_prod_ny}
    \langle \bn^{(l)}, \by \rangle = \frac{1}{\| \bx \|_{2}} \langle \bn^{(l)}, \bx \rangle = \frac{1}{\| \bx \|_{2}}.
\end{equation}
We obtain an explicit form of $(T_{2}^{(l)})^{-1}(\by)$ by following:
\begin{equation*}
    (T_{2}^{(l)})^{-1}(\by) = \bx = \| \bx \|_{2} \by = \frac{\by}{\langle \bn^{(l)}, \by \rangle}.
\end{equation*}

Note that, however, we cannot directly use the fundamental theorem of calculus since a line segment between two points in $\mathcal{S}_{R}^{(l)}$ is not guaranteed to be included in $\mathcal{S}_{R}^{(l)}$. Therefore, we define an open half-space $\mathcal{U}^{(l)}$ and a transformation $F^{(l)}$ on $\mathcal{U}^{(l)}$ by following:
\begin{equation*}
    \mathcal{U}^{(l)} = \{ \bu \in \bbR^{K}: \langle \bn^{(l)}, \bu \rangle > 0 \}
\end{equation*}
and
\begin{equation*}
    F^{(l)}(\bu) = \frac{\bu}{\langle \bn^{(l)}, \bu \rangle}.
\end{equation*}
It is clear that $\mathcal{S}_{R}^{(l)} \subset \mathcal{U}^{(l)}$ and $F^{(l)} = (T_{2}^{(l)})^{-1}$ on $\mathcal{S}_{R}^{(l)}$. Moreover, $F^{(l)} \in C^{1}(\mathcal{U}^{(l)}, \bbR^{K})$ since both numerator and denominator are $C^{1}$. Hence we can show Lipschitz continuity of $F^{(l)}$ by the fundamental theorem of calculus. 

Let $\bu_1,\bu_2 \in \mathcal{S}_{R}^{(l)}$ be given. By definition of $\mathcal{S}_{R}^{(l)}$, there exist $\bx_1,\bx_2 \in \mathcal{C}_{R}^{(l)}$ such that
\[
\bu_j = \frac{\bx_j}{\|\bx_j\|_2}, \quad j=1,2.
\] 
We first bound the operator norm of the Jacobian $DF^{(l)}$ along the line segment between $\bu_1$ and $\bu_2$. 
For $t\in[0,1]$, let
\[
\bu_t := \bu_1 + t(\bu_2-\bu_1),
\]
which parametrizes the line segment between $\bu_1$ and $\bu_2$. 
Using the triangle inequality and the identity $\|\bu_{1} \bu_{2}^\top\|=\|\bu_{1}\|_2\|\bu_{2}\|_2$, we obtain
\begin{align}\label{eq:C_Jacobian_of_Fl}
\|DF^{(l)}(\bu_t)\|
&=
\left\|
\frac{\bI_K}{\langle \bn^{(l)},\bu_t\rangle}
-
\frac{\bu_t \bn^{(l)\top}}{\langle \bn^{(l)},\bu_t\rangle^2}
\right\| \notag \\
&\le
\left\|
\frac{\bI_K}{\langle \bn^{(l)},\bu_t\rangle}
\right\|
+
\left\|
\frac{\bu_t \bn^{(l)\top}}{\langle \bn^{(l)},\bu_t\rangle^2}
\right\| \notag \\
&\le
\frac{1}{\langle \bn^{(l)},\bu_t\rangle}
+
\frac{\|\bu_t\|_2\|\bn^{(l)}\|_2}{\langle \bn^{(l)},\bu_t\rangle^2}.
\end{align}
We bound each term in \eqref{eq:C_Jacobian_of_Fl}. By triangle inequality, we have
\begin{equation}\label{eq:C_upper_bound_of_ut}
    \| \bu_{t} \| = \| (1-t) \bu_{1} + t \bu_{2} \| \leq (1-t) \| \bu_{1} \| + t \| \bu_{2} \| = 1.
\end{equation}
By linearity of inner product,
\begin{align*}
    \langle \bn^{(l)}, \bu_{t} \rangle &= \langle \bn^{(l)}, (1-t)\bu_{1} + t \bu_{2} \rangle \notag \\
    &= (1-t) \langle \bn^{(l)}, \bu_{1} \rangle + t \langle \bn^{(l)}, \bu_{2} \rangle \notag \\
    &= (1-t) \frac{1}{\| \bx_{1} \|_{2}} + t \frac{1}{\| \bx_{2} \|_{2}} \notag \\
    &\geq \frac{1}{\displaystyle\sup_{\bx \in \mathcal{C}_{R}^{(l)}} \| \bx \|} \notag \\
    &= \frac{1}{\displaystyle \max_{1 \leq k \leq K} \| \bR_{k, \cdot}^{(l)} \|_{2}} \notag \\
    &\geq \frac{1}{\| \bR^{(l)} \|}.
\end{align*}
Moreover, by \eqref{eq:D_upper_bound_of_smaxR}, we have
\begin{equation}\label{eq:C_lower_bound_of_innerprod_nut}
    \langle \bn^{(l)}, \bu_{t} \rangle \gtrsim \frac{1}{\sqrt{K}}.
\end{equation}
By substituting each bound \eqref{eq:C_upper_bound_of_n}, \eqref{eq:C_upper_bound_of_ut}, and \eqref{eq:C_lower_bound_of_innerprod_nut} to \eqref{eq:C_Jacobian_of_Fl}, we have
\begin{equation}\label{eq:C_upper_bound_of_Jacobian_Fl}
    \sup_{t \in [0, 1]}\| DF^{(l)}(\bu_{t}) \| \lesssim \sqrt{K} + \frac{K^{3/2}}{\sqrt{\sigma_{B}^{(l)}}} \lesssim \frac{K^{3/2}}{\sqrt{\sigma_{B}^{(l)}}}.
\end{equation}
We show $(T_{2}^{(l)})^{-1}$ is Lipschitz continuous. By the fundamental theorem of calculus and \eqref{eq:C_upper_bound_of_Jacobian_Fl}, we have
\begin{align}\label{eq:C_Lipschitz_of_T2inv}
    \| (T_{2}^{(l)})^{-1}(\bu_{2}) - (T_{2}^{(l)})^{-1}(\bu_{1}) \|_{2} &= \| F^{(l)}(\bu_{2}) - F^{(l)}(\bu_{1}) \|_{2} \notag \\
    &= \| \int_{0}^{1} DF^{(l)}(\bu_{t}) (\bu_{2} - \bu_{1}) dt \|_{2} \notag \\
    &\leq \int_{0}^{1} \| DF^{(l)}(\bu_{t}) (\bu_{2} - \bu_{1}) \| dt \notag \\
    &\leq \int_{0}^{1} \| DF^{(l)}(\bu_{t}) \| \| \bu_{2} - \bu_{1} \|_{2} dt \notag \\
    &\lesssim \int_{0}^{1} \frac{K^{3/2}}{\sqrt{\sigma_{B}^{(l)}}} \| \bu_{2} - \bu_{1} \|_{2} dt \notag \\
    &= \frac{K^{3/2}}{\sqrt{\sigma_{B}^{(l)}}} \| \bu_{2} - \bu_{1} \|_{2}.
\end{align}
By \eqref{eq:C_Lipschitz_of_T2} and \eqref{eq:C_Lipschitz_of_T2inv}, we have
\begin{equation}\label{eq:C_bi_Lipschitz_of_T2}
    \frac{\sqrt{\sigma_{B}^{(l)}}}{K^{3/2}} \| \bx_{2} - \bx_{1} \|_{2} \lesssim \| T_{2}^{(l)}(\bx_{2}) - T_{2}^{(l)}(\bx_{1}) \| \lesssim \frac{\sqrt{K}}{\sqrt{\sigma_{B}^{(l)}}} \| \bx_{2} - \bx_{1} \|_{2}.
\end{equation}

\paragraph{Bi-Lipschitz continuity of $T^{(l)}$.}
Let $\bz_1,\bz_2\in\Delta^{K-1}$ be given, and set
\[
\bx_j := T_1^{(l)}(\bz_j)=\bR^{(l)\top}\bz_j,
\qquad j=1,2.
\]
Then $\bx_1,\bx_2\in\mathcal C_R^{(l)}$, and
\[
T^{(l)}(\bz_j) = T_2^{(l)}(\bx_j), \quad j=1,2.
\]
By \eqref{eq:bi-Lipschitz_constants_T1} and the singular-value bounds for $\bR^{(l)}$ in Section~\ref{sec:Prop_of_R},
\[
\sqrt{\sigma_{B}^{(l)}}\|\bz_2-\bz_1\|_2 \lesssim \|\bx_2-\bx_1\|_2 \lesssim \sqrt K \|\bz_2-\bz_1\|_2.
\]
Combining this with \eqref{eq:C_bi_Lipschitz_of_T2}, we obtain
\begin{align}
\|T^{(l)}(\bz_2)-T^{(l)}(\bz_1)\|_2 &= \|T_2^{(l)}(\bx_2)-T_2^{(l)}(\bx_1)\|_2 \notag\\
&\gtrsim \frac{\sqrt{\sigma_{B}^{(l)}}}{K^{3/2}} \|\bx_2-\bx_1\|_2 \gtrsim \frac{\sigma_{B}^{(l)}}{K^{3/2}} \|\bz_2-\bz_1\|_2,
\end{align}
and similarly,
\begin{align}
\|T^{(l)}(\bz_2)-T^{(l)}(\bz_1)\|_2 &= \|T_2^{(l)}(\bx_2)-T_2^{(l)}(\bx_1)\|_2 \notag\\
&\lesssim \frac{\sqrt K}{\sqrt{\sigma_{B}^{(l)}}} \|\bx_2-\bx_1\|_2 \lesssim \frac{K}{\sqrt{\sigma_{B}^{(l)}}} \|\bz_2-\bz_1\|_2.
\end{align}
Therefore,
\begin{equation}\label{eq:C_bi_Lipschitz_of_Tl}
\frac{\sigma_{B}^{(l)}}{K^{3/2}} \|\bz_2-\bz_1\|_2 \lesssim \|T^{(l)}(\bz_2)-T^{(l)}(\bz_1)\|_2 \lesssim \frac{K}{\sqrt{\sigma_{B}^{(l)}}} \|\bz_2-\bz_1\|_2.
\end{equation}
Thus, each layer-wise map $T^{(l)}$ is bi-Lipschitz on $\Delta^{K-1}$.

\paragraph{Bi-Lipschitz continuity of $T$.}
We now combine the layer-wise bounds. For any $\bz_1,\bz_2\in\Delta^{K-1}$, by the definition of the concatenated map,
\[
\|T(\bz_2)-T(\bz_1)\|_2^2
=
\sum_{l=1}^L
\|T^{(l)}(\bz_2)-T^{(l)}(\bz_1)\|_2^2.
\]
Applying \eqref{eq:C_bi_Lipschitz_of_Tl} to each layer yields
\begin{align}
\|T(\bz_2)-T(\bz_1)\|_2 &\gtrsim \left( \sum_{l=1}^L \left(\frac{\sigma_{B}^{(l)}}{K^{3/2}}\right)^2 \right)^{1/2} \|\bz_2-\bz_1\|_2, \notag\\ \|T(\bz_2)-T(\bz_1)\|_2
&\lesssim \left( \sum_{l=1}^L \frac{K^2}{\sigma_{B}^{(l)}} \right)^{1/2} \|\bz_2-\bz_1\|_2.
\end{align}
Equivalently,
\begin{equation}\label{eq:C_bi_Lipschitz_of_T}
\left( \sum_{l=1}^L \frac{(\sigma_{B}^{(l)})^2}{K^3} \right)^{1/2} \|\bz_2-\bz_1\|_2 \lesssim \|T(\bz_2)-T(\bz_1)\|_2 \lesssim K \left( \sum_{l=1}^L \frac{1}{\sigma_{B}^{(l)}} \right)^{1/2} \|\bz_2-\bz_1\|_2.
\end{equation}
Since $\sigma_{B}^{(l)}>0$ for all $l\in[L]$, this proves that $T$ is bi-Lipschitz on $\Delta^{K-1}$.

%% file: Sections/Appendix-corner_estimation.tex
\section{Corner estimation}\label{sec:appendix_corner_estimation}
Successful corner estimation is a key task of the proposed method. Under the regularity assumptions stated in Section~\ref{sec:Theory}, Theorem~\ref{thm:VH_almost_pure_recovery} establishes that increasing the number of observed layers can improve the probability of accurate corner recovery. Moreover, Corollary~\ref{cor:VH_exact} states that the estimated corner contains exactly one pure node from each community with high probability under the condition~\ref{cond:denoising_purity_separation}(b). In this section, we provide a deterministic error radius of the corner estimation, proof of Theorem~\ref{thm:VH_almost_pure_recovery}, Corollary~\ref{cor:VH_exact} and required preliminary lemmas. 

Throughout this section, we use the following notion to describe whether a candidate index set contains a sufficiently pure representative from each community. Let $\mathcal{I}_{k}(r) = \{ i \in [n]: \bZ_{ik} \geq 1-r \}$ for any $r \in [0, 1)$. Given an index set \(\mathcal J \subseteq [n]\), we say that \(\mathcal J\) is \(r\)-good if
\[
\mathcal J\cap\mathcal I_k(r)\neq\varnothing \quad\text{for every }1\leq k\leq K.
\]


\subsection{Preliminary lemmas}
We present the lemmas required for the corner estimation and the proof of the main theorem. First, the following lemma provides a convenient way to compute the Euclidean distance from a point to a spherical simplex defined in \eqref{eq:def_of_vh_objective}.
\begin{lemma}[Computing the Euclidean distance to a spherical simplex]\label{lem:distance_bw_point_ss}
Let $\mathbf v_1,\ldots,\mathbf v_K\in\mathbb S^{K-1}$, and define
\[
\mathcal C:=\operatorname{cone}\{\mathbf v_1,\ldots,\mathbf v_K\},
\qquad
\mathcal S:=\mathcal C\cap\mathbb S^{K-1}.
\]
For $\mathbf y\in\mathbb S^{K-1}$, let $\mathbf p:=\operatorname{Proj}_{\mathcal C}(\mathbf y).$
If $\mathbf p\neq\mathbf0$, then $\operatorname{Proj}_{\mathcal S}(\mathbf y) = \frac{\mathbf p}{\|\mathbf p\|_2}.$
If $\mathbf p=\mathbf0$, then at least one generating vertex is a Euclidean
projection of $\mathbf y$ onto $\mathcal S$. In particular, any
\[
j^\star\in\operatorname*{arg\,min}_{1\leq j\leq K}
\|\mathbf y-\mathbf v_j\|_2
\]
satisfies
\[
\mathbf v_{j^\star}\in
\operatorname*{arg\,min}_{\mathbf u\in\mathcal S}
\|\mathbf y-\mathbf u\|_2.
\]
Consequently,
\[
d(\mathbf y,\mathcal S)
=
\begin{cases}
\sqrt{2-2\|\mathbf p\|_2}, & \mathbf p\neq\mathbf0,\\[1mm]
\displaystyle\min_{1\le j\le K}\|\mathbf y-\mathbf v_j\|_2,
& \mathbf p=\mathbf0.
\end{cases}
\]
\end{lemma}
Lemma~\ref{lem:distance_bw_point_ss} shows that the distance between a point on the unit sphere and a spherical simplex can be computed by first projecting the point onto the generating cone. If the cone projection is nonzero, we normalize it to unit length; otherwise, it suffices to compare the distances to the \(K\) generating vertices.
\begin{proof}
    Consider the first case $\mathbf{p} \neq \mathbf{0}$. For any $\mathbf{u} \in \mathcal{S}$, the squared Euclidean distance between $\mathbf{y}$ and $\mathbf{u}$ can be written as
    \begin{align*}
        \| \mathbf{y} - \mathbf{u} \|_{2}^{2} = \| \mathbf{y} \|_{2}^{2} + \| \mathbf{u} \|_{2}^{2} - 2 \langle \mathbf{y}, \mathbf{u} \rangle = 2 - 2 \langle \mathbf{y}, \mathbf{u} \rangle.
    \end{align*}
    Hence we aim to find the minimizer $\mathbf{u} \in \mathcal{S}$ such that $\mathbf{u} \in \arg\max_{\mathbf{x} \in \mathcal{S}} \langle \mathbf{y}, \mathbf{x} \rangle$.
    
    By the first-order optimality of projection, for any $\mathbf{x} \in \mathcal{C}$, we have
    \begin{equation}\label{eq:E_fo_optimality_of_proj}
        \langle \mathbf{y} - \mathbf{p}, \mathbf{x} - \mathbf{p} \rangle \leq 0.
    \end{equation}
    By definition of a cone, we know $\mathbf{0}$, $2 \mathbf{p} \in \mathcal{C}$. By substituting $\mathbf{x} = \mathbf{0}, 2\mathbf{p}$ in \eqref{eq:E_fo_optimality_of_proj}, we have $\langle \mathbf{y} - \mathbf{p}, \mathbf{p} \rangle = 0$ and $\langle \mathbf{y} - \mathbf{p}, \mathbf{x} \rangle \leq 0$. Then for any $\mathbf{t} \in \mathcal{S} \subseteq \mathcal{C}$, we have the following upper bound.
    \begin{align*}
        \langle \mathbf{y}, \mathbf{t} \rangle = \langle \mathbf{y} - \mathbf{p}, \mathbf{t} \rangle + \langle \mathbf{p}, \mathbf{t} \rangle \leq  \langle \mathbf{p}, \mathbf{t} \rangle \leq \| \mathbf{p} \|_{2} \| \mathbf{t} \|_{2} = \| \mathbf{p} \|_{2}.
    \end{align*}
    Let $\mathbf{u} = \mathbf{p} / \| \mathbf{p} \|_{2} \in \mathcal{S} \subseteq \mathcal{C}$. Then, 
    \begin{align*}
        \langle \mathbf{y}, \mathbf{u} \rangle = \langle \mathbf{y}, \frac{\mathbf{p}}{\| \mathbf{p}\|_{2}} \rangle = \frac{1}{\| \mathbf{p} \|_{2}} \langle \mathbf{y}, \mathbf{p} \rangle = \frac{1}{\| \mathbf{p} \|_{2}} \left( \langle \mathbf{y} - \mathbf{p}, \mathbf{p} \rangle + \langle \mathbf{p}, \mathbf{p} \rangle \right) = \| \mathbf{p} \|_{2}.
    \end{align*}
    That is, $\mathbf{u} = \mathbf{p} / \| \mathbf{p} \|_{2} \in \arg\max_{\mathbf{x} \in \mathcal{S}} \langle \mathbf{y}, \mathbf{x} \rangle$ and 
    \begin{align*}
        d(\mathbf{y}, \mathcal{S}) = \| \mathbf{y} - \frac{\mathbf{p}}{\| \mathbf{p} \|_{2}} \|_{2} = \sqrt{2 - 2 \langle \mathbf{y}, \frac{\mathbf{p}}{\| \mathbf{p} \|_{2}} \rangle} = \sqrt{2 - \frac{2}{\| \mathbf{p} \|_{2}} \left( \langle \mathbf{y} - \mathbf{p} + \mathbf{p}, \mathbf{p} \right)} = \sqrt{2 - 2 \| \mathbf{p} \|_{2}}.
    \end{align*}

    Next, consider the second case $\mathbf{p} = \mathbf{0}$. Similarly, we aim to find $\mathbf{u} \in \mathcal{S}$ such that $\mathbf{u} \in \arg\max_{\mathbf{x} \in \mathcal{S}} \langle \mathbf{y}, \mathbf{x} \rangle$. We show $\langle \mathbf{y}, \mathbf{v}_{j} \rangle \leq 0$ for all $j \in [K]$. Fix $j \in [K]$. For any $t > 0$, it is known that $t \mathbf{v}_{j} \in \mathcal{C}$. Since $\mathbf{p}= \mathbf{0} \in \mathcal{C}$ is the projected point of $\mathbf{y}$ onto $\mathcal{C}$, we have
    \[
    \|\mathbf y\|_2^2 \le \|\mathbf y-t\mathbf v_j\|_2^2 = \| \mathbf{y} \|_{2}^{2} + t^{2} \| \mathbf{v}_{j} \|_{2}^{2} - 2t \langle \mathbf{y}, \mathbf{v}_{j} \rangle \quad\Longleftrightarrow\quad \langle \mathbf y,\mathbf v_j\rangle\le \frac t2.
    \]
    By taking $t \to 0$, we show $\langle \mathbf{y}, \mathbf{v}_{j} \rangle \leq 0$ for all $j \in [K]$. Let $\mathbf{t} \in \mathcal{S}$ be given. Then there exists a constant $\mathbf{c} = (c_1, \cdots, c_K)^{\top} \in \mathbb{R}^{K}$ with $\sum_{i=1}^{K} c_{i} = 1$ and $c_{i} \geq 0$ for all $i \in [K]$ such that $\mathbf{t} = \frac{\sum_{i=1}^{K} c_{i} \mathbf{v}_{i}}{\| \sum_{i=1}^{K} c_{i} \mathbf{v}_{i} \|_{2}}$. Let $m := \max_{1 \leq i \leq K} \langle \mathbf{y}, \mathbf{v}_{i} \rangle$. Then,
    \begin{align*}
        \langle \mathbf{y}, \mathbf{t} \rangle = \langle \mathbf{y}, \frac{\sum_{i=1}^{K} c_{i} \mathbf{v}_{i}}{\| \sum_{i=1}^{K} c_{i} \mathbf{v}_{i} \|_{2}} \rangle = \frac{1}{\| \sum_{i=1}^{K} c_{i} \mathbf{v}_{i} \|_{2}} \sum_{i=1}^{K} c_{i} \langle \mathbf{y}, \mathbf{v}_{i} \rangle \leq \frac{m}{\| \sum_{i=1}^{K} c_{i} \mathbf{v}_{i} \|_{2}}.
    \end{align*}
    By triangle inequality, $\| \sum_{i=1}^{K} c_{i} \mathbf{v}_{i} \|_{2} \leq \sum_{i=1}^{K} c_{i} \| \mathbf{v}_{i} \|_{2} = 1$. Hence we obtain the following upper bound of $\langle \mathbf{y}, \mathbf{t} \rangle$: since $m \leq 0$,
    \begin{align*}
        \langle \mathbf{y}, \mathbf{t} \rangle \leq \frac{m}{\| \sum_{i=1}^{K} c_{i} \mathbf{v}_{i} \|_{2}} \leq m.
    \end{align*}
    Let $i^{*} \in \arg\max_{i \in [K]} \langle \mathbf{y}, \mathbf{v}_{i} \rangle$. Then $\mathbf{u} = \mathbf{v}_{i^{*}} \in \arg\max_{\mathbf{x} \in \mathcal{S}} \langle \mathbf{y}, \mathbf{x} \rangle$ and
    \begin{align*}
        d(\mathbf{y}, \mathcal{S}) = \| \mathbf{y} - \mathbf{v}_{i^{*}} \|_{2} = \min_{1 \leq j \leq K} \| \mathbf{y} - \mathbf{v}_{j} \|_{2}.
    \end{align*}
\end{proof}

\begin{lemma}[Pullback representation of a spherical simplex]
\label{lem:pullback_SS}
Fix $1 \le l \le L$, and let $\mathcal{J}=\{j_1,\dots,j_K\}\subset [n]$.  
For each $j\in\mathcal{J}$, define
\[
\bz_j := \bZ_{j,\cdot}\in\Delta^{K-1},
\quad
\by_{j}^{(l)} := \mathbf{Y}_{j, \cdot}^{(l)} = T^{(l)}(\bz_j)
= \frac{\bR^{(l)\top}\bz_j}{\|\bR^{(l)\top}\bz_j\|_2}.
\]
Let
\[
C_{\mathcal J}:=\operatorname{conv}\{\bz_j: j\in\mathcal J\}\subset \Delta^{K-1}.
\]
Then
\[
\mathcal{SS}(\mathbf{Y}^{(l)}, \mathcal{J}) = T^{(l)}(C_{\mathcal J}).
\]
\end{lemma}
This lemma shows that the spherical simplex can be equivalently represented via its pullback to the convex set $C_{\mathcal J}\subset\Delta^{K-1}$, allowing us to analyze its geometry in the simplex domain.

\begin{proof}
We first show that $T^{(l)}(C_{\mathcal J}) \subset \mathcal{SS}(\mathbf{Y}^{(l)}, \mathcal{J})$.
Let $\by^{(l)} \in T^{(l)}(C_{\mathcal J})$. Then there exists
\[
\bz=\sum_{j\in\mathcal J} u_j \bz_j \in C_{\mathcal J},
\quad
u_j\ge 0,\quad \sum_{j\in\mathcal J}u_j=1,
\]
such that
\[
\by^{(l)}=T^{(l)}(\bz)=\frac{\bR^{(l)\top}\bz}{\|\bR^{(l)\top}\bz\|_2}.
\]
Since
\[
\bR^{(l)\top}\bz
=
\sum_{j\in\mathcal J} u_j \bR^{(l)\top}\bz_j
=
\sum_{j\in\mathcal J} u_j \|\bR^{(l)\top}\bz_j\|_2 \by_j^{(l)},
\]
define
\[
d_j
:=
\frac{u_j \|\bR^{(l)\top}\bz_j\|_2}
{\sum_{m\in\mathcal J} u_m \|\bR^{(l)\top}\bz_m\|_2},
\qquad j\in\mathcal J.
\]
Then $d_j\ge 0$ and $\sum_{j\in\mathcal J}d_j=1$, and hence
\[
\by^{(l)}
=
\frac{\sum_{j\in\mathcal J} d_j \by_j^{(l)}}
{\left\|\sum_{j\in\mathcal J} d_j \by_j^{(l)}\right\|_2}
\in
\mathcal{SS}(\mathbf{Y}^{(l)}, \mathcal{J}).
\]
This proves $T^{(l)}(C_{\mathcal J}) \subset \mathcal{SS}(\mathbf{Y}^{(l)}, \mathcal{J})$.

Next, we show that $\mathcal{SS}(\mathbf{Y}^{(l)}, \mathcal{J}) \subset T^{(l)}(C_{\mathcal J})$.
Let $\by^{(l)} \in \mathcal{SS}(\mathbf{Y}^{(l)}, \mathcal{J})$. Then there exists
\[
\bu=(u_j)_{j\in\mathcal J}\in\Delta^{K-1}
\]
such that
\[
\by^{(l)}
=
\frac{\sum_{j\in\mathcal J} u_j \by_j^{(l)}}
{\left\|\sum_{j\in\mathcal J} u_j \by_j^{(l)}\right\|_2}.
\]
Define
\[
\gamma:=\sum_{j\in\mathcal J} \frac{u_j}{\|\bR^{(l)\top}\bz_j\|_2},
\qquad
d_j:=\frac{u_j/\|\bR^{(l)\top}\bz_j\|_2}{\gamma},
\qquad j\in\mathcal J,
\]
and let
\[
\bz:=\sum_{j\in\mathcal J} d_j \bz_j \in C_{\mathcal J}.
\]
Then
\[
\bR^{(l)\top}\bz
=
\sum_{j\in\mathcal J} d_j \bR^{(l)\top}\bz_j
=
\sum_{j\in\mathcal J} d_j \|\bR^{(l)\top}\bz_j\|_2 \by_j^{(l)}
=
\frac{1}{\gamma}\sum_{j\in\mathcal J} u_j \by_j^{(l)}.
\]
Therefore,
\[
T^{(l)}(\bz)
=
\frac{\bR^{(l)\top}\bz}{\|\bR^{(l)\top}\bz\|_2}
=
\frac{\sum_{j\in\mathcal J} u_j \by_j^{(l)}}
{\left\|\sum_{j\in\mathcal J} u_j \by_j^{(l)}\right\|_2}
=
\by^{(l)}.
\]
Hence $\by^{(l)}\in T^{(l)}(C_{\mathcal J})$, which proves
\[
\mathcal{SS}(\mathbf{Y}^{(l)}, \mathcal{J}) \subset T^{(l)}(C_{\mathcal J}).
\]
Combining the two inclusions completes the proof.
\end{proof}

\begin{lemma}[Open hemisphere containment of embeddings]\label{lem:hemisphere_with_margin}
    For each $1 \le l \le L$, there exists a unit vector $\bu^{(l)} \in \mathbb{S}^{K-1}$ such that
    \[
    \langle \by_i^{(l)}, \bu^{(l)} \rangle \geq \frac{c_{\mathrm{stab}} \sqrt{\sigma_{B}^{(l)}}}{K}, \quad \forall i \in [n],
    \]
    where $\bu^{(l)}$ is given by
    \[
    \bu^{(l)} = \frac{\bn^{(l)}}{\|\bn^{(l)}\|_2}, 
    \quad 
    \bn^{(l)} = (\bR^{(l)})^{-1} \mathbf{1}_K.
    \]
    Equivalently, all embeddings $\{\by_i^{(l)}\}_{i=1}^n$ lie in an open hemisphere and are separated from its boundary by at least $\frac{c_{\mathrm{stab}} \sqrt{\sigma_{B}^{(l)}}}{K}$. Moreover, if $\hat{h}^{(l)} \leq \frac{c_{\mathrm{stab}} \sqrt{\sigma_{B}^{(l)}}}{2K}$, then 
    \begin{equation*}
        \langle \widetilde{\by}_i^{(l)}, \bu^{(l)} \rangle \geq \frac{c_{\mathrm{stab}} \sqrt{\sigma_{B}^{(l)}}}{2K}, \quad \forall i \in [n],
    \end{equation*}
    where $\widetilde{\by}_{i}^{(l)} := \widetilde{\mathbf{Y}}_{i, \cdot}^{(l)}$.
    In particular, the empirical embeddings also lie in the same open hemisphere with margin at least $\frac{c_{\mathrm{stab}} \sqrt{\sigma_{B}^{(l)}}}{2K}.$
\end{lemma}
This lemma gives the geometric interpretation of the event \(G_l\) used in the main text. On \(G_l\), the empirical row-normalized embeddings \(\{\widetilde \by_i^{(l)}\}_{i=1}^n\) preserve the population hemisphere margin: they remain in the same open hemisphere, with margin at least $\frac{c_{\mathrm{stab}}\sqrt{\sigma_{B}^{(l)}}}{2K}.$
\begin{proof}
    For any \(i\in[n]\), since \(T_1^{(l)}(\mathbf z_i)=(\mathbf R^{(l)})^\top\mathbf z_i\in\mathcal C_R^{(l)}\) and
    \[
    \left\langle \mathbf n^{(l)},T_1^{(l)}(\mathbf z_i)\right\rangle=\left\langle(\mathbf R^{(l)})^{-1}\mathbf 1_K,(\mathbf R^{(l)})^\top\mathbf z_i\right\rangle=\mathbf 1_K^\top\mathbf z_i=1,
    \]
    we have
    \[
    \begin{aligned}
        \left\langle\frac{\mathbf n^{(l)}}{\|\mathbf n^{(l)}\|_2},\mathbf y_i^{(l)}\right\rangle
        &=\frac{1}{\|\mathbf n^{(l)}\|_2\|T_1^{(l)}(\mathbf z_i)\|_2} \\
        &\ge\frac{1}{\|\mathbf n^{(l)}\|_2\sup_{\mathbf x\in\mathcal C_R^{(l)}}\|\mathbf x\|_2} \\
        &=\frac{1}{\|\mathbf n^{(l)}\|_2\max_{1\le k\le K}\|\mathbf R_{k,\cdot}^{(l)}\|_2} \\
        &\ge\frac{1}{\|\mathbf n^{(l)}\|_2\sigma_{\max}(\mathbf R^{(l)})} \ge\frac{c_{\mathrm{stab}}\sqrt{\sigma_{B}^{(l)}}}{K}. \end{aligned}
    \]
    
    Next, assume $\hat{h}^{(l)} \leq \frac{c_{\mathrm{stab}} \sqrt{\sigma_{B}^{(l)}}}{2K}$. Then, we have
    \begin{equation*}
        \begin{aligned}
            \langle \frac{\bn^{(l)}}{\| \bn^{(l)} \|_{2}}, \widetilde{\by}_{i}^{(l)} \rangle &= \langle \frac{\bn^{(l)}}{\| \bn^{(l)} \|_{2}}, \by_{i}^{(l)} - \by_{i}^{(l)} + \widetilde{\by}_{i}^{(l)} \rangle \\
            &= \langle \frac{\bn^{(l)}}{\| \bn^{(l)} \|_{2}}, \by_{i}^{(l)} \rangle + \langle \frac{\bn^{(l)}}{\| \bn^{(l)} \|_{2}}, \widetilde{\by}_{i}^{(l)} - \by_{i}^{(l)} \rangle \\
            &\geq \frac{c_{\mathrm{stab}} \sqrt{\sigma_{B}^{(l)}}}{K} - \left \| \frac{\bn^{(l)}}{\| \bn^{(l)} \|_{2}} \right \|_{2} \| \widetilde{\by}_{i}^{(l)} - \by_{i}^{(l)} \|_{2} \\
            &\geq \frac{c_{\mathrm{stab}} \sqrt{\sigma_{B}^{(l)}}}{K} - \hat{h}^{(l)} \\
            &\geq \frac{c_{\mathrm{stab}} \sqrt{\sigma_{B}^{(l)}}}{2K}.
        \end{aligned}
    \end{equation*}
\end{proof}

\begin{lemma}[Stability of point-to-set distance]\label{lem:stability_of_distance}
Let $\bu_1, \bu_2 \in \mathbb{R}^K$, and let $A,B \subset \mathbb{R}^K$ be nonempty sets. Then
\[
\bigl| d(\bu_1,A)-d(\bu_2,B) \bigr|
\le
\|\bu_1-\bu_2\|_2 + d_H(A,B),
\]
where
\[
d(\bu,A):=\inf_{\ba \in A}\|\bu-\ba\|_2
\]
denotes the point-to-set distance, and $d_H(A,B)$ denotes the Hausdorff distance between $A$ and $B$.
\end{lemma}

\begin{proof}
By the triangle inequality,
\[
\bigl| d(\bu_1,A)-d(\bu_2,B) \bigr|
\le
\bigl| d(\bu_1,A)-d(\bu_1,B) \bigr|
+
\bigl| d(\bu_1,B)-d(\bu_2,B) \bigr|.
\]
It therefore suffices to show that
\[
\bigl| d(\bu_1,A)-d(\bu_1,B) \bigr| \le d_H(A,B)
\quad\text{and}\quad
\bigl| d(\bu_1,B)-d(\bu_2,B) \bigr| \le \|\bu_1-\bu_2\|_2.
\]

We first prove that
\[
\bigl| d(\bu_1,A)-d(\bu_1,B) \bigr| \le d_H(A,B).
\]
Let $\epsilon>0$. Choose $\mathbf{b}_\epsilon \in B$ such that
\[
\|\bu_1-\mathbf{b}_\epsilon\|_2 \le d(\bu_1,B)+\epsilon.
\]
Then
\[
d(\mathbf{b}_\epsilon,A)\le \sup_{\mathbf{b}\in B} d(\mathbf{b},A)\le d_H(A,B).
\]
Hence, there exists $\ba_\epsilon \in A$ such that
\[
\|\mathbf{b}_\epsilon-\ba_\epsilon\|_2 \le d_H(A,B)+\epsilon.
\]
Therefore,
\[
d(\bu_1,A)
\le \|\bu_1-\ba_\epsilon\|_2
\le \|\bu_1-\mathbf{b}_\epsilon\|_2+\|\mathbf{b}_\epsilon-\ba_\epsilon\|_2
\le d(\bu_1,B)+d_H(A,B)+2\epsilon.
\]
Letting $\epsilon\to 0$ yields
\[
d(\bu_1,A)\le d(\bu_1,B)+d_H(A,B).
\]
Interchanging the roles of $A$ and $B$, we also obtain
\[
d(\bu_1,B)\le d(\bu_1,A)+d_H(A,B).
\]
Combining the two inequalities gives
\[
\bigl| d(\bu_1,A)-d(\bu_1,B) \bigr| \le d_H(A,B).
\]

Next, for any $\mathbf{b}\in B$, the triangle inequality gives
\[
\|\bu_1-\mathbf{b}\|_2 \le \|\bu_1-\bu_2\|_2+\|\bu_2-\mathbf{b}\|_2.
\]
Taking the infimum over $\mathbf{b}\in B$ yields
\[
d(\bu_1,B)\le \|\bu_1-\bu_2\|_2+d(\bu_2,B).
\]
Interchanging $\bu_1$ and $\bu_2$ gives
\[
d(\bu_2,B)\le \|\bu_1-\bu_2\|_2+d(\bu_1,B).
\]
Hence,
\[
\bigl| d(\bu_1,B)-d(\bu_2,B) \bigr| \le \|\bu_1-\bu_2\|_2.
\]
This completes the proof.
\end{proof}

\begin{lemma}[Hausdorff bound under a shared parametrization]\label{lem:Hausdorff_shared_parametrization}
Let $U$ be a nonempty set, and let $f,g:U\to\mathbb{R}^K$ be two functions. Define
\[
A := f(U)
\quad\text{and}\quad
B := g(U).
\]
Then,
\[
d_H(A,B) \le \sup_{\bu\in U} \|f(\bu)-g(\bu)\|_2,
\]
where $d_H(A,B)$ denotes the Hausdorff distance between $A$ and $B$.
\end{lemma}

\begin{proof}
By definition of the Hausdorff distance, it suffices to show that
\[
\sup_{\ba\in A} d(\ba,B)
\le
\sup_{\bu\in U} \|f(\bu)-g(\bu)\|_2
\quad\text{and}\quad
\sup_{\mathbf{b}\in B} d(\mathbf{b},A)
\le
\sup_{\bu\in U} \|f(\bu)-g(\bu)\|_2.
\]

We first prove the first inequality. Let $\ba\in A$ be arbitrary. Then there exists some $\bu\in U$ such that $\ba=f(\bu)$. Since $g(\bu)\in B$, we have
\[
d(\ba,B)
=
\inf_{\mathbf{b}\in B}\|\ba-\mathbf{b}\|_2
\le
\|\ba-g(\bu)\|_2
=
\|f(\bu)-g(\bu)\|_2.
\]
Taking the supremum over $\ba\in A$ yields
\[
\sup_{\ba\in A} d(\ba,B)
\le
\sup_{\bu\in U} \|f(\bu)-g(\bu)\|_2.
\]

The second inequality follows analogously. Indeed, for any $\mathbf{b}\in B$, there exists $\bu\in U$ such that $\mathbf{b}=g(\bu)$, and since $f(\bu)\in A$,
\[
d(\mathbf{b},A)
=
\inf_{\ba\in A}\|\mathbf{b}-\ba\|_2
\le
\|\mathbf{b}-f(\bu)\|_2
=
\|g(\bu)-f(\bu)\|_2.
\]
Taking the supremum over $\mathbf{b}\in B$ gives
\[
\sup_{\mathbf{b}\in B} d(\mathbf{b},A)
\le
\sup_{\bu\in U} \|f(\bu)-g(\bu)\|_2.
\]

Combining the two bounds proves the result.
\end{proof}

\subsection{Deterministic almost-pure recovery}
For every \(l\in[L]\), define an embedding-stability event
\[
    G_l := \left\{ \hat h^{(l)} \le \frac{c_{\mathrm{stab}} \sqrt{\sigma_{B}^{(l)}}}{2K} \right\}.
\]
We state a deterministic result of the corner estimation procedure. The following theorem is used in the proof of Theorem~\ref{thm:VH_almost_pure_recovery}.
\begin{theorem}[Deterministic almost-pure recovery]\label{thm:VH_deterministic}
Suppose that Assumptions~\ref{asm:gram_matrix}, \ref{asm:min_singular_value_of_B}, and \ref{asm:embedding_regular}, together with Condition~\ref{cond:identifiability}, hold. Define the layer-aggregated embedding-error quantities and the block-signal summary
\[
\bar h := \frac{1}{L}\sum_{l=1}^L \hat h^{(l)}, \quad \bar p := \frac{1}{L}\sum_{l=1}^L \mathbf 1_{G_l^c}, \quad \bar\sigma_B := \frac{1}{L}\sum_{l=1}^L \sigma_B^{(l)}.
\]

Moreover, let
\[
\rho_{\mathrm{VH}} := \frac{2K^{3/2}}{c_{\mathrm{Lip}}\bar\sigma_{B}} \left[ \left( 1+\frac{4K}{c_{\mathrm{stab}}\sqrt{\sigma_{B, \inf}}} \right)\bar h + 2\bar p \right]
\]
denote the vertex-hunting error radius. Suppose the corner candidate set $\mathcal{I}$ contains at least one pure node from each community.
Then, every minimizer
\[
\widehat{\mathcal J}\in \operatorname*{arg\,min}_{|\mathcal J|=K}\widetilde\Phi_{L, \mathcal{I}}(\mathcal J)
\]
satisfies
\[
\min_{1\le k\le K} \max_{j\in\widehat{\mathcal J}} Z_{jk} \geq 1 - \rho_{\mathrm{VH}}.
\]
\end{theorem}
\begin{proof}
    Given any corner candidate set $\mathcal{I} \subseteq [n]$, we define the population averaged vertex-hunting objective by
    \[
    \Phi_{L, \mathcal{I}}(\mathcal J) := \max_{i \in \mathcal{I}} \frac{1}{L}\sum_{l=1}^L d\bigl(\bY_{i, \cdot}^{(l)},\,\mathcal{SS}(\mathbf{Y}^{(l)}, \mathcal J)\bigr).
    \]
    For each \(k\in[K]\), choose a pure node \(i_k\) from community \(k\), and define a pure node set
    \[
    \mathcal J^\star:=\{i_1,\ldots,i_K\}.
    \]
    Let
    \[
    \widehat{\mathcal J} \in \operatorname*{arg\,min}_{|\mathcal J|=K} \widetilde\Phi_{L, \mathcal{I}}(\mathcal J)
    \]
    be an arbitrary minimizer of the empirical averaged vertex-hunting objective define in \eqref{eq:def_of_vh_objective}. 
    
    We argue by contradiction. Suppose there exists \(k\in[K]\) such that
    \[
    \mathbf{Z}_{jk}<1-r \quad \text{for all }j\in\widehat{\mathcal J}.
    \]
    Under this assumption, we will show that
    \[
    \widetilde\Phi_{L, \mathcal{I}}(\mathcal J^\star) < \widetilde\Phi_{L, \mathcal{I}}(\widehat{\mathcal J}),
    \]
    which contradicts the optimality of \(\widehat{\mathcal J}\). Fix $k$. First, we obtain a lower bound of $\Phi_{L, \mathcal{I}}(\widehat{\mathcal{J}})$. 
    By Lemma~\ref{lem:pullback_SS},
    \[
    \mathcal{SS}(\mathbf{Y}^{(l)}, \widehat{\mathcal J}) = T^{(l)} \left( \operatorname{conv} \{\bZ_{j,\cdot}:j\in\widehat{\mathcal J}\} \right).
    \] 
    Since the convex hull is compact and \(T^{(l)}\) is continuous on \(\Delta^{K-1}\), the set \(\mathcal{SS}(\mathbf{Y}^{(l)}, \widehat{\mathcal J})\) is compact. Therefore, the infimum in $d(\bY_{i_{k}, \cdot}^{(l)}, \mathcal{SS}(\mathbf{Y}^{(l)}, \widehat{\mathcal{J}}))$ is attained at some point in $\mathcal{SS}(\mathbf{Y}^{(l)}, \widehat{\mathcal{J}})$ for every $l \in [L]$. Let $\by^{\ast(l)} \in \mathcal{SS}(\mathbf{Y}^{(l)}, \widehat{\mathcal{J}})$ such that $$d(\bY_{i_{k}, \cdot}^{(l)}, \mathcal{SS}(\mathbf{Y}^{(l)}, \widehat{\mathcal{J}})) = \| \bY_{i_{k}, \cdot}^{(l)} - \by^{\ast(l)} \|_{2}.$$ By Lemma \ref{lem:pullback_SS}, there exists $\bz^{\ast(l)} \in conv\{ \bz_{j}: j \in \widehat{\mathcal{J}} \}$ such that $T^{(l)}(\bz^{\ast(l)}) = \by^{\ast(l)}$. Note that 
    \begin{equation*}
        \bz_{k}^{\ast(l)} = \sum_{j \in \widehat{\mathcal{J}}} c_{j}^{(l)} \bz_{j}(k) \leq \sum_{j \in \widehat{\mathcal{J}}} c_{j}^{(l)} (1-r) = (1-r).
    \end{equation*}
    By bi-Lipschitz continuity of $T^{(l)}$, 
    \begin{equation*}
        d(\bY_{i_{k}, \cdot}^{(l)}, \mathcal{SS}(\mathbf{Y}^{(l)}, \widehat{\mathcal{J}})) = \| T^{(l)}(\bz^{\ast(l)}) - T^{(l)}(\be_{k}) \|_{2} \geq c_{\mathrm{Lip}} \frac{\sigma_{B}^{(l)}}{K^{3/2}} \| \bz^{\ast(l)} - \be_{k} \|_{2} \geq c_{\mathrm{Lip}} \frac{\sigma_{B}^{(l)}}{K^{3/2}} r .
    \end{equation*}
    Therefore, we obtain a lower bound of $\Phi_{L}(\widehat{\mathcal{J}})$ given by:
    \begin{align}\label{eq:VH_lower_bound_of_PhiLJhat}
        \Phi_{L, \mathcal{I}}(\widehat{\mathcal{J}}) &\geq \frac{1}{L} \sum_{l=1}^{L} d(\bY_{i_{k}, \cdot}^{(l)}, \mathcal{SS}(\mathbf{Y}^{(l)}, \widehat{\mathcal{J}})) \notag \\
        &= \frac{1}{L} \sum_{l=1}^{L} \| T^{(l)}(\be_{k}) - T^{(l)}(\bz^{\ast(l)}) \|_{2} \notag \\
        &\geq \frac{1}{L} \sum_{l=1}^{L} c_{\mathrm{Lip}} \frac{\sigma_{B}^{(l)}}{K^{3/2}}r \notag \\
        &= c_{\mathrm{Lip}} \frac{\bar{\sigma}_{B}}{K^{3/2}} r.
    \end{align}
    On the other hand, since $\mathcal{J}^{\star}$ is a true pure node index set, we have $d(\bY_{i, \cdot}^{(l)}, \mathcal{SS}(\mathbf{Y}^{(l)}, \mathcal{J}^{\star})) = 0$ for all $i \in [n]$. Hence we have 
    $$\Phi_{L, \mathcal{I}}(\mathcal{J}^{\star}) = 0 \quad \text{and} \quad \Phi_{L, \mathcal{I}}(\widehat{\mathcal{J}}) - \Phi_{L, \mathcal{I}}(\mathcal{J}^{\star}) \geq c_{\mathrm{Lip}} \frac{\bar{\sigma}_{B}}{K^{3/2}}r.$$

    Next, we bound \(|\Phi_{L, \mathcal{I}}(\mathcal J)-\widetilde\Phi_{L, \mathcal{I}}(\mathcal J)|\) for an arbitrary index set \(\mathcal J=\{j_1,\dots,j_K\}\) with \(|\mathcal J|=K\). Fix \(l\in[L]\). By Lemma~\ref{lem:stability_of_distance},
    \begin{equation}\label{eq:E_bound_of_abs_diff_dist}
        \bigl| d(\bY_{i, \cdot}^{(l)}, \mathcal{SS}(\mathbf{Y}^{(l)}, \mathcal{J})) - d(\widetilde \bY_{i, \cdot}^{(l)}, {\mathcal{SS}}(\widetilde{\mathbf{Y}}^{(l)}, \mathcal{J})) \bigr| \le \|\bY_{i, \cdot}^{(l)}-\widetilde \bY_{i, \cdot}^{(l)}\|_2 + d_H\!\bigl(\mathcal{SS}(\mathbf{Y}^{(l)}, \mathcal{J}),\mathcal{SS}(\widetilde{\mathbf{Y}}^{(l)}, \mathcal{J})\bigr).
    \end{equation}
    We bound the Hausdorff distance in \eqref{eq:E_bound_of_abs_diff_dist}. Define 
    two matrices $\bS_{\mathcal J}^{(l)},\widetilde{\bS}_{\mathcal J}^{(l)}\in\mathbb{R}^{K\times K}$ by
    \[
        (\bS_{\mathcal J}^{(l)})_{k,\cdot}=(\by_{j_k}^{(l)})^\top,
        \quad (\widetilde{\bS}_{\mathcal J}^{(l)})_{k,\cdot} =(\widetilde{\by}_{j_k}^{(l)})^\top,
        \quad k\in[K].
    \] 
    By Lemma~\ref{lem:Hausdorff_shared_parametrization}, we have
    \begin{equation}\label{eq:E_bound_of_hausdorff_distance}
        d_H \! \bigl( \mathcal{SS}(\mathbf{Y}^{(l)}, \mathcal{J}), \mathcal{SS}^{(l)}(\widetilde{\mathbf{Y}}^{(l)}, \mathcal{J}) \bigr) \leq \sup_{\bz\in\Delta^{K-1}} \left\| \frac{\bS_{\mathcal J}^{(l)\top} \bz} {\|\bS_{\mathcal J}^{(l)\top} \bz\|_2} - \frac{\widetilde \bS_{\mathcal J}^{(l)\top} \bz} {\|\widetilde \bS_{\mathcal J}^{(l)\top} \bz\|_2} \right\|_2.
    \end{equation}
    First, we show the denominators in \eqref{eq:E_bound_of_hausdorff_distance} is bounded away from $0$ on the event \(G_l\). For every \(\bz\in\Delta^{K-1}\),
    \[
    \begin{aligned}
        \|\bS_{\mathcal J}^{(l)\top} \bz\|_2 &\ge \left\langle \frac{\bn^{(l)}}{\|\bn^{(l)}\|_2},\, \bS_{\mathcal J}^{(l)\top} \bz \right\rangle \\
        &= \sum_{m=1}^K \bz_m \left\langle \frac{\bn^{(l)}}{\|\bn^{(l)}\|_2},\, \by_{j_m}^{(l)} \right\rangle \ge c_{\mathrm{stab}} \frac{\sqrt{\sigma_{B}^{(l)}}}{K},
    \end{aligned}
    \]
    and similarly,
\[
\begin{aligned}
\|\widetilde \bS_{\mathcal J}^{(l)\top} \bz\|_2
&\ge
\left\langle \frac{\bn^{(l)}}{\|\bn^{(l)}\|_2},\, \widetilde \bS_{\mathcal J}^{(l)\top} \bz \right\rangle \\
&=
\sum_{m=1}^K \bz_m
\left\langle \frac{\bn^{(l)}}{\|\bn^{(l)}\|_2},\, \widetilde \by_{j_m}^{(l)} \right\rangle
\ge
\frac{c_{\mathrm{stab}} \sqrt{\sigma_{B}^{(l)}}}{2K}.
\end{aligned}
\]
Hence,
\[
\min\bigl\{\|\bS_{\mathcal J}^{(l)\top} \bz\|_2,\ \|\widetilde \bS_{\mathcal J}^{(l)\top} \bz\|_2\bigr\}
\ge
\frac{c_{\mathrm{stab}} \sqrt{\sigma_{B}^{(l)}}}{2K}.
\]
Next, we bound numerator in \eqref{eq:E_bound_of_hausdorff_distance}, we have
\[
\begin{aligned}
\| (\bS_{\mathcal J}^{(l)}-\widetilde \bS_{\mathcal J}^{(l)})^\top \bz \|_2
&=
\left\| \sum_{m=1}^K \bz_m \bigl( \by_{j_m}^{(l)}-\widetilde \by_{j_m}^{(l)} \bigr) \right\|_2 \\
&\le
\sum_{m=1}^K \bz_m \| \by_{j_m}^{(l)}-\widetilde \by_{j_m}^{(l)} \|_2 \\
&\le
\hat h^{(l)}.
\end{aligned}
\]
Therefore, by Lemma~\ref{lem:Hausdorff_shared_parametrization},
\[
\begin{aligned}
d_H\!\bigl(\mathcal{SS}(\mathbf{Y}^{(l)}, \mathcal{J}),\mathcal{SS}^{(l)}(\widetilde{\mathbf{Y}}^{(l)} \mathcal{J})\bigr)
&\le
\sup_{\bz\in\Delta^{K-1}}
\left\|
\frac{\bS_{\mathcal J}^{(l)\top} \bz}{\|\bS_{\mathcal J}^{(l)\top} \bz\|_2}
-
\frac{\widetilde \bS_{\mathcal J}^{(l)\top} \bz}{\|\widetilde \bS_{\mathcal J}^{(l)\top} \bz\|_2}
\right\|_2 \\
&\le
\sup_{\bz\in\Delta^{K-1}}
\frac{2\|(\bS_{\mathcal J}^{(l)}-\widetilde \bS_{\mathcal J}^{(l)})^\top \bz\|_2}
{\min\{\|\bS_{\mathcal J}^{(l)\top} \bz\|_2,\ \|\widetilde \bS_{\mathcal J}^{(l)\top} \bz\|_2\}} \\
&\le
\frac{4 K \hat h^{(l)}}{c_{\mathrm{stab}} \sqrt{\sigma_{B}^{(l)}}}.
\end{aligned}
\]

Hence, on \(G_l\),
\begin{equation}\label{eq:E_bound_of_diff_distance_Gl}
    \bigl| d(\by_i^{(l)}, \mathcal{SS}(\mathbf{Y}^{(l)}, \mathcal{J}))
      - d(\widetilde \by_i^{(l)}, \mathcal{SS}^{(l)}(\widetilde{\mathbf{Y}}^{(l)} \mathcal{J})) \bigr|
\le
\left(1+\frac{4K}{c_{\mathrm{stab}} \sqrt{\sigma_{B}^{(l)}}}\right)\hat h^{(l)}.
\end{equation}

On the complementary event \(G_l^c\), both \(\by_i^{(l)}\) and \(\widetilde \by_i^{(l)}\) lie on the unit sphere, while both \(\mathcal{SS}(\mathbf{Y}^{(l)}, \mathcal{J})\) and \(\mathcal{SS}(\widetilde{\mathbf{Y}}^{(l)}\mathcal{J})\) are nonempty subsets of the unit sphere. Therefore,
\[
0 \le d(\by_i^{(l)}, \mathcal{SS}(\mathbf{Y}^{(l)}, \mathcal{J})) \le 2,
\quad
0 \le d(\widetilde \by_i^{(l)}, \mathcal{SS}(\widetilde{\mathbf{Y}}^{(l)}, \mathcal{J})) \le 2,
\]
and hence
\begin{equation}\label{eq:E_bound_of_diff_distance_Glc}
    \bigl| d(\by_i^{(l)}, \mathcal{SS}(\mathbf{Y}^{(l)}, \mathcal{J}))
      - d(\widetilde \by_i^{(l)}, \mathcal{SS}^{(l)}(\widetilde{\mathbf{Y}}^{(l)}, \mathcal{J})) \bigr|
\le 2.
\end{equation}
Combining \eqref{eq:E_bound_of_diff_distance_Gl} and \eqref{eq:E_bound_of_diff_distance_Glc} yields
\[
\bigl| d(\by_i^{(l)}, \mathcal{SS}(\mathbf{Y}^{(l)}, \mathcal{J}))
      - d(\widetilde \by_i^{(l)}, \mathcal{SS}(\widetilde{\mathbf{Y}}^{(l)}, \mathcal{J})) \bigr|
\le
\left(1+\frac{4K}{c_{\mathrm{stab}} \sqrt{\sigma_{B}^{(l)}}}\right)\hat h^{(l)} \mathbf{1}_{G_l}
+
2\mathbf{1}_{G_l^c}.
\]
Averaging over \(l\) and using \(\sigma_{B}^{(l)} \ge \sigma_{B, \min}\) gives
\[
\begin{aligned}
&\biggl|
\frac{1}{L}\sum_{l=1}^L d(\by_i^{(l)}, \mathcal{SS}(\mathbf{Y}^{(l)}, \mathcal{J}))
-
\frac{1}{L}\sum_{l=1}^L d(\widetilde \by_i^{(l)}, \mathcal{SS}(\widetilde{\mathbf{Y}}^{(l)}, \mathcal{J}))
\biggr| \\
&\qquad\le
\frac{1}{L}\sum_{l=1}^L
\left[
\left(1+\frac{4K}{c_{\mathrm{stab}}\sqrt{\sigma_{B}^{(l)}}}\right)\hat h^{(l)} \mathbf{1}_{G_l}
+
2\mathbf{1}_{G_l^c}
\right] \\
&\qquad\le
\left(1+\frac{4K}{c_{\mathrm{stab}}\sqrt{\sigma_{B, \min}}}\right)\bar h + 2\bar p.
\end{aligned}
\]
Since
\[
|\max_i a_i - \max_i b_i| \le \max_i |a_i-b_i|,
\]
it follows that
\[
|\Phi_{L, \mathcal{I}}(\mathcal J)-\widetilde\Phi_{L, \mathcal{I}}(\mathcal J)|
\le
\left(1+\frac{4K}{c_{\mathrm{stab}}\sqrt{\sigma_{B, \min}}}\right)\bar h + 2\bar p
\quad
\text{for every } \mathcal J \subset [n] \text{ with } |\mathcal J|=K.
\]

Applying this bound with \(\mathcal J=\widehat{\mathcal J}\) and using \eqref{eq:VH_lower_bound_of_PhiLJhat}, we obtain
\[
\widetilde\Phi_{L, \mathcal{I}}(\widehat{\mathcal J})
\ge
\Phi_{L, \mathcal{I}}(\widehat{\mathcal J}) - \left(1+\frac{4K}{c_{\mathrm{stab}}\sqrt{\sigma_{B, \min}}}\right) \bar h - 2\bar p
\ge
c_{\mathrm{Lip}} \frac{\bar\sigma_{B}}{K^{3/2}} r - \left(1+\frac{4K}{c_{\mathrm{stab}}\sqrt{\sigma_{B, \min}}}\right) \bar h - 2\bar p.
\]
On the other hand, since \(\mathcal J^\star\) is a pure node index set, \(\Phi_{L, \mathcal{I}}(\mathcal J^\star)=0\), and thus
\[
\widetilde\Phi_{L, \mathcal{I}}(\mathcal J^\star)
\le
\Phi_{L, \mathcal{I}}(\mathcal J^\star) + \left(1+\frac{4K}{c_{\mathrm{stab}}\sqrt{\sigma_{B, \min}}}\right) \bar h + 2\bar p
=
\left(1+\frac{4K}{c_{\mathrm{stab}}\sqrt{\sigma_{B, \min}}}\right) \bar h + 2\bar p.
\]

Therefore, if
\[
c_{\mathrm{Lip}} \frac{\bar\sigma_{B}}{K^{3/2}} r - \left(1+\frac{4K}{c_{\mathrm{stab}}\sqrt{\sigma_{B, \min}}}\right) \bar h - 2\bar p
>
\left(1+\frac{4K}{c_{\mathrm{stab}}\sqrt{\sigma_{B, \min}}}\right) \bar h + 2\bar p,
\]
then
\[
\widetilde\Phi_{L, \mathcal{I}}(\widehat{\mathcal J}) > \widetilde\Phi_{L, \mathcal{I}}(\mathcal J^\star),
\]
which contradicts the definition
\[
\widehat{\mathcal J}\in \arg\min_{|\mathcal J|=K}\widetilde\Phi_{L, \mathcal{I}}(\mathcal J).
\]
This completes the proof.
\end{proof}

\subsection{Proof of Theorem~\ref{thm:VH_almost_pure_recovery}}
We restate Theorem~\ref{thm:VH_almost_pure_recovery} for convenience. 
\VHalmostpure*
\begin{proof}

By Theorem~\ref{thm:VH_deterministic}, for the minimizer $\widehat{\mathcal{J}}(\mathcal{I})$, it follows that 
\begin{equation*}
    \min_{1 \leq k \leq K} \max_{j \in \widehat{\mathcal{J}}(\mathcal{I})} \mathbf{Z}_{jk} \geq 1 - c(\bar{h} + \bar{p})
\end{equation*}
for a universal constant $c>0$. We show $\bar{p} = O(\bar{h})$ and obtain a high probability upper bound of $\bar{h}$ using Bernstein's inequality. On $G_{l}^{c}$, we have $$\hat{h}^{(l)} > \frac{c_{\mathrm{stab}} \sqrt{\sigma_{B}^{(l)}}}{2K} \geq \frac{c_{\mathrm{stab}} \sqrt{\sigma_{B, \min}}}{2K}.$$ Hence, 
\begin{align*}
    \bar{p} = \frac{1}{L} \sum_{l=1}^{L} \mathbf{1}_{G_{l}^{c}} \leq \frac{1}{L} \sum_{l=1}^{L} \frac{2K \hat{h}^{(l)}}{c_{\mathrm{stab}} \sqrt{\sigma_{B, \min}}} = \frac{2K}{c_{\mathrm{stab}} \sqrt{\sigma_{B, \min}}} \bar{h}.
\end{align*}
Under fixed $K$ and assumption~\ref{asm:min_singular_value_of_B}, we showed $\bar{p} = O(\bar{h})$. 

Next, we obtain a high probability upper bound of $\bar{h}$ using the Bernstein's inequality. Recall that on the event $\mathcal{E}^{(l)}$ defined in Appendix~\ref{sec:Closeness_Xhat}, we have $\hat{h}^{(l)} \leq c_{Y} \sqrt{\beta_{n}^{(l)}}$ with probability at least $1 - c_{E}n^{-18}$. We decompose $\hat{h}^{(l)}$ as below:
\begin{equation*}
    \hat{h}^{(l)} = \hat{h}^{(l)} \mathbf{1}_{\mathcal{E}^{(l)}} + \hat{h}^{(l)} \mathbf{1}_{(\mathcal{E}^{(l)})^{c}} \leq c_{Y} \sqrt{\beta_{n}^{(l)}} + 2 \mathbf{1}_{(\mathcal{E}^{(l)})^{c}}.
\end{equation*}
Consequently, we have $\bar{h} \leq \frac{c_{Y}}{L} \sum_{l=1}^{L} \sqrt{\beta_{n}^{(l)}} + \frac{2}{L} \sum_{l=1}^{L} \mathbf{1}_{(\mathcal{E}^{(l)})^{c}}$. Applying Bernstein's inequality to $\sum_{l=1}^{L} \mathbf{1}_{(\mathcal{E}^{(l)})^{c}}$, we have
\begin{equation*}\label{eq:bernstein_bound_of_elc}
    \mathbb{P} \left( \sum_{l=1}^{L} \mathbf{1}_{(\mathcal{E}^{(l)})^{c}} - \sum_{l=1}^{L} \mathbb{E}[\mathbf{1}_{(\mathcal{E}^{(l)})^{c}}] > \sqrt{2 L c_{E} n^{-18} x} + \frac{2x}{3} \right) \leq \exp(-x).
\end{equation*}
Hence, with probability at least $1 - \exp(-x)$, it follows that
\begin{equation}\label{eq:ub_of_avg_of_indicator}
    \frac{1}{L} \sum_{l=1}^{L} \mathbf{1}_{(\mathcal{E}^{(l)})^{c}} \leq c_{E} n^{-18} + \sqrt{\frac{2 c_{E} x}{Ln^{18}}} + \frac{2x}{3L}.
\end{equation}
By AM-GM inequality, we have
\begin{equation*}
    \frac{1}{L} \sum_{l=1}^{L} \mathbf{1}_{(\mathcal{E}^{(l)})^{c}} \leq \frac{3}{2}c_{E} n^{-18} + \frac{5x}{3L}.
\end{equation*}
Therefore, there exists a universal constant $C_{\mathrm{VH}}$ such that
\begin{equation}\label{eq:ub_of_barh}
    \bar{h} \leq C_{\mathrm{VH}} \left( \frac{1}{L} \sum_{l=1}^{L} \sqrt{\beta_{n}^{(l)}} + n^{-18} + \frac{x}{L} \right)
\end{equation}
with probability at least $1 - \exp(-x)$. 
Using $\bar{p} = O(\bar{h})$ and \eqref{eq:ub_of_barh}, 
\begin{equation*}
    \min_{1 \leq k \leq K} \max_{j \in \widehat{\mathcal{J}}(\mathcal{I})} \mathbf{Z}_{jk} \geq 1 - c(\bar{h} + \bar{p}) \geq 1 - C\bar{h} \geq 1 - C_{\mathrm{VH}} \left( \frac{1}{L} \sum_{l=1}^{L} \sqrt{\beta_{n}^{(l)}} + n^{-18} + \frac{x}{L} \right)
\end{equation*}
with probability at least $1 - \exp(-x)$. This completes the proof. 
\end{proof}

\subsection{Proof of Corollary~\ref{cor:VH_exact}}
We restate Corollary~\ref{cor:VH_exact} for convenience. 
\exactrecovery*

\begin{proof}
    For notational convenience, let $r:= r_{n, L}(x)$. By definition of $\mathcal{I}_{k}(r) = \{ i \in [n]: \bZ_{ik} \geq 1-r \}$, It is straightforward to see $\mathcal{P}_{k} \subset \mathcal{I}_{k}(r)$ for any $r \geq 0$. Therefore, we show $\mathcal{I}_{k}(r) \subset \mathcal{P}_{k}$ with $r < \frac{\delta_{\mathrm{mix}}}{\sqrt{2}}$. Let $j \notin \mathcal{P}_{k}$ be given. There are two possible cases for node $j$:
    \begin{enumerate}
        \item $j \in \mathcal{P}_{k^{'}}$ for some $k^{'} \neq k$.
        \item $j$ is a mixed node. 
    \end{enumerate}
    If $j \in \mathcal{P}_{k^{'}}$ for some $k^{'} \neq k$, then $\bZ_{jk} = 0$. Therefore, $j \notin \mathcal{I}_{k}(r)$ for any $r < 1$. 
    
    Assume $j$ is a mixed node. Then by assumption, we have 
    \begin{equation*}
        \| \bZ_{j, \cdot} - \be_{k} \|_{2} \geq \delta_{\mathrm{mix}}.
    \end{equation*}
    Note that for any $i \in \mathcal{I}_{k}(r)$, it satisfies $\bZ_{ik} \geq 1-r$ and
    \begin{equation*}
        \begin{aligned}
            \| \bZ_{i, \cdot} - \be_{k} \|_{2}^{2} &= (1 - \bZ_{ik})^{2} + \sum_{m \neq k} \bZ_{im}^{2} \\
            &\leq (1 - \bZ_{ik})^{2} + (\sum_{m \neq k} \bZ_{im})^{2} \\
            &= 2 (1-\bZ_{ik})^{2} \\
            &\leq 2r^{2}. 
        \end{aligned}
    \end{equation*}
    Assume $r < \frac{\delta_{\mathrm{mix}}}{\sqrt{2}}$. Since
    \begin{equation*}
        \min_{1 \leq k \leq K} \| \bZ_{j, \cdot} - \be_{k} \|_{2} \geq \delta_{\mathrm{mix}} > \sqrt{2}r,
    \end{equation*}
    it is clear that $j \notin \mathcal{I}_{k}(r)$. This completes the proof.
\end{proof}

\subsection{Proof of Proposition~\ref{prop:medoid_preservation}}
We restate Proposition~\ref{prop:medoid_preservation} for convenience.
\medoidpreservation*
\begin{proof}
    Define the \(k\)-medoids objective with $m$ medoids
    \[
    F(\mathcal M) := \frac{1}{n}\sum_{i=1}^n \min_{m\in\mathcal M} \frac{1}{\sqrt L} \|\widetilde{\mathbf y}_i-\widetilde{\mathbf y}_m\|_2, \quad | \mathcal{M}|=m.
    \]
    Let $\widehat{\mathcal{M}}_{m}$ be any one-swap local minimizer of the $m$-medoids objective used in the denoise step in Section~\ref{sec:2_proposed}. That is,  
    \begin{equation}\label{eq:VH_denoise_one_swap}
        F(\widehat{\mathcal{M}}_{m}) \leq F( (\widehat{\mathcal{M}}_{m} \setminus \widehat{m}) \: \cup \: \{ j \} ), \quad \: \forall \widehat{m} \in \widehat{\mathcal{M}}_{m}, \quad \forall  \: j \notin \widehat{\mathcal{M}}_{m}
    \end{equation}
    with $| \widehat{\mathcal{M}}_{m}|=m$. Let $r := r_{n, L}(x)$. 
    Then, to complete the proof, it suffices to show 
    \begin{equation}\label{eq:VH_denoise_goal}
        \widehat{\mathcal{M}}_{m} \text{ is } r \text{-good with } r < \frac{\delta_{\mathrm{mix}}}{\sqrt{2}}
    \end{equation}
    since Corollary~\ref{cor:VH_exact} can then be applied to $m$ medoids indexed by $\widehat{\mathcal{M}}_{m}$. 
    
Now we prove \eqref{eq:VH_denoise_goal} by contradiction. Suppose that $\widehat{\mathcal{M}}_B=\{\widehat{m}_1,\dots,\widehat{m}_B\}$ is not $r$-good for some $r<\frac{\delta_{\mathrm{mix}}}{\sqrt{2}}$. Then there exists $k\in[K]$ such that
\[
\widehat{\mathcal{M}}_B \cap \mathcal{I}_k(r)=\varnothing.
\]
    For each \(i\in[n]\), define
\[
a(i)
:=
\min \left\{
1\le b\le m :
\|\widetilde{\mathbf y}_i-\widetilde{\mathbf y}_{\widehat m_b}\|_2
=
\min_{1\le t\le m}
\|\widetilde{\mathbf y}_i-\widetilde{\mathbf y}_{\widehat m_t}\|_2
\right\},
\]
that is, if the minimum distance is attained by multiple medoids, we assign \(i\) to the medoid with the smallest index. For each \(b\in[m]\), define the corresponding Voronoi cell
\[
\mathcal C_b
:=
\{\, i\in[n] : a(i)=b \,\}.
\]
Then \(\{\mathcal C_b\}_{b=1}^m\) forms a partition of \([n]\), and hence
\[
\sum_{b=1}^m |\mathcal C_b| = n.
\]
Therefore, by an averaging argument, there exists \(b\in[m]\) such that
\begin{equation*}
|\mathcal C_b| \le \frac{n}{m}.
\end{equation*}
    Define another index set 
    \begin{equation*}
        \mathcal{M}_{m}^{\star} := (\widehat{\mathcal{M}}_{m} \setminus \{ \widehat{m}_{b} \}) \cup \{ i_{k} \},
    \end{equation*}
    where $i_{k}$ is an index of a pure node in community $k$.
    Now we show $F(\mathcal{M}_{B}^{\star}) < F(\widehat{\mathcal{M}}_{B})$ by showing
    \begin{equation*}
        \frac{1}{n} \sum_{i=1}^{n} \frac{1}{\sqrt{L}} \left( \min_{m^{\star} \in \mathcal{M}_{m}^{\star}}\| \widetilde{\by}_{i} - \widetilde{\by}_{m^{\star}} \|_{2} - \min_{m \in \widehat{\mathcal{M}}_{m}} \| \widetilde{\by}_{i} - \widetilde{\by}_{m} \|_{2} \right) < 0.
    \end{equation*}
    
    We partition the node set \([n]\) into the following three disjoint groups:
    \begin{enumerate}
        \item \(\mathcal P_k\),
        \item \(\mathcal C_b \setminus \mathcal P_k\),
        \item \([n]\setminus (\mathcal P_k \cup \mathcal C_b)\).
    \end{enumerate}
    We bound the contribution of each group separately. Let $\hat{h} := \frac{1}{\sqrt{L}} \max_{i} \| \by_{i} - \widetilde{\by}_{i} \|_{2}$.
    First, consider the first group. Note that $i_{k} \in \mathcal{M}_{m}^{\star} \cap \mathcal{P}_{k}$. For $i \in \mathcal{P}_{k}$, 
    \begin{align*}
            \min_{m^{\star} \in \mathcal{M}_{m}^{\star}} \frac{1}{\sqrt{L}}\| \widetilde{\by}_{i} - \widetilde{\by}_{m^{\star}} \|_{2} &\leq \min_{m^{\star} \in \mathcal{M}_{m}^{\star}} \frac{1}{\sqrt{L}} \left( \| \widetilde{\by}_{i} - \by_{i} \|_{2} + \| \by_{i} - \by_{m^{\star}} \|_{2} + \| \by_{m^{\star}} - \widetilde{\by}_{m^{\star}} \|_{2} \right) \\
            &\leq \hat{h} + \min_{m^{\star} \in \mathcal{M}_{m}^{\star}} \frac{1}{\sqrt{L}} \| \by_{i} - \by_{m^{\star}} \|_{2} + \hat{h} \\
            &= 2 \hat{h}.
    \end{align*}
    On the other hand, for $i \in \mathcal{P}_{k}$,
    \begin{align*}
            \min_{m \in \widehat{\mathcal{M}}_{m}} \frac{1}{\sqrt{L}} \| \widetilde{\by}_{i} - \widetilde{\by}_{m} \|_{2} &= \min_{m \in \widehat{\mathcal{M}}_{m}} \frac{1}{\sqrt{L}} \left( \| \widetilde{\by}_{i} - \by_{i} + \by_{i} - \by_{m} + \by_{m} - \widetilde{\by}_{m} \|_{2}  \right) \\
            &\geq \min_{m \in \widehat{\mathcal{M}}_{m}} \frac{1}{\sqrt{L}} \left( \| \by_{i} - \by_{m} \|_{2} - \| \widetilde{\by}_{i} - \by_{i} + \by_{m} - \widetilde{\by}_{m} \|_{2} \right) \\
            &\geq \min_{m \in \widehat{\mathcal{M}}_{m}} \frac{1}{\sqrt{L}} \left( \| \by_{i} - \by_{m} \|_{2} - (\| \widetilde{\by}_{i} - \by_{i} \|_{2} + \| \by_{m} - \widetilde{\by}_{m} \|_{2}) \right) \\
            &\geq \min_{m \in \widehat{\mathcal{M}}_{m}} \frac{1}{\sqrt{L}} \| \by_{i} - \by_{m} \|_{2} - 2 \hat{h} \\
            &\geq \Delta_{0} - 2 \hat{h}.
        \end{align*}
    The last inequality follows from the definition of \(\Delta_0\). Indeed, since \(i\in \mathcal P_k\), we have
\[
\mathbf y_i = T(\mathbf e_k).
\]
Moreover, \(\widehat{\mathcal{M}}_m \cap \mathcal I_k(r)=\varnothing\), and \(r<\delta_{\mathrm{mix}}/\sqrt{2}\) implies \(\mathcal I_k(r)=\mathcal P_k\) by Corollary~\ref{cor:VH_exact}. Hence, for every \(j\in[m]\),
\[
\widehat m_j \notin \mathcal P_k.
\]
Therefore, for each \(j\in[m]\), either \(\mathbf Z_{\widehat m_j,\cdot}\)is a mixed node or \(\mathbf Z_{\widehat m_j,\cdot}=\mathbf e_{k'}\) for some \(k'\neq k\). By definition of \(\Delta_0\), it follows that
\[
\frac{1}{\sqrt{L}} \| \mathbf y_i - \mathbf y_{\widehat m_j} \|_2
=
\frac{1}{\sqrt{L}}\| T(\mathbf e_k) - T(\mathbf Z_{\widehat m_j,\cdot}) \|_2
\ge
\Delta_0.
\]
Consequently,
\[
\min_{j\in[B]} \frac{1}{\sqrt{L}} \| \mathbf y_i - \mathbf y_{\widehat m_j} \|_2 \ge \Delta_0.
\]
    Therefore, for the first group, we have
    \begin{equation}\label{eq:VH_denoise_first_group}
        \min_{m^{\star} \in \mathcal{M}_{m}^{\star}} \frac{1}{\sqrt{L}} \| \widetilde{\by}_{i} - \widetilde{\by}_{m^{\star}} \|_{2} - \min_{m \in \widehat{\mathcal{M}}_{m}} \frac{1}{\sqrt{L}} \| \widetilde{\by}_{i} - \widetilde{\by}_{m} \|_{2} \leq 4 \hat{h} - \Delta_{0}.
    \end{equation}
    
    Next, consider the second group. For each $i \in \mathcal{C}_{b} \setminus \mathcal{P}_{k}$, let
    \begin{equation*}
        m_{i}^{\star} := \arg \min_{m^{\star} \in \mathcal{M}_{m}^{\star}} \| \widetilde{\by}_{i} - \widetilde{\by}_{m^{\star}} \|_{2} \quad \text{and} \quad m_{i} := \arg \min_{m \in \widehat{\mathcal{M}}_{m}} \| \widetilde{\by}_{i} - \widetilde{\by}_{m} \|_{2}.
    \end{equation*}
    Then
    \begin{align}
        \min_{m^{\star} \in \mathcal{M}_{m}^{\star}} \frac{1}{\sqrt{L}} \| \widetilde{\by}_{i} - \widetilde{\by}_{m^{\star}} \|_{2} - \min_{m \in \widehat{\mathcal{M}}_{m}} \frac{1}{\sqrt{L}} \| \widetilde{\by}_{i} - \widetilde{\by}_{m} \|_{2} &= \frac{1}{\sqrt{L}} \| \widetilde{\by}_{i} - \widetilde{\by}_{m_{i}^{\star}} \|_{2} - \frac{1}{\sqrt{L}} \| \widetilde{\by}_{i} - \widetilde{\by}_{m_{i}} \|_{2} \notag\\
        &\leq \frac{1}{\sqrt{L}} \| \widetilde{\by}_{m_{i}^{\star}} - \widetilde{\by}_{m_{i}} \|_{2} \notag\\
        &\leq \frac{1}{\sqrt{L}} \| \widetilde{\by}_{m_{i}^{\star}} \|_{2} + \frac{1}{\sqrt{L}} \| \widetilde{\by}_{m_{i}} \|_{2} \notag\\
        &= 2.
\label{eq:VH_denoise_second_group}
\end{align}
    
    Lastly, consider the third group \(i \notin \mathcal P_k \cup \mathcal C_b\). By the definition of \(\mathcal C_b\), the removed medoid \(\widehat m_b\) is not the assigned nearest medoid of \(i\). Hence there exists some \(m_i \in \widehat{\mathcal{M}}_m \setminus \{\widehat m_b\}\) such that
\[
\|\widetilde{\mathbf y}_i-\widetilde{\mathbf y}_{m_i}\|_2
=
\min_{m\in \widehat{\mathcal{M}}_m}\|\widetilde{\mathbf y}_i-\widetilde{\mathbf y}_m\|_2.
\]
Since
\[
m_i \in \widehat{\mathcal{M}}_m \setminus \{\widehat m_b\} \subset \mathcal M_m^\star,
\]
where \(\mathcal M_m^\star := (\widehat{\mathcal{M}}_m \setminus \{\widehat m_b\}) \cup \{i_k\}\), we obtain
\begin{equation}\label{eq:VH_denoise_third_group}
\min_{m^\star \in \mathcal M_m^\star}
\|\widetilde{\mathbf y}_i-\widetilde{\mathbf y}_{m^\star}\|_2
\le
\|\widetilde{\mathbf y}_i-\widetilde{\mathbf y}_{m_i}\|_2
=
\min_{m\in \widehat{\mathcal{M}}_m}
\|\widetilde{\mathbf y}_i-\widetilde{\mathbf y}_{m}\|_2.
\end{equation}
Therefore, replacing \(\widehat m_b\) by \(i_k\) cannot increase the contribution of any node in the third group.

    Since $\widehat{\mathcal M}_m\cap P_k=\varnothing$
and $i_k\in P_k$, we have
$i_k\notin\widehat{\mathcal M}_m$.
Thus,
\[
\mathcal M_m^\star
=
(\widehat{\mathcal M}_m\setminus\{\widehat m_b\})
\cup\{i_k\}
\]
is a feasible one-swap neighbor of
$\widehat{\mathcal M}_m$.

By \eqref{eq:VH_denoise_one_swap},
\eqref{eq:VH_denoise_first_group},
\eqref{eq:VH_denoise_second_group}, and
\eqref{eq:VH_denoise_third_group}, we have
\begin{align*}
0
&\le
F(\mathcal M_m^\star)-F(\widehat{\mathcal M}_m)\\
&\le
\frac{N_k}{n}(4\widehat h-\Delta_0)
+
\frac{2|C_b\setminus P_k|}{n}\\
&\le
\frac{N_k}{n}(4\widehat h-\Delta_0)
+
\frac{2}{m},
\end{align*}
where the first inequality follows from one-swap local optimality and the last inequality uses $|C_b|\le n/m$. By condition~\ref{cond:denoising_purity_separation}(a), we derive a contradiction of one-swap local optimality of $\widehat{\mathcal{M}}_{m}$ if
\begin{equation}\label{eq:prop_contradiction}
    4 \hat{h} + \frac{2}{m p_{\mathrm{pure}}} < \Delta_{0}.
\end{equation}
To this end, we derive a high probability upper bound of $\hat{h}$. Since $\hat{h}^{2} = \frac{1}{L} \max_{i} \| \by_{i} - \widetilde{\by}_{i} \|_{2}^{2} \leq \frac{1}{L} \sum_{l=1}^{L} (\hat{h}^{(l)})^{2}$, we obtain a high probability upper bound of $\frac{1}{L} \sum_{l=1}^{L} (\hat{h}^{(l)})^{2}$. Using a similar decomposition of $\hat{h}^{(l)}$ in the proof of Theorem~\ref{thm:VH_almost_pure_recovery}, we have
\begin{equation*}
    (\hat{h}^{(l)})^{2} = (\hat{h}^{(l)})^{2} \mathbf{1}_{\mathcal{E}^{(l)}} + (\hat{h}^{(l)})^{2} \mathbf{1}_{(\mathcal{E}^{(l)})^{c}} \leq c_{Y}^{2} \beta_{n}^{(l)} + 4 \mathbf{1}_{(\mathcal{E}^{(l)})^{c}}.
\end{equation*}
Using the same upper bound defined in \eqref{eq:ub_of_avg_of_indicator}, it follows that
\begin{align*}
    \frac{1}{L} \sum_{l=1}^{L} (\hat{h}^{(l)})^{2} &\leq \frac{c_{Y}^{2}}{L} \sum_{l=1}^{L} \beta_{n}^{(l)} + 4 \left( c_{E} n^{-18} + \sqrt{\frac{2 c_{E} x}{Ln^{18}}} + \frac{2x}{3L} \right) \\
    &\leq \frac{c_{Y}^{2}}{L} \sum_{l=1}^{L} \beta_{n}^{(l)} + 4 \left( c_{E} n^{-18} + \frac{c_{E}}{2n^{18}} + \frac{x}{L} + \frac{2x}{3L} \right) \\
    &= \frac{c_{Y}^{2}}{L} \sum_{l=1}^{L} \beta_{n}^{(l)} + 6c_{E}n^{-18} + \frac{20x}{3L},
\end{align*}
with probability at least $1 - \exp(-x)$. Hence, on the same event, we have
\begin{equation*}
    \hat{h} \leq \sqrt{\frac{1}{L} \sum_{l=1}^{L} (\hat{h}^{(l)})^{2}} \leq \left( \frac{c_{Y}^{2}}{L} \sum_{l=1}^{L} \beta_{n}^{(l)} + 6c_{E}n^{-18} + \frac{20x}{3L} \right)^{1/2} \leq C \left( \frac{1}{L} \sum_{l=1}^{L} \beta_{n}^{(l)} + n^{-18} + \frac{x}{L} \right)^{1/2}
\end{equation*}
for a sufficiently large constant $C$. Therefore, if 
\begin{equation*}
    4 C \left( \frac{1}{L} \sum_{l=1}^{L} \beta_{n}^{(l)} + n^{-18} + \frac{x}{L} \right)^{1/2} + \frac{2}{m p_{\mathrm{pure}}} < \Delta_{0},
\end{equation*}
then \eqref{eq:prop_contradiction} holds with probability at least $1 - \exp(-x)$ and this completes the proof. 
\end{proof}

%% file: Sections/Appendix-consistency_membership.tex
\section{Proof of Corollary~\ref{cor:init_membership_consistency}}
We restate Corollary~\ref{cor:init_membership_consistency} for convenience. 
\thmInitMembershipConsistency*

\begin{proof}
By relabeling the estimated communities if necessary, we assume throughout the proof that the estimated corner ordering agrees with the population community ordering. Thus, we suppress the permutation matrix \(\mathbf\Pi\) in what follows. 

Let \(\mathcal E_{\mathrm{VH}}\) denote the near-pure recovery event from Theorem~\ref{thm:VH_almost_pure_recovery}, and let \(\mathcal E_X^{(1)}\) denote the rowwise spectral embedding event for the reference layer. By Theorem~\ref{thm:VH_almost_pure_recovery} and the rowwise embedding bound,
\[
\mathbb P(\mathcal E_{\mathrm{VH}})\ge1-\exp(-x), \qquad \mathbb P(\mathcal E_X^{(1)})\ge1-c_{E}n^{-18}.
\]
Hence,
\[
\mathbb P(\mathcal E_{\mathrm{VH}}\cap\mathcal E_X^{(1)})\ge1-\exp(-x) - c_{E}n^{-18}.
\]
We work on this intersection event throughout the proof. Let $\widehat{\mathcal{J}}$ denote the estimated corner index set by the corner estimation method. Then the following statements hold:
\[
\min_{1\le k\le K}\max_{j\in\widehat{\mathcal J}}\bZ_{jk} \ge 1-r_{n,L}(x),
\]
and, for all \(i\in[n]\),
\[
\left\| \be_i^\top \left( \widehat \bX^{(1)} \bW_*^{(1)} - \bX^{(1)} \right) \right\|_2 \le C_X\theta_i^{(1)} \frac{\sigma_{\min}^{(1)}}{K} \sqrt{\beta_n^{(1)}}.
\]

Let
\[
\widetilde \bX^{(1)} = \widehat \bX^{(1)} \bW_*^{(1)}, \quad \widetilde \bS^{(1)} = \widehat \bS^{(1)} \bW_*^{(1)}, \quad \widetilde \by_i^{(1)} = \widehat \by_i^{(1)}\bW_*^{(1)}.
\]
Since
\[
\widehat \bX^{(1)}(\widehat \bS^{(1)})^{-1} = \widetilde \bX^{(1)}(\widetilde \bS^{(1)})^{-1},
\]
we analyze
\[
\widehat \bU^{(1)} - \bU^{(1)} = \widetilde \bX^{(1)}(\widetilde \bS^{(1)})^{-1} - \bX^{(1)}(\bS^{(1)})^{-1}.
\]

First, we control the normalized embedding error on the reference layer. By the elementary normalization inequality,
\[
\left\| \widetilde \by_i^{(1)} - \by_i^{(1)} \right\|_2 \leq \frac{2 \left\| \widetilde \bX_{i,\cdot}^{(1)} - \bX_{i,\cdot}^{(1)} \right\|_2 }{ \| \bX_{i,\cdot}^{(1)}\|_2 }.
\]
By Lemma~\ref{lem:bound_of_spectral}, we have
\[
\|\bX_{i,\cdot}^{(1)}\|_2 \gtrsim \theta_i^{(1)} \frac{\sigma_{\min}^{(1)}}{K}.
\]
Therefore,
\[
\max_{1\le i\le n} \left\| \widetilde \by_i^{(1)} - \by_i^{(1)} \right\|_2 \le C\sqrt{\beta_n^{(1)}}.
\]

Next, we control the corner matrix error. By the almost-pure recovery event, for each community \(k\), there exists \(\hat i_k\in\widehat{\mathcal J}\) such that
\[
\bZ_{\hat i_k,k}\ge 1-r_{n,L}(x).
\]
Since \(\bZ_{\hat i_k,\cdot} \in \Delta^{K-1}\),
\[
\|\bZ_{\hat i_k,\cdot} - \be_k\|_2 \le \sqrt{2}\,r_{n,L}(x).
\]
Let
\[
\Delta_S := \widetilde \bS^{(1)} - \bS^{(1)}.
\]
Then
\[
\|\Delta_S\|_2 \le \|\Delta_S\|_F \le \sqrt K\|\Delta_S\|_{2,\infty}.
\]
For each \(k\in[K]\), let \(i_k\in\mathcal P_k\) denote a pure node representing the \(k\)-th population corner, so that \(\bS_{k,\cdot}^{(1)}=\by_{i_k}^{(1)}\). Then
\begin{align}\label{eq:F_decomposition_of_Delta_S}
\|\widetilde \bS_{k,\cdot}^{(1)} - \bS_{k,\cdot}^{(1)}\|_2 &= \|\widetilde \by_{\hat i_k}^{(1)} - \by_{i_k}^{(1)}\|_2 \nonumber \\
&\le \|\widetilde \by_{\hat i_k}^{(1)} - \by_{\hat i_k}^{(1)}\|_2 + \|\by_{\hat i_k}^{(1)} - \by_{i_k}^{(1)}\|_2.
\end{align}
The first term is bounded by \(C\sqrt{\beta_n^{(1)}}\). For the second term, using the layerwise Lipschitz bound of \(T^{(1)}\),
\[
\|\by_{\hat i_k}^{(1)} - \by_{i_k}^{(1)}\|_2 = \|T^{(1)}(\bZ_{\hat i_k,\cdot}) - T^{(1)}(\be_k)\|_2 \le \frac{C K}{\sqrt{\sigma_{\min}^{(1)}}} \|\bZ_{\hat i_k,\cdot} - \be_k\|_2.
\]
Since \(K\) is fixed and \(\inf_l\sigma_{\min}^{(l)}\ge c_\sigma>0\), this gives
\[
\|\by_{\hat i_k}^{(1)} - \by_{i_k}^{(1)}\|_2 \le C r_{n,L}(x).
\]
Hence
\begin{equation}\label{eq:upper_bound_of_Delta_S}
    \|\Delta_S\|_2 \le C\left( \sqrt{\beta_n^{(1)}}+r_{n,L}(x) \right).
\end{equation}

We now control the inverse terms. By the identifiability conditions, we have \(\bM^{(1)} = \bI_K\). Since $\bR^{(l)} = \bM^{(l)} \bS^{(l)}$, we have \(\bS^{(1)} = \bR^{(1)}\). Recall that $\sigma_{\min}(\bR^{(l)}) \asymp \sqrt{\sigma_{\min}(\bB^{(l)})}$ by (\ref{eq:D_lower_bound_of_sminR}). We obtain an upper bound of $\| (\bS^{(1)})^{-1} \|_{2}$.
\begin{equation*}
    \begin{aligned}
        \| (\bS^{(1)})^{-1} \|_{2} &= \| (\bR^{(1)})^{-1} \|_{2} \\
        &= \frac{1}{\sigma_{\min}(\bR^{(1)})} \\
        &\lesssim \frac{1}{\sqrt{\sigma_{\min}(\bB^{(1)})}} \\
        &\lesssim \frac{1}{\sqrt{\sigma_{\min}^{(1)}}}.
    \end{aligned}
\end{equation*}

Moreover, by Weyl's inequality,
\[
\sigma_{\min}(\widetilde{\bS}^{(1)}) \ge \sigma_{\min}(\bS^{(1)})-\|\Delta_S\|_2.
\]
We show $\| \Delta_{S} \|_{2} \leq \frac{1}{2} \sigma_{\min}(\bS^{(1)})$. By Assumption~\ref{asm:embedding_regular} and the definition of \(r_{n,L}(x)\),
\[
\sqrt{\beta_n^{(1)}}+r_{n,L}\le C\left( \sqrt{c_\beta} + n^{-18} + \frac{x}{L} \right).
\]
Since \(c_\beta\) is sufficiently small, we have \(C\sqrt{c_\beta}\le s_0/4\), where \(s_0\) is a uniform lower bound for \(\sigma_{\min}(\mathbf S^{(1)})\). Moreover, since \(n^{-18} \to 0 \), there exist $n_{0} \in \mathbb{N}$ such that $Cn^{-18} \leq s_{0}/8$ for every $n \geq n_{0}$. Lastly, take $c_{x} := s_{0}/(8C)$ so that $C x/L \leq C c_{x} \leq s_{0}/8$. Then, for every $n \geq n_{0}$ and $0 < x < c_{x}L$, we have
\begin{align*}
    \| \Delta_{S} \|_{2} \leq \frac{1}{2} \sigma_{\min}(\mathbf{S}^{(1)}).
\end{align*}
Therefore,
\[
\sigma_{\min}(\widetilde{\bS}^{(1)})\ge\frac{1}{2}\sigma_{\min}(\bS^{(1)}),
\]
and hence
\[
\|(\widetilde{\bS}^{(1)})^{-1}\|_2\le2\|(\bS^{(1)})^{-1}\|_2\lesssim\frac{1}{\sqrt{\sigma_{\min}^{(1)}}}.
\]

Also, by Lemma~\ref{lem:bound_of_spectral},
\[
\|\bX_{i,\cdot}^{(1)}\|_2 \lesssim \theta_i^{(1)} \sqrt{K}
\]
and
\[
\|\widetilde \bX_{i,\cdot}^{(1)} - \bX_{i,\cdot}^{(1)}\|_2 \le C_{X} \theta_i^{(1)} \frac{\sigma_{\min}^{(1)}}{K} \sqrt{\beta_n^{(1)}}.
\]
Therefore, by triangle inequality,
\[
\|\widetilde \bX_{i,\cdot}^{(1)}\|_2 \le C\theta_i^{(1)} \sqrt{K}
\]
uniformly over \(i\).

Now decompose
\[
\begin{aligned}
\widehat \bU_{i,\cdot}^{(1)} - \bU_{i,\cdot}^{(1)}
&= \be_i^\top \left\{ \widetilde \bX^{(1)}(\widetilde \bS^{(1)})^{-1} - \bX^{(1)}(\bS^{(1)})^{-1} \right\} \\
&= (\widetilde \bX_{i,\cdot}^{(1)} - \bX_{i,\cdot}^{(1)}) (\bS^{(1)})^{-1} + \widetilde \bX_{i,\cdot}^{(1)} \left\{ (\widetilde \bS^{(1)})^{-1} - (\bS^{(1)})^{-1} \right\}.
\end{aligned}
\]
Using
\[
(\widetilde \bS^{(1)})^{-1} - (\bS^{(1)})^{-1} = (\widetilde \bS^{(1)})^{-1} (\bS^{(1)}-\widetilde \bS^{(1)}) (\bS^{(1)})^{-1},
\]
we obtain
\begin{align}\label{eq:F_decomposition_of_U_diff}
\|\widehat \bU_{i,\cdot}^{(1)} - \bU_{i,\cdot}^{(1)}\|_2
&\le \|\widetilde \bX_{i,\cdot}^{(1)} - \bX_{i,\cdot}^{(1)}\|_2 \|(\bS^{(1)})^{-1}\|_2 + \|\widetilde \bX_{i,\cdot}^{(1)}\|_2 \|(\widetilde \bS^{(1)})^{-1}\|_2 \|\Delta_S\|_2 \|(\bS^{(1)})^{-1}\|_2 \\
&\lesssim \theta_{i}^{(1)} \frac{\sqrt{\sigma_{\min}^{(1)}}}{K} \sqrt{\beta_{n}^{(1)}} + \theta_{i}^{(1)} \frac{\sqrt{K}}{\sigma_{\min}^{(1)}} (\sqrt{\beta_{n}^{(1)}} + r_{n, L}) \nonumber \\
&\le C_{U} \theta_i^{(1)} \left( \sqrt{\beta_n^{(1)}}+r_{n,L}(x) \right) \nonumber\\
&\le C_{U}^{\prime} \theta_{i}^{(1)} \left( \sqrt{c_{\beta}} + n^{-18} + \frac{x}{L} \right) \nonumber.
\end{align}
Since \(c_\beta\) is sufficiently small, by choosing \(c_\beta\) and \(c_x\) sufficiently small and \(n_0\) sufficiently large if necessary, for every $n \geq n_0$ and \(0<x\leq c_{x}L\),
\begin{equation}\label{eq:upper_bound_of_diff_U_i}
\|\widehat{\bU}_{i,\cdot}^{(1)}-\bU_{i,\cdot}^{(1)}\|_2 \le C_U^{\prime}\theta_i^{(1)} \left( \sqrt{c_\beta}+ n^{-18} + \frac{x}{L} \right) \le \frac{\theta_i^{(1)}}{2\sqrt K}.
\end{equation}

It remains to translate this \(U\)-error into membership error. Set
\[
\bu_i := \bU_{i,\cdot}^{(1)}, \quad \widehat \bu_i:=\widehat \bU_{i,\cdot}^{(1)}.
\]
Since \(\bM^{(1)} = \bI_K\),
\[
\bu_i = \theta_i^{(1)}\bZ_{i, \cdot}, \quad \mathbf 1_K^\top \bu_i = \theta_i^{(1)}.
\]
Then
\[
\bZ_{i, \cdot} = \frac{\bu_i}{\mathbf 1_K^\top \bu_i}=\frac{\bu_i}{\theta_i^{(1)}}.
\]
By norm equivalence and \eqref{eq:upper_bound_of_diff_U_i}, we have
\[
\|\widehat \bu_i - \bu_i\|_1 \le \sqrt K\|\widehat \bu_i - \bu_i\|_2 \le \frac{\theta_i^{(1)}}{2}.
\]
Hence we can obtain a lower bound of $\mathbf{1}_{K}^{\top} \widehat{\bu}_{i}$ by following:
\begin{equation*}
    \begin{aligned}
        \mathbf{1}_{K}^{\top} \widehat{\bu}_{i} &= \mathbf{1}_{K}^{\top} \bu_{i} + \mathbf{1}_{K}^{\top}(\widehat{\bu}_{i} - \bu_{i}) \\
        &\geq \mathbf{1}_{K}^{\top} \bu_{i} - \abs{\mathbf{1}_{K}^{\top} (\widehat{\bu}_{i} - \bu_{i})} \\
        &\geq \mathbf{1}_{K}^{\top} \bu_{i} - \| \widehat{\bu}_{i} - \bu_{i} \|_{1} \\
        &\geq \theta_{i}^{(1)} - \frac{\theta_{i}^{(1)}}{2} = \frac{\theta_{i}^{(1)}}{2} > 0.
    \end{aligned}
\end{equation*}
Therefore the row-sum normalization is well-defined.

Now,
\[
\begin{aligned}
\left\| \frac{\widehat \bu_i}{\mathbf 1_K^\top \widehat \bu_i} - \frac{\bu_i}{\mathbf 1_K^\top \bu_i} \right\|_1
&\le \frac{\|\widehat \bu_i - \bu_i\|_1}{\mathbf 1_K^\top \widehat \bu_i} + \|\bu_i\|_1 \left| \frac{1}{\mathbf 1_K^\top \widehat \bu_i} - \frac{1}{\mathbf 1_K^\top \bu_i} \right| \\
&\le \frac{4}{\theta_i^{(1)}} \|\widehat \bu_i - \bu_i\|_1.
\end{aligned}
\]
Therefore,
\[
\left\| \frac{\widehat \bu_i}{\mathbf 1_K^\top \widehat \bu_i} - \bZ_{i, \cdot} \right\|_2 \le \frac{4\sqrt K}{\theta_i^{(1)}} \|\widehat \bu_i - \bu_i\|_2.
\]
Since Euclidean projection onto the simplex is non-expansive and \(\bZ_{i, \cdot} \in \Delta^{K-1}\),
\[
\begin{aligned}
\|(\widehat{\bZ}_{0})_{i, \cdot} - \bZ_{i, \cdot} \|_2
&= \left\| \Pi_{\Delta^{K-1}} \left( \frac{\widehat \bu_i}{\mathbf 1_K^\top \widehat \bu_i} \right) - \Pi_{\Delta^{K-1}}(\bZ_{i, \cdot}) \right\|_2 \\
&\le \left\| \frac{\widehat \bu_i}{\mathbf 1_K^\top \widehat \bu_i} - \bZ_{i, \cdot} \right\|_2 \\
&\le \frac{4\sqrt K}{\theta_i^{(1)}} \|\widehat \bu_i - \bu_i\|_2 \\ 
&\le C_{Z} \left( \sqrt{\beta_n^{(1)}}+r_{n,L}(x) \right).
\end{aligned}
\]

Taking the maximum over \(i\) gives
\begin{equation}\label{eq:F_ub_of_two_to_inf_membership}
    \|\widehat{\bZ}_0-\bZ\|_{2,\infty}\le C_Z\left(\sqrt{\beta_n^{(1)}}+r_{n,L}(x)\right).
\end{equation}
Moreover,
\[
\frac{1}{\sqrt n}\|\widehat{\bZ}_0-\bZ\|_F\le\|\widehat{\bZ}_0-\bZ\|_{2,\infty}.
\]
Substituting the definition of \(r_{n,L}(x)\) and restoring the label permutation \(\mathbf\Pi\), we obtain
\[
\frac{1}{\sqrt n}\|\widehat{\bZ}_0\mathbf\Pi-\bZ\|_F\le\|\widehat{\bZ}_0\mathbf\Pi-\bZ\|_{2,\infty}\le C\left(\sqrt{\beta_n^{(1)}}+\frac{1}{L}\sum_{l=1}^L\sqrt{\beta_n^{(l)}}+n^{-18} + \frac{x}{L} \right).
\]
This bound holds with probability at least \(1-e^{-x} - O(n^{-18})\).

Moreover, suppose that $r_{n,L}(x)<\delta_{\mathrm{mix}}/\sqrt{2}$. On the same event, Corollary~\ref{cor:VH_exact} implies $\hat i_k\in\mathcal P_k$ for every $k\in[K]$. Hence
\[
\by_{\hat i_k}^{(1)} = \by_{i_k}^{(1)} = \bS_{k,\cdot}^{(1)}.
\]
Therefore, the second term in \eqref{eq:F_decomposition_of_Delta_S} vanishes, and
\[
\|\Delta_S\|_2 \le \sqrt K\max_{k\in[K]} \|\widetilde\by_{\hat i_k}^{(1)}-\by_{\hat i_k}^{(1)}\|_2 \le C\sqrt{\beta_n^{(1)}}.
\]
Substituting this sharper corner-matrix bound into the inverse-perturbation argument \eqref{eq:F_decomposition_of_U_diff} gives
\[
\|\widehat\bU_{i,\cdot}^{(1)}-\bU_{i,\cdot}^{(1)}\|_2 \le C\theta_i^{(1)}\sqrt{\beta_n^{(1)}}.
\]
The same row-sum normalization and simplex-projection argument then yields, on the same event and with the same permutation,
\[
\frac1{\sqrt n} \|\widehat{\bZ}_0\mathbf\Pi-\bZ\|_F \le \|\widehat{\bZ}_0\mathbf\Pi-\bZ\|_{2,\infty} \le C\sqrt{\beta_n^{(1)}}.
\]
\end{proof}

%% file: Sections/Appendix-minimax.tex
\section{Minimax lower bound for membership estimation}
Let $\mathcal{F}_{n, K, L}$ denote the parameter space of MLDCMM with $n$ nodes, $K$ communities and $L$ layers satisfying identifiability condition~\ref{cond:identifiability} and assumptions~\ref{asm:gram_matrix}, \ref{asm:min_singular_value_of_B} and \ref{asm:embedding_regular}. In this section, we provide a minimax lower bound of membership estimates under MLDCMM. That is, we obtain a lower bound of 
$$\inf_{\widehat{\mathbf{Z}}} \sup_{\bm{\vartheta} \in \mathcal{F}_{n, K, L}} \mathbb{E}_{\bm{\vartheta}}[\min_{\bm{\Pi} \in \mathcal{Q}_{K}} \|\widehat{\mathbf{Z}} \bm{\Pi} - \mathbf{Z} \|_{2, \infty}],
$$
where $\mathcal{Q}_{K}$ is the set of $K \times K$ permutation matrices. We introduce a main challenge on multilayer \textit{mixed-membership} model and provide a minimax lower bound for membership estimation. 

\subsection{A motivating example}\label{sec:Appendix-minimax-example}
In MLDCMM, given a parameter point $\vartheta$ that satisfies the identifiability condition~\ref{cond:identifiability}, one can construct another parameter point $\vartheta^{'}$ such that both $\vartheta$ and $\vartheta^{'}$ yield the same edge probability matrix $\mathbf{P}^{(l)}$ for all non-reference layers but different edge probability matrix for the reference layer. Without loss of generality, let $l=1$ denote the reference layer. 

Consider the following parameter point $\vartheta_{0}:=(\bm{\Theta}_{0}^{(l)}, \mathbf{Z}_{0}, \mathbf{B}_{0}^{(l)})_{l=1}^{l}$: 
\begin{equation*}
    \bm{\Theta}_{0}^{(l)} = \sqrt{\rho_{n}} \mathbf{I}_{n}, \quad \mathbf{B}_{0}^{(l)} = \begin{pmatrix}
        1 & 1/2 \\
        1/2 & 1
    \end{pmatrix}, \quad \forall l \in [L],
\end{equation*}
and $\mathbf{Z}_{0}$ be a \textit{mixed} membership matrix satisfying the identifiability condition. Then $\vartheta_{0}$ satisfies the identifiability condition~\ref{cond:identifiability}. We construct another such parameter point $\vartheta_{t}:=(\bm{\Theta}_{t}^{(l)}, \mathbf{Z}_{t}, \mathbf{B}_{t}^{(l)})_{l=1}^{L}$. Suppose $0 < t \leq 1$. Let $\mathbf D_t:=\operatorname{diag}(e^t,1)$ be a diagonal matrix that, when multiplied on the right of a membership matrix, changes the relative composition of mixed membership vectors while preserving the community direction of pure membership vectors. Since there exists at least one row with $\| (\mathbf{Z}_{0} \mathbf{D}_{t})_{i, \cdot} \|_{1} \neq 1$, we need another normalization matrix $\mathbf{H}_{t} = \operatorname{diag}(\mathbf{Z}_{0} \mathbf{D}_{t} \mathbf{1}_{2})$ so that $\mathbf{Z}_{t} := \mathbf{H}_{t}^{-1} \mathbf{Z}_{0} \mathbf{D}_{t}$ has unit $\ell_{1}$-norm rows. For $l \geq 2$, define $\bm{\Theta}_{t}^{(l)} := \sqrt{\rho_{n}} \mathbf{H}_{t}$ and $\mathbf{B}_{t}^{(l)} := \mathbf{D}_{t}^{-1} \mathbf{B}_{0}^{(l)} \mathbf{D}_{t}^{-1}$. Then, for all non-reference layers $l \geq 2$, 
\[
\begin{aligned}
\mathbf P_t^{(l)} &= \mathbf{\Theta}_t^{(l)} \mathbf Z_t \mathbf B_t^{(l)} \mathbf Z_t^\top \mathbf{\Theta}_t^{(l)} \\
&= \rho_n \mathbf H_t \mathbf Z_t \mathbf D_t^{-1} \mathbf B_0 \mathbf D_t^{-1} \mathbf Z_t^\top \mathbf H_t \\
&= \rho_n \mathbf Z_0\mathbf B_0\mathbf Z_0^\top = \mathbf P_0^{(l)}.
\end{aligned}
\]
For the reference layer $l=1$, define $\bm{\Theta}_{t}^{(1)}:= \sqrt{\rho_{n}}\mathbf{I}_{n}$ and $\mathbf{B}_{t}^{(1)} = \mathbf{B}_{0}^{(1)}$. Since $\mathbf{Z}_{0} \neq \mathbf{Z}_{t}$, we have
\begin{equation*}
    \mathbf{P}_{t}^{(1)} = \rho_{n} \mathbf{Z}_{t} \mathbf{B}_{0}^{(1)} \mathbf{Z}_{t}^{\top} \neq \rho_{n} \mathbf{Z}_{0} \mathbf{B}_{0}^{(1)} \mathbf{Z}_{0}^{\top} = \mathbf{P}_{0}^{(1)}.
\end{equation*}

Note that the two parameter points \(\vartheta_0\) and \(\vartheta_t\) induce identical edge-probability matrices in all layers except the reference layer. Since both $\vartheta_0$ and $\vartheta_t$ satisfy Condition~\ref{cond:identifiability}, either collection of edge-probability matrices, $(\mathbf P_0^{(l)})_{l=1}^L$ or $(\mathbf P_t^{(l)})_{l=1}^L$, uniquely identifies its corresponding parameter point. However, in this construction, all information that distinguishes $\mathbf Z_0$ from $\mathbf Z_t$ is contained in the reference layer, since the two parameter points induce identical edge-probability matrices in every non-reference layer. Consequently, observing additional non-reference layers does not provide further information for distinguishing the two membership matrices. In this sense, membership estimation can be bottlenecked by a single reference layer even as the number of observed layers increases.

\subsection{A minimax lower bound for membership estimation}
\label{sec:Appendix-minimax-lower-bound}

We now show that, under the MLDCMM parameterization, the membership estimation error can be bottlenecked by a single reference layer. In particular, even when the number of layers increases, there exist parameter configurations for which the non-reference layers contain no information for distinguishing two nearby membership matrices. Throughout this subsection, we fix $K=2$ and let $l=1$ denote the reference layer. The lower bound below therefore also applies to any larger parameter space containing this class. We restate Theorem~\ref{thm:minimax-membership-lower-bound} for convenience.

\thmMinimaxLowerbound*
\begin{proof}
We prove the result using Le Cam's two-point method \citep{LeCam1973, Yu1997}. We show the proof when $K=2$ and extend for $K>2$ at the end of the proof.

\paragraph{Step 1: Construction of two parameter points.}

In this step, we construct two parameter points in $\mathcal{F}_{n, 2, L}$ that satisfy the required conditions and assumptions. Fix $\rho_{n} \in (0, 1]$ such that $n \rho_{n} \geq C_{\rho} \log n$ for a sufficiently large $C_{\rho}$ to be specified below. We construct one point $\vartheta_{0} := (\mathbf{\Theta}_{0}^{(l)}, \mathbf{Z}_{0}, \mathbf{B}_{0}^{(l)})_{l=1}^{L} \in \mathcal{F}_{n, 2, L}$ by following: let $\mathbf B_0^{(l)} := \mathbf{B}_{0} = \begin{pmatrix}
1 & 1/2\\
1/2 & 1 
\end{pmatrix}$ and $\mathbf{\Theta}_{0}^{(l)} := \sqrt{\rho_{n}} \mathbf{I}_{n}$ for all $l \in [L]$. Let $\mathbf Z_0$ contain exactly one mixed node $i_*$ satisfying $(\mathbf Z_0)_{i_*,\cdot} = \left(\frac12,\frac12\right),$ and let all remaining nodes be $n_{1}$ pure nodes for the first community and $n_{2}$ pure nodes for the second community with $n_{1} + n_{2} = n - 1$ and $\min(n_{1}, n_{2}) \geq \frac{n}{2}-1$. 

Next, we construct $\vartheta_{t} := (\mathbf{\Theta}_{t}^{(l)}, \mathbf{Z}_{t}, \mathbf{B}_{t}^{(l)})_{l=1}^{L} \in \mathcal{F}_{n, 2, L}$. For $0<t\le1$, define $\mathbf{D}_{t} = \operatorname{diag}(e^{t}, 1)$, $\mathbf H_t = \operatorname{diag}(\mathbf Z_0\mathbf D_t\mathbf 1_2)$ and $\mathbf Z_t = \mathbf H_t^{-1}\mathbf Z_0\mathbf D_t.$ For every non-reference layer $l\ge2$, set $\mathbf{\Theta}_t^{(l)} = \sqrt{\rho_n}\mathbf H_t$ and $\mathbf B_t^{(l)} = \mathbf D_t^{-1}\mathbf B_0\mathbf D_t^{-1}.$
For the reference layer, set $\mathbf{\Theta}_t^{(1)} = \sqrt{\rho_n}\mathbf I_n,$ and $\mathbf B_t^{(1)} = \mathbf B_0.$ By construction, we have $\mathbf{P}_{0}^{(1)} \neq \mathbf{P}_{t}^{(1)}$ and $\mathbf{P}_{0}^{(l)} = \mathbf{P}_{t}^{(l)}$ for all $l \geq 2$.

We show that the two parameter points $\vartheta_0$ and $\vartheta_t$ satisfy Conditions~\ref{cond:identifiability} and \ref{cond:denoising_purity_separation}, as well as Assumptions~\ref{asm:gram_matrix}, \ref{asm:min_singular_value_of_B}, and \ref{asm:embedding_regular}.
\begin{itemize}
    \item \textit{Condition~\ref{cond:identifiability}.} Both $\mathbf{Z}_{0}$ and $\mathbf{Z}_{t}$ have unit $\ell_{1}$-norm rows and there exists at least $\frac{n}{2}-1$ pure nodes for each community. Moreover, $i_*$-th node of $\mathbf{Z}_{0}$ and $\mathbf{Z}_{t}$ satisfies the second condition of Condition~\ref{cond:identifiability}. Lastly, we have the same block connectivity matrix for the reference layer: $\mathbf{B}_{0}^{(1)} = \mathbf{B}_{t}^{(1)} = \begin{pmatrix}
        1 & 1/2 \\
        1/2 & 1
    \end{pmatrix}.$ Since $\mathbf{B}_{0}$ is positive definite, the condition $\| (\mathbf{R}^{(1)})_{k, \cdot} \|_{2} = 1$ for all $k \in [K]$ is equivalent to $\operatorname{diag}(\mathbf{B}^{(1)}) = \mathbf{1}_{K}$, which is true for both $\mathbf{B}_{0}^{(1)}$ and $\mathbf{B}_{t}^{(1)}$. It is clear that $\mathbf{B}_{0}^{(l)}$ is full rank and $\| \mathbf{B}_{0}^{(l)} \|_{\max} = 1$ for all $l \in [L]$. For $\mathbf{B}_{t}^{(l)}$, we have
    \[
    \mathbf B_t^{(l)} = \begin{pmatrix} e^{-2t} & \frac12e^{-t}\\ \frac12e^{-t} & 1 \end{pmatrix}, \quad l\ge2.
    \]
    Therefore, $\mathbf{B}_{t}^{(l)}$ is full rank and $\| \mathbf{B}_{t}^{(l)} \|_{\max} = 1$ for all $l \in [L]$.

    \item \textit{Condition~\ref{cond:denoising_purity_separation}.} By construction, 
    \begin{equation*}
        \min(n_{1}, n_{2}) = \begin{cases}
            \frac{n}{2} - 1 & \text{if } n \text{ is even,} \\
            \frac{n-1}{2} & \text{if } n \text{ is odd}.
        \end{cases}
    \end{equation*}
    Hence, when $n \geq 4$ is even, $\min_{1 \leq k \leq 2} \frac{n_{k}}{n} = \frac{1}{2} - \frac{1}{n} \geq \frac{1}{4}$. When $n \geq 3$ is odd, $\min_{1 \leq k \leq 2} \frac{n_{k}}{n} = \frac{n-1}{2n} \geq \frac{1}{3} > \frac{1}{4}$. Therefore, we can fix $p_{\mathrm{pure}} = \frac{1}{4}$. It is clear that we can choose $\delta_{\mathrm{mix}} = \frac{1}{\sqrt{2}}$ for $\mathbf{Z}_{0}$. The mixed node of $\mathbf{Z}_{t}$ is computed as $(\mathbf{Z}_{t})_{i_{*}, \cdot} = (\frac{e^{t}}{1+e^{t}}, \frac{1}{1+e^{t}})^{\top}$. For any $t \in (0, 1]$, we have
    \begin{equation*}
        \min_{k=1, 2} \| (\mathbf{Z}_{t})_{i_{*}, \cdot} - \mathbf{e}_{k} \|_{2} = \frac{\sqrt{2}}{1+e^{t}} \geq \frac{\sqrt{2}}{1+e}.
    \end{equation*}
    Therefore, we can choose $\delta_{\mathrm{mix}} = \frac{\sqrt{2}}{1+e}$ for $\mathbf{Z}_{t}$.

    \item \textit{Assumption~\ref{asm:gram_matrix}.} Since $\mathbf{G}_{0}^{(l)}$ and $\mathbf{G}_{t}^{(l)}$ are symmetric positive definite, we check Assumption~\ref{asm:gram_matrix} by showing boundedness of the eigenvalues of $(\mathbf{G}_{0}^{(l)})^{2} = \frac{K}{\|\bm{\theta}_{0}^{(l)}\|_{2}^{2}} \mathbf{Z}_{0}^{\top} (\bm{\Theta}_{0}^{(l)})^{2} \mathbf{Z}_{0}$ and $(\mathbf{G}_{t}^{(l)})^{2} = \frac{K}{\|\bm{\theta}_{t}^{(l)}\|_{2}^{2}} \mathbf{Z}_{t}^{\top} (\bm{\Theta}_{t}^{(l)})^{2} \mathbf{Z}_{t}$. Consider $(\mathbf{G}_{0}^{(l)})^{2}$. We have 
    \begin{equation*}
        (\mathbf{G}_{0}^{(l)})^{2} = \frac{2}{n\rho_{n}} \mathbf{Z}_{0}^{\top} (\bm{\Theta}_{0}^{(l)})^{2} \mathbf{Z}_{0} = \frac{2}{n} \mathbf{Z}_{0}^{\top} \mathbf{Z}_{0} = \frac{2}{n} \left( \begin{pmatrix}
            n_{1} & 0 \\
            0 & n_{2}
        \end{pmatrix} + \begin{pmatrix}
            1/4 & 1/4 \\
            1/4 & 1/4
        \end{pmatrix} \right) =: \frac{2}{n} \mathbf{C}_{0}^{(l)}, \ \forall l \in [L].
    \end{equation*}
    Note that $\mathbf{C}_{0}^{(l)}$ is positive definite. Hence, $\lambda_{\max}(\mathbf{C}_{0}^{(l)}) \leq \operatorname{tr}(\mathbf{C}_{0}^{(l)}) = n - \frac{1}{2} < n$ and $\lambda_{\max}\left((\mathbf{G}_{0}^{(l)})^{2} \right) = \frac{2}{n} \lambda_{\max}(\mathbf{C}_{0}^{(l)}) < 2$. On the other hand, by Weyl's inequality, we have 
    \begin{equation*}
        \lambda_{\min}(\mathbf{C}_{0}^{(l)}) \geq \lambda_{\min} \left( \begin{pmatrix}
            n_{1} & 0 \\
            0 & n_{2}
        \end{pmatrix} \right) + \lambda_{\min} \left( \begin{pmatrix}
            1/4 & 1/4 \\
            1/4 & 1/4
        \end{pmatrix} \right) = \min(n_{1}, n_{2}) \geq \frac{n}{2} - 1.
    \end{equation*}
    Therefore, $\lambda_{\min}\left( (\mathbf{G}_{0}^{(l)})^{2} \right) = \frac{2}{n} \lambda_{\min}(\mathbf{C}_{0}^{(l)}) \geq 1 - \frac{2}{n} \geq \frac{1}{3}$. 

    We show $(\mathbf{G}_{t}^{(l)})^{2}$ satisfies the boundedness eigenvalue condition. Consider the reference layer $l=1$. Let $q_{t} := \frac{e^{t}}{1+e^{t}}$ so that $(\mathbf{Z}_{t})_{i_{*}, \cdot} = (q_{t}, 1-q_{t})^{\top}$. Then
    \begin{equation*}
        (\mathbf{G}_{t}^{(1)})^{2} = \frac{2}{n\rho_{n}} \rho_{n} \mathbf{Z}_{t}^{\top} \mathbf{Z}_{t} = \frac{2}{n} \left( \begin{pmatrix}
            n_{1} & 0 \\
            0 & n_{2}
        \end{pmatrix}  + (\mathbf{Z}_{t})_{i_{*}, \cdot} (\mathbf{Z}_{t})_{i_{*}, \cdot}^{\top} \right).
    \end{equation*}
    Similar to the case of $(\mathbf{G}_{0}^{(l)})^{2}$, one can obtain $\frac{1}{3} \leq \lambda_{\min}\left( (\mathbf{G}_{t}^{(1)})^{2} \right) \leq \lambda_{\max}\left( (\mathbf{G}_{t}^{(1)})^{2} \right) <2$. Next, consider $l \geq 2$. For non-reference layers, we have $\bm{\Theta}_{t}^{(l)} = \sqrt{\rho_{n}} \mathbf{H}_{t}$ and $\mathbf{Z}_{t} = \mathbf{H}_{t}^{-1} \mathbf{Z}_{0} \mathbf{D}_{t}$. Hence, 
    \begin{equation*}
        \mathbf{Z}_{t}^{\top} (\bm{\Theta}_{t}^{(l)})^{2} \mathbf{Z}_{t} = \rho_{n} \mathbf{D}_{t} \mathbf{Z}_{0}^{\top} \mathbf{Z}_{0} \mathbf{D}_{t} \quad \text{and} \quad (\mathbf{G}_{t}^{(l)})^{2} = \frac{2 \rho_{n}}{\|\bm{\theta}_{t}^{(l)}\|_{2}^{2}} \mathbf{D}_{t} \mathbf{Z}_{0}^{\top} \mathbf{Z}_{0} \mathbf{D}_{t}.
    \end{equation*}
    Since $0 < t \leq 1$, $1 \leq (H_{t})_{ii} \leq e$ and hence $n \rho_{n} \leq \| \bm{\theta}_{t}^{(l)} \|_{2}^{2} \leq n \rho_{n} e^{2}$. Moreover, $\sigma_{\max}(\mathbf{D}_{t}) = e^{t} \leq e$ and $\sigma_{\min}(\mathbf{D}_{t}) = 1$. We obtain an upper bound of $\lambda_{\max} \left( (\mathbf{G}_{t}^{(l)})^{2} \right)$. 
    \begin{align*}
        \lambda_{\max} \left( (\mathbf{G}_{t}^{(l)})^{2} \right) &= \frac{2 \rho_{n}}{\|\bm{\theta}_{t}^{(l)}\|_{2}^{2}} \lambda_{\max}(\mathbf{D}_{t} \mathbf{Z}_{0}^{\top} \mathbf{Z}_{0} \mathbf{D}_{t}) \\
        &= \frac{2 \rho_{n}}{\|\bm{\theta}_{t}^{(l)}\|_{2}^{2}} (\sigma_{\max}(\mathbf{Z}_{0} \mathbf{D}_{t}))^{2} \\
        &\leq \frac{2 \rho_{n}}{\|\bm{\theta}_{t}^{(l)}\|_{2}^{2}} (\sigma_{\max}(\mathbf{Z}_{0}))^{2} (\sigma_{\max}(\mathbf{D}_{t}))^{2} \\
        &= \frac{2 \rho_{n}}{\|\bm{\theta}_{t}^{(l)}\|_{2}^{2}} \lambda_{\max}(\mathbf{Z}_{0}^{\top} \mathbf{Z}_{0}) (\sigma_{\max}(\mathbf{D}_{t}))^{2} \\
        &\leq \frac{2 \rho_{n}}{n \rho_{n}} (n - 1/2) e^{2} \\
        &\leq 2e^{2}.
    \end{align*}
    We obtain a lower bound of $\lambda_{\min}  \left( (\mathbf{G}_{t}^{(l)})^{2} \right)$. Since $\mathbf{Z}_{0}$ is full column rank and $\mathbf{D}_{t}$ is invertible, $\sigma_{\min}(\mathbf{Z}_{0} \mathbf{D}_{t}) \geq \sigma_{\min}(\mathbf{Z}_{0}) \sigma_{\min}(\mathbf{D}_{t})$. Therefore,
    \begin{align*}
        \lambda_{\min} \left( (\mathbf{G}_{t}^{(l)})^{2} \right) &= \frac{2 \rho_{n}}{\|\bm{\theta}_{t}^{(l)}\|_{2}^{2}} \lambda_{\min}(\mathbf{D}_{t} \mathbf{Z}_{0}^{\top} \mathbf{Z}_{0} \mathbf{D}_{t}) \\
        &= \frac{2 \rho_{n}}{\|\bm{\theta}_{t}^{(l)}\|_{2}^{2}} (\sigma_{\min}(\mathbf{Z}_{0} \mathbf{D}_{t}))^{2} \\
        &\geq \frac{2 \rho_{n}}{\|\bm{\theta}_{t}^{(l)}\|_{2}^{2}} (\sigma_{\min}(\mathbf{Z}_{0}))^{2} (\sigma_{\min}(\mathbf{D}_{t}))^{2} \\
        &= \frac{2 \rho_{n}}{\|\bm{\theta}_{t}^{(l)}\|_{2}^{2}} \lambda_{\min}(\mathbf{Z}_{0}^{\top} \mathbf{Z}_{0}) (\sigma_{\min}(\mathbf{D}_{t}))^{2} \\
        &\geq \frac{2 \rho_{n}}{n \rho_{n} e^{2}} (n/2 - 1) 1^{2} \\
        &\geq \frac{1}{e^{2}} - \frac{2}{ne^{2}} \\
        &\geq \frac{1}{3e^{2}} \qquad \forall n \geq 3.
    \end{align*}
    Consequently, Assumption~\ref{asm:gram_matrix} holds for both parameter points with the common constant $C_G=\sqrt{3}e$.
    \item \textit{Assumption~\ref{asm:min_singular_value_of_B}.} Since $\sigma_{\min}(\mathbf{B}_{0}^{(l)}) = 0.5$, one can choose $\sigma_{B_{0}}^{(l)} = 0.5$ for all $l \in [L]$. For the reference layer $l=1$, it follows that $\mathbf{B}_{t}^{(1)} = \mathbf{B}_{0}$. Hence one can choose $\sigma_{B_{t}}^{(1)} = 0.5$. For non-reference layers $l \geq 2$, we have $\mathbf{B}_{t}^{(l)} = \mathbf{D}_{t}^{-1} \mathbf{B}_{0} \mathbf{D}_{t}^{-1}$. Since $\mathbf{D}_{t}^{-1}$ and $\mathbf{B}_{0}$ are invertible, it follows that $\sigma_{\min}(\mathbf{B}_{t}^{(l)}) = \sigma_{\min}(\mathbf{D}_{t}^{-1} \mathbf{B}_{0} \mathbf{D}_{t}^{-1}) \geq (\sigma_{\min}(\mathbf{D}_{t}^{-1}))^{2} \sigma_{\min}(\mathbf{B}_{0})$. For $0 < t \leq 1$, $\sigma_{\min}(\mathbf{D}_{t}^{-1}) = e^{-t} \geq e^{-1}$. Therefore, 
    \begin{equation*}
        \sigma_{\min}(\mathbf{B}_{t}^{(l)}) \geq (\sigma_{\min}(\mathbf{D}_{t}^{-1}))^{2} \sigma_{\min}(\mathbf{B}_{0}) \geq \left( \frac{1}{e^{1}} \right)^{2} \cdot \frac{1}{2} = \frac{1}{2e^{2}}.
    \end{equation*}
    Thus, one can choose $\sigma_{B_{t}}^{(l)} = \frac{1}{2e^{2}}$ for all $l \in [L]$.

    \item \textit{Assumption~\ref{asm:embedding_regular}.} From the point $\vartheta_{0}$, for all $l \in [L]$, we have
    \begin{equation*}
        \theta_{\min, 0}^{(l)} \| \bm{\theta}_{0}^{(l)}\|_{1} = \sqrt{\rho_{n}} \cdot n\sqrt{\rho_{n}} \geq C_{\rho} \log(n) \quad \text{and} \quad \beta_{n, 0}^{(l)} = \left( \frac{\sqrt{\rho_{n}}}{\sqrt{\rho_{n}}} \right) \frac{2^{8}\sqrt{\rho_{n}} n \sqrt{\rho_{n}} \log(n)}{(1/2)^{3} (n \rho_{n})^{2}} = \frac{2^{11} \log(n)}{n\rho_{n}} \leq \frac{2^{11}}{C_{\rho}}.
    \end{equation*}
    Next, consider $\vartheta_{t}$. For the reference layer $l=1$, we have the same results as above. For non-reference layers $l \geq 2$, we have $\theta_{\min, t}^{(l)} = \sqrt{\rho_{n}}$ and $\theta_{\max, t}^{(l)} = e^{t} \sqrt{\rho_{n}}$. Since $1 \leq (\mathbf{H}_{t})_{ii} \leq e^{t}$, we have
    \begin{equation*}
        n\sqrt{\rho_{n}} \leq \| \bm{\theta}_{t}^{(l)} \|_{1} \leq en\sqrt{\rho_{n}} \quad \text{and} \quad n\rho_{n} \leq \| \bm{\theta}_{t}^{(l)} \|_{2}^{2} \leq e^{2} n \rho_{n}.
    \end{equation*}
    For the first condition, it follows that
    \begin{equation*}
        \theta_{\min, t}^{(l)} \| \bm{\theta}_{t}^{(l)} \|_{1} \geq \sqrt{\rho_{n}} \cdot n \sqrt{\rho_{n}} \geq C_{\rho} \log(n). 
    \end{equation*}
    For the second condition, we have
    \begin{align*}
        \beta_{n, t}^{(l)} &= \left( \frac{e^{t} \sqrt{\rho_{n}}}{\sqrt{\rho_{n}}} \right) \frac{2^{8} e^{t} \sqrt{\rho_{n}} \| \bm{\theta}_{t}^{(l)}\|_{1} \log(n)}{(\frac{1}{2e^{2}})^{3} \| \bm{\theta}_{t}^{(l)} \|_{2}^{4}} \\
        &\leq e^{t} \cdot \frac{2^{11} e^{6+t} \sqrt{\rho_{n}} en\sqrt{\rho_{n}} \log(n)}{(n \rho_{n})^{2}} \\
        &= \frac{2^{11} e^{7+2t} \log(n)}{n \rho_{n}} \\
        &\leq \frac{2^{11} e^{7+2t}}{C_{\rho}}.
    \end{align*}
    For $0 < t \leq 1$, it follows that $\beta_{n}^{(l)} \leq \frac{2^{11} e^{9}}{C_{\rho}}$. Let $C_{\rho} \geq \max \{c_{\mathrm{deg}}, \frac{2^{11}e^{9}}{c_{\beta}} \}$. Then both $\vartheta_{0}$ and $\vartheta_{t}$ satisfy the Assumption~\ref{asm:embedding_regular}.
\end{itemize}
\paragraph{Step 2: Separation between the two membership matrices.}

The transformation leaves every pure membership vector unchanged. For the unique mixed node, we have $(\mathbf Z_t)_{i_*,\cdot} = (q_t,1-q_t)$, where $q_t=\frac{e^t}{1+e^t}$. Therefore, $\Delta_t := \|\mathbf Z_t-\mathbf Z_0\|_{2,\infty} = \sqrt{2} \left(q_t-\frac12\right).$
Since $q_t-\frac12 = \frac{e^t-1}{2(e^t+1)}$, for $0<t\le1$,
\begin{equation}\label{eq:G_ub_of_qt-1/2}
    \frac{t}{2(e+1)} \le q_t-\frac12 \le \frac t4.
\end{equation}
Consequently, $\Delta_t\asymp t$.

We next verify that label permutation does not reduce the separation for this two-point construction. There are only two permutations in $\mathcal P_2$. For the identity permutation, the distance is $\Delta_t$. For the nontrivial permutation $\Pi_{\mathrm{swap}}$, consider any pure node in the first community. Since the transformation leaves this node pure in the first community, $(\mathbf Z_0)_{i,\cdot} = (\mathbf Z_t)_{i,\cdot} = (1,0)$.
Thus, $$\|(\mathbf Z_t \bm{\Pi}_{\mathrm{swap}})_{i,\cdot} -(\mathbf Z_0)_{i,\cdot}\|_2 = \sqrt{2}.$$ Since $\Delta_t<\sqrt{2}$ for $0<t\le1$, the identity permutation is the minimizer. Hence
\begin{equation}\label{eq:G_rate_of_Deltat}
    \min_{\bm{\Pi}\in\mathcal P_2} \|\mathbf Z_t\bm{\Pi}-\mathbf Z_0\|_{2,\infty} = \Delta_t \asymp t.
\end{equation}

\paragraph{Step 3: Kullback--Leibler divergence.}

Let $\mathbb P_t$ and $\mathbb P_0$ denote the joint distributions of the multilayer adjacency matrices under the two parameter points. Since $\mathbf P_t^{(l)} = \mathbf P_0^{(l)}$ for all $l \geq 2$, only the reference layer contributes to the KL divergence.

Moreover, within the reference layer, only edges incident to the mixed node $i_*$ are changed. For every $j\neq i_*$, $(\mathbf P_0^{(1)})_{i_*,j} = \frac34\rho_n.$ Regardless of whether $j$ belongs to the first or the second community,
\[
\left| (\mathbf P_t^{(1)})_{i_*,j} - (\mathbf P_0^{(1)})_{i_*,j} \right| =\left| \rho_{n} \left( \frac{1}{2} + \frac{q_{t}}{2} \right) - \frac{3}{4} \rho_{n} \right| = \frac{\rho_n}{2} \left|q_t-\frac12\right|.
\]

For Bernoulli distributions with parameter $p$ and $q$, we have
\[
\operatorname{kl} \bigl( \operatorname{Bern}(p), \operatorname{Bern}(q) \bigr) \le \frac{(p-q)^2}{q(1-q)}.
\]
Let $p := (\mathbf P_t^{(1)})_{i_*,j}$ and $q := (\mathbf P_0^{(1)})_{i_*,j}$. Since $q=\frac34\rho_n$ and $\rho_n\le1$,
\begin{equation}\label{eq:G_lb_of_denom}
    q(1-q) = \frac34\rho_n \left(1-\frac34\rho_n\right) \ge \frac{3}{16}\rho_n.
\end{equation}
By \eqref{eq:G_ub_of_qt-1/2} and \eqref{eq:G_lb_of_denom}, it follows that
\[
\operatorname{kl} \left( \operatorname{Bern} \bigl((\mathbf P_t^{(1)})_{i_*,j}\bigr), \operatorname{Bern} \bigl((\mathbf P_0^{(1)})_{i_*,j}\bigr) \right) \le \frac43 \rho_n \left(q_t-\frac12\right)^2 \le \frac1{12}\rho_n t^2.
\]
Summing over the $n-1$ affected edges yields
\[
D_{\mathrm{KL}}(\mathbb P_t\|\mathbb P_0) \le \frac{n\rho_n t^2}{12}.
\]

Now choose $t_n := \frac{1}{\sqrt{n\rho_n}}$. Then $0<t_n\le1$ and $D_{\mathrm{KL}}(\mathbb P_{t_n}\|\mathbb P_0) \le \frac1{12}$.
By Pinsker's inequality,
\begin{equation}\label{eq:G_ub_of_TV}
    \operatorname{TV}(\mathbb P_{t_n},\mathbb P_0) \le \sqrt{ \frac12 D_{\mathrm{KL}}(\mathbb P_{t_n}\|\mathbb P_0) } < \frac14.
\end{equation}

\paragraph{Step 4: Le Cam's two-point argument.}
In this step, we use Le Cam's method \citep[Lemma~1]{Yu1997} to obtain a lower bound of $\max_{\vartheta\in\{\vartheta_0,\vartheta_{t_n}\}} \mathbb E_{\vartheta} \left[ d(\widehat{\mathbf Z},\mathbf Z_{\vartheta}) \right]$. Recall that we defined the distance $d(\widehat{\mathbf{Z}}, \mathbf{Z}_{\vartheta}) := \min_{\bm{\Pi}\in\mathcal P_2} \|\widehat{\mathbf Z}\bm{\Pi} - \mathbf Z_{\vartheta}\|_{2,\infty}$. We show $d$ is a pseudo-metric on the space of membership matrices. Nonnegativity and $d(\mathbf Z,\mathbf Z)=0$ are immediate, while symmetry follows from the fact that $\bm{\Pi}^{-1}\in\mathcal P_2$ and from the invariance of the $2\to\infty$ norm under column permutations. It remains to verify the triangle inequality. For any $\mathbf Z_1,\mathbf Z_2,\mathbf Z_3$, we show 
\[
d(\mathbf Z_1,\mathbf Z_3)
\le
d(\mathbf Z_1,\mathbf Z_2)
+
d(\mathbf Z_2,\mathbf Z_3).
\]
Let $\bm{\Pi}_{12}$ and $\bm{\Pi}_{23}$ are minimizing permutations for $d(\mathbf Z_1,\mathbf Z_2)$ and $d(\mathbf Z_2,\mathbf Z_3)$, respectively, then
\[
\begin{aligned}
d(\mathbf Z_1,\mathbf Z_3) &\le \|\mathbf Z_1\bm{\Pi}_{12}\bm{\Pi}_{23}-\mathbf Z_3\|_{2,\infty}\\
&\le \|(\mathbf Z_1\bm{\Pi}_{12}-\mathbf Z_2)\bm{\Pi}_{23}\|_{2,\infty} + \|\mathbf Z_2\bm{\Pi}_{23}-\mathbf Z_3\|_{2,\infty}\\
&= d(\mathbf Z_1,\mathbf Z_2) + d(\mathbf Z_2,\mathbf Z_3),
\end{aligned}
\]
where the last equality follows from the invariance of the $2\to\infty$ norm under column permutations.

Moreover, by \eqref{eq:G_rate_of_Deltat}, the separation between the two membership matrices satisfies
\[
\min_{\bm{\Pi}\in\mathcal P_2} \|\mathbf Z_{t_n}\bm{\Pi} - \mathbf Z_0\|_{2,\infty} = \Delta_{t_n} \ge c_0 t_n
\]
for some universal constant $c_0>0$.

Following the convention
\[
\operatorname{TV}(P,Q) := \sup_A |P(A)-Q(A)| = \frac12\|P-Q\|_1,
\]
the affinity appearing in \citet[Lemma~1]{Yu1997} satisfies $\|P\wedge Q\| = 1-\operatorname{TV}(P,Q).$
Applying Le Cam's method \citep[Lemma~1]{Yu1997} gives, for every estimator
$\widehat{\mathbf Z}$,
\[
\max_{\vartheta\in\{\vartheta_0,\vartheta_{t_n}\}} \mathbb E_{\vartheta} \left[ d(\widehat{\mathbf Z},\mathbf Z_{\vartheta}) \right] \ge \frac{\Delta_{t_n}}{2} \left\{ 1- \operatorname{TV}(\mathbb P_{t_n},\mathbb P_0) \right\}.
\]
Using the total-variation bound \eqref{eq:G_ub_of_TV},
\[
\max_{\vartheta\in\{\vartheta_0,\vartheta_{t_n}\}} \mathbb E_{\vartheta} \left[ d(\widehat{\mathbf Z},\mathbf Z_{\vartheta}) \right] \ge \frac{3}{8}\Delta_{t_n} \gtrsim \frac{1}{\sqrt{n\rho_n}}.
\]
Since $\{\vartheta_0,\vartheta_{t_n}\} \subset \mathcal{F}_{n, 2, L}$, taking the supremum over $\mathcal{F}_{n, 2, L}$ and then the infimum over all estimators completes the proof when $K=2$.

For $K > 2$, construct a membership matrix $\mathbf{Z}_{0}$ with $K-1$ mixed nodes indexed by $\{i_{1}, \cdots, i_{K-1} \}$ such that $(\mathbf{Z}_{0})_{i_{k}, \cdot} := \frac{1}{2} \mathbf{e}_{k} + \frac{1}{2} \mathbf{e}_{k+1}$ and evenly distributed pure nodes. Then one can extend the same proof of $K=2$ with the same $(\bm{\Theta}_{0}^{(l)}, \mathbf{Z}_{0}, \mathbf{B}_{0}^{(l)})_{l=1}^{L}$ and $(\bm{\Theta}_{t}^{(l)}, \mathbf{Z}_{t}, \mathbf{B}_{t}^{(l)})_{l=1}^{L}$.
\end{proof}

%% file: Sections/Appendix-cv.tex
\section{Supplementary Details for the London Bike Data}
In this section, we provide supplementary details for the London bike data analysis in section~\ref{sec:data_analysis}. Specifically, we summarize the compared models and methods in data analysis. Next, we introduce the cross-validation procedure for the model evaluation. 

\subsection{Overview of Compared Models and Methods}\label{sec:appendix_baseline_models}
In data analysis, we compare 4 models: DCBM, DCMM, ML-DCBM, ML-DCMM. Comparing non-overlapping and mixed-membership models in both single-layer and multilayer settings helps illustrate the effect of allowing mixed memberships. On the other hand, comparing single-layer and multilayer models under the same membership structure explains the difference between jointly leveraging layer-specific information and aggregating the layers into a single network $L^{-1} \sum_{l=1}^{L} \mathbf{A}^{(l)}$. We summarize the competing models and methods used in data analysis by the following table~\ref{tab:baseline_methods}.

\begin{table}[!ht]
\centering
\caption{Baseline and ablation methods organized by model class, layer structure, and membership structure.}
\label{tab:baseline_methods}
\small
\setlength{\tabcolsep}{5pt}
\renewcommand{\arraystretch}{1.18}
\begin{tabularx}{\textwidth}{@{}lYY@{}}
\toprule
&
\textbf{Hard membership}
&
\textbf{Mixed membership}
\\
&
\textbf{Non-overlapping communities}
&
\textbf{Overlapping communities}
\\
\midrule
\textbf{Single-layer}
&
\textit{Model:} {DCBM}
\newline
\textit{Method:} \textsc{SCORE} \citep{jin2015fastcommunity}
&
\textit{Model:} {DCMM}
\newline
\textit{Method:} \textsc{Mixed-SCORE} \citep{jin2024mixed}
\\
\midrule
\textbf{Multilayer}
&
\textit{Model:} {ML-DCBM}
\newline
\textit{Method:} \textsc{DCMASE} \citep{Agterberg2025}
&
\textit{Model:} {ML-DCMM}
\newline
\textit{Method:} \textsc{Mixed-SCORE (Avg)},  \textbf{CNASE} 
\\
\bottomrule
\end{tabularx}
\end{table}

\subsection{Cross-validation}\label{sec:appendix-cv}
The cross-validation procedure used in the London bike-sharing network analysis requires reconstructing the layer-specific probability matrices beyond estimating the shared membership matrix. In this section, we describe the reconstruction procedure and evaluate its numerical accuracy at the population level. In particular, we show that the proposed procedure achieves small reconstruction errors even when the estimated membership matrix does not exactly recover the true memberships.

\subsubsection{Cross-validation Procedure}
Following the node-pair cross-validation scheme of \citet{Li2020}, we use a 10-fold cross-validation procedure adapted to multilayer networks.
Let $\mathcal D:=\{(i,j):1\le i<j\le n\}$ denote the set of node pairs in the upper triangle, and randomly partition \(\mathcal D\) into 10 folds \(\mathcal D_1,\ldots,\mathcal D_{10}\). 
For fold \(q\), let \(\Omega_q=\mathcal D\setminus\mathcal D_q\) be the training node-pair set and let \(p_{q}\) be the training proportion. 
We write \(\mathcal P_{\Omega_q}\) for the masking operator that keeps entries in \(\Omega_q\) and sets the held-out entries to zero, symmetrically. For a symmetric matrix \(\bQ\)., let \(S_K(\bQ)\) denote its rank-\(K\) truncated spectral approximation.

For each fold \(q\) and candidate \(K\), we construct both the layerwise
completed training matrices
\[
\widetilde{\bA}_{-q,K}^{(l)}
:=
S_K\!\left(
\frac{1}{p_q}
\mathcal P_{\Omega_q}\bA^{(l)}
\right),
\qquad l=1,\ldots,L,
\]
and the corresponding averaged completed training matrix
\[
\widetilde{\bA}_{-q,K}^{\mathrm{avg}}
:=
S_K\!\left(
\frac{1}{L}
\sum_{l=1}^{L}
\frac{1}{p_q}
\mathcal P_{\Omega_q}\bA^{(l)}
\right).
\]

Let \(\widehat{\bZ}_{-q,K}^{\mathrm{CNASE}}\) denote the shared membership estimate produced by the proposed \textsc{CNASE} estimator, and let \(\widehat{\bZ}_{-q,K}^{\mathrm{MS}}\) denote the membership estimate obtained by applying \textsc{Mixed-SCORE} to \(\widetilde{\bA}_{-q,K}^{\mathrm{avg}}\). Similarly, let \(\widehat{\bZ}_{-q,K}^{\mathrm{DCMASE}}\) and \(\widehat{\bZ}_{-q,K}^{\mathrm{SCORE}}\) denote the membership estimates obtained from \textsc{DCMASE} and \textsc{SCORE}, respectively. The \textsc{Mixed-SCORE (Avg)} method uses \(\widehat{\bZ}_{-q,K}^{\mathrm{MS}}\) as its membership estimate.

For a training matrix \(\bQ\) and a mixed-membership estimate \(\widehat{\bZ}\), let
\(\mathcal R_{\mathrm{mix}}(\bQ,\widehat{\bZ})\) denote probability matrix reconstruction process of mixed-membership models. Similarly, for a non-overlapping membership estimate \(\widehat{\bZ}\), let \(\mathcal R_{\mathrm{hard}}(\bQ,\widehat{\bZ})\) denote the standard DCBM plug-in reconstruction based on the corresponding binary membership matrix and closed-form estimates of the degree-correction and block parameters. Then the raw prediction matrices are defined as follows:
\[
\begin{array}{lll}
\textsc{ML-DCMM}
&
\widetilde{\bP}_{\mathrm{ML},-q,K}^{(l)}
=
\mathcal R_{\mathrm{mix}}
\!\left(
\widetilde{\bA}_{-q,K}^{(l)},
\widehat{\bZ}_{-q,K}^{\mathrm{CNASE}}
\right),
&
l=1,\ldots,L,
\\[0.6em]
\textsc{DCMM}
&
\widetilde{\bP}_{\mathrm{MS},-q,K}^{(l)}
=
\mathcal R_{\mathrm{mix}}
\!\left(
\widetilde{\bA}_{-q,K}^{\mathrm{avg}},
\widehat{\bZ}_{-q,K}^{\mathrm{MS}}
\right),
&
l=1,\ldots,L,
\\[0.6em]
\textsc{ML-DCBM}
&
\widetilde{\bP}_{\mathrm{DCMASE},-q,K}^{(l)}
=
\mathcal R_{\mathrm{hard}}
\!\left(
\widetilde{\bA}_{-q,K}^{(l)},
\widehat{\bZ}_{-q,K}^{\mathrm{DCMASE}}
\right),
&
l=1,\ldots,L,
\\[0.6em]
\textsc{DCBM}
&
\widetilde{\bP}_{\mathrm{SCORE},-q,K}^{(l)}
=
\mathcal R_{\mathrm{hard}}
\!\left(
\widetilde{\bA}_{-q,K}^{\mathrm{avg}},
\widehat{\bZ}_{-q,K}^{\mathrm{SCORE}}
\right),
&
l=1,\ldots,L.
\end{array}
\]

For prediction, we use
\[
\widehat{\bP}_{m,-q,K}^{(l)}
=
\Pi_{[0,1]}\!\left(
\widetilde{\bP}_{m,-q,K}^{(l)}
\right),
\]
where \(\Pi_{[0,1]}\) denotes entrywise projection onto \([0,1]\) with zero
diagonal. This projection is used only for prediction and cross-validation. All procedures are evaluated on the same held-out node pairs. For procedure
\(m\), fold \(q\), and candidate \(K\), the held-out prediction error is
\[
\mathrm{CV}_{m,q}(K)
=
\frac{1}{L|\mathcal D_q|}
\sum_{l=1}^{L}
\sum_{(i,j)\in\mathcal D_q}
\left(
\bA_{ij}^{(l)}
-
\widehat{\bP}_{m,-q,K,ij}^{(l)}
\right)^2.
\]
We then compute
\[
\mathrm{CV}_{m}(K)
=
\frac{1}{10}\sum_{q=1}^{10}\mathrm{CV}_{m,q}(K),
\qquad
\mathrm{SE}_{m}(K)
=
\frac{\mathrm{sd}\{\mathrm{CV}_{m,q}(K):q=1,\ldots,10\}}{\sqrt{10}}.
\]
For each procedure, \(K\) is selected by the one-standard-error rule: among the
values of \(K\) whose cross-validation error is within one standard error of
the minimum, we choose the smallest \(K\). 

\subsubsection{Probability Matrix Reconstruction}
Recall that the probability matrix can be written as $\mathbf{P}^{(l)} = \mathbf{X}^{(l)} \mathbf{I}_{p^{(l)}, q^{(l)}} \mathbf{X}^{(l)\top}$, where $\mathbf{X}^{(l)} = \bm{\Theta}^{(l)} \mathbf{Z} \mathbf{R}^{(l)}$. The main idea for reconstruction is to estimate $\mathbf{X}^{(l)}$ by $\widetilde{\mathbf{X}}^{(l)} := \widehat{\bm{\Theta}}^{(l)} \widehat{\mathbf{Z}} \widehat{\mathbf{R}}^{(l)}$ and reconstruct the probability matrix by $\widehat{\mathbf{P}}^{(l)} := \widetilde{\mathbf{X}}^{(l)} \mathbf{I}_{p_{A}^{(l)}, q_{A}^{(l)}} \widetilde{\mathbf{X}}^{(l)\top}$, where $p_{A}^{(l)}$ and $q_{A}^{(l)}$ is the number of positive and negative eigenvalues among largest $K$ eigenvalues of $\mathbf{A}^{(l)}$ in magnitude. Given the estimated membership matrix $\widehat{\bZ}$, we estimate the layer-specific degree matrices $\{\bTheta^{(l)}\}_{l=1}^L$ and transformation matrices $\{\bR^{(l)}\}_{l=1}^L$ by minimizing the following least-squares objective:
\[
\mathcal L = \sum_{l=1}^{L} \left\| \widehat{\bX}^{(l)} - \bTheta^{(l)}\widehat{\bZ}\bR^{(l)} \right\|_F^2,
\]
where $\widehat{\mathbf{X}}^{(l)}$ is the adjacency spectral embedding of $\mathbf{A}^{(l)}$. We employ an alternating least-squares (ALS) algorithm that iteratively updates $\bR^{(l)}$ and $\bTheta^{(l)}$, in that order, while holding the other parameter fixed at each update. The procedure is summarized in Algorithm~\ref{alg:probability_reconstruction}.

\begin{algorithm}[htbp]
\caption{Probability matrix reconstruction}
\label{alg:probability_reconstruction}
\begin{algorithmic}[1]
\Input Adjacency matrices $\{\bA^{(1)},\ldots,\bA^{(L)}\}$, estimated membership matrix $\widehat{\mathbf{Z}}$, maximum number of iterations $T_{\max}$, convergence criteria $\epsilon$
\Ensure Reconstructed probability matrices $\{ \widehat{\mathbf{P}}^{(1)}, \ldots, \widehat{\mathbf{P}}^{(L)} \}$

\State \textbf{For} $l = 1, \ldots, L$, obtain the adjacency spectral embedding \(\widehat{\bX}^{(l)}\).
\State For the initial points, set $\widehat{\bTheta}_{0}^{(l)} = \bI_n = \mathbf{I}_{n}$ for all $l \in [L]$.
\For{$t=1$ to $T_{\max}$}
    \State Perform one ALS iteration to update 
    $\{\widehat{\bR}^{(l)}\}_{l=1}^L$ and $\{\widehat{\bTheta}^{(l)}\}_{l=1}^L$.
    \If{the convergence criterion is satisfied}
        \State \textbf{break}
    \EndIf
\EndFor
\State Reconstruct $\widetilde{\mathbf{P}}^{(l)} := \widetilde{\mathbf{X}}^{(l)} \mathbf{I}_{p_{A}^{(l)}, q_{A}^{(l)}} \widetilde{\mathbf{X}}^{(l)\top}$, where $\widetilde{\mathbf{X}}^{(l)} := \widehat{\bm{\Theta}}^{(l)} \widehat{\mathbf{Z}} \widehat{\mathbf{R}}^{(l)}$ for all $l \in [L]$.
\State Obtain $\widehat{\mathbf{P}}^{(l)} := \Pi_{[0, 1]}\left( \widetilde{\mathbf{P}}^{(l)} \right)$, where $\Pi_{[0, 1]}$ denotes entrywise projection onto $[0, 1]$ with zero diagonal.
\end{algorithmic}
\end{algorithm}

\subsubsection{Simulation on the Population-Level Reconstruction}\label{sec:additional_simulation_results}

As an additional diagnostic, we apply the proposed estimation and reconstruction procedure directly to the population probability matrices rather than to sampled adjacency matrices. This experiment aims to assess the numerical accuracy of the full pipeline in the absence of sampling noise. In particular, we examine whether the procedure can accurately estimate the shared membership matrix and reconstruct the layer-specific probability matrices at the population level. In particular, it checks whether the proposed pipeline can recover the membership matrix and reconstruct the population matrices when sampling noise is absent.

We use the same six data-generating settings as in Experiment~2, with
\[
(n,n_0,K,L)=(300,30,3,10).
\]
For each target density
\[
\rho\in\{0.04,0.06,0.08,0.10,0.12,0.14,0.16\},
\]
we apply the proposed method to the diagonal-included population matrices
\[
\bP^{(l)}=\bTheta^{(l)} \bZ \bB^{(l)} \bZ^\top \bTheta^{(l)}, 
\qquad 1\le l \le L.
\]
We report the maximum, over the target densities, of the average reconstruction errors across 100 repetitions and layers:
\begin{equation*}
    \begin{aligned}
        E_Z & := \max_{\rho} \frac{1}{100} \sum_{s=1}^{100} \frac{ \left\| \bZ_{\rho,s} - \widehat{\bZ}_{\rho,s} \right\|_F^2}{n} \\
        E_P &:= \max_{\rho} \frac{1}{100\cdot 10} \sum_{s=1}^{100} \sum_{l=1}^{10} \frac{ \left\| \bP_{\rho,s}^{(l)} - \widehat{\bP}_{\rho,s}^{(l)} \right\|_F^2}{n^2}, \\
        E_{P,\mathrm{rel}} &:= \max_{\rho} \frac{1}{100\cdot 10} \sum_{s=1}^{100} \sum_{l=1}^{10} \frac{\left\| \bP_{\rho,s}^{(l)} - \widehat{\bP}_{\rho,s}^{(l)} \right\|_F}{ \left\| \bP_{\rho,s}^{(l)} \right\|_F}, \\
        E_{B,\mathrm{rel}} &:= \max_{\rho} \frac{1}{100\cdot 10} \sum_{s=1}^{100} \sum_{l=1}^{10} \frac{\left\| \bB_{\rho,s}^{(l)} - \widehat{\bB}_{\rho,s}^{(l)} \right\|_F}{ \left\| \bB_{\rho,s}^{(l)} \right\|_F}.
    \end{aligned}
\end{equation*}

\begin{table}[!htbp]
\centering
\caption{
Population-level reconstruction from oracle probability matrices. 
For each density \(\rho\), the membership error is averaged over 100 repetitions. 
The \(\bP\)- and \(\bB\)-reconstruction errors are averaged over 100 repetitions and \(L=10\) layers. 
The table reports the maximum of these density-specific averages over 
\(\rho\in\{0.04,0.06,\ldots,0.16\}\).
}
\label{tab:population_reconstruction}
\begin{tabular}{llcccc}
\hline
Purity & \(\bB\) setting
& \(E_Z\)
& \(E_P\)
& \(E_{P,\mathrm{rel}}\)
& \(E_{B,\mathrm{rel}}\) \\
\hline
\(r=0.2\) & Same \(\bB\)
& \(2.61{\times}10^{-7}\)
& \(4.25{\times}10^{-12}\)
& \(9.60{\times}10^{-6}\)
& \(4.91{\times}10^{-3}\) \\
\(r=0.2\) & Different \(\bB\)
& \(2.72{\times}10^{-7}\)
& \(4.36{\times}10^{-12}\)
& \(9.58{\times}10^{-6}\)
& \(4.94{\times}10^{-3}\) \\
\(r=0.2\) & Alternating \(\bB\)
& \(1.64{\times}10^{-7}\)
& \(2.12{\times}10^{-12}\)
& \(7.06{\times}10^{-6}\)
& \(3.01{\times}10^{-3}\) \\
\(r=0.4\) & Same \(\bB\)
& \(9.22{\times}10^{-9}\)
& \(6.03{\times}10^{-13}\)
& \(4.04{\times}10^{-6}\)
& \(6.92{\times}10^{-4}\) \\
\(r=0.4\) & Different \(\bB\)
& \(9.37{\times}10^{-9}\)
& \(6.15{\times}10^{-13}\)
& \(4.04{\times}10^{-6}\)
& \(6.92{\times}10^{-4}\) \\
\(r=0.4\) & Alternating \(\bB\)
& \(7.15{\times}10^{-9}\)
& \(4.63{\times}10^{-13}\)
& \(3.57{\times}10^{-6}\)
& \(4.71{\times}10^{-4}\) \\
\hline
\end{tabular}
\end{table}

Table~\ref{tab:population_reconstruction} demonstrates that the proposed procedure achieves highly accurate reconstruction at the population level. Across all purity levels and block-structure settings, the averaged probability matrix reconstruction error is on the order of \(10^{-12}\) or smaller after normalization by \(n^2\), while the relative reconstruction error is on the order of \(10^{-6}\). These results indicate that the proposed procedure can accurately reconstruct the population probability matrices even when the estimated membership matrix does not exactly match the true membership matrix.

%% file: Manuscript.bbl
\begin{thebibliography}{58}
\providecommand{\natexlab}[1]{#1}
\providecommand{\url}[1]{\texttt{#1}}
\expandafter\ifx\csname urlstyle\endcsname\relax
  \providecommand{\doi}[1]{doi: #1}\else
  \providecommand{\doi}{doi: \begingroup \urlstyle{rm}\Url}\fi

\bibitem[Agterberg et~al.(2025)Agterberg, Lubberts, and Arroyo]{Agterberg2025}
Joshua Agterberg, Zachary Lubberts, and Jesús Arroyo.
\newblock Joint spectral clustering in multilayer degree-corrected stochastic blockmodels.
\newblock \emph{Journal of the American Statistical Association}, 120\penalty0 (551):\penalty0 1607--1620, 2025.

\bibitem[Aguiar et~al.(2024)Aguiar, Taylor, and Ugander]{Aguiar2024}
Izabel Aguiar, Dane Taylor, and Johan Ugander.
\newblock A tensor factorization model of multilayer network interdependence.
\newblock \emph{Journal of Machine Learning Research}, 25\penalty0 (282):\penalty0 1--54, 2024.

\bibitem[Airoldi et~al.(2008)Airoldi, Blei, Fienberg, and Xing]{airoldi2008mixed}
Edoardo~M. Airoldi, David~M. Blei, Stephen~E. Fienberg, and Eric~P. Xing.
\newblock Mixed membership stochastic blockmodels.
\newblock \emph{Journal of Machine Learning Research}, 9\penalty0 (65):\penalty0 1981--2014, 2008.

\bibitem[Anandkumar et~al.(2013)Anandkumar, Ge, Hsu, and Kakade]{Anandkumar2013}
Animashree Anandkumar, Rong Ge, Daniel Hsu, and Sham Kakade.
\newblock A tensor spectral approach to learning mixed membership community models.
\newblock In \emph{Proceedings of the 26th Annual Conference on Learning Theory}, volume~30 of \emph{Proceedings of Machine Learning Research}, pages 867--881, 2013.

\bibitem[Araújo et~al.(2001)Araújo, Saldanha, Galvão, Yoneyama, Chame, and Visani]{Araujo001}
Mário César~Ugulino Araújo, Teresa Cristina~Bezerra Saldanha, Roberto Kawakami~Harrop Galvão, Takashi Yoneyama, Henrique~Caldas Chame, and Valeria Visani.
\newblock The successive projections algorithm for variable selection in spectroscopic multicomponent analysis.
\newblock \emph{Chemometrics and Intelligent Laboratory Systems}, 57\penalty0 (2):\penalty0 65--73, 2001.

\bibitem[Arroyo et~al.(2021)Arroyo, Athreya, Cape, Chen, Priebe, and Vogelstein]{arroyo2021inference}
Jes{\'u}s Arroyo, Avanti Athreya, Joshua Cape, Guodong Chen, Carey~E Priebe, and Joshua~T Vogelstein.
\newblock Inference for multiple heterogeneous networks with a common invariant subspace.
\newblock \emph{Journal of Machine Learning Research}, 22\penalty0 (142):\penalty0 1--49, 2021.

\bibitem[Athreya et~al.(2018)Athreya, Fishkind, Tang, Priebe, Park, Vogelstein, Levin, Lyzinski, Qin, and Sussman]{Athreya2017}
Avanti Athreya, Donniell~E. Fishkind, Minh Tang, Carey~E. Priebe, Youngser Park, Joshua~T. Vogelstein, Keith~D. Levin, Vince Lyzinski, Yichen Qin, and Daniel~L. Sussman.
\newblock Statistical inference on random dot product graphs: A survey.
\newblock \emph{Journal of Machine Learning Research}, 18\penalty0 (226):\penalty0 1--92, 2018.

\bibitem[Ball and Newman(2013)]{BALLNEWMAN2013}
Brian Ball and M.E.J. Newman.
\newblock Friendship networks and social status.
\newblock \emph{Network Science}, 1\penalty0 (1):\penalty0 16--30, 2013.

\bibitem[Bhattacharyya and Chatterjee(2018)]{Battacharyya2018}
Sharmodeep Bhattacharyya and Shirshendu Chatterjee.
\newblock Spectral clustering for multiple sparse networks: {I}, 2018.
\newblock arXiv preprint arXiv:1805.10594.

\bibitem[Bickel and Chen(2009)]{Bickel2009}
Peter~J. Bickel and Aiyou Chen.
\newblock A nonparametric view of network models and {N}ewman–{G}irvan and other modularities.
\newblock \emph{Proceedings of the National Academy of Sciences}, 106\penalty0 (50):\penalty0 21068--21073, 2009.

\bibitem[Boccaletti et~al.(2014)Boccaletti, Bianconi, Criado, {del Genio}, Gómez-Gardeñes, Romance, Sendiña-Nadal, Wang, and Zanin]{BOCCALETTI2014}
Stefano Boccaletti, Ginestra Bianconi, Regino Criado, Charo~I. {del Genio}, Jesús Gómez-Gardeñes, Miguel Romance, Irene Sendiña-Nadal, Zheng Wang, and Massimiliano Zanin.
\newblock The structure and dynamics of multilayer networks.
\newblock \emph{Physics Reports}, 544\penalty0 (1):\penalty0 1--122, 2014.

\bibitem[Bro(1998)]{Bro1998}
Rasmus Bro.
\newblock \emph{Multi-way Analysis in the Food Industry: Models, Algorithms, and Applications}.
\newblock PhD thesis, University of Amsterdam, Amsterdam, The Netherlands, 1998.

\bibitem[Chen et~al.(2021)Chen, Chi, Fan, and Ma]{chen2021spectral}
Yuxin Chen, Yuejie Chi, Jianqing Fan, and Cong Ma.
\newblock Spectral methods for data science: A statistical perspective.
\newblock \emph{Foundations and Trends in Machine Learning}, 14\penalty0 (5):\penalty0 566--806, 2021.

\bibitem[De~Domenico(2017)]{DeDomenico2017}
Manlio De~Domenico.
\newblock Multilayer modeling and analysis of human brain networks.
\newblock \emph{GigaScience}, 6\penalty0 (5):\penalty0 1--8, 2017.

\bibitem[Diaconis and Sturmfels(1998)]{diaconis1998algebraic}
Persi Diaconis and Bernd Sturmfels.
\newblock Algebraic algorithms for sampling from conditional distributions.
\newblock \emph{The Annals of statistics}, 26\penalty0 (1):\penalty0 363--397, 1998.

\bibitem[Fan et~al.(2022)Fan, Fan, Han, and Lv]{Fan2022}
Jianqing Fan, Yingying Fan, Xiao Han, and Jinchi Lv.
\newblock Simple: Statistical inference on membership profiles in large networks.
\newblock \emph{Journal of the Royal Statistical Society Series B: Statistical Methodology}, 84\penalty0 (2):\penalty0 630--653, 04 2022.

\bibitem[Fortunato(2010)]{FORTUNATO201075}
Santo Fortunato.
\newblock Community detection in graphs.
\newblock \emph{Physics Reports}, 486\penalty0 (3--5):\penalty0 75--174, 2010.

\bibitem[Gao et~al.(2018)Gao, Ma, Zhang, and Zhou]{Gao2018}
Chao Gao, Zongming Ma, Anderson~Y. Zhang, and Harrison~H. Zhou.
\newblock {Community detection in degree-corrected block models}.
\newblock \emph{The Annals of Statistics}, 46\penalty0 (5):\penalty0 2153--2185, 2018.

\bibitem[Hagmann et~al.(2008)Hagmann, Cammoun, Gigandet, Meuli, Honey, Wedeen, and Sporns]{Hagmann2008}
Patric Hagmann, Leila Cammoun, Xavier Gigandet, Reto Meuli, Christopher~J Honey, Van~J Wedeen, and Olaf Sporns.
\newblock Mapping the structural core of human cerebral cortex.
\newblock \emph{PLOS Biology}, 6\penalty0 (7):\penalty0 e159, 2008.

\bibitem[Han et~al.(2015)Han, Xu, and Airoldi]{Han2015}
Qiuyi Han, Kevin Xu, and Edoardo Airoldi.
\newblock Consistent estimation of dynamic and multi-layer block models.
\newblock In \emph{Proceedings of the 32nd International Conference on Machine Learning}, volume~37 of \emph{Proceedings of Machine Learning Research}, pages 1511--1520, 2015.

\bibitem[Han et~al.(2023)Han, Yang, and Fan]{Han2023}
Xiao Han, Qing Yang, and Yingying Fan.
\newblock {Universal rank inference via residual subsampling with application to large networks}.
\newblock \emph{The Annals of Statistics}, 51\penalty0 (3):\penalty0 1109 -- 1133, 2023.

\bibitem[Harshman and Lundy(1996)]{Harshman_Lundy_1996}
Richard~A. Harshman and Margaret~E. Lundy.
\newblock Uniqueness proof for a family of models sharing features of {T}ucker’s three-mode factor analysis and {PARAFAC/Candecomp}.
\newblock \emph{Psychometrika}, 61\penalty0 (1):\penalty0 133--154, 1996.

\bibitem[Holland et~al.(1983)Holland, Laskey, and Leinhardt]{HollandLaskeyLeinhardt1983}
Paul~W. Holland, Kathryn~Blackmond Laskey, and Samuel Leinhardt.
\newblock Stochastic blockmodels: First steps.
\newblock \emph{Social Networks}, 5\penalty0 (2):\penalty0 109--137, 1983.

\bibitem[Jin(2015)]{jin2015fastcommunity}
Jiashun Jin.
\newblock Fast community detection by {SCORE}.
\newblock \emph{The Annals of Statistics}, 43\penalty0 (1):\penalty0 57--89, 2015.

\bibitem[Jin et~al.(2023)Jin, Ke, Luo, and Wang]{Jin2023optimal}
Jiashun Jin, Zheng~Tracy Ke, Shengming Luo, and Minzhe Wang.
\newblock Optimal estimation of the number of network communities.
\newblock \emph{Journal of the American Statistical Association}, 118\penalty0 (543):\penalty0 2101--2116, 2023.

\bibitem[Jin et~al.(2024)Jin, Ke, and Luo]{jin2024mixed}
Jiashun Jin, Zheng~Tracy Ke, and Shengming Luo.
\newblock Mixed membership estimation for social networks.
\newblock \emph{Journal of Econometrics}, 239\penalty0 (2):\penalty0 105369, 2024.

\bibitem[Jing et~al.(2021)Jing, Li, Lyu, and Xia]{Jing2021}
Bing-Yi Jing, Ting Li, Zhongyuan Lyu, and Dong Xia.
\newblock {Community detection on mixture multilayer networks via regularized tensor decomposition}.
\newblock \emph{The Annals of Statistics}, 49\penalty0 (6):\penalty0 3181--3205, 2021.

\bibitem[Jones and Rubin-Delanchy(2020)]{jones2020multilayer}
Andrew Jones and Patrick Rubin-Delanchy.
\newblock The multilayer random dot product graph.
\newblock \emph{arXiv preprint arXiv:2007.10455}, 2020.

\bibitem[Karrer and Newman(2011)]{karrer2011stochastic}
Brian Karrer and Mark~EJ Newman.
\newblock Stochastic blockmodels and community structure in networks.
\newblock \emph{Physical Review E}, 83\penalty0 (1):\penalty0 016107, 2011.

\bibitem[Kaufman and Rousseeuw(1990)]{KaufmanRousseeuw1990}
Leonard Kaufman and Peter~J. Rousseeuw.
\newblock Partitioning around medoids (program pam).
\newblock In \emph{Finding Groups in Data: An Introduction to Cluster Analysis}, chapter~2, pages 68--125. 1990.

\bibitem[Ke and Wang(2025)]{ke2025optimal}
Zheng~Tracy Ke and Jingming Wang.
\newblock Optimal network membership estimation under severe degree heterogeneity.
\newblock \emph{Journal of the American Statistical Association}, 120\penalty0 (550):\penalty0 948--962, 2025.

\bibitem[Kivelä et~al.(2014)Kivelä, Arenas, Barthelemy, Gleeson, Moreno, and Porter]{Kivela2014}
Mikko Kivelä, Alex Arenas, Marc Barthelemy, James~P. Gleeson, Yamir Moreno, and Mason~A. Porter.
\newblock Multilayer networks.
\newblock \emph{Journal of Complex Networks}, 2\penalty0 (3):\penalty0 203--271, 2014.

\bibitem[Le~Cam(1973)]{LeCam1973}
Lucien Le~Cam.
\newblock Convergence of estimates under dimensionality restrictions.
\newblock \emph{The Annals of Statistics}, 1\penalty0 (1):\penalty0 38--53, 1973.

\bibitem[Lei and Lin(2023)]{lei2023biasadjusted}
Jing Lei and Kevin~Z. Lin.
\newblock Bias-adjusted spectral clustering in multi-layer stochastic block models.
\newblock \emph{Journal of the American Statistical Association}, 118\penalty0 (544):\penalty0 2433--2445, 2023.

\bibitem[Lei and Rinaldo(2015)]{Lei2015}
Jing Lei and Alessandro Rinaldo.
\newblock {Consistency of spectral clustering in stochastic block models}.
\newblock \emph{The Annals of Statistics}, 43\penalty0 (1):\penalty0 215--237, 2015.

\bibitem[Lei et~al.(2020)Lei, Chen, and Lynch]{Lei2020}
Jing Lei, Kehui Chen, and Brian Lynch.
\newblock Consistent community detection in multi-layer network data.
\newblock \emph{Biometrika}, 107\penalty0 (1):\penalty0 61--73, 2020.

\bibitem[Li et~al.(2020)Li, Levina, and Zhu]{Li2020}
Tianxi Li, Elizaveta Levina, and Ji~Zhu.
\newblock Network cross-validation by edge sampling.
\newblock \emph{Biometrika}, 107\penalty0 (2):\penalty0 257--276, 2020.

\bibitem[Lukemire et~al.(2021)Lukemire, Kundu, Pagnoni, and Guo]{Lukemire2021}
Joshua Lukemire, Suprateek Kundu, Giuseppe Pagnoni, and Ying Guo.
\newblock Bayesian joint modeling of multiple brain functional networks.
\newblock \emph{Journal of the American Statistical Association}, 116\penalty0 (534):\penalty0 518--530, 2021.

\bibitem[Mao et~al.(2017)Mao, Sarkar, and Chakrabarti]{Mao2017}
Xueyu Mao, Purnamrita Sarkar, and Deepayan Chakrabarti.
\newblock On mixed memberships and symmetric nonnegative matrix factorizations.
\newblock In \emph{Proceedings of the 34th International Conference on Machine Learning}, volume~70 of \emph{Proceedings of Machine Learning Research}, pages 2324--2333, 2017.

\bibitem[Mao et~al.(2018)Mao, Sarkar, and Chakrabarti]{mao2018overlapping}
Xueyu Mao, Purnamrita Sarkar, and Deepayan Chakrabarti.
\newblock Overlapping clustering models, and one (class) {SVM} to bind them all.
\newblock \emph{Advances in Neural Information Processing Systems}, 31, 2018.

\bibitem[Mao et~al.(2021)Mao, Sarkar, and Chakrabarti]{Mao2021}
Xueyu Mao, Purnamrita Sarkar, and Deepayan Chakrabarti.
\newblock Estimating mixed memberships with sharp eigenvector deviations.
\newblock \emph{Journal of the American Statistical Association}, 116\penalty0 (536):\penalty0 1928--1940, 2021.

\bibitem[Mucha et~al.(2010)Mucha, Richardson, Macon, Porter, and Onnela]{mucha2010community}
Peter~J Mucha, Thomas Richardson, Kevin Macon, Mason~A Porter, and Jukka-Pekka Onnela.
\newblock Community structure in time-dependent, multiscale, and multiplex networks.
\newblock \emph{Science}, 328\penalty0 (5980):\penalty0 876--878, 2010.

\bibitem[Munoz-Mendez et~al.(2018)Munoz-Mendez, Klemmer, Han, and Jarvis]{Munoz-Mendez2018}
Fernando Munoz-Mendez, Konstantin Klemmer, Ke~Han, and Stephen Jarvis.
\newblock {Community Structures, Interactions and Dynamics in London's Bicycle Sharing Network}.
\newblock In \emph{UbiComp/ISWC 2018 Adjunct}, pages 1015--1023, 2018.

\bibitem[Paul and Chen(2016)]{PaulChen2016}
Saptarshi Paul and Yuguo Chen.
\newblock Consistent community detection in multi-relational data through restricted multi-layer stochastic blockmodel.
\newblock \emph{Electronic Journal of Statistics}, 10\penalty0 (2):\penalty0 3807--3870, 2016.

\bibitem[Peixoto(2015)]{Peixoto2015}
Tiago~P. Peixoto.
\newblock Inferring the mesoscale structure of layered, edge-valued, and time-varying networks.
\newblock \emph{Physical Review E}, 92\penalty0 (4):\penalty0 042807, Oct 2015.

\bibitem[Qin and Rohe(2013)]{Qin2013}
Tai Qin and Karl Rohe.
\newblock Regularized spectral clustering under the degree-corrected stochastic blockmodel.
\newblock In \emph{Advances in Neural Information Processing Systems}, volume~26, pages 3120--3128, 2013.

\bibitem[Rohe et~al.(2011)Rohe, Chatterjee, and Yu]{Rohe2011}
Karl Rohe, Sourav Chatterjee, and Bin Yu.
\newblock {Spectral clustering and the high-dimensional stochastic blockmodel}.
\newblock \emph{The Annals of Statistics}, 39\penalty0 (4):\penalty0 1878--1915, 2011.

\bibitem[Rubin-Delanchy et~al.(2022)Rubin-Delanchy, Cape, Tang, and Priebe]{Rubin2022}
Patrick Rubin-Delanchy, Joshua Cape, Minh Tang, and Carey~E. Priebe.
\newblock A statistical interpretation of spectral embedding: The generalised random dot product graph.
\newblock \emph{Journal of the Royal Statistical Society Series B: Statistical Methodology}, 84\penalty0 (4):\penalty0 1446--1473, 2022.

\bibitem[Sussman et~al.(2012)Sussman, Tang, Fishkind, and Priebe]{Sussman2012}
Daniel~L. Sussman, Minh Tang, Donniell~E. Fishkind, and Carey~E. Priebe.
\newblock A consistent adjacency spectral embedding for stochastic blockmodel graphs.
\newblock \emph{Journal of the American Statistical Association}, 107\penalty0 (499):\penalty0 1119--1128, 2012.

\bibitem[Szell et~al.(2010)Szell, Lambiotte, and Thurner]{Szell2010}
Michael Szell, Renaud Lambiotte, and Stefan Thurner.
\newblock Multirelational organization of large-scale social networks in an online world.
\newblock \emph{Proceedings of the National Academy of Sciences}, 107\penalty0 (31):\penalty0 13636--13641, 2010.

\bibitem[Usevich(2023)]{usevich2023}
Konstantin Usevich.
\newblock A lifting approach to {ParaTuck-2} tensor decompositions, 2023.
\newblock arXiv preprint arXiv:2302.00922.

\bibitem[Wang et~al.(2026)Wang, Li, Madrid~Padilla, Yu, and Rinaldo]{wang2026multilayer}
Fan Wang, Wanshan Li, Oscar~Hernan Madrid~Padilla, Yi~Yu, and Alessandro Rinaldo.
\newblock Multilayer random dot product graphs: Estimation and online change point detection.
\newblock \emph{Journal of the Royal Statistical Society Series B: Statistical Methodology}, 88\penalty0 (1):\penalty0 282--312, 2026.

\bibitem[Xie and Huang(2025)]{Xie2025}
Fangzheng Xie and Hsin-Hsiung Huang.
\newblock Uncertainty quantification for mixed membership in multilayer networks with degree heterogeneity using {G}aussian variational inference, 2025.
\newblock arXiv preprint arXiv:2512.08146.

\bibitem[Young and Scheinerman(2007)]{Young2007}
Stephen~J. Young and Edward~R. Scheinerman.
\newblock Random dot product graph models for social networks.
\newblock In \emph{Algorithms and Models for the Web-Graph}, pages 138--149, 2007.

\bibitem[Yu(1997)]{Yu1997}
Bin Yu.
\newblock {Assouad}, {Fano}, and {Le Cam}.
\newblock In \emph{Festschrift for Lucien Le Cam: Research Papers in Probability and Statistics}, pages 423--435. Springer, 1997.

\bibitem[Zhang et~al.(2020)Zhang, Levina, and Zhu]{zhang2020detecting}
Yuan Zhang, Elizaveta Levina, and Ji~Zhu.
\newblock Detecting overlapping communities in networks using spectral methods.
\newblock \emph{SIAM Journal on Mathematics of Data Science}, 2\penalty0 (2):\penalty0 265--283, 2020.

\bibitem[Zhao et~al.(2012)Zhao, Levina, and Zhu]{Zhao2011}
Yunpeng Zhao, Elizaveta Levina, and Ji~Zhu.
\newblock {Consistency of community detection in networks under degree-corrected stochastic block models}.
\newblock \emph{The Annals of Statistics}, 40\penalty0 (4):\penalty0 2266--2292, 2012.

\bibitem[Zniyed and {de Almeida}(2025)]{zniyed2025}
Yassine Zniyed and André~L.F. {de Almeida}.
\newblock A stochastic algorithm for the {ParaTuck} decomposition.
\newblock \emph{Digital Signal Processing}, 156:\penalty0 104767, 2025.

\end{thebibliography}
